\documentclass[12pt,oneside,a4paper]{book}

\pdfoutput=1
\usepackage[T1]{fontenc}
\usepackage[english]{babel} 
\usepackage[latin1]{inputenc}
\usepackage{verbatim}                  
\usepackage[pdftex]{graphicx}           
\usepackage{setspace}                   
\usepackage{indentfirst}                
\usepackage{makeidx}                    
\usepackage[nottoc]{tocbibind}          
\usepackage{courier}                    
\usepackage{type1cm}                    
\usepackage{listings}                   
\usepackage{titletoc}
\usepackage[font=small,format=plain,labelfont=bf,up,textfont=it,up]{caption}
\usepackage[usenames,svgnames,dvipsnames]{xcolor}
\usepackage[a4paper,top=3.0cm,bottom=3.0cm,left=3.0cm,right=2.0cm]{geometry} 
\usepackage[pdftex,plainpages=false,pdfpagelabels,pagebackref,colorlinks=true,citecolor=DarkGreen,linkcolor=NavyBlue,urlcolor=DarkRed,filecolor=green,bookmarksopen=true] 
{hyperref}
\usepackage[all]{hypcap}                
\fontsize{60}{62}\usefont{OT1}{cmr}{m}{n}{\selectfont}

\usepackage{epstopdf}
\usepackage{enumerate}
\usepackage{amsmath}
\usepackage{amssymb}
\usepackage{slashed}
\usepackage{caption}
\usepackage{subcaption}

\usepackage{cite}

\usepackage{fancyhdr}
\renewcommand{\chaptermark}[1]{\markboth{\MakeUppercase{#1}}{}}

\graphicspath{{./figures/}}             
\makeindex                              
\fontsize{60}{62}\usefont{OT1}{cmr}{m}{n}{\selectfont}
\cleardoublepage
\normalsize

\begin{document}
\frontmatter 
\fancyhead[RO]{{\footnotesize\rightmark}\hspace{2em}\thepage}
\setcounter{tocdepth}{2}
\fancyhead[LE]{\thepage\hspace{2em}\footnotesize{\leftmark}}
\fancyhead[RE,LO]{}
\fancyhead[RO]{{\footnotesize\rightmark}\hspace{2em}\thepage}

\onehalfspacing  

%
%
%
%
%
%
%
%

\begin{titlepage}

\setlength{\voffset}{0pt}
\setlength{\hoffset}{0pt}

\centering

\Large{University of S\~ao Paulo\\
Institute of Physics}

\vspace{\stretch{3}}

\LARGE{\bf Strongly coupled non-Abelian plasmas in a magnetic field}

\vspace{\stretch{1}}

\Large{Renato Anselmo Judica Critelli
}

\vspace{\stretch{1}}

\begin{flushright}

\vspace{\stretch{3.5}}

\begin{minipage}{.6\textwidth}
\normalsize{Advisor: Prof. Dr. Jorge Jos\'e Leite Noronha Junior}\\

\vspace{\stretch{0.5}}

\normalsize{Dissertation presented to the Institute of Physics of University
of S\~ao Paulo as a requirement to the title of
Master in Science.}
\end{minipage}
\end{flushright}

\vspace{\stretch{1.5}}

\begin{flushleft}
\normalsize
Examining committee:\\
\vspace{\stretch{.4}}
\hspace{.003\textwidth}
\begin{minipage}{.97\textwidth}

Prof. Dr. Jorge José Leite Noronha Junior (IF-USP) \\
Prof. Dr. Eduardo Souza Fraga (IF-UFRJ)\\
Prof. Dr. Ricardo D'Elia Matheus (IFT-UNESP)

\end{minipage}
\end{flushleft}

\vspace{\stretch{2}}

S\~ao Paulo\\
2016
\end{titlepage}

\pagenumbering{roman}     

\chapter*{Acknowledgements}
First, I would like to thank my advisor Jorge for accepting me as his student and for the guidance so far. It has been a pleasure to share his enthusiasm with physics, since his great classes of mathematical methods for physicists some years ago.

I want to thank my collaborators, Stefano and Maicon, for their help in Chapter 5. Another special thanks goes to Romulo for the hard work done in Chapter 7 and 8, and, more than that, for being my unofficial co-advisor during the last year. Not even half of this work would have been done without their support.

A very special thanks goes to all the great teachers that I had (and will have!).

The working environment at the GRHAFITE has been nice. Consequently, I want to thank the professors Fernando, Marina, Renato, Alberto and Manuel, for their kindness and support. Inherently, my colleagues have been nice too: Luiz, Davi, Bruno, Hugo, Daniel, Jorgivan, Rafael, Diego, Samuel and Andre. Also, I would like to thank my favorite member of the DFMA, Anderson, whose friendship, from the very beginning, has been refreshing.

I thank CNPq for the financial support.

I also thank professors Eduardo Fraga and Ricardo Matheus for comments and suggestions for this dissertation.

Last but not least, I want to thank my family for all their love and support. My brother in special, with whom all this science thing began.

\chapter*{Abstract}
\noindent Critelli, R. \textbf{Strongly coupled non-Abelian plasmas in a magnetic field} 
2016.
Dissertation (M.Sc.) - Instituto de F\'isica,
Universidade de S\~ao Paulo, S\~ao Paulo, 2016.
\\

In this dissertation we use the gauge/gravity duality approach to study the dynamics of strongly coupled non-Abelian plasmas. Ultimately, we want to understand the properties of the quark-gluon plasma (QGP), whose scientifc interest by the scientific community escalated exponentially after its discovery in the 2000's through the collision of ultrarelativistic heavy ions.

One can enrich the dynamics of the QGP by adding an external field, such as the baryon chemical potential (needed to study the QCD phase diagram), or a magnetic field. In this dissertation, we choose to investigate the magnetic effects. Indeed, there are compelling evidences that strong magnetic fields of the order $eB\sim 10 m_\pi^2$ are created in the early stages of ultrarelativistic heavy ion collisions.

The chosen observable to scan possible effects of the magnetic field on the QGP was the viscosity, due to the famous result $\eta/s=1/4\pi$ obtained via holography. In a first approach we use a caricature of the QGP, the $\mathcal{N}=4$ super Yang-Mills plasma to calculate the deviations of the viscosity as we add a magnetic field. We must emphasize, though, that a magnetized plasma has a priori seven viscosity coefficients (five shears and two bulks). In addition, we also study in this same model the anisotropic heavy quark-antiquark potential in the presence of a magnetic field.

In the end, we propose a phenomenological holographic QCD-like model, which is built upon the lattice QCD data, to study the thermodynamics and the viscosity of the QGP with an external strong magnetic field.
\\

\noindent \textbf{Keywords:} gauge-gravity duality, non-Abelian plasmas, transport phenomena, viscosity.

\chapter*{List of Publications}

This dissertation is based on the papers:

\begin{enumerate}
\item R. Critelli, S. I. Finazzo, M. Zaniboni and J. Noronha, \textit{Anisotropic shear viscosity of a strongly coupled non-Abelian plasma from magnetic branes}. Phys. Rev. D \textbf{90}, 066006 (2014), [arXiv:1406.6019 [hep-th]].

\item R. Rougemont, R. Critelli and J. Noronha, \textit{Anisotropic heavy quark potential in strongly-coupled N=4 SYM in a magnetic field}. Phys. Rev. D \textbf{91}, 066001 (2015), [arXiv:1409.0556 [hep-th]].

\item R. Rougemont, R. Critelli and J. Noronha,\textit{Holographic calculation of the QCD crossover temperature in a magnetic field}.  Phys.\ Rev.\ D {\bf 93}, 045013 (2016), [arXiv:1505.07894 [hep-th]].
\end{enumerate}

\tableofcontents    



\listoffigures            
\listoftables            

\mainmatter

\fancyhead[RE,LO]{\thesection}

\onehalfspacing            

\chapter{Introduction}
\label{Chap1.0}

The so-called \emph{Standard Model} (SM) is the state-of-the-art description for the basic constituents of matter, having a striking success in describing physics phenomena up to distances  $\sim 10^{-17}$ m. The SM is built within the quantum field theory (QFT) framework, whose physical objects are the fields and their respective excitations, i.e. the \emph{particles}, which can be either fermions (half-integer spin) or bosons (integer spin).

Putting aside the Higgs mechanism \cite{Higgs1,Higgs2,Higgs3,Higgs4}, responsible to give mass for the elementary particles, one can separate, for practical and pedagogical reasons, the SM into two: The electroweak sector and the strong sector. The electroweak sector embraces the leptons (e.g. electron), the neutrinos, the massive vector bosons, and the photon; the strong sector is concerned with the quarks and gluons. Furthermore, this idea can be formalized in terms of group theory since the SM contains the following set of internal \emph{gauge symmetries}
\begin{equation}
SU(3)\times SU(2)\times U(1),
\end{equation} 
where $SU(N)$ denotes the special unitary group of rank $N$. Above, the $SU(3)$ represents the strong interaction, while $SU(2)\times U(1)$ represents the electroweak sector\footnote{Note that the SM does not contain information about the gravity. Indeed, an experimental sign of some \emph{quantum gravity} effect is far beyond our reach once it requires energies to the order of the Planck mass ($M_P\sim 10^{19}GeV/c^2$)}. Since the focus of this dissertation is to unveil aspects in the strong interaction domain, we will omit further explanations regarding the weak interactions. 

The strong interaction gives rise to \emph{hadronic matter}, formed basically by quarks and gluons. Inside a hadron, such as the proton, one has an intricate interplay between quarks and gluons, which is also responsible to maintain an atom cohesive. We call \emph{baryons} the hadrons formed by three quarks (e.g. proton), whereas \emph{meson} is the designation for hadrons with a pair quark-antiquark (e.g. pion)\footnote{\textit{A priori}, there is no reason to limit the numbers of quarks inside an hadron, but only these two classes of hadrons are stable. We remark, though, the recent discovery of the so-called \emph{pentaquark} \cite{Aaij:2015tga}.}.

The \emph{quantum chromodynamics} (QCD) is the theory of strong interactions and it gives the rules for how the quark and gluon fields interact. For example, one great triumph of QCD is the correct calculation of a broad variety of states in the hadronic spectrum \cite{Durr:2008zz}. However, due to its non-perturbative nature, the hadronic spectrum can only be accurately calculated using  \emph{lattice QCD} \cite{Creutz}, which is a computational method to explore the non-perturbative aspects of  QCD. On the other hand, using the celebrated properties of \emph{asymptotic f
reedom} \cite{Gross:1973id}, one can access analytically, via perturbation theory, high energy processes.

Another amazing feature of QCD is color confinement. Quarks and gluons have color charge, the fingerprint of the strong interaction and, for some reason, nature forbids free colored particles to exist. We can only observe their bound states, the hadrons. This puzzle is also difficult to tackle because confinement is a non-perturbative property of QCD. 

In our way to understand how the basic constituents of matter behave, we also want to understand what happens when one increases the temperature $T$ and density, i.e. we wants to unfold the \emph{phase diagram} of hadronic matter. In the context of hadronic matter, the density is given by the baryonic chemical potential $\mu_B$ (see Refs. \cite{Aarts:2015tyj,fodorreview} for a review). The experimental exploration of this phase diagram is done by the means of a heavy ion collision (HIC), in which the kinetic energy of the ultrarelativistic ions is converted in temperature; nowadays the main operational facilities performing HICs are the Relativistic Heavy Ion Collider (RHIC), and the Large Hadron Colider (LHC).  Alternatively, one could extract some data from the early universe (very high $T$, $T\gg m_\pi$), or from the dense stars (very high $\mu_B$, around $\mathcal{O}\left(1\,\textrm{GeV}\right)$ \cite{Kurkela:2009gj,Kurkela:2016was,Fraga:2015xha}), but it is a tougher task, naturally.

It turns out that the phase diagram of the hadronic matter is quite rich. As one increases the temperature, one will eventually end up with a hadron gas. Increasing even more the temperature, this hadronic matter undergoes a (pseudo) phase transition (crossover \cite{Aoki:2006we}), where the hadrons ``melt'' and one has a quark-gluon plasma (QGP). Additionally, by increasing $\mu_B$, one suspects the existence of a critical ending point (CEP) along with the first order transition line. One also suspects that, for extremely large values of $\mu_B$, one has the so-called color superconductor \cite{Alford:2007xm}.

The possibility of the quark-gluon plasma phase raised several theoretical questions and answers. The experimental way to reach it, the community concluded, was colliding two ultrarelativistic heavy ions, as mentioned before. A major result accumulated from decades of efforts came in 2004, as in this year all the experimental collaborations at RHIC made an announcement claiming that the QGP was formed in the heavy ion collisions \cite{expQGP1,expQGP2,expQGP3,expQGP4}.Theoretical support was also released as well \cite{QGP}.

The most startling feature of this new state of matter is, perhaps, its extremely low shear viscosity to entropy ratio $\eta/s$, now supported by solid measurements and theoretical predictions \cite{Heinz:2013th}. This value is also coherent with a naive estimate of what would be the lowest possible value for $\eta/s$ using kinetic theory and the uncertainty principle \cite{Danielewicz:1984ww} (cf. Sec. \ref{Chap2.2.1}). For this reason, one often says that the QGP is the most ``perfect fluid'' ever created. Furthermore, there has been great success in describing the strongly coupled QGP using relativistic hydrodynamic evolution \cite{Shuryak:2004cy,Schaefer1,Schaefer2}.

However, QCD perturbation techniques (pQCD) are not able to obtain such small viscosity \cite{Arnold:2000dr,Arnold:2003zc,Asakawa:2006tc,Meyer:2007ic,Xu:2007ns}. This is a compelling sign of the non-perturbative nature of the QGP formed in these experiments. Also, lattice QCD is not suited for calculations of nonequilibruim phenomena \cite{Meyer:2011gj}. It is in this daunting scenario that the Anti-de Sitter/Conformal field (AdS/CFT) correspondence \cite{Maldacena:1997re,Gubser:1998bc,Witten:1998qj} flourished because in 2004 a calculation performed within this framework gave the following result for the shear viscosity of the maximally supersymmetric $SU(N_c)$ Yang-Mills theory (a.k.a. $\mathcal{N}=4$ SYM) in the strongly coupled regime \cite{Kovtun:2004de}\footnote{The number of colors is also infinite.}
\begin{equation}
\frac{\eta}{s} = \frac{\hbar}{4\pi k_B},
\end{equation}
which is close to what was estimated in RHIC, and later at the LHC. Such astonishing result served as motivation for the enormous efforts made towards a better understanding of the QGP using holographic dualities.

Just to emphasize how a ``simple'' heavy collision may reveal some of the most recondite secrets of nature, we list briefly some of its possibilities:
\begin{itemize}
\item One can study a quantum field theory (QCD) at finite temperature in a laboratory. Unfortunately, there still no means to access the thermal electroweak sector, basically because the energy required is just too high. However, a novel $100$ TeV  $pp$ collider may shed some light in the electroweak phase transition - See Ref. \cite{NimaRev} for a review. 

\item It might connects us with the origin of the Universe. Indeed, after the Big-Bang (the first few seconds), the visible matter was a soup of quark-gluon plasma. In this sense, one often refers to a heavy ion collision as being a \emph{little bang} - although this is misleading name since the energy scales of a heavy ion collision vastly differ from the early universe.

\item Relativistic hydrodynamics is far from being a natural extension of the Navier-Stokes equation - see Ref. \cite{Romatschke:2009im} for a review. It has many subtleties and (apparent) flaws, mainly on its dissipative aspect - we explain this briefly in Sec. \ref{Chap2.1.1}. Therefore, the QGP formed in HIC represents a great opportunity to reveal how a relativistic dissipative fluid behaves.

\item The applications of the gauge/gravity duality may lead to some progress in string theory.
\end{itemize}

In more recent years, it was perceived that in a peripheral heavy ion collision, there may be the formation, for a short period of time, of the strongest magnetic field ever created in laboratory with an upper limit around $\mathcal{O}(10^{19}G)$ - or $\mathcal{O}\left(0.3\,\textrm{GeV}^2\right)$ in natural units\footnote{To translate the magnetic field expressed in natural units to the CGS system of units, one may use the fact that  $B^{\text{(CGS)}}\simeq 1.69\times 10^{20}$ Gauss for $(eB)^{\text{(natural)}}=1 \ \text{GeV}^2$.}, at the LHC \cite{noncentralB1,noncentralB2,noncentralB3,noncentralB4,noncentralB5,noncentralB6,noncentralB7,noncentralB8,noncentralB9}.  Moreover, extreme magnetic fields are found in dense neutron stars known as magnetars \cite{magnetar}, and is very likely to have existed in the early universe \cite{universe1, universe2,Green:2015fss}. Extreme magnetic fields can considerably change the thermodynamics of the QGP, and, in this sense, one is effectively adding a new $B$-axis on the phase diagram \cite{latticedata0,latticedata2,latticedata3,Endrodi:2015oba,cohen}. Another characteristic effect of strong magnetic fields is the breaking of the spatial isotropy, due to the appearance of a preferred direction along the $B$-axis; this feature may have profound impact on transport coefficients, as we shall see in this work. In summary, it is an auspicious time  to investigate these magnetic effects, either with lattice QCD, effective models, or the gauge/gravity duality \cite{Gursoy:2014aka,Agasian:2008tb,Mizher:2010zb,Evans:2010xs,Preis:2010cq,Fukushima:2012xw,Bali:2012zg,Fukushima:2012kc,Blaizot:2012sd,Callebaut:2013ria,Bali:2013esa,Bonati:2014ksa,Fukushima:2013zga,Machado:2013rta,Fraga:2013ova,Andersen:2013swa,Bali:2013owa,Ferreira:2013oda,Ruggieri:2014bqa,Ferreira:2014kpa,Farias:2014eca,Ayala:2014iba,Ayala:2014gwa,Ferrer:2014qka,Kamikado:2014bua,Yu:2014xoa,Braun:2014fua,Mueller:2015fka,Fraga:2012rr,reviewfiniteB1,reviewfiniteB2,reviewfiniteB3, DK1,DK2,DK3,DK-applications1,DK-applications2,DK-applications3,DK-applications4,DK-applications5,Rougemont:2015oea,Drwenski:2015sha,Endrodi:2013cs,Miransky:2002rp,Bloczynski:2012en,bookdima,Bonati:2013lca,Bonati:2013vba,Fraga:2012ev,Albash:2007bk,Johnson:2008vna,Ballon-Bayona:2013cta,Mamo:2015dea,Dudal:2015wfn,Fraga:2012fs,Mamo:2015aia}.

Therefore, following the holographic spirit, \emph{we investigate in this dissertation the interplay between the hot and dense matter, i.e. the QGP, with extreme magnetic fields} - That is our goal. More specifically, we investigate the dependence of shear and bulk viscosities, and the potential between a quark-antiquark with respect the magnetic field. In the end, we propose a holographic \emph{bottom-up} model that emulates the QCD equation of state (EoS) at $\mu_{B}=0$ and $B\neq 0$. A detailed resume of this dissertation is presented below \\

\subsection{Dissertation's briefing}
Here we present how this dissertation is organized, and give a short summary of each chapter.

We continue this introduction with the basics of the strong interactions involving the QCD Lagrangian at zero temperature. Then, we review the formation and the basic features of the QGP, with a focus on its low viscosity since we want to exploit this feature in presence of a magnetic field in Chapters \ref{Chap5.0} and \ref{Chap6.0}.

In Chapter \ref{Chap 2} we perform a study of the shear viscosity and bulk viscosity, aiming possible applications in strongly coupled non-Abelian plasmas, such as the QGP. For sake of completeness, we also briefly discuss the kinetic theory's formulation of shear and bulk viscosities. This chapter serves as preparation for a more detailed study made in Chapter \ref{Chap5.0} and Chapter \ref{Chap6.0}, in which we calculate the viscosities as functions of the magnetic field.

Chapter \ref{Chap 3.0} is dedicated to introduce in some detail the gauge/gravity duality, which will be our tool to deal with strongly coupled non-Abelian plasmas. As an instructive exercise, we computed the isotropic shear viscosity from two different ways in Sec. \ref{Chap3.4}; the bulk viscosity is examined in Sec. \ref{Chap3.5}. 

In Chapter \ref{Chap4.0} we introduce the effects of a magnetic field on the QGP. Also, we introduce here the important magnetic brane solution found by D'Hoker and Kraus \cite{DK1,DK2,DK3} - the gravity dual of magnetic $\mathcal{N}=4$ SYM, which is used in Chapters \ref{Chap5.0}, \ref{Chap6.0} and \ref{Chap7.0}. Moreover, this Chapter contains the discussion of how we deal with viscosity when one has an anisotropy induced by the magnetic field, i.e. we learn that now one has seven viscosity coefficients, being five shears and two bulks; this will be important for the subsequent chapters.

In Chapter \ref{Chap5.0} we calculate the anisotropic shear viscosities of the strongly coupled $\mathcal{N}=4$ SYM plasma in presence of a magnetic field, using the magnetic brane solution developed in the previous Chapter. This Chapter is based on Ref. \cite{DK-applications2}.

In Chapter \ref{Chap6.0} we calculate the two bulk viscosities of the strongly coupled $\mathcal{N}=4$ SYM plasma in presence of a mangetic field using the magnetic brane background. Although we argue that the non-vanishing trace of the magnetic brane could induce a bulk viscosity, we found that both bulk viscosities vanish.

In Chapter \ref{Chap7.0} we calculated the anisotropic heavy quark-antiquark potential in the presence of a magnetic field. Again, we have used the magnetic brane solution. This chapter is based on Ref. \cite{DK-applications2}.

The Chapter \ref{Chap8.0} is devoted to present a novel bottom-up non-conformal  holographic model, which is constructed upon the lattice results for the QCD EoS in order to emulates the effect of an external magnetic field on the non-confromal strongly interacting QGP. At the present stage, we have calculated some thermodynamic variables, such as entropy density and pressure, and the anisotropic shear viscosity. This Chapter is based on Ref. \cite{Rougemont:2015oea}. 

We close this dissertation in Chapter \ref{Chap9} where we present our conclusions and an outlook.

\subsection{Notation and conventions}
To avoid possible misunderstandings, we define here our notation and conventions used throughout this dissertation, if not otherwise specified.

We adopt the natural units system, i.e. $c=k_b=\hbar=1$. The signature of the metric is mostly plus, i.e. $(-++\cdots)$. We also adopt the Einstein summation notation, which means that two repeated indices are being summed, e.g. $\sum_{j}a_j b^j=a_j b^j$.  

The greek indices $(\mu,\nu,\dots)$ run through all the space dimensionality. The latin indices $(i,j,k,\dots)$ are reserved for the spatial dimensions, such as $x$, $y$, and so on.

Our Riemann curvature tensor is given by
\begin{equation}
R^{\alpha}_{ \ \beta\mu\nu} = \partial_{\mu}\Gamma^{\alpha}_{\beta\nu}- \partial_{\nu}\Gamma^{\alpha}_{\beta\mu}+\Gamma^{\alpha}_{\mu\sigma}\Gamma^{\sigma}_{\beta\mu} -\Gamma^{\alpha}_{\nu\sigma}\Gamma^{\sigma}_{\beta\mu},
\end{equation}
where $\Gamma^{\alpha}_{\mu\nu}$ is the Christoffel tensor, defined as
\begin{equation}
\Gamma^{\alpha}_{\mu\nu}=\frac{1}{2}g^{\alpha\sigma}\left( \partial_\nu g_{\sigma\mu} +\partial_\mu g_{\sigma\nu}-\partial_\sigma g_{\mu\nu} \right).
\end{equation}

Regarding the AdS$_5$ space, whenever we use $u$ as the ``extra'' radial coordinate, it is understood that the conformal boundary is located at $u=0$. On the other hand, if $r$ is used for the radial coordinate, the boundary is located at $r\rightarrow\infty$.

The physical magnetic field on the magnetic brane context, following the previous literature, is represented by $\mathcal{B}$. However, following Ref. \cite{Rougemont:2015oea}, we denote as $B$ the physical magnetic field in Chapter \ref{Chap8.0}.

\pagebreak

\section{Strong Interactions at zero temperature}
\label{Chap1.1}

This section reviews the basic aspects of QCD, which defines the interactions among gluons (spin-1 bosons) and quarks (fermions with spin 1/2), and gluons among themselves. The material covered here can be found in any QFT textbook \cite{Peskin}.

The strong interactions are ruled by the QCD Lagrangian, which is obtained from the SU($N_c$) non-Abelian Yang-Mills theory, where $N_c=3$ for the QCD, but is often enlightening to leave the number of colors free.

The QCD Lagrangian is defined as being a Lorentz scalar in $(3+1)$ dimensions, and it is given by
\begin{equation}
\mathcal{L}=-\frac{1}{4}G^{a}_{\mu\nu}G^{a\mu\nu}+\sum_{f}^{N_f}\bar{\psi}_f\left(\gamma^\mu D_{\mu}-m_f\right)\psi_f,
\end{equation}
where $\psi_f$ ($\bar{\psi}_f$) denotes the quark (antiquark) Dirac field, and $m_f$ its respective mass. The number of flavors is represented by $N_f$. So far, there are six flavors for the QCD (quarks up, down, strange, charm, bottom and top), but we usually consider only the first three in general, since the rest of them are very heavy\footnote{The masses of the quarks are: $m_u=2.3$ MeV, $m_d=4.8$ MeV, $m_s=95$ MeV, $m_c=1275$ MeV, $m_b=4180$ MeV, $m_t=173$ GeV \cite{Agashe:2014kda}.}. The covariant derivative $D_\mu$ is given by
\begin{equation}
D_\mu=\partial_\mu +i g A_\mu^a t^a,
\end{equation}
where $A_{\mu}^{a}$ are the gluon fields in the adjoint representation, $t^a$ are the generators of the $SU(N_c)$ group, and $g$ is the coupling constant among quarks and gluons.

The structure $G^{a}_{\mu\nu}$ is the non-Abelian Yang-Mills field strength, defined as
\begin{equation}
G^{a}_{\mu\nu}=\partial_\mu A_{\nu}+\partial_\nu A_{\mu}+gf^{abc}A^aA^b,
\end{equation}
where $f^{abc}$ denotes the structure constants of the group $SU(N_c)$, $[t^a,t^b]=f^{abc}t^c$. From $G^{a}_{\mu\nu}$, we also deduce that the gluons (bosons) interact directly with each other, in opposition of what happens in QED, whose photons do not interact directly among themselves. Although we can condense in one equation the essence of the strong interactions, it is extremely difficult to deal with it. For instance, for the gluon interaction $gluon+gluon \rightarrow 8\,gluon$, at \emph{tree level}, we need to take into account more than one million Feynman diagrams \cite{Mangano:1990by}!

QCD, as well as the whole SM, is renormalizable. The beta function $\beta(\mu)\equiv \mu\partial g/\partial\mu$ tells us how the coupling $g(\mu)$ evolves with the energy scale. For QCD ($N_c=3$), at the \emph{1-loop level}, it is given by
\begin{equation}\label{eq:betaQCD}
\beta(\mu)=-\left(11-\frac{2N_f}{3}\right)\frac{g^3}{16\pi^2}.
\end{equation}

Using the beta function, at 1-loop level, we have
\begin{equation}\label{eq:QCDg}
g(\mu)\sim \frac{1}{\ln \frac{\mu}{\Lambda_{QCD}}},
\end{equation}
where $\Lambda_{QCD}\approx 200$ MeV is the intrinsic energy scale of the strong interactions, and $\mu$ is the energy scale of the specific process.

From Eq. \eqref{eq:QCDg}, one concludes that the interactions among quarks and gluons, represented by the coupling $g$, become weaker at high energies for $N_f<33/2$ - this is the property of asymptotic freedom \cite{Gross:1973id}. With asymptotic freedom at hand, we can derive the potential felt by the pair quark-antiquark $V_{Q\bar{Q}}$ for short distances. The expression for $V_{Q\bar{Q}}$ is \cite{Peskin}
\begin{equation}\label{eq:QCDPotePert}
V_{Q\bar{Q}}=-\frac{4}{3}\frac{\alpha_s}{r}, \ \ \ \text{(short distances)}.
\end{equation}

Naturally, the above potential does not hold for long distances, i.e. when one has to deal with the non-perturbative regime of QCD, which is evidenced by the increase of $g(\mu)$ as we diminish the energy scale. Fortunately, lattice QCD is able to capture this static potential between the $Q\bar{Q}$ pair and the result is generally parametrized by the so-called Cornell potential \cite{Cornell}
\begin{equation}
V_{Q\bar{Q}} = -\frac{4}{3}\frac{\alpha_s}{r}+\sigma r,
\end{equation}
where the linear factor $\sigma r$ is responsible for color confinement. Also, we say that $\sigma$ is the \emph{string tension}, because of the string flux-tube of the chromo-eletromagnetic charge.
Moreover, there are some effective models to deal with the non-perturbative aspect of the QCD, such as the MIT bag model \cite{MITBag}, the Nambu-Jona-Lasinio \cite{NJL1, NJL2} model, etc.

Since the Chapter \ref{Chap7.0} is devoted to the study of the $\bar{Q}Q$ potential immersed in a magnetic field, it is worth to give some further theoretical details about the potential $V_{\bar{Q}Q}$. Using standard tools in QFT, we can obtain Eq. \eqref{eq:QCDPotePert} in, at least, two different ways. The first one is to consider a simple tree-level Feynman diagram interaction of the $Q\bar{Q}$ pair intemediated by a gluon; by comparing the result of this diagram, i.e. its $S-$Matrix, with the Born-level potential for the nonrelativistic scattering, we are led to the result \eqref{eq:QCDPotePert} \cite{Peskin}.

The other way to obtain $V_{\bar{Q}Q}$ is using the so-called \emph{Wilson loop} \cite{Wilson:1974sk}. The Wilson loop is a \emph{non-local} but gauge invariant observable, whose structure is given in terms of the holonomy of the gauge connection. The explicit formula for the Wilson loop is
\begin{equation}\label{eq:WilsonLoop}
\langle W(C) \rangle = \text{Tr}_{\mathcal{R}} P e^{i\,\oint_C A_\mu dx^\mu},
\end{equation}
where $\text{Tr}=$Trace, $\mathcal{R}$ is the representation of the group $SU(N_c)$, and $P$ is the path-ordering operator. Notice that \eqref{eq:WilsonLoop} resembles the Aharanov-Bohm phase in quantum mechanics, which is not a coincidence since the Wilson loop is the phase of a charged particle moving through the contour $C$.

To investigate further the physical meaning of the Wilson loop \eqref{eq:WilsonLoop}, we take a rectangular contour in, say, the $tx-$plane  with sides $T$ ($t-$axis) and $D$ ($x-$axis). Taking the limit $T\rightarrow\infty$, we have
\begin{equation}
\lim_{T\rightarrow\infty}\langle W(C) \rangle = e^{-iTV_{\bar{Q}Q}(D)},
\end{equation}
where $D$ now is the distance between the $\bar{Q}Q$ pair. In a confining theory, such as the QCD, as we increase the distance $D$ the Wilson loop behaves like
\begin{equation}
\langle W(C) \rangle \sim e^{-i\sigma D T}.
\end{equation}
Notice that the energy of the interaction is proportional to the loop's area, i.e. there is an \emph{area law} for confining theories. Moreover, in his seminal paper \cite{Wilson:1974sk}, Wilson tried to explain confinement ($g\rightarrow \infty$) arguing that the links (pieces of the loop) in one direction do not compensate links in opposite direction, but the flaw is that this is valid even for QED. Thus, analytical approaches for the confinement problem are certainly an open question.

The importance of the Wilson loop exceeds the mere $Q\bar{Q}$ potential calculation since it also can be defined as an order parameter for phase transitions. We postpone further discussions about this subject to the next section when we introduce temperature effects. Moreover, in the Appendix \ref{AppE} we revise the holographic calculation of this observable for  $\mathcal{N}=4$ super Yang-Mills at strong coupling \cite{maldacena,rey,sonne,yaffe,jorge}.

Incidentally, the QCD Lagrangian has some additional symmetries. One very important symmetry is the \emph{chiral symmetry}. Decomposing the (lightest) quarks in left-handed (L) and right-handed (R), and considering that $m_u\cong m_d\approx 0$, we have the global symmetry $U(2)_L\times U(2)_R=SU(2)_L\times SU(2)_R\times U(1)_V \times U(1)_A$, with the part $SU(2)_L\times SU(2)_R$ denoting the chiral symmetry; the $U(1)$ symmetries are the vector and axial symmetries, respectively\footnote{Actually, the vector and axial symmetries are only exact, at the classical level, in the chiral limit $m_u=m_d=0$, once $\partial_\mu J_{V}^{\mu}\propto m$ and $\partial_\mu J_{A}^{\mu}\propto m$. However quantum effects implies that $\partial_\mu J_{A}^{\mu}=\frac{g^2}{16\pi^2}\epsilon^{\alpha\beta\gamma\delta}G^{a}_{\alpha\beta}G^{a}_{\gamma\delta}$ (chiral anomaly) \cite{Peskin}.}. Furthermore, chiral symmetry was spontaneously broken in the early universe as the temperature cooled down below a certain critical temperature  $T_\chi$, which is very close to the critical temperature ($T_c$) of the deconfinement phase transition. Therefore, one may probe the restoration of the chiral symmetry in heavy ion collisions.

The pion is a Nambu-Goldstone boson associated with the spontaneous symmetry breaking of chiral symmetry. However, since the masses of the $u$ and $d$ quarks are not identically zero, the pion actually has a mass, which is
\begin{equation}
m_{\pi}^{2}=\frac{(m_u+m_d)}{f_\pi}\langle\bar{\psi}\psi \rangle,
\end{equation}
where $f_\pi=92$ MeV is the pion's decay constant. Hence, we usually say that the pion is a \emph{pseudo} Nambu-Goldstone boson. The term denoted by $\langle\bar{\psi}\psi \rangle$ is known as the chiral condensate, a non-perturbative observable per se, which spontaneously breaks the chiral symmetry $SU(2)_L\times SU(2)\rightarrow SU(2)_V$ to the isospin symmetry. Furthermore, the chiral condensate is an intrinsic property of quarks in the fundamental representation.

\section{The hot and dense QCD matter}
\label{Chap1.2}

In this section we begin to heat up ordinary hadronic matter until we observe a phase transition leading to the QGP, which is the object of our studies. Also, we intend to pave the way to Chapter \ref{Chap8.0} where we deal with the QGP thermodynamics in the presence of a magnetic field. 

The usual treatment in thermal QFT \cite{Das} is to Wick rotate the time coordinate, i.e. $t\rightarrow i\tau$, so that the path integral formulation becomes a partition function,
\begin{equation}
Z = \int \mathcal{D}\phi \, \text{exp}\left[ \int_{0}^{1/T}d\tau\int d^{d}x \mathcal{L}(\phi) \right].
\end{equation}
Since $Z$ is the partition function, we can obtain information about thermodynamics using standards identities of the partition function. However, none of this will be done in this work. As will be clear along the dissertation, the gauge/gravity duality provides the same information from a gravitational point of view.  \\

\begin{figure}[h]
\centering
\includegraphics[width=12.0cm,height=7cm]{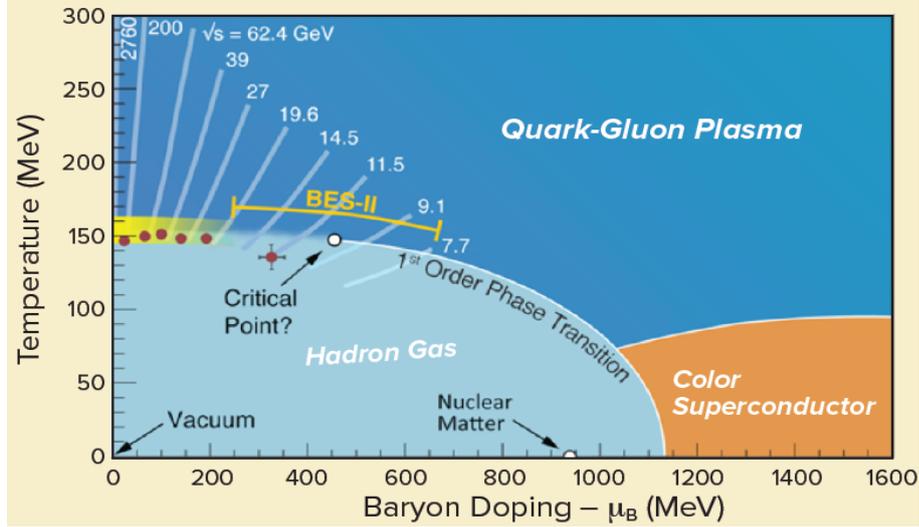}
\caption{A sketch of the phase diagram of the hadronic matter. The yellow band near $T=150$ MeV and $\mu_B=0$ MeV corresponds to the crossover region between confined/deconfined matter probed so far. The \emph{BES-II} refers to the \emph{Beam Energy Scan II} RHIC's program to search the critical ending point (CEP) of QCD, planned to start soon \cite{BES2}. Notice that we can reach higher values of $\mu_B$ by decreasing the energy of the beam collider \cite{BES1}. Also, for extremely high values of $\mu_B$, we have an intriguing phase known as \emph{Color Superconductor} (CSC) phase, which is likely to happen in very dense neutron stars \cite{Alford:2007xm}. Figure adapted from \cite{FutureNuke}.}
\label{fig:QDCPhase}
\end{figure}

We know that quarks and gluons are confined inside hadrons and there is no hope to see them freely. Nevertheless, it was realized some decades ago that there are some conditions under which quarks and gluons may be observed as the true degrees of freedom. These scenarios are feasible in \emph{extreme conditions}: very high temperature (melted hadrons) or/and very high density (squeezed hadrons). In Fig. \ref{fig:QDCPhase}, which is a sketch of the \emph{phase diagram} for the hadronic matter, we present the current view of what happens in these extreme situations. Therefore, as one increases the temperature we have a (pseudo) phase transition between the gas of hadrons and the QGP; on the other hand, for extremely large baryonic chemical potential, we infer the existence of a color superconducting phase \cite{Rischke:2003mt}.

In order to estimate the critical temperature $T_c$ transition between the hadronic confined matter and the deconfined QGP, we shall use the crude but instructive bag model with $\mu_B=0$. In this model, the QGP is treated as a free gas of fermions (quarks with $m=0$) and bosons (gluons), whilst the hadrons (confined phase) are regarded as ``bags'' with an inward pressure $P_B$ locking the quarks inside the hadron. One can compute the critical temperature by doing $P_{QGP}=P_B$. 

To obtain $P_{QGP}$ we use standard quantum thermodynamics. The starting point is the state density $dn$ in an interval $d^3p$,
\begin{equation}
dn=\frac{d^3p}{(2\pi)^2}g f(p)= \frac{dp}{(2\pi)^2}4\pi p^2 g f(p),
\end{equation}
where $g$ is the degeneracy factor. The distribution function $f(p)$ is given by
\begin{equation}
f(p)=\frac{1}{e^{p/T}\pm 1},
\end{equation}
where the plus sign is for quarks and antiquarks (Fermi-Dirac distribution), and the minus sign if for gluons (Bose-Einstein distribution).

Before we calculate the energy density, by integrating its differential $d\varepsilon=p\,dn$, let us derive what is the degeneracy factor for quarks ($g_q$) and gluons ($g_g$). For quarks (antiquarks have the same degeneracy), we have
\begin{equation}
g_q = \underbrace{N_{spin}}_{=2}\times\underbrace{N_{c}}_{=3}\times\underbrace{N_{f}}_{=2}=12,
\end{equation}
where we assumed contributions only for the lightest quarks, up and down. On the other hand, the gluon degeneracy factor is
\begin{equation}
g_g = \underbrace{N_{spin}}_{=2}\times\underbrace{N_{c}^2-1}_{=8}=16.
\end{equation}
The next step is to calculate the energy density $\varepsilon_{QGP}$,
\begin{align}
\varepsilon_q &= 4\pi g_q\int \frac{dp}{(2\pi)^3}\frac{p^3}{e^{p/T}+1} = \frac{7\pi^2g_q}{240}T^4, \\
\varepsilon_{\bar{q}} &= \varepsilon_q, \\
\varepsilon_g &=4\pi g_g\int \frac{dp}{(2\pi)^3}\frac{p^3}{e^{p/T}-1}= \frac{\pi^2 g_g}{30}T^4 \\
\therefore & \ \varepsilon_{QGP} = \varepsilon_q+\varepsilon_{\bar{q}}+\varepsilon_g. \label{eq:EnergyDensityCrude}
\end{align}

For an ultrarelativistic gas, the relation between the pressure ($P$) and the energy density ($\varepsilon$) is
\begin{equation}
P=\frac{1}{3}\varepsilon.
\end{equation}
Hence, to extract the critical temperature, we equate the above pressure with the bag pressure,
\begin{equation}
\frac{1}{3}\varepsilon_{QGP}=P_B \Rightarrow T_c = \left(\frac{45P_B}{17\pi^2}\right)^{1/4} \sim 140 \text{MeV},
\end{equation}
where we used $P_B^{1/4}\sim 200$ MeV. Although this is a rough estimative, it is in  agreement with realistic calculations \cite{fodorreview}, though it misses the \emph{order} of the transition.

\begin{figure}
\begin{center}
\begin{tabular}{c}
\includegraphics[width=0.55\textwidth]{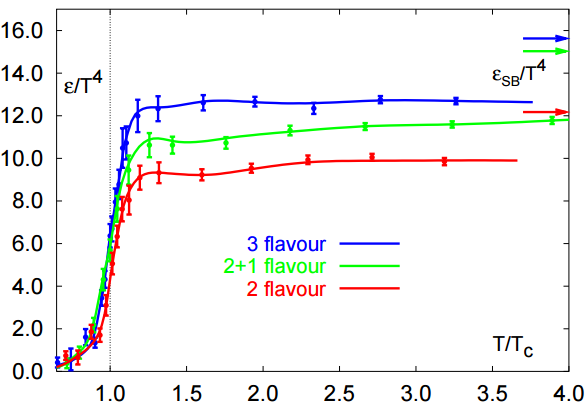} 
\end{tabular}
\begin{tabular}{c}
\includegraphics[width=14.0cm,height=7cm]{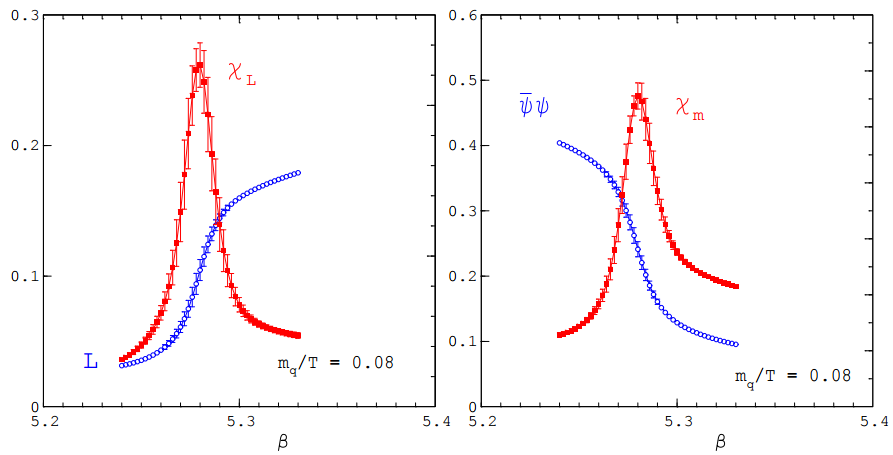} 
\end{tabular}
\end{center}
\caption{\emph{Top figure}: The energy density as function of temperature on lattice \cite{Karsch:2001cy}. Notice that we have an abrupt though smooth increase of the energy density in the crossover region; with the asymptotic behavior ($T>>T_c$) being the Stefan-Boltzmann relation found in Eq \eqref{eq:EnergyDensityCrude}. \emph{Bottom-left figure:} The expected value for the Polyakov loop $\langle L \rangle$ along with its respective susceptibility $\chi_L$ as function of the coupling $\beta=6/g^2$, with a clear sign of the deconfined phase for $T>T_c$. \emph{Bottom-right figure:} The quark condensate is monotonically decreasing, which means the recovering of the chiral symmetry in the same region of the confined/deconfined phase transition given by the other figures. Both figures were taken from \cite{Karsch:2001cy}.}
\label{fig:LQCDthermo}
\end{figure}

Let us go back to the case of the Wilson loop, discussed in the previous section. As mentioned already, the Wilson loop can be used as an order parameter for phase transition. Actually, one defines the Wilson \emph{line}, known as the \emph{Polyakov loop}, which has the following form
\begin{equation}
L(\vec{x})=\frac{1}{N_{c}}\text{Tr} P \,\text{exp}\left[ \int_{0}^{1/T}A_\tau(\vec{x},\tau)d\tau \right],
\end{equation}
where the integral is taken in the compact ``time'' direction with period $1/T$, which is the usual in any thermal quantum field theory. For a pure gauge theory\footnote{This is equivalent to assume quarks with infinite mass.}, we can summarize the important (qualitative) result of the Polyakov loop as being:
\begin{align}
T<T_c: \ \ \langle L \rangle = 0, \notag \\
T>T_c: \ \ \langle L \rangle > 0.
\end{align}
Thus, we can regard $\langle L \rangle$ as an important observable in the crossover region. We can obtain the respective critical temperature from the peak of the susceptibility, $\chi_L\sim (\langle L^2\rangle - \langle L\rangle^2)$, as depicted in Fig. \ref{fig:LQCDthermo}. Moreover, we can infer that $\langle L \rangle \propto e^{-F_{q}(T)/T}$, for $T>T_c$, where $F_q(T)$ is the quark's free energy.

The chiral $\langle\bar{\psi}\psi \rangle$ condensate is also an order parameter that is related with the chiral symmetry breaking, which is restored above some critical temperature $T_\chi$; below this temperature we have the formation of $N_f^2-1$ pions and other hadrons. Although $T_\chi$ is not directly connected with the deconfinement  critical temperature, it seems to be very close to it. The $T_\chi$ can be obtained from the peak of the susceptibility $\chi_m=\partial\langle\bar{\psi}\psi \rangle/\partial m$, cf. Fig. \ref{fig:LQCDthermo}.

Our cutting-edge knowledge about this phase transition comes from lattice QCD \cite{Borsanyi:2013bia}, which furnishes $T_c\sim 150$ MeV. And very important, it gives us a crossover, which is not a \textit{bona fide} phase transition since all functions are smooth and analytical, i.e. there is no discontinuity. We present some lattice results supporting these conclusions, including the Polyakov loop and the quark condensate in Fig. \ref{fig:LQCDthermo}, whose behavior is characteristic for a crossover phase transition.

There is a plethora of observables from which one can extract information about the critical temperature\footnote{It is not a problem that we have some slight difference between different critical temperatures, obtained from different observables. However, it is a necessary condition that they coalesce to the same $T_c$ at the CEP.}. For instance, in Chapter \ref{Chap8.0}, we calculated the critical temperature as function of the magnetic field using the entropy density inflexion point.

Now that we have discussed some of the theoretical aspects of the confined-deconfined phase transition, it is time to discuss the basics of a typical heavy ion collision, which is how we can achieve high temperatures. Extensive reviews can be found in Refs. \cite{Schaefer1, Schaefer2, reviewQGP1,reviewQGP2} - in particular, we indicate Ref. \cite{deSouza:2015ena} for the history of heavy ion collisions. 

The first attempt to study experimentally hot and dense QCD matter began in 1971 with the Bevalac, the first heavy ion collider\footnote{In general, the ions used to perform these experiments currently are lead (Pb) or gold (Au), and their velocity at the collision is very close to the speed of light.} at the Lawrence Berkeley National Laboratory (LBNL); although the motivation at the time was to probe the partonic structure of the nucleons, since in the 60's it was understood that they were not fundamental. Some years later, the theoretical predictions for the fluid-like behavior of the QGP began to appear \cite{Bjorken1}. The next facilities designed to perform heavy ion collisions were the Super Proton Synchrotron (SPS) at CERN in 1981, and the Alternating Gradient Synchrotron Booster (AGS) at the Brookhaven National Laboratory (BNL) in 1991. Currently, we have two operational facilities, the Relativistic Heavy Ion Collider (RHIC) at the BNL, with energy capability of $7.7\, \text{GeV} \lesssim \sqrt{s_{NN}}\lesssim 200$ GeV, and the Large Hadron Collider (LHC) at the CERN, with energy capability of $\sqrt{s_{NN}}=5.02$ TeV\footnote{In his first run (2009-2013), the LHC operated with $\sqrt{s_{NN}}=2.76$ TeV.}. 

\begin{figure}[h]
\centering
\includegraphics[width=13.0cm,height=4.5cm]{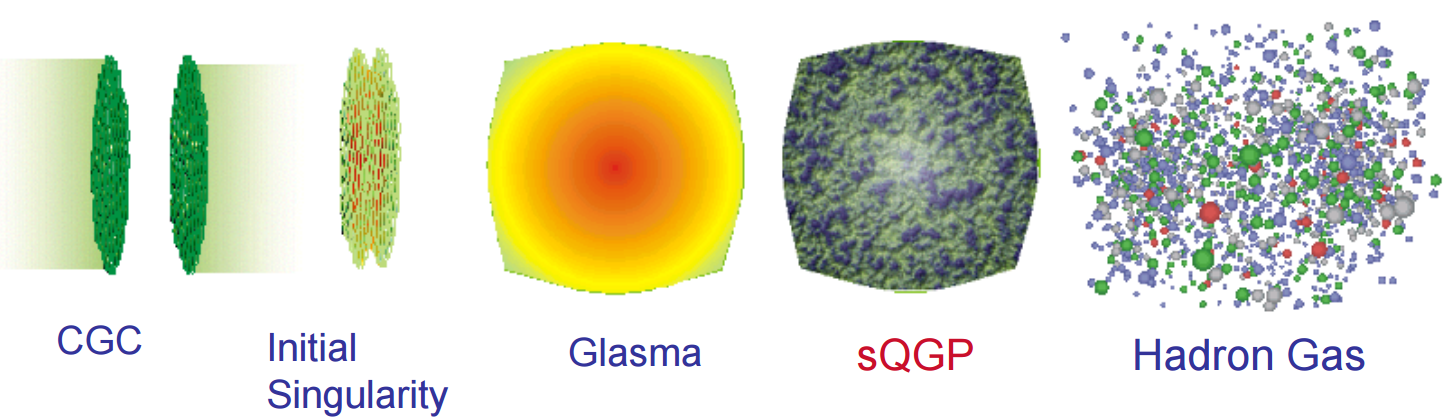}
\caption{The time evolution of a typical heavy ion collision. See the main text for the explanation. Soon after the formation of the QGP, the temperature is about $T\sim 400$ MeV. Figure adapted from \cite{McLerran:2010uc}.}
\label{fig:HICevol}
\end{figure}

We shall outline now what happens in a heavy ion collision and why expect the formation of the QGP. To help the visualization, we sketched in Fig. \ref{fig:HICevol} the time evolution of a typical collision event.

The first highly non-trivial situation is already the \emph{initial state} (the first stage in Fig. \ref{fig:HICevol}). It is not a surprise, since we have $\sim 260$ nucleons per ion at almost the speed of light, and they will eventually interact with the other ion. The simplest way to model this initial condition is using the so-called Glauber Model \cite{Glauber:1970jm}, in which we assume a Woods-Saxon profile for the nuclei. A more sophisticated way to describe the initial state is using the Color Glass Condensate (CGC) - see Ref. \cite{CGC1} for a review, which is an effective theory for high energy QCD where the saturation scale $Q_s$ guarantees the validity of perturbation theory, i.e.  $\alpha_s(Q_s)\ll 1$, though the system is \emph{strongly correlated}, due to its high occupancy level\footnote{This is known as the saturation of the gluon fields \cite{Albacete:2014fwa}. One can infer the importance of the gluon's field from the Balitsky-Fadin-Kuraev-Lipatov (BFKL) evolution equation for the gluon's density. Given the gluon distribution function $\mathcal{G}$, we have that $\mathcal{G}\sim x^{-\frac{4N_c\ln 2}{\pi}\alpha_s}$, where $x$ is the usual Bjorken$-x$. Hence, if we increase the energy (low $x$), the gluon's occupancy grows \cite{Albacete:2014fwa}.} . Moreover, the collision between two nuclei described by the CGC leads to the \emph{glasma}\cite{Glasma1, Glasma2} formation.

The next stage is the \emph{thermalization} of the glasma towards the strongly coupled quark-gluon plasma (QGP). We must emphasize that the thermalization is not completely understood yet, on contrary, it is a very active area of research, with recent studies using tools from QCD \cite{Kurkela:2015qoa}, as well as some holographic approaches \cite{Chesler:2008hg,Chesler:2010bi}. Nevertheless, we do know that thermalization is fast, i.e. $\tau_{\text{therm}}\sim 1$ fm, and the initial temperature of the thermalized QGP is about $T\sim 400$ MeV. Furthermore, the initial conditions for the hydrodynamic evolution of the QGP is provided by matching the energy,
\begin{equation}
T_{\mu\nu}^{initial}(\tau_{\text{therm}}) = T_{\mu\nu}^{hydro}(\tau_{\text{therm}}),
\end{equation}
where ``initial'' refers to some model (e.g. CGC) used to describe the early stages of the collision. The QGP phase is the focus of this dissertation. The existence of the QGP was announced by RHIC in 2004 \cite{expQGP1,expQGP2,expQGP3,expQGP4}\footnote{There were some previous evidences for the QGP at SPS found by looking at the suppression of the $J/\psi$ meson \cite{Heinz:2000bk}.}.

Finally, we have the last stage, the hadronization of the deconfined matter. Concomitant with its (fast) expansion, the QGP cools down and once it reach the transition temperature, we have the formation of the bounded states - the hadrons.  Eventually, this hadron gas will reach a temperature such that all the inelastic collisions cease, which is denoted as being the \emph{chemical freeze-out}, since the hadron's species are maintained after this threshold temperature. As the temperature keeps decreasing, one has the \emph{kinetic freeze-out}, wherein the elastic interactions cease (the gas does not interact any more) and the momentum distribution and the correlation distributions are frozen. After the kinetic freeze-out the remaining unstable hadrons decays and we have the stream of particles measured by detectors. The theoretical description of this hadronization can be described using, for example, the Hadron Resonance Gas (HRG) model \cite{HRG1}.

The experimental evidences for the existence of the QGP in a heavy ion collision, are related to:
\begin{itemize}
\item Jet suppression: In vacuum, a di-jet event has an equal distribution of energy among its jets. However, the QGP acts like a medium that reduces the momentum/energy from the jets, and we can measure this ``jet quenching''. Jet quenching is an important observable, which can be studied from the pQCD point of view (See Ref. \cite{Majumder:2010qh} for a review) assuming a weakly interacting QGP. To tackle the strongly coupled QGP one can resort to holographic techniques \cite{Liu:2006ug, D'Eramo:2010ak}, or lattice calculations \cite{Panero:2013pla}.

\item Elliptic flow: A well-defined elliptic flow is characteristic of the collective behavior. We come back to this issue in Sec. \ref{Chap1.2.1} with further details.\\
\end{itemize}

Notice that we did not try to give further details of how this matter can be formed inside neutron stars; the main reason is because the holographic methods developed in the subsequent chapters are not capable (yet!) to deal with large $\mu_B$. In the next subsection, we will speak more about key issues regarding the viscosity of the QGP.

\subsection{The viscosity of the QGP}
\label{Chap1.2.1}

Let us discuss now, in some detail, the striking feature of the smallness of the QGP $\eta/s$. This discussion will motivate Chapter \ref{Chap 2} which, in turn, will pave the way to tackle the anisotropic viscosities due to an external magnetic field.

\begin{figure}[t]
\centering
\includegraphics[width=12.0cm,height=9cm]{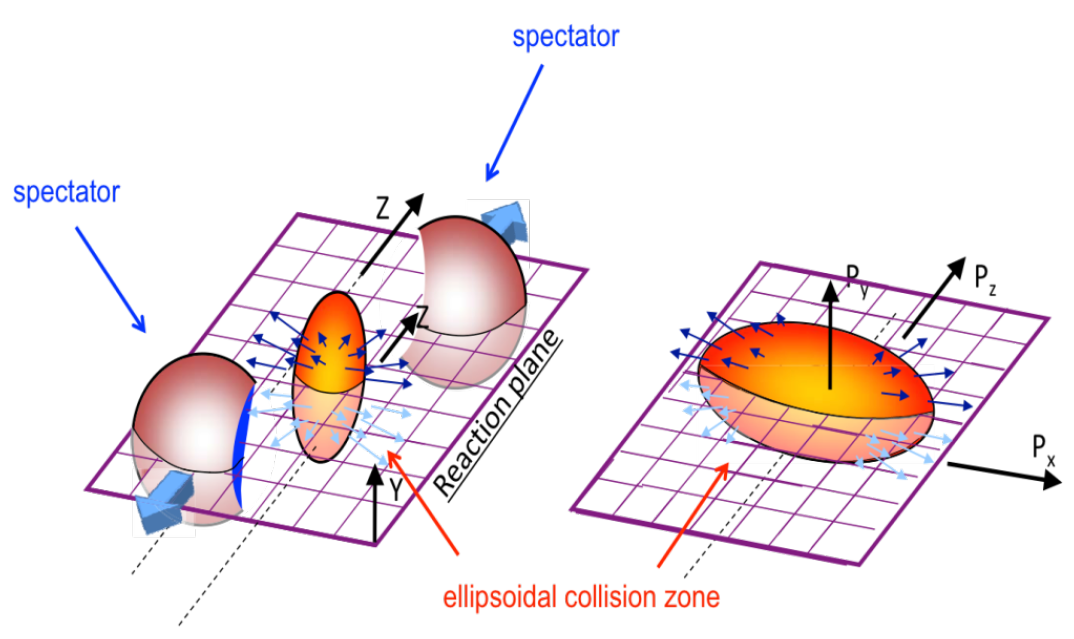}
\caption{A typical peripheral heavy ion collision. The number of spectators nuclei is given by $N_{spec}=2A-N_{partic}$, where $A$ is the mass number of the ion. Figure adapted from \cite{LaceyDraw}.}
\label{fig:HIreact}
\end{figure}

To connect the QGP formed in a heavy ion collision with its viscosity, we need to give some further details about the geometry of the collision. In Figure \ref{fig:HIreact} we have a schematic non-central collision (also called peripheral collision). The parameter that characterizes a peripheral collision is the impact parameter $b$, which is the distance between the centres of two colliding nuclei; we do not measure the impact parameter experimentally, nor $N_{spec.}$ or $N_{partic}$ (cf. Fig. \ref{fig:HIreact}). What is actually measured is the particle multiplicity in momentum space, which is decomposed in terms of Fourier coefficients:
\begin{equation}
E\frac{d^3N}{d^3p}=\frac{1}{2\pi}\frac{d^2N}{p_Tdp_Tdy}\left[ 1+2\sum_{n=1}^{\infty}v_n\cos(n(\phi-\psi_n)) \right],
\end{equation}
where $E$, $p_T$, $\phi$, and $y$, are the particle's energy, transverse momentum, azimuthal angle and rapidity, respectively. The angle $\psi_n$ is the event plane angle. The $v_n$ is the Fourier coefficient associated with the respective mode, with the first having specific names, i.e. $v_1$ is the direct flow, $v_2$ is the elliptic flow, $v_3$ is the triangular flow and so on.

When the QGP is formed in a peripheral heavy ion collision, it has initially an ellipsoidal shape (almond). As time goes by, this formed ellipsoid will expand, faster in the perpendicular direction of the collision (notice the momentum anisotropy on the left of Fig. \ref{fig:HIreact}), generating the elliptic flow. We can formally represent the momentum asymmetry using the eccentricity $\epsilon$,
\begin{equation}
\epsilon = \frac{\langle T_{xx}-T_{yy}\rangle}{\langle T_{xx}+T_{yy}\rangle},
\end{equation}
where $T_{xx}$ and $T_{yy}$ are the components of the stress-energy momentum tensor, with $\langle \dots \rangle$ meaning that we are averaging it on the reaction plane. Intuitively, we can understand the elliptic flow as being originated from the gradient pressure of the QGP formed in the collision, with the large elliptic flow indicating that the partons of the QGP are interacting strongly with small shear viscosity to entropy density (momentum diffusion).

\begin{figure}[tbp]
\begin{center}
\begin{tabular}{cc}
\includegraphics[width=0.5\textwidth]{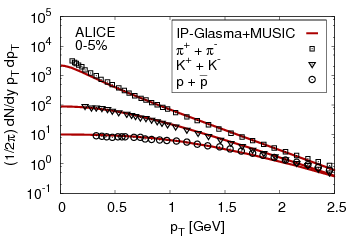} & %
\includegraphics[width=0.5\textwidth]{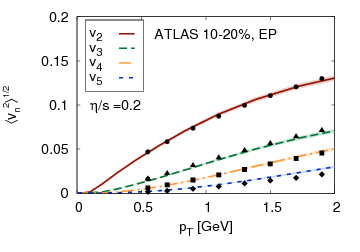} \\
&
\end{tabular}%
\caption{\emph{Left panel}: The success of the hydrodynamic modelling of the QGP (the red line crossing the experimental dots). The \emph{IP+Glasma} is the initial condition model, whilst \emph{MUSIC} is the hydrodynamic code that models the spacetime evolution of the QGP. \emph{Right Panel}: The experimental coefficients $v_n$ are in good agreement with the theoretical model for $\eta/s=0.2$ (very low viscosity). Figure adapted from \cite{Gale:2012rq}.}
\label{Fig:HIhydro}
\end{center}
\end{figure}

The question of whether relativistic hydrodynamics can describe  elliptic flow satisfactorily is ``answered'' in Fig. \ref{Fig:HIhydro}, which shows good agreement of the hydrodynamic model with the experimental data. Notice that, from the data analysis, we have a very small shear viscosity ($\eta/s=0.2$ \cite{Gale:2012rq})\footnote{The value of the shear viscosity depends of the temperature. Consequently, we can have some deviations of $\eta/s$ as we vary the energy of the collision \cite{NoronhaHostler:2008ju,Csernai:2006zz}.}.

\begin{figure}[t]
\centering
\includegraphics[width=12cm,height=7cm]{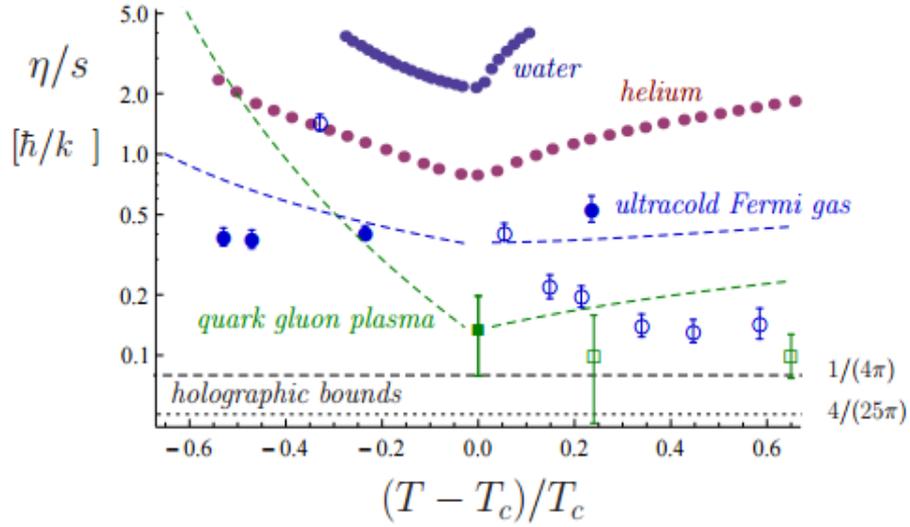}
\caption{The comparison of $\eta/s$ for a variety of substances that admits a hydrodynamic description. The ``holographic bounds'' are related to the KSS bound \eqref{eq: taxa}, and the possible Gauss-Bonnet correction \eqref{eq:EtaSGB}. Figure adapted from \cite{adams}.}
\label{fig:etacompar}
\end{figure}

Using the standard perturbative QCD, at the next-to-leading order, we have the following result for the viscosity \cite{Arnold:2000dr,Arnold:2003zc}:
\begin{equation}
\eta\sim \frac{T^3}{g^4\ln g^{-1}} \underset{g\sim 2}{\implies} \frac{\eta}{s}\sim 1.
\end{equation}
This value found is about one order of magnitude higher than the experimental values of $\eta/s$, cf. Fig. \ref{Fig:HIhydro}. Thus, we have compelling reasons to believe that the QGP formed in these heavy ion collisions is strongly coupled. Moreover, in Figure \ref{fig:etacompar} we show the expected behavior of $(\eta/s)_{QGP}$ and compare it with some other known fluids. 

Therefore, the we can draw the following big picture for the QGP: We can compute properties of the QGP using pQCD whenever the temperature is high enough and we can compute these same properties for low temperatures when we the QGP is already hadronized using some thermal model for hadrons \cite{NoronhaHostler:2008ju,NoronhaHostler:2012ug} (the shear viscosity using the HRG is done in Ref. \cite{Demir:2008tr}); it is the crossover region (strongly coupled regime) the source of great problems.

\begin{figure}[h]
\centering
\includegraphics[width=10cm,height=5cm]{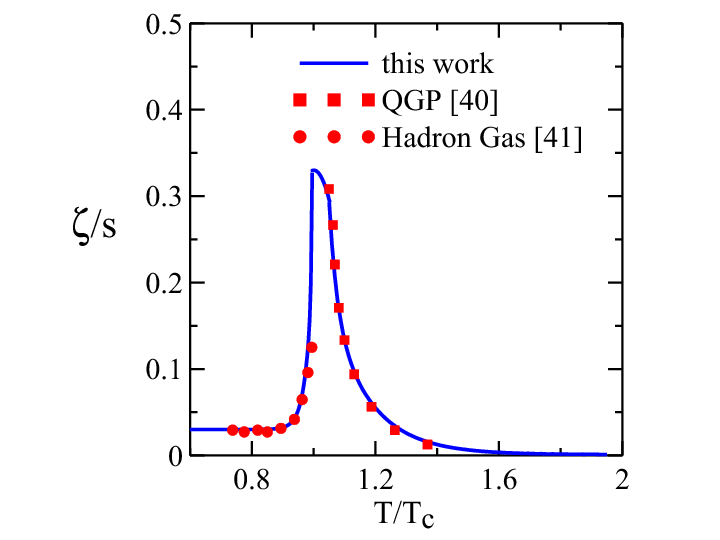}
\caption{The newest estimation for the bulk viscosity of the QGP. The HRG circle-dots were taken from \cite{NoronhaHostler:2008ju}, whilst the ``QGP'' square-dots refers to \cite{Karsch:2007jc} Figure adapted from \cite{Ryu:2015vwa}.}
\label{fig:bulkQGP}
\end{figure}

More recently, physicists became aware of the importance of the bulk viscosity \cite{Noronha-Hostler:2013gga,Noronha-Hostler:2014dqa,Ryu:2015vwa,Karsch:2007jc}. Because QCD is not conformal, though it can be approximately conformal at high temperatures, it is indispensable to build a non-conformal theory from a strong coupling framework to model near crossover region. In addition, bulk viscosity affects directly the value of the shear viscosity: If $\zeta$ grows then $\eta$ has to decrease and vice-versa. Figure \ref{fig:bulkQGP} shows a plot for the bulk viscosity \cite{Ryu:2015vwa} used in hydrodynamic simulations compared with experimental data, which seems to be one order of magnitude above the holographic calculations \cite{GN1,GN2,Benincasa:2005iv,Mas:2007ng,Buchel:2007mf,Buchel:2008uu,Eling:2011ms,Hoyos:2013cba,Finazzo:2014cna,gubser2}. We return to the holographically computed $\zeta$ in Sec. \ref{Chap3.5}.

\section{A bump on the road: The gauge/gravity duality}
\label{Chap1.3}

String theory appeared in the late 1960's as an attempt to describe the strong interactions of mesons \cite{Veneziano}. Despite its first success in describing the Regge trajectories of mesons, it was overcome by the QCD. Nowadays, string theory is seen as a promising theory of \emph{quantum gravity} since it has a massless spin-2 particle in its spectrum.

However, after a Maldacena's paper in 1997 \cite{Maldacena:1997re}, which connects a strongly coupled conformal theory in four dimensions with a string theory in higher dimensions, string theory returned as an attempt to describe the strong interactions. In a few words, Ref. \cite{Maldacena:1997re} conjectured a duality between $\mathcal{N}=4$ Super Yang-Mills theory in (1+3) dimensions with the type IIB super string theory. This duality is known as the AdS/CFT correspondence, since the $\mathcal{N}=4$ SYM is a conformal field theory (CFT), and the Anti-de Sitter (AdS) space is the background solution of the supergravity action originated from  string theory.

Soon after the publication of Maldacena's paper, Witten \cite{Witten:1998qj}, Gubser, Klebanov and Polyakov \cite{Gubser:1998bc} defined the map with more precise statements, allowing to ``easily'' extract  properties of strongly coupled systems. One remarkable result was the derivation of the ratio $\eta/s$ for the $\mathcal{N}=4$ SYM with infinite coupling and infinite number of colors \cite{Kovtun:2004de}
\begin{equation}
\frac{\eta}{s}=\frac{1}{4\pi},
\end{equation}
with $(\eta/s)_{sQGP}$ being in the vicinity of this value. This remarkable result opened a new window to explore non-equilibrium properties of strongly coupled theories, similar to QCD. 

The AdS/CFT correspondence is encompassed in a more general idea that relates field theories to gravitational theories in higher dimensions. To introduce it, we remind the reader that Bekenstein and Hawking \cite{Bek, Hawk} taught us that the entropy of a black hole scales with its horizon area,
\begin{equation}
S= \frac{A_h}{4\pi}.
\end{equation}
The above result is intriguing because the entropy is an extensive quantity, i.e. it should scale with the volume. This result inspired 't-Hooft in his seminal paper \cite{'tHooft:1993gx} to propose that the information is encoded on the boundary of the theory; later, this idea was perfected and vaunted by Susskind \cite{Susskind:1994vu}, giving origin to the \emph{holographic principle}. Therefore, AdS/CFT is the first serious realization of this holographic principle. This also explains why we often refer to the AdS/CFT correspondence as being an holography.

The fact that the original Maldacena's conjecture maps two highly symmetric theories is good, in the sense that we have more control of quantities (``BPSness''), and we can test some aspects of this conjecture more easily \cite{Drukker:2000rr,Faraggi:2016ekd}. However, its very unpleasant to be bounded only to the $\mathcal{N}=4$ SYM, since it is a highly symmetric theory, contrary to the real QCD, which is a non-conformal theory and does not have supersymmetry. This scenario naturally leads us to the pursuit of broader dualities with broken symmetries \cite{Klebanov:1999tb}, going from a more theoretical view (\emph{top-down} constructions) \cite{Klebanov:2000hb,Sakai:2004cn} to a phenomenological approach (\emph{bottom-up} construction) \cite{ihqcd-2,ihqcd-1,hot-ihqcd,gubser1,ihqcd-veneziano}. The agenda of connecting gauge theories with gravitational theories in higher dimensions is also known as the \underline{gauge/gravity duality} \cite{Horowitz:2006ct}. 

In this dissertation we want to apply this gauge/gravity duality idea to strongly coupled non-Abelian plasmas embedded in a magnetic field. We shall discuss its precise formulation in Chapter \ref{Chap 3.0} and apply it in the subsequent chapters in the case of including a magnetic field. Chapters \ref{Chap4.0} (shear viscosity), \ref{Chap5.0} (bulk viscosity) and \ref{Chap6.0} (heavy $\bar{q}q$ potential) utilize the gravitational dual of the $\mathcal{N}=4$ SYM in presence of a magnetic field developed by D'Hoker and Kraus in Refs. \cite{DK1, DK2, DK3}, which is reviewed in Sec. \ref{sec:magbranes}. The final Chapter \ref{Chap8.0} introduces a bottom-up model that we developed which is designed to describe QCD with magnetic field near the crossover temperature.

\chapter{Transport coefficients: the shear and bulk viscosities}
\label{Chap 2}
Now that we are more familiar with the properties of the QGP formed in a heavy ion collision, it is time to perform a thorough study of the so-called \emph{transport coefficients}. The transport coefficients are important observables to fully characterize a medium, or a fluid, in our case of interest. They arise to parametrize the response of the system under a small perturbation: when the system is out of its equilibrium it undergoes dissipative processes to return to the equilibrium, and the dissipation will be proportional to the correspondent transport coefficient. 

In this dissertation we are interested on the transport coefficients that causes dissipations on a fluid, i.e. the shear and bulk viscosities without other conserved charges such as $J^{\mu}$. Just to cite another example of transport coefficient, we also have the conductivity, which is the measure of how well a system  conducts some conserved charge under an external influence; for instance, the \emph{electrical} conductivity measures how well the system conducts an electric current under an external electric field. Just to say the obvious, this is the realm of non-equilibrium statistical physics.

Therefore, the next section will be devoted to analyse the underlying physics of dissipative processes in a fluid from a macroscopic point of view, i.e. hydrodynamics \cite{LandauFluid}. The fluid mechanics (or hydrodynamics), is an effective theory, relying in small departures from the equilibrium (long-wavelength), and trustful whenever the microscopic scale (e.g. the mean free path of the molecules in a gas) is much smaller than the macroscopic scale. Thus, we shall be able to connect the dissipative processes with some coefficients, the transport coefficients. However, the fluid mechanics cannot \emph{derive} these constants once it is not a microscopic description, so they are obtained by experimental measures.

The section \ref{Chap 2.2} introduces kinetic theory \cite{LandauKine}, which allows us to look at the microscopic foundations of (diluted) fluids. Although the calculations become harder, this theory bypasses the limitations of the macroscopic fluid mechanics because in the framework of the kinetic theory, we can actually derive the transport coefficients. Nevertheless, this method relies in how diluted the fluid is and how weakly the particles interact; as outlined in the Introduction, the QGP seems to be a liquid, which severely constraint this method. To circumvent this problem, we will use the gauge/gravity duality introduced in Chapter \ref{Chap 3.0}.

The end of this chapter finishes in Section \ref{Chap 2.3} with \emph{linear response theory}, which relates the transport coefficients with Green's functions (``correlators'' and ``two-point functions'' are synonyms here). This is a very powerful tool to calculate quantities, once it does not rely in assumptions such as weakly coupling and/or how diluted the fluid is. Indeed, this formulation will be used to calculate all the transport coefficients (viscosities) throughout this work.\footnote{Also, one can use kinetic theory to calculate the Green's functions \cite{Denicol:2011fa}.}

\section{Dissipation in fluid mechanics}
\label{Chap 2.1}

Let us start with ideal (non-relativistic) hydrodynamics \cite{LandauFluid}, which is appropriate when the viscosity (internal friction) and the thermal conductivity can be suppressed. In this scenario, according to the standard theory of fluid mechanics, we need, along with the equation of state, three equations to completely describe the fluid's motion,
\begin{equation}
\frac{\partial \rho}{\partial t}+\nabla\cdot (\rho \vec{v})=0,
\end{equation}
\begin{equation}\label{eq:Euler-eq}
\frac{\partial \vec{v}}{\partial t}+\vec{v}\cdot \nabla \vec{v}=-\frac{1}{\rho}\nabla P ,
\end{equation}
\begin{equation}\label{eq:Energy-balance}
\frac{\partial (1/2mv^2+\rho\varepsilon)}{\partial t}+\nabla\cdot \left[\rho\vec{v}(1/v^2+h) \right]=0, 
\end{equation}
where $\rho$ is the fluid's density, $\vec{v}$ is the velocity, $P$ is the pressure, $\varepsilon$ is the energy density, and $h$ is the enthalpy. The first equation is the continuity equation, expressing the conservation of the mass of the fluid. The second set of equations, \eqref{eq:Euler-eq}, is Newton's second law at work, known as the Euler's equations. The third equation takes into account the energy balance of the fluid. Furthermore, all the quantities above should be regarded as fields at some point $(t,\vec{x})$ of the space-time, \emph{not} the fluid itself, i.e. we are in the \emph{Eulerian picture}.

In writing the fundamental equations of the fluid mechanics, we tacitly ignored an external force density $\vec{f}$. To remedy this, one could include this force in the RHS of the Euler's equations \eqref{eq:Euler-eq}. For instance, if the fluid is under the effect of a gravitational field $g$, then, $\vec{f}=\vec{g}$; for plasmas, it is usual to have $\vec{f}=e(\vec{v}\times \vec{B}+\vec{E})$, where $e$ is the ion/electron charge, and $\vec{B}$ ($\vec{E}$) is the magnetic (electric) field. 

To include the effects of the energy dissipation (closely related with the increase of the entropy) in the fluid's equations of motion, we have to alter the equations above. More specifically, we alter the eqs. \eqref{eq:Euler-eq} and \eqref{eq:Energy-balance}.

For the heuristic derivation of the \emph{viscous stress tensor} $T_{ij}$, we first rewrite the Euler's equations in the following way,
\begin{equation}
\frac{\partial (\rho v_i)}{\partial t}= -\frac{\partial T_{ij}}{\partial x^j},
\end{equation}
where $T_{ij}=P\delta_{ij}+\rho v_i v_j$, is the momentum flux density. The question now is how to add a dissipative term $\Pi_{ij}$ for this flux density. For such task, we need some further phenomenological considerations:
\begin{itemize}
\item Internal friction occurs when we have relative motion between the fluid's constituents. We expect then something like $\partial_i v_j$ - a gradient in the fluid's velocity. Also, the friction vanishes for $\vec{v}=\text{constant}$;

\item Assume linearity in the dissipation with respect $\partial_i v_j$, in analogy with the standard classical mechanics. Fluids that obey this law are called \emph{Newtonian} fluids. We mention some non-Newtonian cases by the end of this subsection;

\item For an uniform rotation, with angular velocity $\vec{\Omega}$, there are no frictions too. In this case, the velocity $\vec{v}$ goes like $\vec{\Omega}\times\vec{r}$;

\item For an isotropic fluid (or even with axial symmetry), $\Pi_{ij}$ is a symmetric tensor.
\end{itemize}

Bearing in mind all theses assumptions, we can construct the following viscous stress tensor,
\begin{equation}\label{eq:viscous-stress-tensor}
\Pi_{ij}=\eta \left(\partial_i v_j +\partial_j v_i-\frac{2}{D-1}\delta_{ij}\nabla\cdot \vec{v} \right)+\zeta \delta_{ij}\nabla\cdot \vec{v},
\end{equation}
where $D$ is the number of dimensions of the space and time, $\eta$ is the shear viscosity, $\zeta$ is the bulk viscosity\footnote{The shear viscosity $\eta$ is also known as the dynamical viscosity, whereas $\zeta$ is also known as the second viscosity.}, and they are independent of the fluid's velocity. These two viscosity coefficients certainly depends of some parameters, such as the temperature (see Fig. \ref{fig:etacompar} for the case of the QGP) but, as mentioned before, we cannot (yet) determine their values. Moreover, notice that we arranged the tensor $\Pi_{ij}$ in such a way that $\eta$ is related with the vanishing trace of  $\Pi_{ij}$ whereas $\zeta$ does not vanish if we take the trace of the viscous stress tensor - this will always be the case, even for the magnetic scenario in Section \ref{Chap4.3}.

With the viscous stress tensor \eqref{eq:viscous-stress-tensor} at hand, we can finally modify Euler's equations ($D=3$ hereafter),
\begin{equation}
\frac{\partial \vec{v}}{\partial t}+\vec{v}\cdot \nabla \vec{v}=-\frac{1}{\rho}\nabla p + \frac{\eta}{\rho}\nabla^2\vec{v}+\frac{1}{\rho}\left(\frac{1}{3}\eta+\zeta\right)\nabla(\nabla\cdot\vec{v}).
\end{equation}
The above set of equations are the famous \emph{Navier-Stokes equations}. Analytical solutions for the Navier-Stokes are very challenging, as one can easily guess by looking at it, mostly because of its non-linearity\footnote{Indeed, the proof of existence and smoothness of the Navier-Stokes equations stands as one of the millenium problems - see Clay Mathematics Institute \cite{Clay}.}; usually, one tries to perform some sort of approximation. Last but not least, the dissipation effects enter as a scalar function in the energy equation \eqref{eq:Energy-balance}.

Another important feature of the shear and bulk viscosities is their positiveness, i.e. $\eta>0$ and $\zeta>0$. To arrive at this conclusion, we just need to check the rate of entropy increasing due to the internal friction, whose formula is given by
\begin{equation}
T\partial_t (\rho s) = \frac{\eta}{2}\left(\partial_i v_j +\partial_j v_i-\frac{2}{3}\delta_{ij}\nabla\cdot \vec{v} \right)^2+\zeta(\nabla\cdot \vec{v})^2.
\end{equation}
Thus, taking for granted the second law of thermodynamics ($s$ is the entropy density), we conclude that $\eta>0$ and $\zeta>0$. Moreover, the increase of the entropy (irreversibility) is consonant with the fact that $\Pi_{ij}$ breaks the time reversal symmetry.

After the discussion of the viscous stress tensor, it is time to think about what is the physical meaning of the viscosities (see also \cite{ViscoLi}). Let us begin by studying the effects of the shear viscosity.

\begin{figure}[h]
\centering
  \includegraphics[width=.58\linewidth]{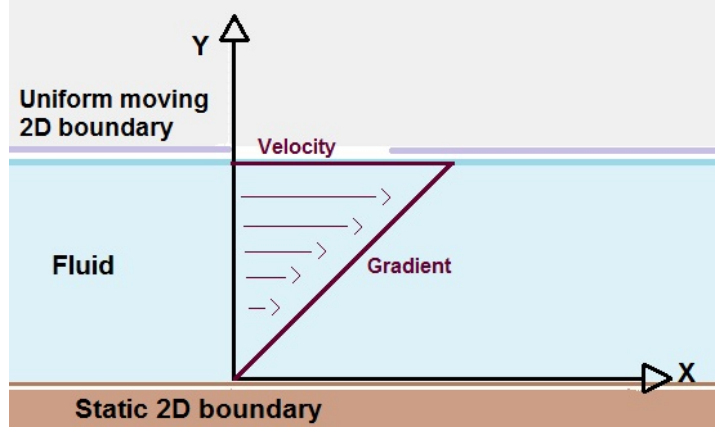}
  \caption{The schematic representation of a laminar flux induced by a moving boundary. The gradient of the fluid's velocity along the $y-$axis, induced by the friction with the moving plate, will result in an internal friction between the ``layers" of the fluid. This is known as Couette flow.}
  \label{fig:lami-flow}      
\end{figure}

For the shear viscosity, it is convenient to think of a laminar flow, as sketched in Fig. \ref{fig:lami-flow}. In this case, when the fluid is \emph{Newtonian}, the force per unit of area obeys the following relation
\begin{equation}\label{eq:newtonian-fluid}
\frac{F}{A}=\eta \frac{v_x}{d},
\end{equation}
where  $v_x$ is the $x-$component of the moving plate's velocity, and $d$ is the separation between the plates in Fig. \ref{fig:lami-flow}. Thus, the fluid's shear viscosity is a measure of the resistance to flow or shear.

Additionally, shear viscosity (and bulk viscosity as well) describes \emph{momentum diffusion}. To see this, consider a fluid's infinitesimal layer in the Couette flow (Fig. \ref{fig:lami-flow}), and then apply Newton's laws along with \eqref{eq:newtonian-fluid}. The result is
\begin{equation}
\partial_t p_x -\frac{\eta}{\rho}\partial^{2}_{x}p_x=0,
\end{equation}
where $p_x$ is the $x-$component of the layer's momentum . The above equation is the well known form of the diffusion equation.

To see whether the shear viscosity is important or not, we cannot perform a naive analysis by looking at only its absolute value; instead, we must analyse all the variables. A simple  way to do this is defining the Reynolds number ($Re$). If one neglects $\nabla\cdot\vec{v}=0$ (incompressible fluid), the Navier-Stokes equations become\footnote{This is achieved by a simple rescalying: $\vec{x}\rightarrow \frac{\vec{x}}{L}$, $t\rightarrow \frac{V}{L}t$, $\vec{v}\rightarrow \frac{\vec{v}}{V}$, $p\rightarrow \frac{p}{\rho V^2}$.}
\begin{equation}
\frac{\partial \vec{v}}{\partial t}+\vec{v}\cdot \nabla \vec{v}=-\nabla p + \frac{1}{Re}\nabla^2\vec{v},
\end{equation}
where
\begin{equation}
Re \equiv \frac{\rho|\vec{v}|L}{\eta},
\end{equation}
with $L$ being some macroscopic characteristic scale of the flow (e.g. the distance between two plates). Thus, for $Re\gg 1$, we can treat, in a good approximation, the fluid as being inviscid.

In table \ref{table-visc} we provide some experimental values for $\eta$ obtained from various elements. The cgs physical units for the viscosity is the \emph{poise} (P, $1P=0.1  kg.m^{-1}.s^{-1}$), originated from Jean Leonard Marie Poiseuille.

\begin{center}
\begin{table}[h]
\centering
\caption{Some selected fluids and their respective shear viscosity $\eta$. The values were measured under temperature and pressure conditions of $T=300$ K and $P=1$ atm \cite{Tabeta}, respectively.}
\label{table-visc}
\begin{tabular}{l|l}
Element       & $\eta$ (cP) \\ \hline
air         & 18.5           \\
hydrogen & 9.0       \\
helium     & 20        \\
Honey (non-Newtonian)         & 2000-10000       
\end{tabular}
\end{table}
\end{center}

We turn our attentions to the bulk viscosity now. The bulk viscosity plays a major role whenever we have an expansion of the fluid. This is evident once one notes that $\zeta$ is always associated with $\nabla\cdot \vec{v}$, with the latter being different than zero for fluids being compressed ($\partial_t\rho\neq 0$, from the continuity equation). Furthermore, this closely connects  the bulk viscosity with the sound speed $c_s$, $c_s^2\equiv \frac{\partial P}{\partial\rho}$ (or $c_s^2=\frac{\partial P}{\partial\varepsilon}$ for relativistic fluids). 

Another very important aspect of the bulk viscosity is that it vanishes in conformal field theories (CFT), as shown in Appendix \ref{appB}. In field theory, a conformal theory does not have a characteristic energy scale (e.g. the particle masses), and this is represented by the vanishing trace of the stress-energy tensor $T^{\mu}_{\mu}=0$; once one recalls that $T^{\mu}_{\mu}\propto\zeta$ it becomes obvious that $\zeta$ has to vanish for a CFT.

In general, bulk viscosity can be of the same magnitude as the shear viscosity, and, frequently, we have expressions relating both. For instance, a simple kinetic model of the expanding (viscous) universe gives that $\zeta/\eta\sim (1/2-c_s^2)^2$ \cite{Weinberg:1971mx}. For the case of the strongly coupled non-Abelian plasma, calculations within the gauge/gravity correspondence, indicate the following relation between bulk and shear viscosity \cite{Buchel:2007mf,Buchel:2008uu,Eling:2011ms,Hoyos:2013cba}
\begin{equation}
\frac{\zeta}{\eta} \sim \frac{1}{2}\left(\frac{1}{3}-c_s^2 \right)
\end{equation}
We shall return to the holographic case in more detail in Chapter \ref{Chap 3.0}.

\subsubsection{Extreme viscosities}
\begin{figure}[t]
\centering
  \includegraphics[width=.5\linewidth]{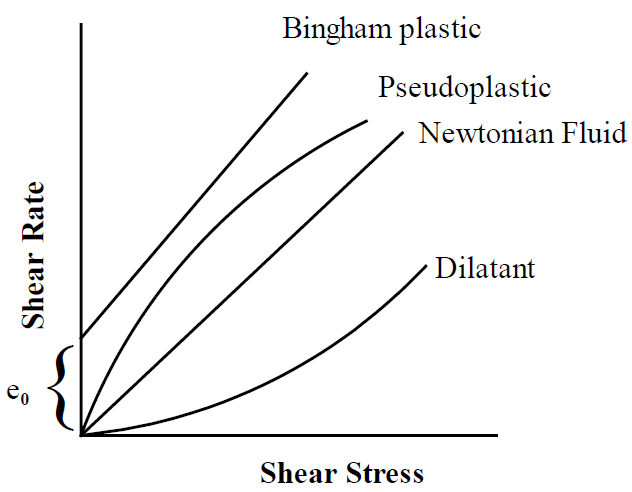}
  \caption{Classification of some materials according to their response under a shear stress. Recent holographic attempts to emulate a solid behavior in the context of the AdS/CMT is done in Refs. \cite{Alberte:2016xja,Hartnoll:2016tri}. This image was taken from \cite{ViscoLi}.}
  \label{fig:NonNewt}      
\end{figure}

It is worth mentioning some further ``extremal'' cases. By extremal cases we mean fluids with very high or low viscosity. In general, highly viscous fluids (liquids) deviate from the Newtonian behavior. Moreover, as we increase the viscosity, the fluid begin to behave as a type of plastic, or solid \cite{LandauEla}; we have some grey zone between liquids, plastics and solids. Fig. \ref{fig:NonNewt} illustrates the behavior of non-Newtonian fluids. 

One prominent example which fits in the above description, is the \emph{asthenosphere}. When one studies the Earth's structure, it is useful to divide it in different rheological layers (e.g. crust, mantle and core); the asthenosphere is located just above the mantle and below the lithosphere (very solid). Remarkably, the asthenosphere, though solid at first sight, behaves like a highly viscous fluid ($\eta\sim 10^{20}$ poises!) over geological times \cite{Asthe}, working as a type of ``grease'' between the mantle and the crust.

\begin{figure}[h]
\centering
  \includegraphics[width=.3\linewidth]{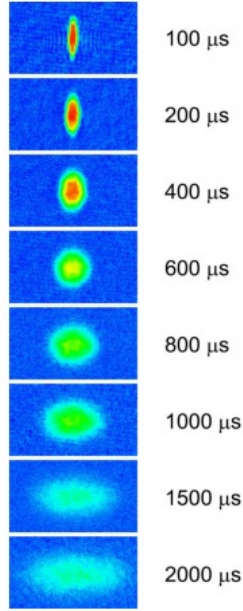}
  \caption{Images of the time evolution of a ultracold fermi gas. The initial condition (anisotropic pressure) resembles the one found in a non-central heavy ion collision. Moreover, the sharp elliptic flow, which arises in the course of the time, is only possible for small values of the viscosity $\eta/s$, like the QGP. The typical time scale of this experiment is about 1 mili second. This image was taken from \cite{O'Hara:2002zz}.}
  \label{fig:ultracold}      
\end{figure}

Now, let us cool down the temperature until the nano Kelvin scale, and mention some novel fluids which possess a tiny viscosity. As one cool down certain alkaline metals, through some laser beam trapping mechanism, we may have the formation of quantum gases, such as ultracold fermi gas \cite{Giorgini:2008zz,adams}, which are the prototype of a many-body quantum system. An ultracold Fermi gas can be brought to a strongly coupled phase as it condensates displaying a small value for the viscosity (superfluidity), in analogy with the QGP. We ilustrate the elliptic flow of the fermi gas in Fig. \ref{fig:ultracold}.

\subsection{The relativistic generalization}
\label{Chap2.1.1}

The relativistic generalization, in the sense of the special relativity\footnote{One could also consider the general relativity but, in order to simplify the discussion, we shall not consider curved spacetimes here.}, of the viscous fluid dynamics is not easy. If one tries to use a naive relativistic version of the Navier-Stokes equations, one finds some very undesirable features, because now the theory is plagued with instabilities (for boosted frames) and it does not respect causality \cite{Hiscock1, Hiscock2}. Here we briefly discuss how these problems can be circumvented. For a complete discussion we suggest Ref. \cite{Romatschke:2009im}.

For the relativistic case, we also start with the inviscid case. In this case, we have the following equations for the fluid motions (supplemented by an equation of state, $\varepsilon=\varepsilon(\rho, P)$, where $\varepsilon$ is the energy density)
\begin{equation}
\partial_\mu (\rho u^\mu)=0, \ \ \ \text{(particle number conservation)}
\end{equation}
\begin{equation}\label{eq:rel-flu-cons}
\partial_\mu ((\varepsilon+P)u^\mu u^\nu)+Pg^{\mu\nu})=0,
\end{equation}
where $u^\mu=(\gamma,\gamma\vec{v})$ is the fluid's four-velocity, with $\gamma$ being the Lorentz factor and $u^\mu u_\mu=-1$. Since the stress energy tensor is given by\footnote{Note that the stress tensor $T_{ij}$ is just the spatial components of the stress-energy tensor $T_{\mu\nu}$.} 
\begin{equation}
T^{\mu\nu}=(\varepsilon+P)u^\mu u^\nu +Pg^{\mu\nu}.
\end{equation}
The second set of equations \eqref{eq:rel-flu-cons} expresses the local conservation of energy and momentum.

The relativistic version of the viscous stress tensor for he Navier-Stokes theory is given by
\begin{equation}
\Pi_{\mu\nu}=-2\eta\left(w_{\mu\nu}-\Delta_{\mu\nu}\frac{\theta}{3} \right)-\zeta\theta,
\end{equation}
where $w_{\mu\nu}=\frac{1}{2}\left(D_{\mu}u_{\nu}+D_{\nu}u_{\mu}\right)$, $D_{\mu}=\Delta_{\mu\alpha}\partial^{\alpha}$, $\Delta_{\mu\nu}=g_{\mu\nu}+u_{\mu}u_{\nu}$ (orthogonal projector), and $\theta=\partial_\mu u^\mu$.

Now, if one employs this viscous tensor in Eq. \eqref{eq:rel-flu-cons}, just as done in the non-relativistic case, the theory will suffer from instabilities and acausality \cite{Hiscock1, Hiscock2}. For instance, the relativistic theories for viscous fluids, such as the Eckart and Landau-Lifshitz theories, predict that water, in room temperature, should explode in $10^{-34}$ s \cite{Hiscock2}!

So far, the most well succeeded way to fix these problems, is the so-called Israel-Stewart theory \cite{Israel1, Israel2}, derived from some entropy argument that guesses correctly the entropy current out of the equations. In this framework, we have a \emph{relaxation equation} for the viscous tensor rather than a simple algebraic relation. From the kinetic theory point of view, one can obtain a relativistic hydrodynamic system from the truncation of the \emph{gradient expansion}, in which the Knudsen number ($K_n\equiv l_{micro}/L_{macro}$) is the small parameter \cite{Cerci} - this is the so-called Chapman-Enskog method; though, this expansion leads to the (acausal and unstable) NS equations. A better way to proceed in kinetic theory is to use the moments method \cite{Denicol:2010xn}. For strongly coupled theories, the fluid/gravity duality can provide useful insights to construct the gradient expansion in terms of spacetime parameters \cite{FluidGravity}\footnote{Let us comment on how the stress-energy tensor of fluids can be constructed from gravitational arguments in the light of the AdS/CFT correspondence. Firstly, we write the thermal AdS$_5$ metric in the Fefferman-Graham coordinates,
\begin{equation}
ds^2 = -\frac{(1-u^4/u_h^4)^2}{(1+u^4/u_h^4)u^2}dt^2+(1+\frac{u^4}{u_h^4})d\vec{x}^2+\frac{du^2}{u^2}.
\end{equation}.

The next step is to use the formula \eqref{eq:stress-energy-tensor} \cite{ren1,ren2,ren3,ren4,ren5} for the expected value of the stress-energy tensor of the dual theory,
\begin{equation}
\langle T_{\mu\nu}\rangle=\frac{g^{(4)}_{\mu\nu}}{4\pi G_5} = \text{diag}\left(\varepsilon,\varepsilon/3,\varepsilon/3,\varepsilon/3 \right), \ \ \ \text{where} \ \varepsilon\equiv \frac{3}{16\pi G_5 u_h^4}.
\end{equation}
Hence, we obtained the stress-energy tensor of a conformal ($\varepsilon=3P$) ideal fluid in the rest frame. Performing a rigid boost $u^\mu$ in the metric, we obtain the ideal part for the stress-energy tensor,
\begin{equation}
\langle T_{\mu\nu}\rangle = (\varepsilon+p)u^\mu u^\nu +Pg^{\mu\nu}.
\end{equation}

We can gradually include higher gradient terms in the gravity side in order to obtain the dissipative (higher gradient expansion) part of the fluid stress-energy tensor \cite{FluidGravity}.
}

One can separate the shear/bulk (traceless/non-traceless) contributions for $\Pi^{\mu\nu}$ in the following way
\begin{equation}\label{eq:Pi-Sep}
\Pi^{\mu\nu}=\underbrace{\pi^{\mu\nu}}_{traceless}+\underbrace{\Pi}_{g^{\mu\nu}\Pi_{\mu\nu}/3}\Delta_{\mu\nu}.
\end{equation}
Thus, in Israel-Stewart theory, the equation for the shear channel becomes
\begin{equation}
\tau_\pi\left( D\pi^{\langle \mu\nu \rangle}+\frac{4}{3}\theta\pi^{\mu\nu} \right)+\pi^{\mu\nu}=-2\eta \sigma^{\mu\nu}+\dots,
\end{equation}
where $\tau_{\pi}$ is the relaxation time. We have defined also $D\equiv u^\mu \partial_\mu$, and $A^{\langle \mu\nu \rangle}\equiv\Delta^{\mu\nu\alpha\beta}A_{\alpha\beta}$, with $\Delta^{\mu\nu\alpha\beta}=(\Delta^{\mu\alpha}\Delta^{\nu\beta}+\Delta^{\mu\beta}\Delta^{\nu\alpha})/2-1/3\Delta^{\mu\nu}\Delta^{\alpha\beta}$, for any second rank tensor $A^{\mu\nu}$. In this equation, the time dependent variable is the tensor $\pi^{\mu\nu}=\Delta^{\mu\nu\alpha\beta}T_{\alpha\beta}$. The dots represent higher order corrections of the theory; for instance, for an holographic calculation of the second order transport coefficients, see Ref. \cite{Finazzo:2014cna}.

For the non-vanishing trace contribution of $\Pi_{\mu\nu}$, one has
\begin{equation}
\tau_{\Pi}(D\Pi+\theta\Pi)+\Pi=-\zeta\theta+\dots,
\end{equation}
where $\tau_{\Pi}$ is the relaxation time, and $\Pi$ defined in Eq. \eqref{eq:Pi-Sep}.

\subsection{Estimating the shear viscosity of liquids}

So far, we were not able to infer some value for the shear and bulk viscosities. Indeed, even nowadays we do not have a theory which enables us to derive them for liquids - see \cite{ViscoLi} for a complete review. Of course, we do not have a quasiparticle description for liquids, so it is really an astonishing fact that we are, perhaps, closer to compute $\eta(T)$ for the QGP than we are to obtain $\eta(T)$ for water from first principles. The analysis for diluted gases is done in Section \ref{Chap 2.2}.

The complexity of the interactions between the molecules in a liquid stands as a great challenge to derive $\eta(T)$. Each type of liquid has its own peculiarities and it would be a ludicrous task to tackle them individually. What is done, in the vast majority of the cases, to find the analytical expression for $\eta(T)$ of some liquid, is to resort to some empirical method. For example, one takes the data for $\eta(T)$ of a liquid and then fits this data to a function (e.g. $\eta(T)=A e^{-B T}$).

However, one can learn at least one lesson about $\eta(T)$ from  Eyring's pioneer work \cite{Eyring}. In this work, the dissipation rate comes from the filling of some vacancy (hole) by a molecule; in this picture the liquid is like a crystal because the molecules can freely fill (and leave) the holes. Thus, the shear viscosity depends on the activation energy $E$ of this process,
\begin{equation}
\eta\simeq hne^{\frac{E}{k_B T}},
\end{equation}
where $n$ is the molecule density, $h$ is the Plack's constant and $T$ is the temperature. The important feature of this estimate is that $\eta$ varies greatly with the temperature, which is the opposite of what occurs in gases.

\section{The kinetic theory's point of view}
\label{Chap 2.2}

In order to take a step further towards understanding the shear and bulk viscosities of a fluid, one may look at short distance behavior, i.e. the microscopic foundations of hydrodynamics \cite{LandauKine}. As already emphasized in previous sections, this is mainly suited for gases, as will be clearer below. Furthermore, this approach is a \textit{quasiparticle} method, since we consider the granulations (the molecules) to formulate the equations.

The basic quantity to be considered here is the distribution function $f(t,\vec{x},\Gamma)$, where $\Gamma$ covers the dependence on some other(s) possible(s) variable(s)\footnote{The most common dependence is the momentum $p$. Indeed, to form the phase space, we need all the conjugated momenta of the generalized coordinates.}. The distribution function gives us the statistics of the gas. For instance, the distribution for a classical diluted gas at rest is given by the well-known Boltzmann distribution,
\begin{equation}\label{eq:Bolt-dis}
f_0=exp\left(\frac{\mu-\varepsilon(\Gamma)}{T} \right),
\end{equation}
where $\mu$ is the chemical potential, $T$ is the temperature, and $\varepsilon(\Gamma)$ is the energy per molecule. For a quantum gas, we can have either the Fermi-Dirac distribution for fermions or the Bose-Einstein distribution for bosons.

One extracts macroscopic (measurable) quantities from kinetic theory by taking averages (moments). For instance, the spatial distribution density of molecules is
\begin{equation}
N(t,\vec{x})= \int f(t,\vec{x},\Gamma) d\Gamma,
\end{equation}
while the macroscopic mean velocity of the gas is
\begin{equation}
\vec{V}(t,\vec{x})=\frac{1}{N}\int \vec{v}(t,\vec{x},\Gamma) f(t,\vec{x},\Gamma) d\Gamma,
\end{equation}
among others.

Kinetic theory is concerned also with the evolution of the system through the course of the time, i.e. how the distribution function evolves with time. This information is obtained form the Boltzmann transport equation,
\begin{equation}\label{eq:Boltzmann-eq}
\frac{df}{dt}=\partial_t f+\vec{v}\cdot\nabla f=C[f].
\end{equation}
The factor $C[f]$ appears on the RHS of the Boltzmann equation is the so-called \emph{collision term}. This collision term tells us how the molecules of the fluid interact, and, depending of the interaction, the distribution function evolves differently with time. Assuming that the molecular interactions are fast\footnote{By fast, we mean that the interaction occurs in one point of the space-time.} binary collision, and two molecules collide elastically, one can write the collision term explicitly,
\begin{equation}
C[f]=\int w(f'f_1'-ff_1)d\Gamma_1d\Gamma'd\Gamma_1,
\end{equation}
where we assumed collisions of the kind $\Gamma,\Gamma_1\rightarrow\Gamma',\Gamma_1'$ (with the respective distributions $f',f_1',f,f_1$),  meaning that we have inhomogeneities in the gas once $\Gamma\neq\Gamma_1$. Of course, in equilibrium $\Gamma=\Gamma_1$ and $C[f]=0\Rightarrow df/dt=0$. The term $w'$ is related to the differential cross section of the molecule's interactions, $d\sigma=w(\Gamma,\Gamma_1;\Gamma',\Gamma_1')/|\vec{v}-\vec{v}_1|d\Gamma'd\Gamma_1'$. Thus, the Boltzmann equation \eqref{eq:Boltzmann-eq} is a non-linear integro-differential equation.

For extensive reviews and studies of the Boltzmann transport equation, we suggest Ref. \cite{LandauKine}. Here, we shall only pinpoint the basics in order to extract the shear and bulk viscosity in non-relativistic gases. 

To solve exactly \eqref{eq:Boltzmann-eq} is a tough task. Usually, we consider small departures from equilibrium,
\begin{equation}\label{eq:small-corr-distr}
f=f_0+\delta f,
\end{equation}
where $f_0$ is the equilibrium distribution \eqref{eq:Bolt-dis} and $\delta f$ is a small correction. A useful parametrization for the correction is $\delta f = f_0\chi/T$, with $\chi$ being the unknown function.

Before we plug the correction in the Boltzmann equation, we recast its LHS in the following way (see $§6$ of \cite{LandauKine})\footnote{The assumptions behind this rearrangement involve the equation of state of the ideal gas and the enthalpy $h=c_p T$, which is valid for classical gases with no vibrational modes.}
\begin{equation}\label{eq:LHSBolt}
\frac{T}{f_0}\frac{df}{dt}=  \frac{\varepsilon(\Gamma)-c_p}{T}\vec{v}\cdot\nabla T +\left[ mv_i v_j-\delta_{ij}\frac{\varepsilon(\Gamma)}{c_v} \right]w_{ij},
\end{equation}
where $c_p$ $(c_v)$ is the thermal capacity with constant pressure (volume), $m$ the molecule's mass, and $w_{ij}=1/2\partial_{(i}v_{j)}$ (we already met this structure in the viscous stress tensor \eqref{eq:viscous-stress-tensor}).
The above structure for the RHS of Boltzmann equation is very enlightening because we have written it in terms of a (first) gradient expansion, with the thermal conductivity $\kappa$ being related with the gradient of temperature, and the viscosity coefficients, $\eta$ and $\zeta$, being related with the gradient of velocity.

Substituting \eqref{eq:small-corr-distr} into \eqref{eq:Boltzmann-eq}, and using the form \eqref{eq:LHSBolt}, we have
\begin{equation}
 \frac{\varepsilon(\Gamma)-c_p}{T}\vec{v}\cdot\nabla T +\left[ mv_i v_j-\delta_{ij}\frac{\varepsilon(\Gamma)}{c_v} \right]w_{ij} = I[\chi],
\end{equation}
where
\begin{equation}
I[\chi]=\int w f_{01}(\chi'+\chi_1'-\chi-\chi_1)d\Gamma_1d\Gamma'd\Gamma_1',
\end{equation}
is the linear operator for the collisions. Since we are not interested on the thermal conduction of the gas, we omit the temperature gradient contribution hereafter.

To calculate the viscosities, we split the traceless (shear channel) and non-traceless (bulk channel) contributions of the velocity in the Boltzmann equation. This can be easily achieved with the following procedure 
\begin{align}\label{eq:kine-shear-bulk-channel}
\left[mv_i v_j-\delta_{ij}\frac{\varepsilon(\Gamma)}{c_v} \right]w_{ij} &= mv_i v_jw^{ij}-\frac{\varepsilon(\Gamma)}{c_v}\nabla\cdot\vec{v} \notag \\
    &= \underbrace{mv_iv_j\left(w^{ij}-\frac{1}{3}\delta^{ij}\nabla\cdot\vec{v}  \right)}_{\text{shear channel}} + \underbrace{\left(\frac{1}{3}mv^2-\frac{\varepsilon(\Gamma)}{c_v} \right)\nabla\cdot\vec{v}}_{\text{bulk channel}}.
\end{align}
Also, notice the similarity of the above equation with the viscous stress tensor \eqref{eq:viscous-stress-tensor}. Indeed, in kinetic theory, we define the visoucs tensor $\Pi_{ij}$ as being
\begin{equation}\label{eq:mom-flux-kine}
\Pi_{ij}=\int mv_iv_j f d\Gamma.
\end{equation}

When calculating the shear viscosity, we neglect the bulk channel. Then, we end up with the equation\footnote{More precisely, one can divide the contributions of the transport coefficients $\kappa$, $\eta$ and $\zeta$, in the collision integral as $I[\chi]=I_\kappa[\chi]+I_\eta[\chi]+I_\zeta[\chi]$.}
\begin{equation}
m\left(v_i v_j-\frac{1}{3}\delta_{ij}v^2 \right)w^{ij} = I[\chi].
\end{equation}

We will search for solutions of the equation above adopting the Ansatz
\begin{equation}
\chi = A_{ij}w^{ij},
\end{equation}
where $A_{ij}(\Gamma)$ is a symmetric second rank tensor. Moreover, for a monoatomic gas, the tensor $A_{ij}$ must depend exclusively of the velocity. Thus, the general form for this symmetric tensor is given by
\begin{equation}
 A_{ij}=\left(v_i v_j-\frac{1}{3}\delta_{ij}v^2 \right)A(v),
 \end{equation} 
where $A(v)$ is some unknown scalar function of $v$.

The equation for the shear channel reduces to
\begin{equation}
m(v_iv_j-\frac{1}{3}\delta_{ij}v^2)=I[A_{ij}].
\end{equation}

Concerning the viscous stress tensor, if we plug the distribution function $\chi$ in \eqref{eq:mom-flux-kine}, we obtain its traceless dissipative part $\sigma_{ij}$,
\begin{align}
\sigma_{ij}&=-\frac{m}{T}\int v_iv_jf_0\chi d\Gamma = \eta_{ijkl}w^{kl}, \\
\eta_{ijkl}&=-\frac{m}{T}\int f_0v_iv_jA_{kl}d\Gamma, \label{eq:rank4Visc}
\end{align}
At this point, we introduced the rank four tensor $\eta_{ijkl}$; this object will be very important in defining the viscosity within the context of anisotropic media - we discuss its properties in Sec. \ref{Chap4.3}. Because of the isotropic nature of the gas, the tensor $\eta_{ijkl}$ is symmetric under the index exchanges: $i\leftrightarrow j$,  $k\leftrightarrow l$, and  $ij \leftrightarrow jl$. Thus, we can construct it as follows (the traceless part)
\begin{equation}
\eta_{ijkl}=\eta\left(\delta_{ik}\delta_{jl}+\delta_{il}\delta_{jl}-\frac{2}{3}\delta_{ij}\delta_{kl} \right),
\end{equation}
so that $\sigma_{ij}=2\eta w_{ij}$ and, consequently, $\eta$ is the desired shear viscosity coefficient. To calculate $\eta$, we contract the the tensor $\eta_{ijkl}$ with respect to the pairs of suffixes $(ij)$ and $(kl)$. Therefore ,the expression for the shear viscosity becomes
\begin{equation}
\eta = -\frac{m}{10T}\int v^iv^jA_{ij}f_0d\Gamma.
\end{equation}

Instead of solving the equation above\footnote{\label{ft12} One can obtain a fairly accurate solution by expanding the scalar function $A(v)$ in terms of the Laguerre's polynomials \cite{LandauKine}.}, let us here only analyze its physical content. For such a task, we shall digress about key concepts of the kinetic theory.

A fundamental concept in kinetic theory is the \emph{mean free path}, which we denote by $l_{mfp}$. The mean free path tells us how much, in average, a molecule travels in space before colliding again with another molecule. Intuitively, $l_{mfp}$ should be small for a dense gas and for molecules with large interactions. A simple dimensional analysis estimate gives the relation: $l_{mfp}\sim 1/(N\sigma)$, where $\sigma$ is the collision cross-section; if we consider the molecular gas of hard spheres, then $\sigma=\pi d^2$, with $d$ being the molecule's diameter.

Aside the mean free path, one may also define a relaxation time $\tau\sim l_{mfp}/\langle v\rangle$, called mean free time. Bearing this in mind, we introduce the so-called Bhatnagar-Gross-Krook (BGK) model \cite{BGK}. In this approach, we approximate the collision integral by the expression
\begin{equation}
C[f]=\left(\frac{\partial f}{\partial t}\right)_{coll.}=-\frac{f-f_0}{\tau},
\end{equation}
where $\tau$ is the relaxation time. Although this approach can be a good qualitative description of transport coefficients, it is not precise enough to determine an overall factor. Using the BGK operator, the shear viscosity is
\begin{equation}
\eta_{BGK} = N T \tau,
\end{equation}
with $\tau \sim 1/(n \langle v\rangle\sigma)$. So equivalently, one can write
\begin{equation}
\eta\sim m\langle v\rangle N l_{mfp},
\end{equation}

Using $l_{mfp}\sim 1/(N\sigma)$ and $\langle v\rangle\sim \sqrt{T/m}$, one also obtains
\begin{equation}\label{eq:visc-gas-dens}
\eta \sim \frac{\sqrt{mT}}{\sigma}.
\end{equation}

We can compare quantitatively the BGK operator method with the solution of the Boltzmann's equations described in the footnote \ref{ft12} \cite{Zwan}, 
\begin{align}
\left(\frac{\eta}{\kappa} \right)_{Bolt.}=\frac{4}{15}m, \ \ \ \ \left(\frac{\eta}{\kappa} \right)_{BGK}=\frac{2}{5}m.
\end{align}
where $\kappa$ is the thermal conductivity of the gas. As aforementioned, we see a significant numerical disagreement due to accuracy limitations of the BGK method.

The result obtained by James C. Maxwell in 1860 \cite{Brush} for the shear viscosity, which is carried out in detail in Appendix \ref{appA}, is
\begin{equation}\label{eq:ViscMaxw}
\eta=\frac{1}{3}mn\langle v\rangle l_{mfp},
\end{equation}
which is in agreement (up to some overall constant) with the previous discussion. The most startling fact about the shear viscosity for dilute gases is its dependence with the density - This fact is explicit in Eq. \eqref{eq:visc-gas-dens}. This result had great importance to establish confidence in kinetic theory \cite{Brush}.

We now come to the bulk viscosity. Returning to the expression \eqref{eq:kine-shear-bulk-channel}, one considers now the bulk channel,
\begin{equation}
\left( \frac{1}{3}mv^2-\frac{\varepsilon(\Gamma)}{c_v} \right)\nabla\cdot\vec{v}=I[\chi].
\end{equation}
In a similar way to what was done for the shear, we shall seek for solutions with the form
\begin{equation}
\chi = A(v)\nabla\cdot\vec{v},
\end{equation}
so that
\begin{equation}\label{eq:kine-gas-bulk}
\frac{1}{3}mv^2-\frac{\varepsilon(\Gamma)}{c_v} = I[A].
\end{equation}
Thus,
\begin{equation}
\zeta = -\frac{m}{3T}\int v^2Af_0d\Gamma.
\end{equation}
For monoatomic gases, we have $\varepsilon(\Gamma)=1/2mv^2$ and $c_v=3/2$, therefore, the LHS of Eq. \eqref{eq:kine-gas-bulk} is zero. Consequently, we have that $\zeta=0$ for non-relativistic monoatomic gases\footnote{However, if we perform the virial expansion, which is some correction in the gas' EoS in terms of the gaseousness parameter $Nd^3$, one obtains a nonzero bulk viscosity \cite{LandauKine}.}. In the next subsection we shall mention what happens in the relativitstic case, but it is convenient to examine the ultrarelativistic (massless) case now. Using the fact that $3P=\varepsilon$ for ultrarelativistic gases, we see that \eqref{eq:kine-gas-bulk} vanishes in this limit too, though, as we shall see below, it does not in the purely relativistic case.

\subsection{Relativistic Boltzmann equation}

So far, we have only dealt with the non-relativistic case for the diluted gas since that was enough to develop our intuition about the viscosity coefficients. However, the case of bulk viscosity has some appeal, once it does not vanish in the relativistic case. Also, the high energy physics requires the usage of the relativistic version of the Boltzamnn equation. 

The relativistic generalization of the Boltzmann distribution \eqref{eq:Bolt-dis} is (see \cite{Hakim} for an extensive review)
\begin{equation}
f_0(p)=exp\left(\frac{u_\mu p^\mu-\mu}{T} \right),
\end{equation}
where $u^\mu$ ($p^\mu$) is the four-velocity (momentum). This is also known as the Juttner-Synge distribution function.

The relativistic version of the Boltzmann equation \eqref{eq:Boltzmann-eq} is (in flat spacetime)
\begin{equation}\label{eq:Rel-Boltz}
p^\mu\partial_\mu f = C[f].
\end{equation}

In this relativistic scenario, and using the BGK operator method, the shear and bulk viscosities for a monoatomic (classical) gas are respectively given by (see section 2 of Ref. \cite{Hakim})
\begin{equation}
\eta = \frac{\tau}{15}\beta m^5 4\pi e^{\beta\mu}\left[3\frac{K_3(\beta m)}{(\beta m)^2}-\frac{K_2(\beta m)}{\beta m} +K_1(\beta m)-K_{i1}(\beta m)\right],
\end{equation}
\begin{align}
\zeta=-\tau m^44\pi & e^{\beta\mu}\left[\frac{K_2(\beta m)}{(\beta m)^2}\frac{(\beta m)^2h'(\beta m)+\beta mh(\beta m)}{(\beta m)^2h'(\beta m)+1} \right. \notag \\
 & \left. -\frac{K_3(\beta m)}{\beta m}\frac{1}{(\beta m)^2h'(\beta m)+1}-\frac{\beta m}{9}\left( \frac{3K_2(\beta m)}{(\beta m)^2}-\frac{K_3(\beta m)}{\beta m}+K_1(\beta m)-K_{i1}(\beta m) \right)  \right],
\end{align}
where $\tau$ is the relaxation time, $\beta=1/T$, and $h(\beta m)=K_3(\beta m)/K_2(\beta)$, with $K_n(x)$ being the modified Bessel function of the second kind
\footnote{The modified Bessel function of the second kind may be defined as
\begin{equation}
K_n(x)\equiv \int_{0}^{\infty} ds \, e^{-s\cosh s}\cosh\left(n x\right).
\end{equation}}.

Moreover, as we mention ahead in Sec. \ref{Chap4.3.1}, Ref. \cite{Tuchin:2011jw} used this relativistic formalism, with the addition of an external magnetic field, to derive the anisotropic viscosity coefficients that arise in anisotropic media.

There are some recent developments regarding the relativistic kinetic theory, with possible applications to  heavy ion collisions. We highlight a novel study on the analyticity of the Green's function, from which one can extract the transport coefficients analysing its poles; the reference \cite{DenicolPhd} offers a good summary of recent developments. Similar philosophy is found in the holographic context, wherein one can compute the transport coefficients via the quasinormal modes (QNM) of the black branes which also correspond to poles of the retarded Green's function in the gauge theory \cite{Kovtun:2005ev}. 

\subsection{Minimal shear viscosity to entropy density ratio from the uncertainty principle}
\label{Chap2.2.1}

We now present a simple argument, based on the uncertainty principle ($\Delta p\Delta x\geq\hbar/(2\pi)$), of the minimal ratio $\eta/s$ that one could find in nature \cite{Danielewicz:1984ww}. In this sense,one argues that the particle momentum $\langle p \rangle$ cannot be measured with precision higher than $\sim \hbar/\langle p \rangle$. Oh the other hand, the mean free path $l_{mfp}$ must balances this accuracy in a way that $\langle p \rangle l_{mfp} \gtrsim \hbar$. As for the entropy density, recovering the Boltzmann constant $k_B$, we have $s\sim k_B$. Therefore, from Maxwell's formula \eqref{eq:ViscMaxw}, we have that
\begin{equation}
\frac{\eta}{s} \gtrsim \frac{\hbar}{k_B}.
\end{equation}
This supposed minimum is still larger than the ratio $\eta/s$ obtained from the AdS/CFT correspondence \cite{Kovtun:2004de}.

\section{Linear response theory}
\label{Chap 2.3}

It is time to develop an important tool to tackle the calculation of transpot coefficients in dense fluids. As we saw above, in section \ref{Chap 2.2}, there are some very standard ways to derive the shear and bulk viscosities of diluted gases. However, we need to surpass this dilute limitation and the way to achieve this is via \emph{linear response theory} \cite{LeStat}, from which one can derive the \emph{Green-Kubo relations}. This is, by far, the most used method to extract the transport coefficients of the QGP, without \cite{Demir:2008tr,Kovtun:2004de,Kadam2} or with external magnetic field \cite{DK-applications2,Kadam1}.

We have a classical formulation of this problem, but let us bypass it and go straightforward to the quantum case. Suppose now that we have a quantum theory and we perform some small time-dependent fluctuation. The effect of this disturbance is seen as a change in the original Hamiltonian, and, if we relate the fluctuation with some operator $\mathcal{O}$, we have the following correction to the original Hamiltonian
\begin{equation}
H'(t)=\lambda\mathcal{O}(t),
\end{equation}
where $\lambda$ is a small parameter.

We calculate now the expectation value of some operator $\mathcal{A}$ with this new correction. Here, we will always work in the canonical ensemble ($\rho=e^{-\beta H}$), if not otherwise specified. Thus, we have
\begin{align}
\langle \mathcal{A}(t)\rangle &= \text{Tr}\lbrace \rho(t)\mathcal{A}(t) \rbrace.
\end{align}

To proceed with the calculation, it is useful to work in the \emph{interaction picture}, which gives us the following rule to evolve the density matrix operator
\begin{equation}\label{eq:dens-matr-evo}
\rho(t)=U(t)\rho_0 U^{-1}(t),
\end{equation}
where $U(t)$ is the time evolution operator, and $\rho_0\equiv\rho(t=0)$ is the density matrix just before the perturbation. The time evolution operator is defined as
\begin{equation}
U(t)=T \text{exp}\left(-i\int_{0}^{t}H'(t')dt' \right),
\end{equation}
where $T$ is the time ordering operator. The above equation is solution of the evolution equation $dU/dt=H'U$.

Using Eq. \eqref{eq:dens-matr-evo}, we write the expectation value as
\begin{align}
\langle \mathcal{A}(t)\rangle &= \text{Tr}\lbrace \rho_0 U^{-1}(t)\mathcal{A}(t)U(t) \rbrace \notag \\
                          &= \text{Tr} \lbrace \rho_0\left( \mathcal{A}(t)+i\int_{-\infty}^{t}dt'[H'(t'),\mathcal{A}(t)]+\cdots \right)  \rbrace \notag \\
                          &\approx \langle\mathcal{A}(t)\rangle_{\mathcal{O}=0}+i\int_{0}^{t}dt'\langle [H'(t'),\mathcal{A}(t)]\rangle,
\end{align}
where in the second line we used the Baker-Campbell-Hausdorff formula\footnote{This formula is defined by
\begin{equation}
e^{A}B e^{A}=B+[A,B]+\frac{1}{2}[A,[A,B]]+\cdots \notag,
\end{equation}
with $A$ and $B$ being two distinct operators.}. Also, the small parameter $\lambda$ inside $H'(t)$ allows us to make the approximation above. Defining $\delta\langle\mathcal{A}\rangle=\langle \mathcal{A}\rangle-\langle\mathcal{A}\rangle_{\mathcal{O}=0}$, we have
\begin{align}
\delta \langle\mathcal{A} (t)\rangle &= \int_{0}^{\infty}dt'\langle [\lambda\mathcal{O}(t'),\mathcal{A}(t)]\rangle \notag \\
         &=\int_{-\infty}^{\infty}dt'\theta(t)\langle [\lambda\mathcal{O}(t'),\mathcal{A}(t)]\rangle.
\end{align}

We can trivially generalize this result for some space dependent operator, $\mathcal{O}(t)\rightarrow\mathcal{O}(t,\vec{x})$. Moreover, if we assume also that the disturbance happened in a very short time scale, i.e. $\mathcal{O}(t)=\mathcal{O}(0)\delta(t)$, we obtain
\begin{equation}\label{eq:Green-Kubo-rel1}
\delta \langle\mathcal{A} (t,\vec{x})\rangle = -i\theta(t)\langle [\mathcal{A}(t,\vec{x}),\lambda\mathcal{O}(0,\vec{0})]\rangle,
\end{equation}
or, equivalently, in Fourier space
\begin{equation}\label{eq:Green-Kubo-rel2}
\delta \langle \mathcal{A}(\omega,\vec{k}) \rangle = -i \int d^4x e^{i(\omega t-\vec{k}\cdot \vec{x})}\theta(t)\langle\left[\mathcal{A}(t,\vec{x}),\lambda \mathcal{O}(0,\vec{0})  \right] \rangle.
\end{equation}
The above equations, \eqref{eq:Green-Kubo-rel1} and \eqref{eq:Green-Kubo-rel2}, are known as the Green-Kubo relations, or Kubo formulas, for short. The $\theta(t)$ term ensures causality: only for $t>0$ we can have an effect from a perturbation made at $t=0$. Therefore, it is natural to write the relation
\begin{equation}\label{eq:kubo-fourier-app}
\delta \langle \mathcal{\mathcal{O}}(\omega,\vec{k}) \rangle = - \phi(\omega,\vec{k})G^{R}(\omega,\vec{k}),
\end{equation}
where $\phi(\omega,\vec{k})$ is the source of the operator $\mathcal{O}$, and  $G^{R}(\omega,\vec{k})$ is the retarded Green's function (also know as correlator, or two-point function) defined as
\begin{equation}
G^{R}(\omega)= -i \int d^4x e^{i(\omega t-\vec{k}\cdot \vec{x})}\theta(t)\langle\left[\mathcal{O}(t,\vec{x}),\mathcal{O}(0,\vec{0})  \right] \rangle.
\end{equation}

The transport coefficient $\chi(\omega)$ associated to the system's response for the original disturbance is given by the Green's function in the low frequency regime ($\vec{k}\rightarrow\vec{0}$), i.e.
\begin{equation}
\chi(\omega)=-\frac{G^{R}(\omega,\vec{k}=0)}{i\omega} = -\frac{\text{Im}\,G^R(\omega,\vec{k}=0)}{\omega},
\end{equation}
where we used the property that $\text{Re}\,G^R$ ($\text{Im}\,G^R$) is odd (even) with respect to $\omega$. We emphasize that the dissipative information is all encoded in the imaginary part of the Green's function, as it occurs in the damped harmonic oscillator or in the Drude's model for the conductivity.

For example, the conductivity tensor $\sigma_{ij}(\omega)$ can be seen as the response of the system to some electromagnetic disturbance, $H'=A_{\mu} J^\mu$. In this case, the conductivity is expressed as
\begin{equation}
\sigma_{ij}=-\frac{\text{Im}G^R_{ij}(\omega,\vec{k}=0)}{\omega},
\end{equation}
where
\begin{equation}
G^R_{ij}(\omega,\vec{k}=0)=-i \int d^4x e^{i(\omega t-\vec{k}\cdot \vec{x})}\theta(t)\langle\left[J_i(t,\vec{x}),\lambda J_{j}(0,\vec{0})  \right] \rangle.
\end{equation}

\subsection{The Kubo formulas for the viscosities}
\label{Chap2.3.1}

The goal now is to obtain the Kubo formula for both the shear and bulk viscosities. In this chapter we deal only with the isotropic case while the generalization for the anisotropic case induced by a magnetic field is done in Sec. \ref{Chap4.3}. So firstly, we have to specify what is the relevant operator to extract the formulas of the viscosities; turns out that the stress-energy tensor $T^{\mu\nu}$ is the required one, as evidenced by Eq. \eqref{eq:viscous-stress-tensor}. In fact, the metric field $g_{\mu\nu}$ couples with $T^{\mu\nu}$ in the interaction Hamiltonian so we have (in a linearised level)
\begin{equation}
H' =-\frac{1}{2} h_{\mu\nu}\delta(t)\delta^{(3)}(\vec{x})T^{\mu\nu}(t,\vec{x}),
\end{equation}
where $h_{\mu\nu}$ is a small deviation of the background metric $g_{\mu\nu}$ and $T^{\mu\nu}$ is the stres-energy tensor. With this disturbance, if we identify $\mathcal{A}=T^{\mu\nu}$, the eq. \eqref{eq:kubo-fourier-app} becomes
\begin{equation}\label{eq:var-T-}
\delta \langle T^{\mu\nu}(\omega,\vec{k}) \rangle =-\frac{1}{2}h^{\rho\sigma}G_{T^{\mu\nu}T^{\rho\sigma}}^{R}(\omega,\vec{k}),
\end{equation}
with
\begin{equation}
G^{R}_{T^{\mu\nu}T^{\rho\sigma}}(\omega,\vec{k}) \equiv - i\int d^4x e^{i(\omega t-\vec{k}\cdot \vec{x})}\theta(t) \langle \left[T_{\mu\nu}(t,\vec{x}), T_{\rho\sigma}(0,\vec{0}) \right] \rangle,
\end{equation}
being the retarded Green function.

We need some extra equation for $\delta T_{\mu\nu}$ to compare with \eqref{eq:var-T-}, and then, extract the Green's functions. The way to do this is generalizing $T^{\mu\nu}$ to a curved spacetime by introducing the covariant derivative $\partial_\mu\rightarrow\nabla_\mu$, and performing a small fluctuation of the metric. Because we are dealing with a field theory in flat spacetime, we assume that the background is flat; also, we work in the rest frame of the fluid where $u^{\mu}=(1,0,0,0)$\footnote{In other words, we will work in the Landau-Lifshitz frame, where $u_\mu\Pi^{\mu\nu}=0$, and all the information about the viscosities are in the components $\lbrace i, j, k, l \rbrace$ of the retarded Green function.}, and we assume a homogeneous perturbation, which means that we can work only with the spatial indeces, i.e.  $g_{ij}= \eta_{ij}+h_{ij}(t)$, with $h_{00}=h_{0i}=0$. Lastly, we can set $\vec{k}=0$ in the very beginning simplifying the intermediate steps.

To clarify, let us rewrite the expression for $T_{\mu\nu}$,
\begin{equation}
T_{\mu\nu}=T^{(0)}_{\mu\nu}+\Pi_{\mu\nu},
\end{equation}
where
\begin{align}
T^{(0)}_{\mu\nu} = (\varepsilon+P)u^\mu u^\nu +Pg^{\mu\nu}, \ \ \ \text{and} \ \ \ \Pi_{\mu\nu}=-2\eta\left(w_{\mu\nu}-\Delta_{\mu\nu}\frac{\theta}{3} \right)-\zeta\theta.
\end{align}

Thus, after we fluctuate the metric within $T^{\mu\nu}$, we have
\begin{equation}
\delta T_{ij}=ph_{ij}-\frac{1}{2}K\delta_{ij}h^{k}_{k}+\eta\left(-\partial_th_{ij}+\frac{1}{3}\delta_{ij}\partial_t h^{k}_{k} \right)-\frac{1}{2}\zeta\partial_t h^{k}_{k},
\end{equation}
where $K=-VdP/dV$ is the bulk modulus.

Taking $(ij)=(xy)$ in $\delta T_{ij}$, one obtains
\begin{equation}\label{eq:deltaTxy_iso}
\delta T_{xy} = (p+i\omega\eta)h_{xy}.
\end{equation}

Comparing Eq. \eqref{eq:deltaTxy_iso} with Eq. \eqref{eq:var-T-}, we arrive at the Kubo formula for the shear viscosity,
\begin{equation}
\eta=-\lim_{\omega\rightarrow 0}\frac{1}{\omega}\text{Im}\,G^{R}_{T^{xy}T^{xy}}(\omega,\vec{0}).
\end{equation}

In order to obtain the bulk formula, we take the trace of $\delta T_{ij}$, and compare it with the trace of \eqref{eq:var-T-}. The result is
\begin{equation}\label{eq:Bulk-Kubo-iso}
\zeta=-\frac{1}{9}\lim_{\omega\rightarrow 0}\frac{1}{\omega}\text{Im}\,G^{R}_{T^{i}_{i}T^{j}_{j}}(\omega,\vec{0}).
\end{equation}

\chapter{The Gauge/Gravity duality}
\label{Chap 3.0}

In the section \ref{Chap1.3} of this dissertation, it was mentioned how the dualities coming from string theory helped us to understand strongly coupled theories since they provide a map between the strong and the weak coupling regime. The most studied and understood duality is the Anti-de Sitter/Conformal field theory (AdS/CFT) correspondence, in which it is conjectured that there is a map between type IIB string theory in AdS$_5\times S^5$ and the maximally supersymetric $SU(N)$ theory, $\mathcal{N}=4$ SYM. In the correspondence, the low energy limit of the string theory corresponds to the strong coupled regime of the field theory, providing a unique way to study quantum field theories beyond weak coupling; the opposite direction of the duality, i.e. the weakly interacting CFT, is more challenging because it  a theory of quantum gravity\footnote{The usual superstring theory is equivalent to the ``first quantization'' of particles, in the sense that we have individual strings being quantized.} theory, and we did not achieve it in string theory yet. Therefore, in the remaining of this Chapter, we shall review this gauge/gravity duality and apply it to obtain hydrodynamic transport coefficients of strongly coupled field theories. Furthermore, as explained in Sec. \ref{Chap1.3}, one may use the terms AdS/CFT, gauge/gravity and holography interchangeably.

The first glimpse of a possible connection between gauge theories and  string theory came from the seminal paper of 't  Hooft \cite{Planar} in 1974 in which he considered the large $N_c$ limit of $SU(N_c)$ gauge theories (see Ref. \cite{Lucini:2012gg} for a review). In this large $N_c$ expansion, perturbation theory is governed by 't  Hooft's coupling $\lambda\equiv N_c g^2$ rather than just the gauge coupling $g$. In the 't Hooft limit, the beta function for $\lambda$ is given by
\begin{equation}
\mu \frac{d\lambda}{d \mu} =-\frac{11}{24\pi^2}\lambda^2 +\mathcal{O}(\lambda^3).
\end{equation}

Notice that the above flow equation tells us that the large $N_c$ theory is still asymptotically free. In fact, one can include effects of flavors by setting $N_f\rightarrow\infty$ but keeping $N_f/N_c$ small; this is the Veneziano limit \cite{Veneziano}, which is also being used recently in some holographic models \cite{ihqcd-veneziano,Drwenski:2015sha}.

Remarkably, as $N_c\rightarrow \infty$, there is a great simplification of the gauge theory at the perturbative level since only the \emph{planar diagrams}\footnote{A planar diagram is one that can be drawn without crossing lines.} will contribute in the calculations; the non-planar diagrams are suppresed by powers of $1/N_c$. This fact can be seen directly from the general rule to calculate some Feynman amplitude $\mathcal{M}$ in large $N_c$ theories (for the gluonic sector):
\begin{equation}\label{eq:largeN}
\mathcal{M}=\sum_{h, b=0}^{\infty}N_{c}^{2-2h-b}\sum_{n=0}^{\infty}c_n(h,b)\lambda^n,
\end{equation}
where $h$ denotes the number of ``handles'', and $b$ the number of boundaries. It turns out that the form of the expansion \eqref{eq:largeN} is the same one encounters in string perturbation theory. This fact is a strong indication of some deeper relation between gauge theories and string theories.

The last key ingredient within string theory towards the formulation of the AdS/CFT correspondence came in 1995 when it was shown that string theory also admits extended objects, called Dirichlet branes, or just D-branes for short \cite{Polchinski:1995mt}. A Dp-brane is like a $(p+1)$ dimensional membrane moving through the spacetime where the open strings endpoints are allowed to end (Dirichlet boundary condition). Another striking feature of D-branes is the existence of gauge theories in their worldvolume, thanks to the spectrum of the open-strings living on it; the worldvolume of a D-brane carries a U(1) SUSY gauge theory in $(p+1)$ dimensions \cite{Witten:1995im}. 

Therefore, with the concept of D-branes at hand, we shall outline in the next subsection the underlying motivations behind Maldacena's conjecture. The material covered in this Chapter can be found in Refs. \cite{Ramallo:2013bua,KiriBook,NastaseBook,ErdBook}.

\section{The Conjecture}
\label{Chap 3.1}

Maldacena's original conjecture \cite{Maldacena:1997re} relies on the examination of the same physical system from two rather distinct points of view. Assuming that we are on the type IIB superstring theory framework, we can add the theory a stack of $N$ coincident D3-branes and depending of certain limits involving the string coupling $g_s$ and the number of D3-branes, we may use distinct effective theories. For instance, we can face this situation from the gauge theory induced on the D3-branes (open string picture). The other situation is obtained for $g_sN\gg 1$, in which the D3-branes bend the space and one can use the supergravity limit (closed string picture).

Let us begin with the open string picture. In this scenario, we can decompose the contributions of the open/closed strings in the action as follows
\begin{equation}\label{eq:DecOpe}
S= S_{\text{branes}}+S_{\text{bulk}}+S_{\text{int}},
\end{equation}
where $S_{\text{branes}}$ contains the contributions of the gauge theory generated by the open strings living within the stack of $N$ D3-branes. Turns out that this gauge theory is $SU(N_c=N)$ $\mathcal{N}=4$ SYM in four dimensions \cite{Witten:1995im}, an exact conformal field theory with vanishing beta function, whose Lagrangian can be written in terms of the string coupling as (the bosonic part)
\begin{equation}
\mathcal{L}=\frac{1}{4\pi g_s}\text{Tr} \left[\frac{1}{4}F^{\mu\nu}F_{\mu\nu}+\frac{1}{2}D_\mu\phi^i D^{\mu\nu}\phi^i +[\phi^i,\phi^j]^2 \right].
\end{equation}
Thus, if we compare it with the Lagrangian of the $\mathcal{N}=4$ SYM with coupling $g$, we infer the following relation between the couplings
\begin{equation}
g=4\pi g_s^2.
\end{equation}

Returning to Eq. \eqref{eq:DecOpe}, $S_{\text{bulk}}$ is related to the closed strings distributed around the space and $S_{\text{int}}$ denotes the interaction between open/closed strings. Knowing that $S_{\text{int}}\propto \alpha'^2$, we conclude that $S_{\text{int}}\rightarrow 0$ in the low energy limit $(\alpha'\rightarrow 0)$, and the closed string modes decouple from the open string modes. Moreover, in the limit where $\alpha'\rightarrow 0$, the closed strings behaves as free gravity in $\mathbb{R}^{9,1}$.

\begin{equation}\label{eq:OpenPic}
\underline{\text{Open string picture:} \ \  \mathcal{N}=4 \ SU(N_c) \  \text{SYM in four dimensions} + \text{Free gravity}}.
\end{equation}
\\

Now, let us analyse the \emph{closed string picture} involving the stack of $N_c$ D3-branes. Let us also assume that we have a large number of D3-branes, i.e. $g_s^2 N\gg 1$, still in the low energy limit ($\alpha'\rightarrow 0$). Intuitively, the large number of D3-branes will bend spacetime, so we need a general relativity plus supersymmetry. We can describe this scenario using an effective theory called \emph{supergravity} (SUGRA). Since this will give rise to the AdS$_5$ space, we elaborate a little bit more the detail below.

Roughly speaking, these supergravity actions are obtained by equating the bosonic and fermionic degrees of freedom while demanding supersymmetry. The general SUGRA action of the type IIB (not necessarily a D3-brane system) is given by \cite{KiriBook,NastaseBook,ErdBook}
\begin{equation}
S_{\text{IIB Sugra}}= S_{NS}+S_{R}+S_{CS} + \text{fermions},
\end{equation}
where $S_{NS}$ describes the Neveu-Schwarz sector, $S_{R}$ is for Ramond sector while $S_{CS}$ is the Chern-Simons (topological) term. Also, commonlly the fermionic part (gravitino and dilatino) of the supergravity action are neglected since it vanishes in the classical limit. 

The explicit bosonic part of $S_{\text{IIB Sugra}}$, in the \emph{string-frame}\footnote{The Einstein-frame metric $g_{\mu\nu}^{E}$, is related to the string-frame metric $g_{\mu\nu}^{s}$, by a Weyl rescaling, i.e.
\begin{equation}
g_{\mu\nu}^{s}=e^{\Phi/2}g_{\mu\nu}^{E}.
\end{equation}}, is given by
\begin{align}
S_{NS} &= \frac{1}{16\pi G_{10}}\int d^{10} x\sqrt{-g} e^{-2\Phi}\left( R+4(\partial\Phi)^2-\frac{1}{2}|H_3|^2 \right), \\
S_{R} &=-\frac{1}{32\pi G_{10}}\int d^{10} x\sqrt{-g}\left(|F_1|^2+|\tilde{F}_3|^2+\frac{1}{2}|\tilde{F}_5|^2\right), \\
S_{CS} &=  -\frac{1}{32\pi G_{10}}\int d^{10}x \ C_4\wedge H_3\wedge F_3,
\end{align}
where $\Phi$ is the dilaton, and the field strengths $F_p$ are obtained from some potential $C_{p-1}$ such that $F_{p}=dC_{p-1}$; in type IIB (IIA) $p$ is odd (even). In the equations above, we also make use of the definitions
\begin{align}
|F_p|^2\equiv\frac{1}{p ! }F^{\mu_1\dots \mu_p}F_{\mu_1\dots \mu_n}, \ \ \ \tilde{F}_3 & =F_3-C_0\wedge H_3, \ \ \ \tilde{F}_5=F_5-\frac{1}{2}C_2\wedge H_3+\frac{1}{2}B_2\wedge F_3, \notag \\
 & H_3=dB_2.
\end{align}

The Dp-branes are the sources of the field strengths $F_p$. For instance, a D3-brane will source the (self-dual) field $F_5$, just as a charged particle is the source of the a U(1) gauge field. Furthermore, this imposes the Dirac quantization condition for some integer charge $Q$,
\begin{equation}
Q\equiv \int_{S^{8-p}}\star F_{2+p} = N_c,
\end{equation}
where we already identify the charge as being the rank of the $SU(N_c)$ group. We can also rewrite this in terms of some other constant $L$ in the following manner
\begin{equation}
L^4 = 4\pi l_{s}^{4} g_s N_c,
\end{equation}
and we say, in advance, that $L$ will be the AdS radius.

For a complete guide in solving the equations of motions for Dp-branes, we recommend Ref. \cite{Argurio:1998cp}. Here, we just give the final result of the extremal black hole metric created by a stack of $N_c$ D3-branes ($F_1=F_3=H_3=0$):
\begin{align}
ds_{s}^{2} &= \frac{1}{\sqrt{H(r)}}\left(-dt^2+\delta_{ij}dx^idx^j \right)+\sqrt{H(r)}(dr^2+d\Omega_{5}^{2}), \notag \\
      F_5 &= (1+\star)dH(r)^{-1}\wedge dx \wedge dy\wedge dz, \notag \\
      e^{\Phi} &= g_s = \text{constant},
\end{align}
where
\begin{equation}
H(r)=1+\frac{L^4}{r^4}.
\end{equation}

\begin{figure}[h]
\centering
\includegraphics[width=12cm,height=7cm]{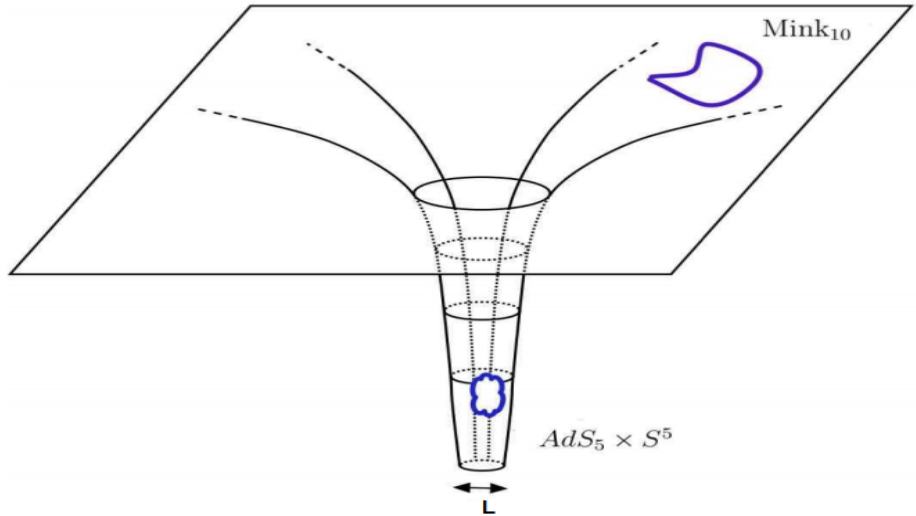}
\caption{\emph{The closed string picture:} The strings in the flat space are long wavelength excitations and do not see the throat. On the other hand, the strings near the throat still interact as they are pushed down the throat. Figure adapted from \cite{Mateos:2007ay}.}
\label{fig:ads5s5}
\end{figure}

Notice that the case of D3-branes is special in the sense that the dilaton profile is trivial. Taking the limit $r\gg 1$ (near the ``throat'', cf. Fig \ref{fig:ads5s5}), one obtains
\begin{equation}\label{eq:limAdsCon}
ds^2=\frac{r^2}{L^2}(-dt^2+dx^2+dy^2+dz^2)+\frac{L^2dr^2}{r^2}+L^2d\Omega_{5}^{2}.
\end{equation}

The metric \eqref{eq:limAdsCon} is $AdS_5\times S^5$ -  we are close now to fully state the conjecture. To finish the closed string picture, we still have to notice that the closed strings far from the throat are weakly interacting, so we have a free gravity theory; this is justified because energies near of horizon $E_{hor}$ is small with respect to an observer far from the horizon, i.e. $E\sim  E_{hor}/r\rightarrow 0$. This situation is depicted in Fig. \ref{fig:ads5s5}.

In summary, from the closed string picture, we have
\begin{equation}\label{eq:CloPic}
\underline{\text{Closed string picture:} \ \ \text{type IIB in} AdS^5\times S^5 + \text{Free gravity}, \ \ \ (gN_c\gg 1).}
\end{equation}

Now, if we compare Eq. \eqref{eq:OpenPic} with Eq. \eqref{eq:CloPic}, we are lead to the conjecture made by Maldacena \cite{Maldacena:1997re}:
\begin{equation}\label{eq:Con-dual-Mal}
\boxed{ \mathcal{N}=4 \ SU(N_c) \ \text{SYM in four dimensions} = \text{type IIB in} AdS^5\times S^5.}
\end{equation}

We expressed the duality in its \emph{strong} form, i.e. the statement above is supposed to be valid at any value of the coupling, though the derivation assumed $g N_c \gg 1$ in the closed string picture. In his \emph{weak} form, the conjecture is assumed to hold only between strongly coupled $\mathcal{N}=4$ SYM and the weakly coupled type IIB string theory (classical approximation).

The parameters $L$ (AdS radius), $l_s$ (string length), $g_s$ (string coupling), $g$ (Yang-Mills coupling), and $\lambda$ ('t Hooft coupling) are related by
\begin{equation}
 \frac{L^4}{l_s^4}=4\pi g_s N_c = 4\pi g^2 N_c = 4\pi \lambda.
 \end{equation} 
 
For the classical (super)gravity approximation on the AdS side to be valid, one must have
\begin{equation}
L\gg l_s \Rightarrow \lambda \gg 1 \  \text{('t Hoof limit)}.
\end{equation}
The relation above makes evident the weak/strong duality implied by the AdS/CFT correspondence.

\subsubsection{Basic Checks}

Now that we stated the conjecture, one may perfome some checks to verify it. For the purposes of this dissertation, to analyse the symmetries of both sides suffices. Thus, let us 

\begin{itemize}
\item The global symmetries on both sides of the duality must match since they reflect the physical properties of the system. The global symmetry of $\mathcal{N}=4$ SYM (CFT side) is provided by conformal symmetry $SU(2,2)$ plus its R-charge symmetry, i.e. one has $SU(2,2)\times SU(4)_R\sim SO(2,4)\times SO(6)$, which is the bosonic subgroup of the supergroup $SU(2,2|4)$. On the gravity side, the symmetry group of the AdS$_5$ space is $SO(2,4)$, while the 5-sphere possess a $SO(6)$ rotational symmetry; so again, one has $SO(2,4)\times SO(6)$ symmetry. For the interested reader, in Appendix \ref{appA} we discuss more about the conformal group.

\item The $AdS^5\times S^5$ has 32 Killing spinors and  $\mathcal{N}=4$ SYM has 32 supercharges.

\item There is a $SL(2,\mathbb{Z})$ symmetry related with an S-duality on both sides,
\begin{equation}
\tau = \frac{4\pi i}{g_s}+\frac{\theta}{2\pi} = \frac{i}{g}+\frac{\chi}{2\pi}.
\end{equation}
\end{itemize}

Another very important match is the comparison of the spectra of operators (CFT side) and the fields (AdS side). This check can be verified in Table 7 of Ref. \cite{D'Hoker:2002aw}.

\subsection{The renormalization group argument}
\label{Chap 3.1.1}

Although the first great d\'ebut of the holographic principle took place within the framework of  string theory, it is a widespread belief that it may be well defined without any mention to string theory. Thus we outline here how one can arrive at the AdS/CFT correspondence using solely concepts of QFT \cite{adams}. 

We consider then some generic lattice field theory in $d$ dimensions, such as Ising model, whose Hamiltonian is given by
\begin{equation}
H = \sum_{x} J_i(x) \mathcal{O}^i(x),
\end{equation}
where $J_i(x)$ is the coupling of the correspondent operator $\mathcal{O}^i(x)$ at the lattice $x$. Following the usual renormalization group approach, we examine the behavior of the couplings as one varies the energy, which means that one can coarse-grains the system, similarly to a block-spin, as we vary the lattice scale $u$ (e.g. $u=a,2a,\dots$). Mathematically, this idea is expressed by the beta function $\beta(J(x),u)$,  
\begin{equation}
u\frac{\partial}{\partial u}J_i(x)=\beta(J(x),u).
\end{equation}

\begin{figure}[t]
\centering
\includegraphics[width=12cm,height=7cm]{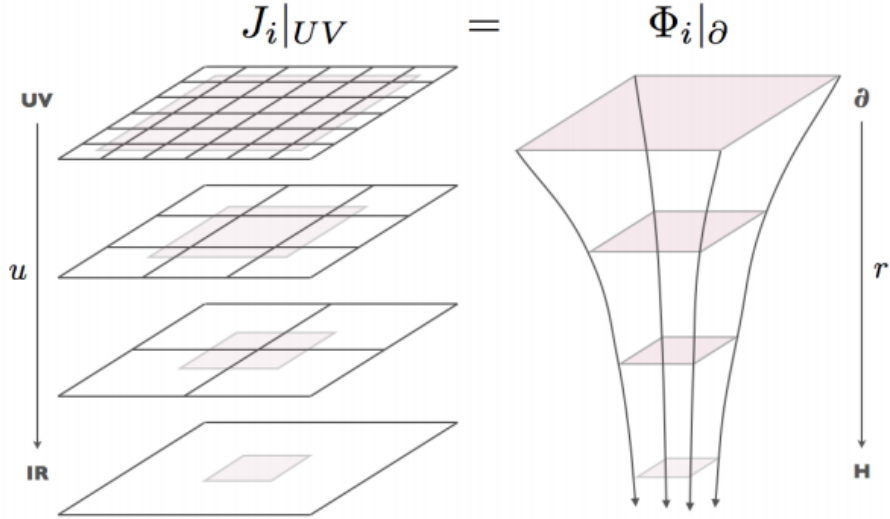}
\caption{The renormalization picture of the AdS/CFT. This illustration was taken from \cite{adams}.}
\label{fig:renargu}
\end{figure}

Now, one can think of a stack of different coarse-grained lattices obeying the energy hierarchy, i.e. from the IR to the UV, as depicted in Fig. \ref{fig:renargu}. Taking the continuum limit, now one has $(d+1)$ dimensional theory, with $u$ being the extra (energy) coordinate. Furthermore, one can associate some bulk field $\Phi$ of this new higher dimensional theory to the coupling of the old theory in the UV as follows
\begin{equation}\label{eq:field/op}
\Phi_i (r=\text{boundary})= J_i(x,u\rightarrow 0).
\end{equation}

Hence, we arrived at some field/operator relation, i.e. the boundary value of the bulk field is the source of the original field theory - this is the key idea behind the holographic dictionary that we will present in Sec.  \ref{Chap3.3}. We can infer that the higher dimensional theory is a gravity theory by wondering what would be the bulk field associated with stress-energy operator $T^{\mu\nu}$; it turns out that the only admissible spin-2 that couples with $T^{\mu\nu}$ is the metric field $g_{\mu\nu}$. Assuming now that we have a CFT as the original field theory, this choice implies the scaling symmetry $u\rightarrow \lambda u$ (besides Poincar\'e invariance), which is accomplished, on the gravity side, by the AdS space \eqref{eq:limAdsCon}.

As we said above one can understand the radial coordinate $u$ as an energy scale. With \eqref{poincare-patch} at hand, we can put this statement in a mathematical form. To do this, we consider a test particle with four-momentum $P^{\mu}=(0,-E,\vec{p})$, and a static observer at the conformal boundary with four-velocity given by $U^{\mu}=(0,1,\vec{0})/\sqrt{-g_{tt}}$. The observer at the boundary measures the following energy ($E_{bdry}$) for the test particle
\begin{equation}
E_{bdry}=-g_{\mu\nu}U^{\mu}P^{\nu}=\frac{E}{\sqrt{-g_{tt}}}=\frac{E}{u},
\end{equation}
with $1/\sqrt{-g_{tt}}$ being the red shift factor. This is nothing else than the UV/IR correspondence \cite{Peet:1998wn,Susskind:1998dq}: the deep bulk of AdS space is associated with the low energy limit of the theory (IR), whilst the near boundary region is identified with the high energy branch (UV).

\section{General properties of AdS space and its black holes}
\label{Chap3.2}

In this dissertation, all the calculations are related to the gravity side of the gauge/gravity duality in the sense that we perform some calculation in the curved spacetime, and then we extract information about gauge theory in flat spacetime. Bearing this in mind, in this section we comment some properties  of AdS spacetime, and other aspects of general relativity as well, which are fundamental for this dissertation. There is plenty of material on this subject ranging from the pure general relativity viewpoint \cite{Wald} to some stringy reviews \cite{KiriBook, NastaseBook, ErdBook}.

The AdS space belongs to a category of manifolds entitled \emph{maximally symmetric}. As the name indicates, a maximally symmetric manifold possess the maximum number of Killing vectors\footnote{Recall that a Killing vector $\xi^\mu$ represents a symmetry of the manifold. They are characterized by the equation
\begin{equation}
\nabla_\mu\xi_\nu+\nabla_\nu\xi_\mu=0.
\end{equation}} allowed. For instance, the Euclidian space $\mathbb{R}^{n}$ is a maximally symmetric manifold. To count the number of Killing vectors of a maximally symmetric manifold, one takes a generic point $P$ and its neighbourhood (assuming local flatness) and considers the following symmetry operations: translations and rotations. For an Euclidian maximally symmetric manifold in $n$ dimensions, we have $n$ independent translations, and  $\frac{1}{2}n(n-1)$ independent rotations; now we sum both to get the number of Killing vectors:
\begin{equation}\label{eq:killing}
n + \frac{1}{2}n(n-1) = \frac{1}{2}(n+2)(n+1).
\end{equation}
If the manifold has Lorentzian signature, as Minkowski spacetime does, some of the rotations are boosts as well.

A maximally symmetric manifold has constant curvature, which can be easily deduced from the following relation (valid for maximally symmetric spaces)
\begin{equation}
R_{\alpha\beta\mu\nu} = \frac{R}{n(n-1)} (g_{\alpha\mu}g_{\beta\nu}- g_{\alpha\nu}g_{\beta\nu} ),
\end{equation}
where $R_{\alpha\beta\mu\nu}$ is the Riemann curvature tensor and $R$ is the curvature scalar. In general relativity, a manifold with constant negative curvature is called Anti-de Sitter (AdS), whereas manifolds with zero and positive constant curvatures, are called Minkowski and de Sitter (dS), respectively. 

There is an elegant way to begin the study of the AdS$_{p+2}$ space.  In this way we first consider the Minkowski manifold $\mathbb{R}^{2,p+1}$, whose metric is given by the line element
\begin{equation}
ds^2 = -dX_{0}^2 - dX_{p+2}^2 + \sum_{i=1}^{p+1} dX_{i}^{2},
\end{equation}
with isometry group SO(2,p+1). Now, to see the AdS$_{p+2}$ space to rise and shine, we embed a hyperboloid into Minkowski  space
\begin{equation}
-X_{0}^2 - X_{p+2}^2 +  \sum_{i=1}^{p+1} X_{i}^{2} = -L^2,
\end{equation}
where the element $L$ is the curvature radius.

We introduce now a parametrization known as \emph{global coordinates}, which is useful to map the causal structure and the topology of the AdS space since it covers all the manifold,
 \begin{align}
&X_0 = L \cosh\rho \cos \tau \notag \\
&X_{p+2} =   L \cosh\rho \sin \tau\notag \\
&X_i = L \sinh\rho \, \Omega_i, \ \ ( \Sigma_{i=1}^{p+1} \Omega_{i}^{2} =1, i= 1, ..., p+1).
 \end{align}
The induced metric of this embedding is given by
 \begin{equation}\label{eq:lineAdS}
 ds^2 = L^2 \left( -\cosh^2\rho d\tau^2 + d\rho^2 + \sinh^2\rho d\Omega_{p}^2  \right)
 \end{equation}
where $d\Omega_{p}^2$ is the line-element of a p-sphere with unity radius. We emphasize also that the domain of the coordinates, $\rho \in [0, \infty)$ and $\tau \in [0,2\pi)$, covers all the AdS space. The topology of the  $AdS_{p+2}$ manifold is $\mathbb{S}^{1}\times \mathbb{R}^{p+1}$, with $\mathbb{S}^{1}$ being related to the temporal coordinate. A peculiar fact of this construction is that the ``time'' coordinate $\tau$ is cyclic and, consequently, this allows for a closed time-like curve - it cause problems with causality. To remedy this, we unwrap  the AdS manifold and ``glue" it with some copy; after this procedure, the domain of the temporal coordinate becomes $t \in (-\infty, \infty) $. 

Another important feature of the vacuum AdS$_{p+2}$ space is that it is a solution of the vacuum Einstein's equations with a negative cosmological constant $(\Lambda<0)$
\begin{equation}\label{eq:eq. einstein. vacuo}
R_{\mu\nu} =  \frac{2\Lambda}{p}g_{\mu\nu}.
\end{equation}
To verify the statement above, we write the metric solution in $p+2$ dimensions of the vacuum,
\begin{equation}
ds^2 = - \left( 1-\frac{2\Lambda r^2}{(p+1)p} \right)dt^2 + \left( 1-\frac{2\Lambda r^2}{(p+1)d} \right)^{-1}dr^2 +r^2 d\Omega^2_{p},
\end{equation}
and recalling that $\Lambda<0$, we have
\begin{equation}\label{eq: rel. lambda L^2}
\frac{2\Lambda}{(p+1)p} = - \frac{1}{L^2} <0,
\end{equation}
which leads us to
\begin{equation}
ds^2 = - \left(1-\frac{r^2}{L^2}\right)dt^2 +\left(1-\frac{r^2}{L^2}\right)^{-1}dr^2 +r^2d\Omega^2_{p}.
\end{equation}
After the following change of variables
\begin{equation}
t = L \tau; \ \ \ r = L\sinh\rho
\end{equation}
we recover the line element of \eqref{eq:lineAdS}.

\begin{figure}[h]
\centering
\includegraphics[width=7cm,height=7cm]{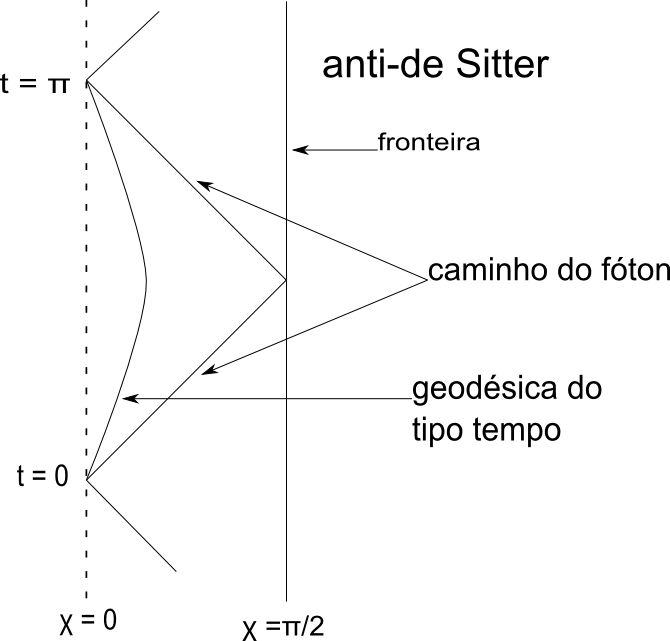}
\caption{Penrose diagram for the $AdS$ space. Notice that a massive particle, represented by the time-like path, can never reach the boundary. On the other hand, a photon can eventually reach the boundary and take its way back in finite time.}
\label{fig:diagramaAdS}
\end{figure}

To gain information about the causal structure of AdS space, one constructs its causal map, i.e. the Penrose diagram (also known as Carter-Penrose, or Conformal diagram) for the space. The main idea of these diagrams is to consider some nice coordinate transformations such that null-geodesics, i.e. the light path, can be drawn as straight diagonal lines, as in Minwkoswki space.  The first step to build this map is to consider the global coordinates \eqref{eq:lineAdS} and perform a change of variable with respect to the radial coordinate \cite{Wald}
\begin{equation}
\cosh\rho = \frac{1}{\cos \chi}, \ \ \chi \in [0,\pi/2[,
\end{equation}
so that
\begin{equation}
ds^2 = \frac{L^2}{\cos^2\chi}(- dt^2 + d\chi^2 +\chi^2d\Omega^2_{p}).
\end{equation}
Note that term within the parenthesis is just the Einstein's static universe. Also, the conformal factor multiplying the static universe does not alter the null geodesics - this explains the epithet \emph{conformal} diagram. In Fig. \ref{fig:diagramaAdS} the Penrose diagram for AdS space.

The domain of the radial coordinate of the Einstein's static universe is $[0,2\pi)$, one says that the AdS manifold is conformally related to \emph{half} of the Einstein's static universe. Consequently, a timelike slice\footnote{To categorize some hypersurface (spacelike, null, or timelike), one just needs to look at its normal vector $n^{\mu}$. For instance, a null hypersurface has a null normal, $n^\mu n_\mu=0$.} of this space has the topology of the inner hemisphere of $\mathbb{S}^{p}$, including the boundary. To draw the geodesics of some particle one needs to solve the geodesic equations, as usual.

As drawn in Fig. \ref{fig:diagramaAdS}, massless particles within the AdS space have the striking capability of reaching the boundary and return to the bulk in finite time. This feature enables the thermal equilibrium between a black hole inside the bulk and the boundary since the former will not effectively evaporate by radiation emission. We discuss more about the thermodynamics in the next subsection.

Another patch of coordinates, $\left\lbrace t, u, \vec{x} \right\rbrace$, often employed in calculations\footnote{Indeed, this patch is, by far, the most employed in holographic calculations.}, is the so-called \emph{Poincar\'e patch}. It is defined as follows
\begin{align}
&X_0 = \frac{u}{2} \left[ 1 +\frac{1}{u^2}\left( L^2+\vec{x}^2-t^2 \right) \right], \notag\\
&X_{p+2} = \frac{Lt}{u}, \notag\\
&X_i = \frac{Lx^{i}}{u}, \ \ i=1,\dots, p \notag \\
&X_{p+1} = \frac{u}{2} \left[ 1 -\frac{1}{u^2}\left( L^2 -\vec{x}^2+t^2 \right) \right]. 
\end{align}
The domain of the radial coordinate $u$ is $(0,\infty)$. By doing this, we cut the AdS hyperboloid in half (the other half is located in $u<0$). To see this more explicitly, we write
\begin{equation}
\frac{X_0 - X_{p+1}}{L^2} = \frac{1}{u},
\end{equation}
confirming the statement that we cut the hyperboloid with the condition that $X_0 > X_{p+1}$. 

In these coordinates, the metric has the following form,
\begin{equation}\label{poincare-patch}
ds^2 = \frac{L^2}{u^2} \left( -dt^2 + du^2 +d\vec{x}^2_p \right)
\end{equation}
with a manifest dilation symmetry $(t,u,\vec{x}) \rightarrow (\lambda t,\lambda u, \lambda\vec{x})$. Another self-evident characteristic is that slices of the manifold to $u=cte$ are conformally related with the Minkowski space.

For the patch \eqref{poincare-patch}, one sees that the time-like Killing vector $\partial_t$ goes to zero as $u\rightarrow \infty$, which does not occur in global coordinates. Performing some analogy with the usual Schwarzschild metric, whose time-like Killing vector vanishes near its horizon, one can say that $u=\infty$ is a Killing horizon. Naturally, the \textit{locus} of the conformal boundary is $u=0$.

\begin{figure}[h]
\centering
\includegraphics[width=7.3cm,height=7.3cm]{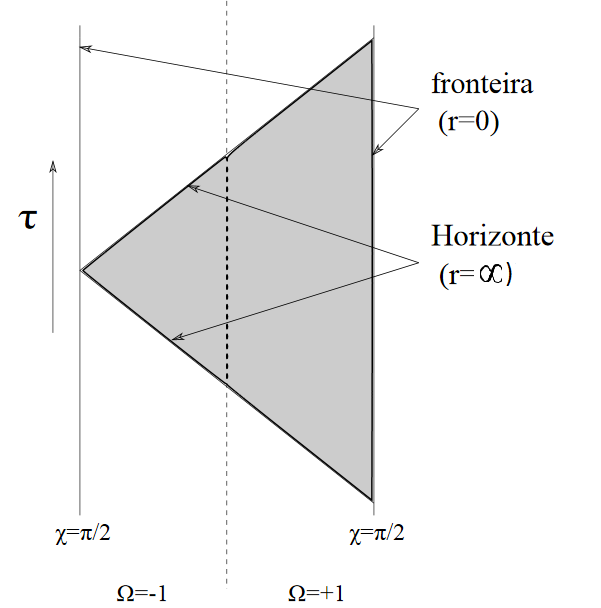}
\caption{The conformal diagram for the AdS$_{p+2}$ space using the Poincar\'e patch \eqref{poincare-patch}. In this patch, we can only access the dark region.}
\label{fig:poincpatch}
\end{figure}

\subsection{Thermodynamics}

We discussed in some detail how massless particles are ``locked'' inside the AdS space. Indeed, topologically, the AdS space is like a closed box. Thanks to this, if we insert a black hole in  AdS space, the former will be in thermal equilibrium with the boundary of AdS and will not evaporate by thermal radiation emission (Hawking radiation); black branes have positive specific heat and are thermodynamically stable. Black branes are black holes with extended translational symmetry, though we use both terms interchangeably here. Therefore, to include temperature on the field theory, we add a black hole in the bulk. 

The usual AdS$_{5}$-Schwarzschild metric of the black brane is given by (the $S^5$ piece is discarded from now on)
\begin{equation}
ds^2 = \frac{r^2}{L^2}\left[ -(1-r^4/r_{h}^{4})dt^2+d\vec{x}^2 \right]+\frac{L^2dr^2}{r^2(1-r^4/r^{4}_{h})},
\end{equation}
where we adopted $r$ as being the extra radial coordinate with $r\rightarrow\infty$ being the locus of the conformal boundary.

The task now is to obtain the temperature of a generic five dimensional black hole, whose value will be the same as in the dual field theory.  To obtain the explicit formula for the temperature of a black hole, we follow the standard procedure in thermal field theories (briefly mentioned in the beginning of Sec. \ref{Chap1.2}) and Wick rotate the time coordinate to its Euclidean version, i.e. $t\rightarrow i \tau$. Let us assume that the Euclidean metric of the black brane has the form
\begin{equation}
ds^2 = g(r)[f(r)d\tau^2+d\vec{x}^2]+\frac{dr^2}{h(r)},
\end{equation}
with $f(r_h)=h(r_h)=0$. From a general relativity point of view, we still have to compactify the $\tau$ coordinate, $\tau = \tau + \beta$, to avoid a conical singularity. The temperature of the black hole will be proportional to the length circle, just as in field theory \cite{Das}.

To make the above statement more precise and explicit, let us take $r\approx r_h$. Then, we have
\begin{equation}
f(r)\approx f'(r_h)(r-r_h),\ \ \ h(r)\approx h'(r_h)(r-rh),
\end{equation}
and we define a new coordinate $\rho$ (the radius of the circle) such that
\begin{equation}
\frac{1}{r-r_h}\frac{dr^2}{h'(r_h)}=d\rho^2 \Rightarrow \rho = 2\sqrt{\frac{r-r_r}{h'(r_h)}}.
\end{equation}
Next, we define the the coordinate $\theta$, the angle of the circle, in the following way
\begin{equation}
g(r_h)f'(r_h)(r-r_h)d\tau^2 =\rho^2 d\theta^2 \Rightarrow \theta = \frac{1}{2}\sqrt{g(r_h)f'(r_h)h'(r_h)}\theta.
\end{equation}
With these definitions, the near horizon metric becomes
\begin{equation}
ds^2\approx \left( d\rho^2+\rho^2d\theta^2\right)+g(r_h)d\vec{x}^2.
\end{equation}
Notice that the term inside the brackets corresponds to flat space in polar coordinates. In order for $ds^2$ to be well defined at the horizon ($\rho=0$), we need to compactify the angle $\theta$ to avoid the conical singularity (angle defect); the periodicity of $\theta$ is $2\pi$. This procedure also implies the periodicity $\tau=\tau+\frac{1}{T}$, where $T$ is the Hawking temperature. Therefore, the explicit formula for the temperature (for the bulk and boundary) is
\begin{equation}
T = \frac{\sqrt{g(r_h)f'(r_h)h'(r_h)}}{4\pi},
\end{equation}
or, more generally,
\begin{equation}
T = \frac{\sqrt{|g_{tt}'(r_h)g'^{rr}(r_h)|}}{4\pi},
\end{equation}
for a generic line element of the form
\begin{equation}
ds^2=g_{tt}(r)dt^2+g_{rr}(r)dr^2+\dots .
\end{equation}

The entropy density $s=S/V_3$, where $V_3=\int dxdydz$, is another important thermodynamic quantity but we will not derive it here.\footnote{This can be done in, at least, two different ways. The first one is using the path integral formalism, in which we plug the Gibbons-Hawking-York (GHY) boundary term in the on-shell action to obtain the partition function, and then extract the entropy \cite{Bek, Hawk} using thermodynamic identities. The second way to derive it, is using the Wald entropy formula \cite{Wald:1993nt}
\begin{equation}
S = -2\pi \int_{H}\sqrt{h}\frac{\delta \mathcal{L}}{\delta R_{\alpha\beta\mu\nu}}d^{D-1}x,
\end{equation}
where $H$ is the slice of the horizon and $h$ the induced metric on it.}. We just give the final result, which is the famous Bekenstein-Hawking formula \cite{Bek, Hawk}
\begin{equation}
s = \frac{A_h}{4G_5},
\end{equation}
where $A_h$ is the are of the black hole, and $G_5$ is the Newton's constant for a five dimensional space.

Therefore, given the geometry of the bulk space (the hardest part most of the times), it is quite easy to extract the temperature and the entropy density.

\subsection{The scalar field in AdS: The BF bound and scalings dimensions}
\label{Chap3.2.2}

Now that we are more familiar with AdS space, let us take a free massive scalar field in this space. From this apparently simple exercise, we shall learn some valuable lessons about the bulk fields according to the renormalization group perspective. Consider then the AdS$_{d+1}$ space with no temperature and zero density, whose metric is
\begin{equation}
ds^2 =\frac{L^2}{u^2}(-dt^2+du^2+d\vec{x}^2_{d-1}).
\end{equation}
The scalar field Lagrangian is
\begin{equation}
S=\int d^{d+1}x \sqrt{-g} \left[-\frac{1}{2}(\partial\Phi)^2-\frac{m^2\Phi^2}{2} \right],
\end{equation}
where $m$ is the mass of the scalar field. For sake of simplicity, we assume no backreaction of the scalar field in the background. Taking the plane-wave Ansatz $\Phi(t,u,x) = \phi(u)e^{i k\dot x}$, the equation of motion of the scalar field becomes
\begin{align}
&  \frac{1}{\sqrt{-g}}\partial_{\mu}(\sqrt{-g}\partial^\mu\Phi)-m\Phi=0, \notag \\
& \Rightarrow u^2\phi''(u)+r(d-1)\phi'(u)-(k^2r^2-m^2L^2)\phi(u)=0,
\end{align}
whose solution is given in terms of the Bessel's modified functions $K_\nu$ and $I_{\nu}$,
\begin{align}
&\phi(u)  = \phi_{reg}r^{d/2}K_{\Delta-\frac{d}{2}}(\sqrt{k^2} u)+\phi_{irreg}r^{d/2}I_{\Delta-\frac{d}{2}}(\sqrt{k^2} u), \\ \text{with}  \ \  & \Delta(\Delta-d)=m^2L^2. 
\end{align}
The constants $\phi_{reg}$ and $\phi_{irreg}$  are related to the regularity of the correspondent solution at the boundary ($u=0$); remember that $K_{\nu}(0)=0$ and $I_{\nu}(0)\rightarrow\infty$. Thus, if we impose regularity at the boundary we should discard the irregular solution ($\phi_{irreg}=0$).  Demanding the reality condition for the \emph{scaling dimension} $\Delta$, i.e. $\Delta \in \mathbb{R}$, we find the Breitenlohner-Freedman (BF) bound \cite{Breitenlohner:1982bm}
\begin{equation}
m^2 \geq -\frac{d^2}{4L^2},
\end{equation}
or, for the five dimensional AdS space,
\begin{equation}\label{eq:BFbound5}
m^2 \geq -\frac{4}{L^2}.
\end{equation}
The BF bound means that a (small) negative mass does not induce an instability. Naturally, the bottom-up model developed in Chapter \ref{Chap8.0} respects this bound.

It is instructive to further examine the near boundary behavior of the scalar field. As $\phi(u)$ approaches the boundary, one can write
\begin{equation}
\phi(u\rightarrow 0)\sim \phi_{d-\Delta}(k)u^{d-\Delta}+\phi_{\Delta}(k)u^{\Delta}+\dots 
\end{equation}
so that
\begin{equation}\label{eq:scalingscalar}
\phi (u=0,x)= \lim_{u\rightarrow 0} u^{\Delta-d}\phi(u,x).
\end{equation}

Although we arrived at Eq. \eqref{eq:scalingscalar} for the scalar field, this is a very general way to express the behavior of some field in the bulk. Moreover, with the scaling dimension $\Delta$ (the relation \eqref{eq:scalingscalar} is valid only for scalar fields) we can study how the introduction of a new operator breaks the conformality of the original theory. Suppose that we introduce some Lorentz scalar operator $\mathcal{O}$ in the old CFT theory,
\begin{equation}\label{eq:DefCFT}
S_{CFT} \rightarrow S_{CFT}+ \int_x \mathcal{O}(x).
\end{equation}
Then, one may categorize this deformation as
\begin{itemize}
\item Relevant operator, $d-\Delta>0$: This deformation is weak in the UV ($\phi(u\rightarrow 0)\rightarrow 0$) and strong in the IR .

\item Irrelevant operator, $d-\Delta<0$: This deformation is strong in the UV ($\phi(u\rightarrow 0)\rightarrow \infty$) and weak in the IR.

\item Marginal operator, $d-\Delta=0$: Does not break conformal invariance at the leading order of the deformation. However, the operator can be marginally relevant or marginally irrelevant after the inclusion of string corrections.
\end{itemize}

\section{The holographic dictionary: Extracting Green's functions}
\label{Chap3.3}

After our brief detour through the basics aspects of holography and the AdS spacetime, it is time to stablish a precise duality, from which we can calculate the Green's functions of the strongly coupled field theory, erstwhile impossible via standard methods. 

From the conjecture of the AdS/CFT correspondence \eqref{eq:Con-dual-Mal}, given the generating function $Z[J]$ of the $\mathcal{N}=4$ SYM in four dimensions supplied by some source $J(x)$ and remembering the relation between the source and the bulk field ($J=\Phi_0$) in Eq. \eqref{eq:field/op}, one can write the Euclidean partition function as follows
\begin{equation}
Z[J] = \int \mathcal{D}[\dots]\text{exp}\left(S+ \int_{x}\Phi_0 \mathcal{O}\right) = \left\langle \text{exp}\left( \int_{x}\Phi_0 \mathcal{O} \right) \right\rangle,
\end{equation}
where the ellipsis denotes the field content of the field theory.

Following \cite{Witten:1998qj,Gubser:1998bc}, one states the \emph{field/operator correspondence}, where 
\begin{equation}
\left\langle \text{exp}\left( \int_{x}\Phi_0 \mathcal{O} \right) \right\rangle_{CFT} = Z_{String}[\Phi],
\end{equation}
where $Z_{String}$ is the partition function of the IIB superstring theory. As aforementioned, we are rather clueless of what $Z_{string}$ is for quantum gravity, which is needed to check the AdS/CFT correspondence in some arbitrary coupling strength. Fortunately, for strongly coupled $\mathcal{N}=4$ SYM, we can approximate the partition function of the string theory by the (super)gravity action
\begin{equation}
Z_{String}[\Phi]\approx \text{exp}\left( -S_{clas}[\Phi(u=0,x)=\Phi_0]\right).
\end{equation}

In the Euclidean signature, one can extract the $n-$point function of the field theory by taking functional derivatives as follows
\begin{equation}\label{eq:EuclPres}
\langle \mathcal{O}_1(x_1)\dots\mathcal{O}_n(x_n)\rangle = \left.\frac{\delta^n Z_{QFT}[J]}{\delta J_1(x_1)\dots \delta J_n(x_n)}\right|_{J=0}.
\end{equation}
Therefore, one can calculate the Euclidean correlators of the field theory by taking functional derivatives of the boundary value of the respective bulk field in the classical (on-shell) gravitation action! However, this procedure is valid \emph{only} for the Euclidian signature. If we want to compute real-time correlators (e.g.: needed for the shear viscosity) using the relation \eqref{eq:Con-dual-Mal} we find incongruences on the correlators. In the next subsection we shall outline how one can circumvent this and extract the 2-point function (Green's function) in real-time; we shall employ this recipe to calculate the shear and bulk viscosities late in this chapter.

If by chance we need to obtain higher point functions in real-time, such as 3-point function, the discussion becomes more complicated. Instead, we have to solve the bulk-to-bulk, bulk-to-boundary, and boundary-to-boundary propagators of the AdS side of the theory   \cite{Skenderis:2008dg}.

\subsection{The Euclidian one-point function the holographic renormalization}
\label{Chap3.3.1}

Before we tackle the problem of the retarded two-point function, from which most of the quantities come from, it is important and instructive to discuss some aspects of the one-point function. We shall specialize our discussion to the expectation value of stress-energy tensor $\langle T^{\mu\nu}\rangle$,and we shall comment on the divergences of the on-shell action and how one can deal with them. 

We begin by defining the Brown-York (BY) \cite{ghy1} stress-energy tensor, defined at the boundary of the manifold via
\begin{equation}\label{eq:BYtensor}
T^{\mu\nu}_{BY} =-\frac{2}{\sqrt{-\gamma}}\frac{\delta S}{\delta\gamma^{\mu\nu}}, 
\end{equation}
where $\gamma_{\mu\nu}$ is the induced metric at the boundary. The BY tensor is called \emph{quasilocal} since it is defined at the boundary of a given manifold; the definition of a local energy density in general relativity is flawed. Also, the BY tensor is often divergent. However, this is very suitable for the AdS case: regarding the boundary of the AdS space as being the CFT, we can assign a stress-energy tensor for the latter $\langle T^{\mu\nu} \rangle$ by using the BY tensor at the boundary; the (gravitational) divergences correspond to the original ultraviolet divergences of the field theory.

Just as in field theory, one can cancel the divergences of the AdS space by introducing the so-called \emph{counter term action} $S_{ct}$, which is defined at the boundary. Therefore, the complete gravitational action, i.e. the renormalized action, for the AdS space becomes
\begin{equation}
S_{ren} = S_{EH}+S_{GHY}+S_{ct},
\end{equation}
where $S_{EH}$ is the Hilbert-Einstein action (plus cosmological constant), and $S_{GHY}$ is the Gibbons-Hawking-York (GHY) boundary term needed for a well posed variational problem with boundary \cite{ghy1, ghy2}. For a (d+1)-dimensional AdS space, the boundary terms are
\begin{equation}\label{eq:GHYterm}
S_{GHY}= \frac{1}{8\pi}\int_{\partial M} d^dx \sqrt{-\gamma}K,
\end{equation}
\begin{equation}
S_{ct}=-\int _{\partial M} d^dx  \sqrt{-\gamma}\frac{3}{L}\left(1-\frac{L^2}{12}R(\gamma) \right),
\end{equation}
where $\gamma$ is the induced metric on the boundary and $K$ is its extrinsic curvature.

With the renormalized action at hand, the Brown-York tensor \eqref{eq:BYtensor} obtained is
\begin{equation}
T^{\mu\nu}=\frac{1}{8\pi G_{5}} \left[K^{\mu\nu}-K\gamma^{\mu\nu}-\frac{3}{L}\gamma^{\mu\nu}-\frac{L}{2}\left(R^{\mu\nu}-\frac{\gamma^{\mu\nu}}{2}R \right) \right].
\end{equation}

The trace of $T^{\mu\nu}$ is given by
\begin{equation}
T^{\mu}_{\mu} = -\frac{L^3}{8\pi G_{5}}\left( -\frac{1}{8}R^{\mu\nu}R_{\mu\nu}-\frac{1}{24}R^2 \right),
\end{equation}
which is zero for the case of AdS space in remarkably agreement with the CFT (see Eq. (23) of Ref. \cite{ren1} and subsequent discussion). 

The discussion done here is also connected with the bulk viscosity since it vanishes when the trace of the stress-energy momentum tensor vanishes (see the discussion in Appendix \ref{appA}).

\subsection{The retarded two-point function}
\label{Chap3.1.2}

In this subsection we shall show how to compute retarded two-point functions. Originally, this recipe was proposed in \cite{Son:2002sd} and was put on firmer grounds correct later in \cite{Herzogcorre}. Moreover, this procedure is generalized in Chapter \ref{Chap6.0} where we have a \emph{mixing} of operators, i.e. we have to work with matrices. For this reason, this section is not a ``step-by-step'' calculation since this is done in Chapter \ref{Chap6.0}.

Let $\phi(u,x)$ be the dual bulk field of the operator $\mathcal{O}(x)$. The field $\phi$ will obey some equation of motion, from the Einstein's equations for instance. Performing the following Fourier transformation
\begin{equation}
\phi(u,x) = \int\frac{d^4k}{(2\pi)^4}e^{ik\cdot x}\phi(u,k),
\end{equation}
we can decompose the solution as
\begin{equation}\label{eq:phifouri}
\phi(u,x)=f_k(u) \phi_0(k),
\end{equation}
with the Dirichlet boundary condition
\begin{equation}
\lim_{u\rightarrow 0}u^{4-\Delta}f_k(u)=1.
\end{equation}

Now we substitute the on-shell field \eqref{eq:phifouri} into the action. We then recast the on-shell action in the following way
\begin{equation}
S_{on-shell} = \left. \int\frac{d^4k}{(2\pi)^4} \phi_0(k)\mathcal{F}(k,u)\phi_0(k)\right|_{u=0}^{u=u_h},
\end{equation}
where $\mathcal{F}(k,u)$ is some flux whose imaginary part is conserved, i.e. $\partial_u (\mathcal{F}-\mathcal{F}^*)=0$. 

If we try the naive Euclidean prescription \eqref{eq:EuclPres}, we find
\begin{equation}\label{eq:EucliPresRT}
G(k)\stackrel{?}{=}-\mathcal{F}(k,u)\vert_{u=0}^{u=u_h}-\mathcal{F}(k,u)\vert_{u=u_h}^{u=0}.
\end{equation}

However, as mentioned in the previous section, this result does not hold. The fact that the result \eqref{eq:EucliPresRT} does not possess an imaginary part already rules it out. To circumvent this problem we take just the boundary contribution, in the following way \cite{Son:2002sd,Herzogcorre}
\begin{equation}\label{eq:StarinetsRecip}
G^R(k)=-2\mathcal{F}(k,u)\vert_{u=0},
\end{equation}
The above expression gives us the correct retarded two-point function.

Now, we pass to some applications of the formalism developed in this chapter. We shall focus on the shear and bulk viscosities.

\section{The holographic shear viscosity}
\label{Chap3.4}

After discussing the viscosities in Chapter \ref{Chap 2} and the AdS/CFT correspondence, it is time to link these two concepts in order to extract the shear viscosity from the AdS/CFT correspondence. Loosely speaking, it is common to use the word ``holographic'' to designate quantities calculated via the duality; e.g. the holographic shear viscosity. Thus, this subsection is devoted to calculate the shear viscosity of strongly coupled non-Abelian theories in the large $N_c$ limit.

For sake of completeness, and for pedagogical reasons, in this dissertation we shall calculate the usual holographic shear viscosity using three distinct methods: 
\begin{itemize}
\item Associating the absorption cross section with the imaginary part of the retarded Green's function;

\item Using the standard recipe described in Sec. \ref{Chap3.1.2}. This method is also exploited in Chapter \ref{Chap6.0} in order to calculate the bulk viscosity (from the magnetic brane) and in Sec. \ref{SecAnisoShear} where we calculated the anisotropic shear viscosity for the bottom-up magnetic model;

\item Adopting the so-called \emph{membrane paradigm} approach \cite{Iqbal:2008by,Thorne:1986iy}. The discussion of this method is postponed to Chapter \ref{Chap5.0}, where we calculate the anisotropic shear viscosity using the magnetic brane solution.
\end{itemize}

Before we go straight to tackle the first item, we need to establish the field/operator relation for the shear viscosity. To guess what is the bulk field associated with the shear viscosity, let us rewrite the Kubo formula for the shear viscosity derived in Sec. \ref{Chap2.3.1}, as follows
\begin{equation}\label{eq:KuboShearAg}
\eta=-\lim_{\omega\rightarrow 0}\frac{1}{\omega}\text{Im}\,G^{R}_{T_{xy},T_{xy}}(\omega,\vec{0}).
\end{equation}
From the above formula, we learn that the important operator is the stress-energy tensor, i.e. $\mathcal{O}=T^{\mu\nu}$. The stress-energy tensor operator is sourced by the metric field and, at the linearised level, we have the following interaction term
\begin{equation}
S_{int} = \frac{1}{2}\int d^4x h_{\mu\nu}T^{\mu\nu} \supset h_{xy}T^{xy},
\end{equation}
where in the last step we emphasized the important part to calculate \eqref{eq:KuboShearAg}. Therefore, in order to calculate the shear viscosity one needs to perform small fluctuations of the metric, $g_{\mu\nu}\rightarrow g_{\mu\nu}+h_{\mu\nu}$.

\subsubsection{From the absorption cross section}

In the early ages of the AdS/CFT correspondence it was a common exercise to compare the absorption rates of gravitons (closed strings) on the D-brane world volume (AdS side) with the perturbative side \cite{Callan:1996dv, Das:1996wn}. It turns out that in the low energy limit  both views of the absorption rate agree \cite{Klebanov:1997kc}. Motived by these facts, let us consider some graviton wave (metric disturbance) of frequency $\omega$, which propagates along a perpendicular direction to the black brane ($h\sim h(r)e^{-i\omega t}$). From the field theory perspective, the graviton absorption cross section is related to the component $T_{xy}$ of the stress-energy tensor in the following way\footnote{To connect the absorption cross section with the imaginary part of the Green's function one first realize that, according to Fermi's golden rule, the net absorption rate of the graviton is
\begin{equation}
\Gamma = V_3 \sum_{i,f}\frac{e^{-\beta E_i}}{Z} |\langle f| T_{x,y}(0) |i \rangle|^2 (2\pi)^4\delta^{(3)}(\vec{p}_f-\vec{p}_i)\left[ \delta(E_f-E_i-\omega)-\delta(E_f-E_i+\omega) \right].
\end{equation}
Thus, if one compares $\Gamma$ with the spectral decomposition of the Green's function, i.e. $\Gamma=-2V_3 \text{Im}\,G^{R}(\omega)$, we arrive at the formula \eqref{eq:etaOldAbs}.}
\begin{equation} \label{eq: absor-dual}
\sigma_{abs}(\omega) = - \frac{2\kappa_5^2}{\omega} \text{Im}\,G^R(\omega) = \frac{\kappa_5^2}{\omega} \int dt d\vec{x} e^{i\omega t} \left\langle \left[ T_{xy}(t,\vec{x}), T_{xy}(0,\vec{0}) \right] \right\rangle
\end{equation}
with $\kappa_5 = \sqrt{8\pi G_5}$. Comparing \eqref{eq: absor-dual} with \eqref{eq:KuboShearAg}, we have
\begin{equation}\label{eq:etaOldAbs}
\eta = \frac{\sigma_{abs}(0)}{2\kappa^2} = \frac{\sigma_{abs}(0)}{16\pi G}.
\end{equation}

The calculation of $\sigma_{abs}(0)$ was carried out in Ref. \cite{Policastro:2001yc}. The result is 
\begin{equation}
\eta = \frac{\pi}{8}N_c^2 T^3.
\end{equation}

In this subsection, however, we shall follow Ref. \cite{Kovtun:2004de} in which the general result for the ratio $\eta/s=\frac{1}{4\pi}$ was obtained using some few assumptions. The first assumption is about symmetry of the black brane, whose metric we assume to be
\begin{equation}\label{eq:BackMetrShear}
ds^2 = f(\xi)(dx^2+dy^2)+\dots,
\end{equation}
where $\xi$ represents the dependence of some variable except $x$ or $y$, i.e. we have a $SO(2)$ symmetry, valid for isotropic theories.

With the background metric \eqref{eq:BackMetrShear} at hand, we perform some disturbance on the metric (graviton scattering)
\begin{equation}
g_{\mu\nu} = g^{(0)}_{\mu\nu} + h_{\mu\nu},
\end{equation}
where $g^{(0)}_{\mu\nu}$ denotes the background metric and $h_{\mu\nu}$ is the disturbance. To calculate the shear viscosity, it is enough to work only with $h_{xy}$ different than zero, since this mode decouples from the others. This can be viewed as being a gravitational wave with $\times$ (times) polarization. Moreover, let us assume that $h_{xy}$ does not depend of $x$ or $y$, i.e. $h_{xy}=h_{xy}(\xi)$.

In a $(d+1)$ dimensional manifold, we can write the Einstein's equations as follows\footnote{To get rid of the scalar curvature $R$ in Einstein's equations we just take their trace and write $R$ as function of $T^{\lambda}_{\lambda}$.}
\begin{equation} \label{eq:einst. invert.}
R_{\mu\nu} = \kappa^2_5 \left(T_{\mu\nu} - \frac{T^{\lambda}_{\lambda}}{d-1}g_{\mu\nu}  \right),
\end{equation}
where $T^{\mu\nu}$ is the stress-energy tensor supplied by some matter field (e.g. dilaton field). To calculate $\sigma_{abs}$, one still has to massage the equations obtained from \eqref{eq:einst. invert.} in order to arrive in a nice expression for the equations of motion for $h_{xy}$. Assuming that
\begin{equation}
T_{\mu\nu} = - g_{\mu\nu}\mathcal{L} + \dots,
\end{equation}
where the ellipsis denotes higher order corrections, we obtain the following expressions for the components $R_{xx}$ and $R_{xy}$ of \eqref{eq:einst. invert.}, respectively
\begin{equation}\label{eq: visc. lin. 1}
\frac{1}{2} \left[ \square f + \frac{\partial_\mu f\partial^\mu f}{f} \right]   = \kappa_5^2 \left( \mathcal{L} + \frac{T^{(0)\lambda}_{\lambda}}{d-1} \right),
\end{equation}
\begin{equation}\label{eq: visc. lin. 2}
- \square h_{xy}+ \frac{2\partial^\mu f\partial_\mu h_{xy}}{f} - \frac{\partial_\mu f\partial^\mu f}{f^2} h_{xy} = -2 \kappa^2 \left( \mathcal{L} + \frac{T^{(0)\lambda}_{\lambda}}{d-1} \right) h_{xy}.
\end{equation}

The fundamental observation of Ref. \cite{Kovtun:2004de} is that $h_x^y=h_{xy}/f$ obeys the same equation of a massless scalar field,
\begin{equation}
\square h_{x}^{y} = 0.
\end{equation}
Hence, in order to obtain the holographic shear viscosity, we need to calculate the absorption cross section of a massless scalar field. The detailed calculation of this cross section is done in the Appendix \ref{appC}. One can show that $\sigma_{abs}(0)$ depends solely on the area of the black hole horizon \cite{DasAbs}, 
\begin{equation}
\sigma_{abs}(\omega=0)=A_h.
\end{equation}

Therefore, if we divide \eqref{eq:etaOldAbs} by the entropy density $s=A_h/4G_5$, we are lead to the celebrated ratio \cite{Kovtun:2004de} (recovering now $\hbar$ and $k_B$)
 \begin{equation}\label{eq: taxa}
  \frac{\eta}{s} = \frac{\hbar}{4\pi k_B}.  
 \end{equation}
 
As we already emphasized throughout this dissertation, the ratio \eqref{eq: taxa} had tremendous consequences in establishing the usefulness of the gauge/gravity duality in the study of strongly coupled systems, such as the QGP near the crossover region. It is also a robust result since it is valid for every isotropic field theory described with a dual gravity theory with at most two derivatives in the action. 

\subsubsection{From the conserved flux}

Now, let us apply the recipe developed in Sec. \ref{Chap3.1.2} to calculate the shear viscosity for an isotropic theory. Once one knows that the metric fluctuation $h_{x}^y$ obeys the same equation of motion as a massless scalar fild field does, one can write the on-shell action as follows
\begin{equation}\label{eq:hxyAct}
	S = \frac{1}{16\pi G_5}\int d^5x\sqrt{-g}\left( -\frac{1}{2}(\partial h_x^y)^2\right)
\end{equation}
where $g$ is the determinant of the background metric. Actually, there is an overall factor that we must fix by perturbing the whole action; we postpone this calculation to Sec. \ref{MetricFluAniso}, where we perform the action perturbation for the magnetic brane case. In this brief calculation, let us adopt the following black 3-brane metric
\begin{equation}
ds^2 = e^{A(r)}(-h(r)dt^2+d\vec{x}^2)+e^{2B(r)}\frac{dr^2}{h(r)},
\end{equation}
with $h(r)$ being the blackening factor.  

The next step is to Fourier transform the fluctuation\footnote{We take the Fourier transform only with respect the time coordinate because we take $\vec{k}=\vec{0}$ in the Green's function.}
\begin{equation}\label{eq:Fourhxy}
h_x^{y}(t,r) = \int \frac{d\omega}{(2\pi)}e^{-i\omega t}\Phi(\omega,r),
\end{equation}
with
\begin{equation}
\Phi(\omega,u) = \phi_\omega (r)\phi_0, \ \ \ \text{and} \ \ \lim_{r\rightarrow \infty} \phi_\omega (r) =1.
\end{equation}

Making the Lagrangian explicitly complex, i.e. $(\partial h_x^y)^2\rightarrow (\partial_\mu h_{x}^{y\,\dagger})\partial^\mu h_x^y$, and plugging \eqref{eq:Fourhxy} into \eqref{eq:hxyAct}, one obtains
\begin{equation}
S = V_3\int \frac{d\omega}{2\pi}\phi_0 \mathcal{F}(\omega,r)\phi_0,
\end{equation}
where
\begin{equation}\label{eq:fluxhxy}
\mathcal{F}(\omega,r)=he^{4A-B}\phi_\omega^*\partial_r\phi_\omega.
\end{equation}

Thus, from Eqs. \eqref{eq:StarinetsRecip} and \eqref{eq:KuboShearAg}, we have the relation
\begin{equation}\label{eq:ShearFluxRec}
\eta=-\frac{1}{16\pi G_5}\lim_{\omega\rightarrow 0}\frac{1}{\omega}\text{Im}\,\mathcal{F}.
\end{equation}

Since $\partial_r\text{Im}\,\mathcal{F}=0$, we can calculate it in the most convenient region, which turns out to be the near horizon limit. Near the horizon, we have that
\begin{equation}\label{eq:phihxynear}
\phi_\omega(r\rightarrow r_h)=c_{-}(r-r_h)^{-\frac{i\omega}{4\pi T}}+c_{+}(r-r_h)^{+\frac{i\omega}{4\pi T}},
\end{equation}
where $c_{-}$ and $c_{+}$ are two integration constants that can be determined via a matching procedure with the boundary; this is done carefully in Sec. \ref{SecAnisoShear} and here we only state the result $c_{-}=1$. We shall take $c_{+}=0$ because this solution is related to the \emph{advanced} Green's function, whilst $c_{-}$ is related to the retaded Green's function.  

Substituting \eqref{eq:phihxynear} into the flux \eqref{eq:fluxhxy}, and the latter in \eqref{eq:ShearFluxRec}, we obtain the following expression for the shear viscosity
\begin{equation}
\eta = \frac{1}{16\pi G_5}e^{3A(r_h)},
\end{equation}
and, dividing it by the entropy density $s =e^{3A(r_h)}/4G_5$, we are lead to the result
\begin{equation}
\frac{\eta}{s} =\frac{1}{4\pi}.
\end{equation}

Lastly, it is time to notice a very important point. To calculate the shear viscosity, the on-shell action, and so on, we tacitly assumed that there were no divergences. This may seem odd since in Sec. \ref{Chap3.3.1} we spoke about these divergences. The answer for this apparent puzzle is that the imaginary part of the Green's function is free from divergences. To see this more clearly, from the gravity side, remember that the counter-terms are defined as boundary terms and if we add the boundary term $\partial_r (\phi_\omega\phi_\omega)$ to the action, the effect of this new term on $\mathcal{F}$ is
\begin{equation}
\mathcal{F}=\alpha |\phi_\omega|^2+\dots.
\end{equation}
Hence, it has no effect on the imaginary part of the Green's function.

\subsubsection{Corrections to $\frac{\eta}{s}=\frac{1}{4\pi}$}

Although the result for shear viscosity to entropy density ratio obtained via the AdS/CFT duality is very robust, there are some situations in which $\eta/s\neq 1/4\pi$. Here, we shall list all known the situations where are deviations from the original result.

The first realization of $\eta/s\neq 1/4\pi$ came from supergravity corrections, which are the so-called $\alpha'$-corrections. Using the first $\alpha'$-correction, the holographic shear viscosity becomes \cite{Kats:2007mq}
\begin{equation}
\frac{\eta}{s} = \frac{1}{4\pi}\left( 1+\frac{135\zeta(3)}{8(2\lambda)^{3/2}} \right),
\end{equation}
where $\lambda$ is the 't Hooft coupling. Notice that this correction \emph{increase} the value of the shear viscosity, giving some support for the conjecture that the value $\eta/s= 1/4\pi$ is a minimum. Moreover, this is in agreement with the QCD cf. Fig. \ref{fig:etacompar}.

The first supergravity correction in type IIB goes like $R^4$ in the gravitational action. Therefore, we can imagine somehow a quadratic correction, going like $R^2$, though without a stringy guide. This correction is accomplished by introducing the so-called Gauss-Bonnet term $\mathcal{L}_{GB}=\lambda_{GB}(R^2-4R^{\mu\nu}R_{\mu\nu}+R^{\alpha\beta\mu\nu}R_{\alpha\beta\mu\nu})$. The Gauss-Bonnet term modifies the shear viscosity in the following manner \cite{Brigante:2007nu,Brigante:2008gz}
\begin{equation}\label{eq:EtaSGB}
\frac{\eta}{s}= \frac{1-4\lambda_{GB}}{4\pi},
\end{equation}
with the condition $-\frac{7}{36}<\lambda_{GB}<0.09$ needed to preserve causality \cite{Brigante:2008gz}.

Another way to modify the result \eqref{eq: taxa} is to consider \emph{anisotropic} theories, such as the one induced by the magnetic brane solution \cite{DK-applications2}, focus of this dissertation. In anisotropic theories though, we have more than one shear (and bulk) viscosity coefficient - the detailed discussion about this fact in Sec. \ref{Chap4.3}.

The first calculation of anisotropic shear viscosities was done in Ref. \cite{Rebhan:2011vd} for the case of an anisotropic plasma created by a spatial dependent axion profile, which was proposed originally in Ref. \cite{Mateos:2011ix} - see also Refs. \cite{Jahnke:2014vwa, Cheng:2014qia} for extensions of this axion+dilaton model; the result resembles some qualitative features of the viscosities obtained from the magnetic brane background as we shall see ahead. By the same token, one can find model with a dilaton driven anisotropy \cite{Jain:2014vka,Jain:2015txa}, an anisotropic $SU(2)$ model used for superfluids \cite{Natsuume:2010ky,Erdmenger:2010xm, Erdmenger:2012zu}, and a black brane whose temperature is modulated by the spatial directions $\vec{x}$ \cite{Moskalets:2014hoa, Ovdat:2014ipa}. Lastly, recent violations of the viscosity result of isotropic theories were found in the context of massive gravity \cite{Alberte:2016xja,Hartnoll:2016tri}.

\section{The holographic bulk viscosity}
\label{Chap3.5}

The bulk viscosity is the remaining transport coefficient to characterize energy dissipation due to internal friction in a strongly coupled non-Abelian plasma. In the same spirit of this introductory chapter, we begin discussing the standard $\mathcal{N}=4$ SYM. However, this is a somewhat brief discussion because the $\mathcal{N}=4$ SYM is an exactly conformal field theory (beta function vanishes identically). The discussion of why a conformal theory has zero bulk viscosity is presented in Appendix \ref{appB}.

Ultimately, we are interested in make contact with the real world and QCD, for instance, does have a trace anomaly \eqref{eq:TraAnQCD}. The the way to deform the four dimensional CFT and obtain an energy scale was sketched in Eq. \eqref{eq:DefCFT}  \cite{Klebanov:1999tb}, which is accomplished by introducing some operator $\mathcal{O}$ such that 
\begin{equation}
\mathcal{L}_{\mathcal{N}=4}\rightarrow \mathcal{L}_{\mathcal{N}=4}+\Lambda^{4-\Delta}\mathcal{O}.
\end{equation}
In the equation above, the operator $\mathcal{O}$ is dimensionless and the scale is represented by $\Lambda^{4-\Delta}$. Moreover, we are interested in relevant deformations $\Delta<4$ because of the dimension of $\text{Tr}F^2<4$.

With the above deformation and the trace of the stress-energy tensor becomes \cite{Yarom:2009mw}
\begin{equation}
T^\mu_\mu = (\Delta-4 )\langle \mathcal{O}_\Delta\rangle\Lambda^{4-\Delta},
\end{equation}
which is congruent with the existence of a bulk viscosity.

Now, let us see how we can effectively implement such deformations. From the top-down perspective, we have several models, such as the Sakai-Sugimoto model \cite{Sakai:2004cn} or the Klebanov-Strassler cascade model \cite{Klebanov:2000hb}. However, none of these top-down models are able to capture the correct thermodynamics of the QCD and, for this reason, we adopt a bottom-up perspective. 

The simplest bottom-up addition to the bulk action to bring the dual CFT closer to the QCD is achieved by adding a scalar field backreacting with the metric field. This is the procedure adopted in the Improved Holographic QCD (IHQCD) \cite{ihqcd-1,ihqcd-2,ihqcd-veneziano}, and also in Gubser's model \cite{GN1,GN2}. Since Chapter \ref{Chap8.0} is an extension of the latter with a magnetic field, our discussion done here is based on Refs. \cite{GN1,GN2,gubser1,gubser2}. The bulk action is given by
\begin{equation}\label{eq:NonConfAcGub}
S = \frac{1}{16\pi G_5}\int d^5 x\sqrt{-g}\left(R-\frac{1}{2}\partial_\mu\phi \partial^\mu \phi - V(\phi) \right)
\end{equation}
where $\phi =\phi(r)$ is the scalar field (dilaton) along with its respective potential $V(\phi)$. The Einstein equation of this dilatonic gravity is
\begin{equation}\label{eq:EinsteinNConfGub}
 R_{\mu\nu}-\frac{1}{3}g_{\mu\nu}V(\phi)-\frac{1}{2}\partial_\mu\phi\partial_\nu\phi=0,
 \end{equation} 
which is supplied by the dilaton equation
\begin{equation}
 (\square-V'(\phi))\phi=0.
 \end{equation} 
 
The most general metric Ansatz for a black hole with $SO(3)$ symmetry in the $\vec{x}$ spatial directions is
\begin{equation}
ds^2 = e^{a(r)}\left( -h(r)dt^2+d\vec{x}^2 \right)+e^{2b(r)}\frac{dr^2}{h(r)}.
\end{equation}
 
Both Einstein and dilaton equations can be solved for a broad variety  of potentials $V(\phi)$; we have a landscape of possible black holes. So now comes the phenomenological aspect of the model: we shall fix the parameters of the potential $V(\phi)$ using the lattice QCD results for the equation of state (EoS). The observable that we choose from the lattice in order to fix $V(\phi)$ is the speed of sound $c_s$ given by
\begin{equation}\label{eq:cs2grav}
c_s^2 =\frac{d \log T }{d\log s}.
\end{equation}
For each $V(\phi)$ that we take, we will have a different background geometry and, consequently, we will also have the correspondent $c_s(T)$ of this geometry given by Eq. \eqref{eq:cs2grav}. Therefore, the idea is to choose the potential in a way that it maximally approaches the lattice data.  The functional form to make this fit (by ``eyeball'') is
\begin{equation}\label{eq:PotFitGub}
L^2V(\phi) = -12\cosh\gamma\phi+b_2\phi^2+b_4\phi^4+b_6\phi^6, 
\end{equation}
where $\gamma$, $b_2$, $b_4$ and $b_6$ are the fit parameters. Although this scalar model does not include fermions in the fundamental representation, we use the lattice EoS with 2+1 flavors from Ref. \cite{latticedata1} to perform the fit. As pointed out in \cite{GN1}, this is a way to mimic QCD at finite temperature near the crossover region. We show the result of this fit in Fig. \ref{fig2} and in Eq. \eqref{2.19}. Furthermore, we assign a value for the constant $G_5$ by fitting the gravitational results of $p/T^4$ found on lattice.

Expanding the potential \eqref{eq:PotFitGub} in the UV (near the boundary), where the dilaton goes to zero, we have
\begin{equation}\label{eq:AsymPotGub}
V(\phi\rightarrow 0)=-\frac{12}{L^2}+\frac{1}{2L^2}m^2\phi^2+\mathcal{O}(\phi^4).
\end{equation}
The first term is the negative cosmological constant, which guarantees an asymptotic AdS$_5$ space. The second term is the dilaton mass and for the parameters given in \eqref{2.19} one finds $m^2\approx-3$; although this is a negative mass, it respect the BF bound \eqref{eq:BFbound5}.
\\

Now that we are more familiar with the non-conformal bottom-up model, let us calculate the bulk viscosity. First, the Kubo formula for the bulk viscosity is
\begin{equation}
\zeta = -\frac{4}{9}\lim_{\omega\rightarrow 0}\frac{1}{\omega}\text{Im}\,G^R_{T^{i}_{i}T^{j}_{j}}(\omega),
\end{equation}
where
\begin{equation}
G^R_{T^{i}_{i}T^{j}_{j}}(\omega) = -i\int dt e^{i\omega t}\left\langle \left[\frac{1}{2} T^{i}_{i}(t,\vec{x}), \frac{1}{2} T^{j}_{j}(0,\vec{0})\right] \right\rangle.
\end{equation}

Therefore, to calculate the bulk viscosity, we must consider fluctuations of the \emph{diagonal} part of the metric field around the background
\begin{equation}
h_{\mu\nu} = \text{diag}\lbrace h_{tt}, h_{rr}, h_{xx}, h_{xx},h_{xx} \rbrace,
\end{equation}
adopting the plane wave Ansatz as usual, i.e. $h_{\mu\nu}=h_{\mu\nu}e^{-i\omega t}$. We also set $h_{xx}=h_{yy}=h_{zz}$ due to $SO(3)$ symmetry.

However, there are some difficulties here. The dilaton fluctuation also couples with this diagonal fluctuated part of the metric, so one cannot ignore it. There are two ways to remedy this complication. The first one is to make a gauge change (using diffeo. invariance) to eliminate the dilaton fluctuation, though giving up the radial gauge $h_{\mu r}=0$; the second way is to work with a gauge invariant quantity involving the fluctuations of the metric and the dilaton \cite{gubser2}. Here, we choose the first path.

Before we proceed with the calculations, it is useful to take the gauge where the dilaton is the radial coordinate\footnote{In this case, we switched of the location of the conformal boundary to $\phi=0$ ($r=0$).},
\begin{equation}
\phi = r.
\end{equation}
Hence, defining $H_{xx}\equiv h^x_x$, the resulting equation of motion derived from \eqref{eq:EinsteinNConfGub} is 
\begin{equation}
H_{xx}''+\left(4a'-b'+\frac{h'}{h}-\frac{2A''}{A'} \right) H_{xx}'+ \left(\frac{h'b'}{h}-\frac{h'}{6hA'}+\frac{e^{2a-2b}}{h^2}\omega^2 \right)H_{xx}=0,
\end{equation}
where the primes denote derivatives with respect to $\phi$.

The conserved flux ($\text{Im} \mathcal{F}$) for the differential equation above  is\footnote{The flux of the second order differential equation
\begin{equation}
y''(x)+p(x)y'(x)+q(x)y(x)=0,
\end{equation}
is given by Abel's identity, 
\begin{equation}
\text{Im}\mathcal{F} = \text{exp}\left( \int^x p(u)du\right)W(y_1,y_2),
\end{equation}
where $W(y_1,y_2)=y_1'y_2-y_2'y_1$ is the Wronskian.}
\begin{equation}\label{eq:ConsFluxGub}
\text{Im} \mathcal{F} = \frac{e^{4a-b}h}{4a'}\text{Im}(H_{xx}^{*}\partial_r H_{xx}).
\end{equation}
With the bulk viscosity being given by
\begin{equation}
\zeta = -\frac{\text{Im}\mathcal{F}}{16\pi G_5}.
\end{equation}
Actually, we skipped the calculation of the on-shell action, which is important to determine the overall factor on the conserved flux. 

As mentioned before, we have the freedom to evaluate the conserved flux in the most convenient region, which is the horizon. The near horizon solution of $H_{xx}$ is
\begin{equation}\label{eq:NearHorFluBul}
H_{xx}(\phi\rightarrow \phi_h) = C e^{i\omega t}|\phi-\phi_h|^{-\frac{i\omega}{4\pi T}},
\end{equation}
where $C$ is some consant, which is determined by imposing the Dirichlet boundary condition $H_{xx}(\phi=0)=1$. But this constant cannot be analytically calculated and one must resorts to numerics. Plugging \eqref{eq:NearHorFluBul} into the conserved flux \eqref{eq:ConsFluxGub}, one obtains
\begin{equation}
\text{Im} \mathcal{F}(\omega,\phi\rightarrow\phi_h)\approx \frac{e^{3a(\phi_h)}\omega h'(\phi_h)|C|^2}{4a'(\phi_h)^2}\frac{e^{a(\phi_h)-b(\phi_h)}}{4\pi T} \approx \frac{\omega e^{2a(\phi_h)}}{4a'(\phi_h)^2}|C|^2.
\end{equation}

\begin{figure}[h]
\centering
\includegraphics[width=7.5cm,height=7cm]{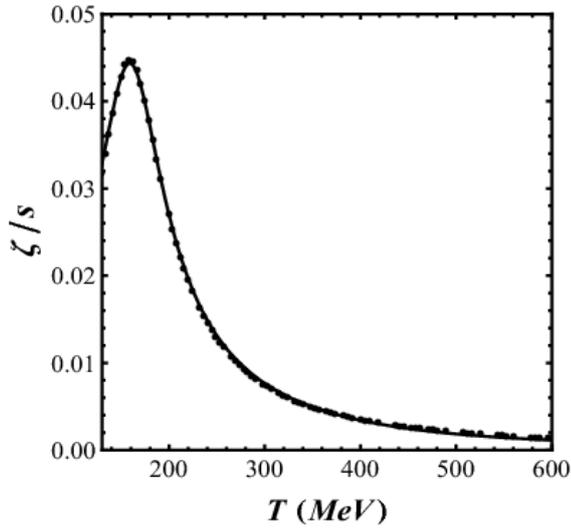}
\caption{The holographic result for the bulk viscosity using the non-conformal action \eqref{eq:NonConfAcGub}. Figure adapted from \cite{Finazzo:2014cna}.}
\label{fig:bulkHolo}
\end{figure}

Using the relation $a'=-V/3V'$ obtained from the Einstein's equations, we finally obtain the expression for the bulk viscosity to the entropy density,
\begin{equation}
\frac{\zeta}{s} = \frac{|C|^2}{4\pi}\frac{V'(\phi_h)}{V(\phi_h)}.
\end{equation}
The result is shown in Fig. \ref{fig:bulkHolo}.

\chapter{Strong magnetic fields in hot and dense matter}
\label{Chap4.0}

In this chapter we begin to include effects of strong magnetic fields in our studies of strongly coupled matter. Our main goal is to better understand the interplay of magnetic fields with the hot and dense matter. By hot and dense matter we mean the quark-gluon plasma as described in the Introduction. Therefore, this section is devoted to provide the big picture behind the generation of intense magnetic fields within the QGP context.\\

The study of the equilibrium and transport properties of the QGP as functions of parameters such as the temperature $T$, chemical potential(s), and (electro)magnetic fields are of great relevance for the characterization and understanding of this new state of QCD matter. In particular, very strong magnetic fields up to $\mathcal{O}\left(0.3\,\textrm{GeV}^2\right)$ are expected to be created in the early stages of noncentral relativistic heavy ion collisions \cite{noncentralB1,noncentralB2,noncentralB3,noncentralB4,noncentralB5,noncentralB6,noncentralB7,noncentralB8,noncentralB9}. and even much larger magnetic fields of $\mathcal{O}\left(4\,\textrm{GeV}^2\right)$ may have been produced in the early stages of the Universe \cite{universe1,universe2} (see also Fig.\ 10 in \cite{latticedata0}). Moreover, magnetic fields up to $\mathcal{O}\left(1\,\textrm{MeV}^2\right)$ are present in the interior of very dense neutron stars known as magnetars \cite{magnetar}. Therefore, the study of the effects of strong magnetic fields on the QGP has sparked a large amount of interest in the community in recent years \cite{cohen,Gursoy:2014aka,Agasian:2008tb,Mizher:2010zb,Evans:2010xs,Preis:2010cq,Fukushima:2012xw,Fukushima:2012kc,Bali:2012zg,Blaizot:2012sd,Callebaut:2013ria,Bali:2013esa,Bonati:2014ksa,Fukushima:2013zga,Machado:2013rta,Fraga:2013ova,Andersen:2013swa,Bali:2013owa,Ferreira:2013oda,Ruggieri:2014bqa,Ferreira:2014kpa,Farias:2014eca,Ayala:2014iba,Ayala:2014gwa,Ferrer:2014qka,Kamikado:2014bua,Yu:2014xoa,Braun:2014fua,Mueller:2015fka,Endrodi:2015oba} (for extensive reviews and other references, see for instance, \cite{Fraga:2012rr,reviewfiniteB1,reviewfiniteB2,reviewfiniteB3})

In the last few years, several works have emphasized that non-central heavy ion collisions are not only characterized by a sizable anisotropic flow but also by the presence of very strong electromagnetic fields formed at the early stages of the collisions \cite{noncentralB1,noncentralB2,noncentralB3,noncentralB4,noncentralB5,noncentralB6,noncentralB7,noncentralB8,noncentralB9}. This has created a lot of interest on the effects of strong electromagnetic fields in strongly interacting QCD matter \cite{bookdima} and, recently, lattice calculations with physical quark masses have determined how a strong external magnetic field changes the thermodynamic properties of the QGP \cite{latticedata0,latticedata2,latticedata3}. Lattice calculations have also been used in \cite{Bonati:2013lca,Bonati:2013vba,Bali:2013owa} to determine the magnetization of QCD matter in equilibrium and the authors of Ref.\ \cite{Bali:2013owa} argued that the paramagnetic behavior \cite{Fraga:2012ev} found in these lattice simulations leads to a sort of paramagnetic squeezing that could contribute to the overall elliptic flow observed in heavy ion collisions. If the magnetic field is still large enough at the time that elliptic flow is building up, it is natural to also consider the effects of strong magnetic fields on the subsequent hydrodynamic expansion of the QGP. 

Since the properties of a strongly coupled QGP cannot be reliably studied using perturbative techniques one has to resort to nonperturbative approaches that are valid at strong coupling. Interestingly enough, contrary to what happens in the case of a nonzero baryon chemical potential where the sign problem of the fermion determinant prevents the application of the Monte Carlo importance sampling method in lattice simulations (for a review see \cite{fodorreview}), in the case of a nonzero magnetic field (at vanishing baryon chemical potential) standard lattice techniques may be employed to study the equilibrium properties of QCD in the $(T,B)$-plane, see for instance, \cite{latticedata0,latticedata2,latticedata3}.

Another nonperturbative method that is suited to study strongly coupled non-Abelian gauge theories is the holographic AdS/CFT correspondence (also known as the gauge/gravity duality) as detailed in the previous chapter. The correspondence has been employed to obtain useful insights into the properties of the strongly coupled QGP, as recently reviewed in \cite{solana,adams}. A very attractive feature of the gauge/gravity duality is that it may be easily employed to compute transport coefficients of strongly coupled non-Abelian gauge theory plasmas, as done in Secs. \ref{Chap3.4} and \ref{Chap3.5}, which is a challenging task to perform on the lattice \cite{Meyer:2011gj}.

\section{The magnetic field generated by a peripheral heavy ion collision}
\label{Chap4.1}

In this section we present some further details about the generating of strong magnetic field in a heavy ion collision.

\begin{figure}
\begin{center}
\begin{tabular}{c}
\includegraphics[width=0.45\textwidth]{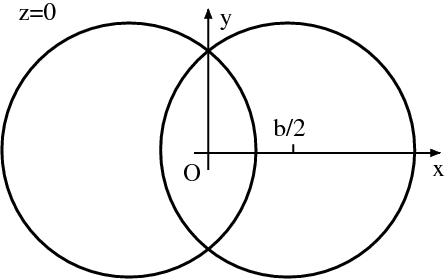} 
\end{tabular}
\begin{tabular}{c}
\includegraphics[width=0.45\textwidth]{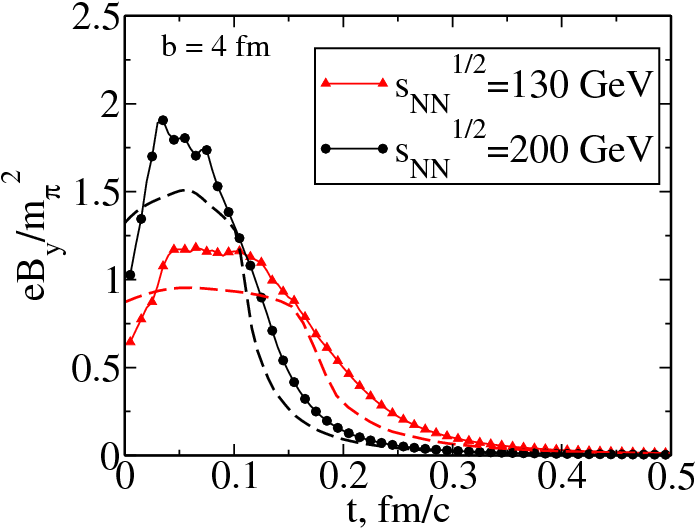} 
\end{tabular}
\begin{tabular}{c}
\includegraphics[width=0.45\textwidth]{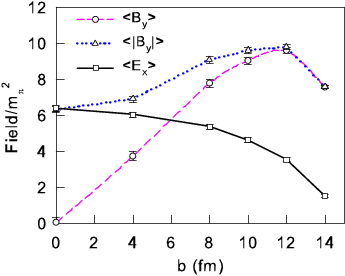} 
\end{tabular}
\begin{tabular}{c}
\includegraphics[width=0.45\textwidth]{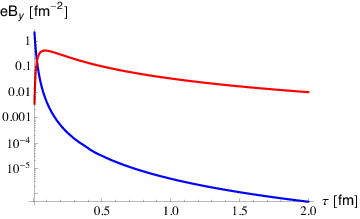} 
\end{tabular}
\end{center}
\caption{\emph{Top-left pannel}: Transverse impact plane of a typical peripheral heavy ion collision that will produce a strong magnetic field (large impact parameter). \emph{Top-right pannel:} Monte-Carlo simulation of the time evolution of the magnetic field (neglecting conductivity effects) created in a Au+Au collision. \emph{Bottom-left pannel:} Monte-Carlo simulation of the initial EM fields using a nonzero electrical conductivity for the QGP; notice that the value of the electric field can be large. \emph{Bottom-right pannel:} Illustration of how a finite electric conductivity (top red line) may enhance the magnetic field duration; the bottom curve is assuming a vacuum media. Figures adapted from Ref. \cite{noncentralB3,noncentralB6,Gursoy:2014aka}, respectively.}
\label{fig:HIMag}
\end{figure}

The first thing that we must consider is the form of the magnetic field generated by relativistic charged particle, since one considers that the magnetic field is produced by the spectator protons of the heavy ion collison (see Fig. \ref{fig:HIreact}). This information is obtained from the Li\'enard-Wiechert potentials,
\begin{equation}
\vec{B}(t,\vec{r})=\alpha_{EM}\sum_{i=1}^{N_{proton}}Z_i \frac{\vec{v}_i\times \vec{R}}{R_i-\vec{R}_i\times\vec{v}_i}(1-v_i^{2}),
\end{equation}
where $\vec{R}=\vec{R}(t)$, $Z_i$ and $\vec{v}_i$ are the proton's position, charge and speed, respectively; the top-left pannel in Fig. \ref{fig:HIMag} provides the coordinate axes. Although the fine-structure constant $\alpha_{EM}\approx 1/137$ is small, it is balanced by the large number of protons. Indeed, we can see how $eB$ scales with the total number of protons $Z$ using that $eB\sim Z/R^{2}$, where $R$ is the nucleus size $R\sim A^{1/3}\sim Z^{1/3}$. Therefore, we have that $eB\sim Z^{1/3}$.

In order to obtain an accurate evolution of the magnetic field one usually proceed with the Monte-Carlo (MC) simulation \cite{noncentralB3,noncentralB5,noncentralB6}. The result of this MC  simulation at the vacuum is given in Fig. \ref{fig:HIMag} (top-right pannel), and we can see a quite strong magnetic field $eB\sim m_{\pi}^{2}$ in the early stages of the collision.

It is not clear at the moment if the electromagnetic fields present in the early stages of heavy ion collisions remain strong enough to directly affect equilibrium and transport properties of the plasma produced at later stages. As shown in \cite{noncentralB4}, the electrical conductivity of the QGP can greatly increase the lifetime of the magnetic field on the QGP. Thus, a precise estimate of the QGP electrical conductivity is imperative. A lattice QCD result for the electrical conductivity of the QGP is \cite{noncentralB8}
\begin{equation}
\sigma_{LQCD}= (5.8 \pm 2.9) \frac{T}{T_c} \text{MeV},
\end{equation}  
with large uncertainty yet. The best lattice calculation of this electrical conductivity, shown in Fig. \ref{figlimitations}, is found in Ref. \cite{Aarts:2014nba}. We illustrate the effects of the electrical conductivity on the bottom-right pannel of Fig. \ref{fig:HIMag}.

Since we have the coexistence of magnetic and electric fields during the early stages of the QGP, it is natural to ask whether we have the formation of a non-trivial topological field configurations. The information about this is given by the Chern-Simons following term 
\begin{equation}
\int d^4x \epsilon^{\alpha\beta\mu\nu}F_{\alpha\beta}F_{\mu\nu}\neq 0,
\end{equation}
whose presence may create a current that will unbalance\footnote{The QCD chiral anomaly is also necessary to provide the initial chirality imbalance.} quarks and antiquarks leading to some chiral imbalance . This effect is called the \emph{chiral magnetic effect} (CME), which is of great interest in recent years \cite{noncentralB1,noncentralB2,Li:2014bha}.

\section{The magnetic brane background}
\label{sec:magbranes}


Now that we are conscious about the very strong magnetic fields created in the early stages of (peripheral) heavy ion collision, we would like to scan possible effects induced by this magnetic field. In this dissertation we choose to work with the gauge/gravity duality to tackle this problem, given its suitability to describe strongly coupled systems in equilibrium, near-the-equilibrium, or even far-from-equilibrium. The holographic approach is one specific approach among many, such as lattice QCD, the NJL model, the chiral Lagrangian, etc. 

There are, so far, four distinct ways to implement a magnetic field in a holographic model aiming possible applications to the sQGP: the magnetic brane solution \cite{DK1, DK2, DK3}; the Sakai-Sugimoto model \cite{Sakai:2004cn,Callebaut:2013ria,Albash:2007bk,Johnson:2008vna,Ballon-Bayona:2013cta}, from which the magnetic field is generated by the Dirac-Born-Infeld action inherent to the $Dp/\bar{Dp}$ branes embedding; a bottom-up model mimicking the QCD EoS \cite{Rougemont:2015oea} - develoved in Chap. \ref{Chap8.0}; and another bottom-up model based on the IHQCD in the Veneziano limit \cite{Drwenski:2015sha}. There are some extensions of the magnetic brane solution as well using the hard/soft wall perspective \cite{Mamo:2015dea,Dudal:2015wfn}. In this Chapter, though, we shall develop the magnetic brane solution for the subsequent applications in Chapters \ref{Chap5.0}, \ref{Chap6.0}, and \ref{Chap7.0}.

The magnetic brane solution is a bona-fide top-down construction \cite{DK1,DK2,DK3} dual to the magnetic $\mathcal{N}=4$ SYM. Of course, it is a caricature of QCD  but it can be very enlightening  and show some universal qualitative behaviors. For instance, the anisotropic shear viscosity obtained from the magnetic brane (Chap. \ref{Chap5.0}) has the same qualitative behavior of the QCD-like model \ref{SecAnisoShear}. Before we present the (super)gravity action for the magnetic brane, let us examine the conformal field theory side in the presence of a magnetic field.

To include the effects of a magnetic field on the maximally supersymmetric $SU(N_c)$ theory, in the large $N_c$ limit and in four dimensions, we must deform the theory by including an external Abelian $U(1)$ gauge field
\begin{equation}
S_{\mathcal{N}=4}\rightarrow S_{\mathcal{N}=4} + \int d^4 x j^\mu(x) A^{ext}_{\mu}(x),
\end{equation}
where
\begin{equation}
A^{ext} = \mathcal{B}dx, \ \ \ \text{(Landau gauge)}
\end{equation}
with $\mathcal{B}$ being the physical magnetic field, and $j^\mu(x)$ is the conserved U(1) current, associated to the four Weyl fermions and the three complex scalars of $\mathcal{N}=4$ SYM. Diving more deep in some technical details, recall that all the matter content of the $\mathcal{N}=4$ SYM theory is in the adjoint representation, with the Weyl fermions in the \textbf{4} of the SO(6), and the scalars in the \textbf{6} of SO(6). The global SO(6) R-symmetry accommodates up to three distinct magnetic fields, i.e. $U(1)^3\subset SO(6)$, though we use only one U(1) Cartan subgroup to create the magnetic field.

From the gravitational point of view, we have originaly the type IIB SUGRA in $AdS_5\times S^5$. The reduction of the five-sphere breaks the group $SO(6)$ into $SO(2)\times SO(2)\times SO(2)=U(1)_a\times U(1)_b\times U(1)_c$; the black brane solutions are charged under these $U(1)$ Cartan subroups. As in the field theory, we are only interested in one of these subgroups.\footnote{For instance, if we want to include the baryonic chemical potential $\mu_B$ we need to turn on another Abelian gauge field.}

A consistent truncation of the 5-dimensional bosonic supergravity is\footnote{We note that our definition for the Riemann tensor possesses an overall minus sign in comparison to the one used in \cite{DK1}.}
\begin{equation}
\label{eq:action}
S = \frac{1}{16 \pi G_5} \int d^5x \, \sqrt{-g}  \left(R + \frac{12}{L^2} - F^{\mu\nu}F_{\mu\nu}\right) + S_{CS} + S_{bdry},
\end{equation}
where $G_5$ is the 5-dimensional gravitational constant, $L$ is the asymptotic $AdS_5$ radius and $F=dA$ is the Maxwell field strength 2-form. The term $S_{CS}$ is the Chern-Simons term, given by
\begin{equation}
S_{CS}=\frac{1}{6\sqrt{3}\pi G_5}\int_{M}A\wedge F\wedge F.
\end{equation}
For the case where there is only the magnetic field, the Chern-Simons term is identically zero and for this reason we shall ignore it hereafter.

The boundary term $S_{bdry}$ encodes the contributions of the Gibbons-Hawking-York action, necessary to define a well posed variational problem, and the counter-term action that eliminates the divergences of the on-shell action. The explicit form of $S_{bdry}$ is
\begin{equation}
S_{bdry} = S_{GHY}+S_{ct},
\end{equation}
where $S_{GHY}$ is given by Eq. \eqref{eq:GHYterm} and 
\begin{equation}
S_{ct}=\frac{1}{8\pi G_5}\int_{\partial M}d^4x\sqrt{-\gamma}\left( \frac{L}{4}R(\gamma)-\frac{3}{L}+\frac{L}{2}\left(\ln\frac{r}{L}\right)F^{\mu\nu}F_{\mu\nu} \right).
\end{equation}

The equations of motion are obtained from the Einstein-Maxwell field equations
\begin{equation}
\label{eq:eom1}
R_{\mu \nu} = -\frac{4}{L^2} g_{\mu \nu} - \frac{1}{3} F_{\rho \sigma} F^{\rho \sigma} g_{\mu \nu} + 2 F_{\mu \rho} F_{\nu}^{\ \rho},
\end{equation}
and also from the Maxwell's field equations for the Abelian field,
\begin{equation}
\label{eq:eom2}
\nabla_{\mu} F^{\mu \nu} = 0.
\end{equation}

If we want a constant magnetic field along the z-direction, which breaks the original SO(3) rotation symmetry, the natural Ansatz for the magnetic brane geometry is 
\begin{equation}
\label{eq:background}
ds^2 = -U(r) dt^2 + \frac{dr^2}{U(r)} + f(r)(dx^2+dy^2) + p(r) dz^2,
\end{equation}
where $U(r)$, $f(r)$ and $p(r)$ are determined by solving the equations of motion. The holographic coordinate $r$ is such that the boundary is located at $r \to \infty$. We want a black brane background and, thus, we require that at a given $r=r_h$ the function $U(r)$ has a simple zero. The Ansatz for the field strength $F$ is given by
\begin{equation}
\label{eq:fieldstrength}
F = B\, dx \wedge dy,
\end{equation}
where the constant $B$ is the bulk magnetic field oriented along the $z$ direction. It can be checked that the equation of motion \eqref{eq:eom2} is trivially satisfied by this Ansatz.

In the absence of a magnetic field $p(r) = f(r)$, which reflects the spatial $SO(3)$ invariance of the boundary gauge theory. However, since the magnetic field establishes a preferred direction in space, it breaks the $SO(3)$ spatial symmetry to only a $SO(2)$ symmetry in the $x,y$ directions. In the bulk theory this is taken into account by the fact that in this case $f(r) \neq p(r)$.

The equations of motion derived from \eqref{eq:background} are (we set $L=1$ from now on)
\begin{align}
\label{eq:eom3}
 U(V''-W'') + \left(U'+U(2V'+W') \right)(V'-W') & =  -2B^2 e^{-4V}, \nonumber \\
2 V'' + W'' + 2 (V')^2 + (W')^2 & =  0, \nonumber \\
\frac{1}{2} U'' + \frac{1}{2} U' (2V'+W') & =  4 + \frac{2}{3} B^2 e^{-4V} \quad \quad \\
2 U' V' + U'V + 2U(V')^2+4UV'W' & =  12 - 2B^2 e^{-4V}, \nonumber
\end{align}
where we defined $V$ and $W$ by $f = e^{2V}$ and $p = e^{2W}$. By Bianchi's identity, the fourth equation of motion can be shown to be a consequence of the three first equations and, thus, it can be taken as a constraint on initial data.

It is well-known that charged systems undergo dimensional reduction in the presence of strong fields due to the projection towards the lowest Landau level \cite{Gusynin:1994re,Gusynin:1994xp,Gusynin:1995nb} (see the recent review in \cite{Shovkovy:2012zn}). Taking that into account, the authors of \cite{DK1} proposed that the background \eqref{eq:background} satisfied two conditions. The first condition is that the geometry must be asymptotically $\mathrm{AdS}_5$, that is, $U(r) \to r^2$, $p(r) \to r^2$ and $f(r) \to r^2$ when $r \to \infty$ since in the UV we must recover the dynamics of $\mathcal{N} = 4$ SYM without the influence of the magnetic field. The second condition is that in the asymptotic IR the geometry becomes a BTZ black hole \cite{Banados:1992wn} times a two dimensional torus $T^2$ in the spatial directions orthogonal to the magnetic field. In fact, deep in the IR the geometry near the horizon of the black brane $r_h$, $r \sim r_h$, is given by
\begin{equation}
\label{eq:BTZ}
ds^2 = \left[ -3(r^2-r_h^2) dt^2 + 3 r^2 dz^2 + \frac{dr^2}{3(r^2-r_h^2)} \right] + \left[ \frac{B}{\sqrt{3}} (dx^2+dy^2)\right].
\end{equation}
This implies that in the IR the dynamics corresponds to a (1+1) dimensional CFT. Thus, imposing that the background interpolates between the BTZ black hole for $r \sim r_h$ and $\mathrm{AdS}_5$ for high $T$ and interpreting the flow along the $r$ direction as a renormalization group flow, this solution flows from a (1+1) dimensional CFT in the IR to a 4 dimensional CFT in the UV \cite{DK1}.

\subsection{Numerical solution and thermodynamics}
\label{sec:num}

No analytic solution which interpolates between $AdS_5$ and the BTZ$\times T^2$ geometry is known and, thus, we must resort to numerics. In this subsection we briefly review the numerical procedure for solving the equations of motion and the thermodynamics, first elaborated in \cite{DK1}. 

The strategy is to first choose the scale for the $t$ and $r$ coordinates to fix the horizon position at $r_h=1$ so that $\tilde{U}(1) = 0$, where the tilde indicates that we are in the rescaled coordinates $\tilde{t}$ and $\tilde{r}$. By using the fact that any physical quantity in this model should depend on the dimensionless ratio $T/\sqrt{B}$, we also fix the temperature at $T=1/(4 \pi)$ - this means that we take $\tilde{U}'(1) = 1$. Also, we rescale the $x$, $y$, and $z$ coordinates to have $\tilde{V}(1)=\tilde{W}(1)=0$. In these new coordinates, the magnetic field is $b$. After these redefinitions, the first and fourth equations in \eqref{eq:eom3} imply that 
\begin{align}
\label{eq:initialdata}
\tilde{V}'(1) = & \,4-\frac{4}{3} b^2 \quad \quad \mathrm{and} \nonumber \\ 
\tilde{W}'(1) = & \, 4+\frac{2}{3}b^2.
\end{align}

This gives a well posed initial value problem for $\tilde{U}(\tilde{r})$, $\tilde{V}(\tilde{r})$, and $\tilde{W}(\tilde{r})$, which can be integrated out from $\tilde{r}=1$ to a large value of $\tilde{r}$. It can be checked numerically that the geometry has the asymptotic behavior
\begin{equation}
\tilde{U}(\tilde{r})\rightarrow \tilde{r}^2, \ \ \ e^{2\tilde{V}(\tilde{r})} \rightarrow v \tilde{r}^2, \ \ \ e^{2\tilde{W}(\tilde{r})} \rightarrow w \tilde{r}^2,
\end{equation}
where $v(b)$ and $w(b)$ are proportionality constants that depend on the rescaled magnetic field $b$. This result implies that, apart from a coordinate rescaling, the geometry is asymptotically $\mathrm{AdS}_5$.  To go back to the original units and have the correct $\mathrm{AdS}_5$ asymptotic behavior, we need to rescale back to our original coordinate system by doing $(\tilde{x},\tilde{y},\tilde{z})\rightarrow(x/\sqrt{v},y/\sqrt{v},z/\sqrt{w})$. The metric is then (in coordinates that are asymptotically $\mathrm{AdS}_5$)
\begin{equation}
\label{eq:resc.metric}
ds^2 = -\tilde{U}(r)dt^2 + \frac{dr^2}{\tilde{U}(r)} + \frac{e^{2\tilde{V}(r)}}{v}(dx^2+dy^2) + \frac{e^{2\tilde{W}(r)}}{w}dz^2,
\end{equation}
where we note that we have taken $r = \tilde{r}$. By the same token, the field strength is now written as
\begin{equation}\label{eq:Non-Phys-Mag}
F = \frac{b}{v}dx \wedge dy.
\end{equation}
Therefore, the rescaled magnetic field is related to the physical field at the boundary by $B = b/v$. Also, note that the first equation \eqref{eq:initialdata} implies that for $b > \sqrt{3}$ we have $V'(1) < 0$, which means that the geometry will not be asymptotically $\mathrm{AdS}_5$. Thus, the rescaled field $b$ has an upper value given by $b_{max} = \sqrt{3}$.

From \eqref{eq:resc.metric} one can obtain the thermodynamics of the gauge theory. The physical field is $\mathcal{B} = \sqrt{3} B$, as argued in \cite{DK1} by comparing the Chern-Simons term in \eqref{eq:action} with the $\mathcal{N} = 4$ SYM chiral anomaly. The dimensionless ratio $T/\sqrt{\mathcal{B}}$ is given by
\begin{equation}
\frac{T}{\sqrt{\mathcal{B}}} = \frac{1}{4 \pi\, 3^{1/4}} \sqrt{\frac{v}{b}}.
\end{equation}
while the dimensionless ratio of the entropy density $s$ by $N^2 \mathcal{B}^3/2$ (using that $G_5 = \pi/2N^2$) is
\begin{equation}
\frac{s}{N^2\mathcal{B}^{3/2}} = \frac{1}{3^{3/4} 2\pi}\sqrt{\frac{v}{b^3 w}}\,.
\end{equation}

\begin{figure}[t]
\centering
  \includegraphics[width=.6\linewidth]{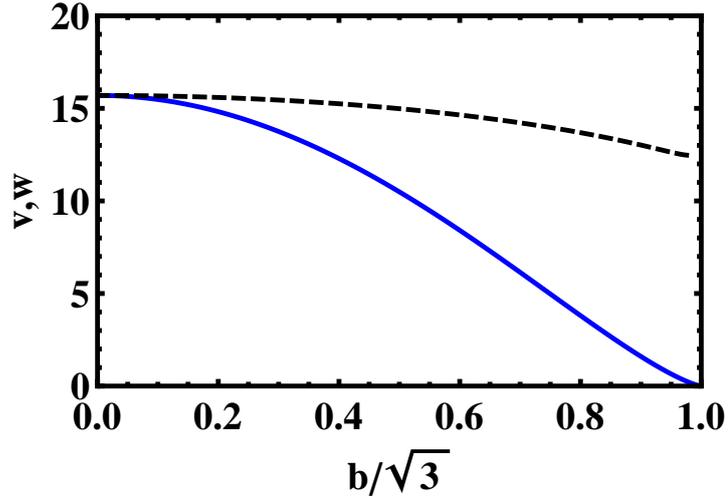}
  \caption{ The rescaling parameters $v$ (solid blue curve) and $w$ (dashed black curve) as a function of $b/\sqrt{3}$.}
  \label{fig:vwb}      
\end{figure}

\begin{figure}[h]
\centering
  \includegraphics[width=.6\linewidth]{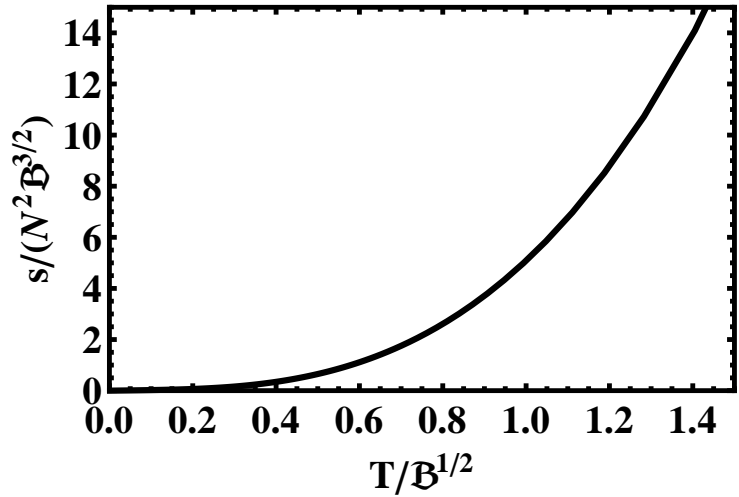}
  \caption{The normalized entropy density $s/(N^2 \mathcal{B}^{3/2})$ as a function of the dimensionless combination $T/\sqrt{\mathcal{B}}$.}
  \label{fig:thermo}      
\end{figure}

The numerical procedure for evaluating the thermodynamics can then be summarized as follows: one chooses a value of the rescaled magnetic field $b$, numerically solves the equations of motion, and obtains the rescaled parameters $v$ and $w$ by fitting the asymptotic data for $\tilde{V}(r)$ and $\tilde{W}(r)$ to the functions $v r^2$ and $w r^2$. By varying $b$, one can obtain the functions $v(b)$ and $w(b)$ and evaluate $T/\sqrt{\mathcal{B}}$ versus $s/(N^2 \mathcal{B}^{3/2})$ by using $b$ as a parameter. In Fig.\ \ref{fig:vwb} we show $v$ and $w$ as a function of $b$. The entropy density is shown in Fig.\ \ref{fig:thermo} and we have checked that our results match those previously found in \cite{DK1}.

\section{Viscous relativistic magnetohydrodynamics}
\label{Chap4.3}

In this section we shall elaborate in detail the consequence of breaking the $SO(3)$ rotation on the fluid's dissipation, i.e. we shall see the rising of anisotropic viscosities. Below, we shall see that we have seven viscosity coefficients, five shear viscosities and two bulk viscosities. Historically, the calculations of the anisotropic transport coefficients in plasmas were carried out in the 1950's, mainly by Braginskii \cite{Bragi}, in the context of the abelian plasmas. In more recent years, we became aware of high energy relativistic systems, like neutron stars \cite{Huang:2009ue,Huang:2011dc}, where the anisotropic nature of the plasma may play an important role.  Although our discussion is about the anisotropic viscosity in a plasma driven by a magnetic field, we stress that this phenomena occurs in various others systems, like plastics and superfluids \cite{LandauEla}; see Refs. \cite{Natsuume:2010ky,Erdmenger:2010xm,Erdmenger:2012zu} for the holographic approach of the later.

Ultimately, we are interested in relativistic viscous plasmas and, consequently, we want a causal and stable theory of magnetohydrodynamics. For the viscous magnetohydrodynamics one has the Navier-Stokes-Fourier-Ohm theory \cite{Huang:2009ue}, which is an extension of the old (acausal and unstable) relativistic Navier-Stokes theory - we shall not exploit this further. There was an attempt to include relativistic effects on the magnetohydrodynamics for the weakly collisional (abelian) plasmas \cite{Chandra:2015iza}; this may be important to study black hole's accretion flows, where the magnetic field is intense. Recently, though, Ref. \cite{Molnar:2016vvu} extended the Israel-Stewart formalism (cf. Sec. \ref{Chap2.1.1}) in order to accommodate anisotropic fluids.

Before we tackle the viscosity part, it is worth to discuss a bit the inviscid (ideal) case, which is well established. Ideal magnetohydrodynamics follows the same idea of the usual hydrodynamics described in Chapter \ref{Chap 2}, i.e. it is an effective theory, valid for long-wavelength and low-frequency excitations. Ideal hydrodynamics is the zeroth order result of the gradient expansion and it can be built using the following quantities:
\begin{align}
T^{\mu\nu}_{ideal} &= T^{\mu\nu}_{FO}+T^{\mu\nu}_{EM}, \\
T^{\mu\nu}_{FO} &= \varepsilon u^\mu u^\nu -P\Delta^{\mu\nu}+M^{\lambda(\mu}F_{\lambda}^{\nu)}, \\
n^\mu&=n u^\mu, \\
s^\mu &= s u^\mu,
\end{align}
where $u^\mu$ is the four-velocity with normalization $u_\mu u^\mu=-1$, $T^{\mu\nu}_{EM}=F^{\mu\alpha}F^{\nu}_{\ \alpha}-1/4\eta^{\mu\nu}F^2$ is the electromagnetic contribution for the stress-energy tensor, $\Delta^{\mu\nu}=g^{\mu\nu}+u^\mu u^\nu$ is the orthogonal projector, and $\varepsilon$, $P$, $s$ are the energy density, pressure, and entropy density, respectively. The symbol $n$ represents any possible charges that the theory may contain; for instance, it could be the baryon number. The antisymmetric tensor $M^{\mu\nu}$ is the polarization tensor, which can be obtained from the thermodynamic potential $\Omega$, $M^{\mu\nu}=\partial\Omega/\partial F_{\mu\nu}$.

The magnetohydrodynamics equations are obtained from the conservation laws
\begin{align}
\partial_\mu T^{\mu\nu}_{ideal}= \partial_\mu n^\mu = 0, \ \ \ \partial_\mu s^\mu \geq 0.
\end{align}
For instance, the one-dimensional magnetic Bjorkern flow is solved in Ref. \cite{Roy:2015kma}.

In order to make contact with the dissipative part of magnetohydrodynamics in it first order formulation, we write
\begin{align}
T^{\mu\nu} &= T^{\mu\nu}_{ideal}+\Pi^{\mu\nu}, \\
n^\mu &= n u^\mu + j_n^\mu, \\
s^\mu &= s u^\mu+j_s^\mu,
\end{align}
where $\Pi_{\mu\nu}$ is the viscous stress tensor, with $j_n^\mu$ and $j_s^\mu$ being the dissipative fluxes.

The task now is to derive the form of the viscous stress tensor. For highly magnetized plasmas, it cannot be the same of the usual isotropic plasmas since it has a reduced axial symmetry around the magnetic vector; from the gravitational side, the magnetic brane \eqref{eq:background} tells us the same. Therefore, to arrive at some expression for $\Pi_{\mu\nu}$, we shall need the rank-4 viscosity tensor $\eta^{\alpha\beta\mu\nu}$ which we already found in Eq. \eqref{eq:rank4Visc} when we discussed the shear viscosity in the context of kinetic theory. To clarify the discussion, let us first define the dissipation function $\mathcal{R}$
\begin{equation}\label{eq:dissipation}
\mathcal{R} = \frac{1}{2}\eta^{\mu\nu\alpha\beta}w_{\mu\nu}w_{\alpha\beta},
\end{equation}
where $w_{\mu\nu}=\frac{1}{2}\left(D_\mu u_\nu+D_\nu u_\mu \right)$, and $D_\mu= \Delta_{\mu\nu}\partial^\nu$.  Taking the derivative of \eqref{eq:dissipation} with respect to $w_{\mu\nu}$, we obtain the usual stress tensor $\Pi^{\mu\nu}$
\begin{equation}
\Pi^{\mu\nu}= \eta^{\mu\nu\alpha\beta}w_{\alpha\beta},
\label{onsagercondition}
\end{equation}
which is the same relation that we found in Eq. \eqref{eq:rank4Visc}.

The construction of the viscosity tensor is based on its symmetry properties. Assuming the existence of an external magnetic field $B$, we have
\begin{equation}
 \eta^{\mu\nu\alpha\beta}(B)= \eta^{\nu\mu\alpha\beta}(B) =\eta^{\mu\nu\beta\alpha}(B).
\end{equation}
Also, the the Onsager principle \cite{LandauKine} tells us that
\begin{equation}
  \eta^{\mu\nu\alpha\beta}(B) =  \eta^{\alpha\beta\mu\nu}(-B).
\end{equation}
Now, we write down all the linear independent objects satisfying the above conditions of symmetry
\begin{align}
\text{(i)}& \ \Delta^{\mu\nu}\Delta^{\alpha\beta}, \notag \\
\text{(ii)}& \ \Delta^{\mu\alpha}\Delta^{\nu\beta}+\Delta^{\mu\beta}\Delta^{\nu\alpha},  \notag \\
\text{(iii)}& \ \Delta^{\mu\nu}b^{\alpha}b^{\beta}+\Delta^{\alpha\beta}b^{\mu}b^{\nu}, \notag \\
\text{(iv)}& \ b^{\mu}b^{\nu}b^{\alpha}b^{\beta}, \notag \\
\text{(v)} & \ \Delta^{\mu\alpha}b^{\nu}b^{\beta}+\Delta^{\mu\beta}b^{\nu}b^{\alpha}+\Delta^{\nu\alpha}b^{\mu}b^{\beta}+\Delta^{\nu\beta}b^{\mu}b^{\alpha},  \notag \\
\text{(vi)}& \ \Delta^{\mu\alpha}b^{\nu\beta}+\Delta^{\mu\beta}b^{\nu\alpha}+\Delta^{\nu\alpha}b^{\mu\beta}+\Delta^{\nu\beta}b^{\mu\alpha},  \notag \\
\text{(vii)}& \ b^{\mu\alpha}b^{\nu}b^{\beta}+b^{\mu\beta}b^{\nu}b^{\alpha}+b^{\nu\alpha}b^{\mu}b^{\beta}+b^{\nu\beta}b^{\mu}b^{\alpha},
\end{align}
where $b^\mu$ is a spacelike vector orthogonal to the magnetic field ($b_\mu b^\mu=1$), and $b^{\mu\nu}=\epsilon^{\mu\nu\alpha\beta}b^\alpha u^\beta$. This means that we have seven coefficients, five shear viscosities and two bulk viscosities. The shear viscosities are related to the traceless part of $\Pi^{\mu\nu}$ while the bulk viscosities are related to the trace of the stress tensor. We note that Onsager's condition in Eq.\ (\ref{onsagercondition}) is responsible for the presence of the two last tensors, (vi) and (vii), involving the Levi-Civita symbol $\epsilon^{\mu\nu\alpha\beta}$. These structures are inherent in magnetized plasmas \cite{LandauKine,Huang:2009ue,Huang:2011dc} but they are not present in the case of anisotropic superfluids, in which we have only five viscosity coefficients altogether.

We say in advance, however, that the form of the line element in Eq. \eqref{eq:background} allows up to five different viscosity coefficients. This is because we have the following independent metric fluctuations: $h_{xy}$, $h_{xz}$, $h_{xx}+h_{yy}$ and $h_{xx}-h_{yy}$. Moreover as we shall see below after the calculation of the Kubo formulas, three shear viscosity coefficients are trivially zero in this case. 

For the sake of convenience, we will adopt the same combination of viscosity coefficients chosen in \cite{Huang:2009ue,Huang:2011dc} \footnote{This is a different convention for the coefficients than the one adopted in \cite{Bragi} and in $Â§13$ of \cite{LandauKine}.}. Thus, using the general linear combination of the structures above, we find the most general form of the viscosity tensor in the presence of a constant magnetic field
\begin{align}\label{eq:r4ViscoTensMag}
\eta^{\mu\nu\alpha\beta} =& (-2/3\eta_0 +1/4\eta_1 +3/2\zeta_\perp)\text{(i)} + (\eta_0 )\text{(ii)} +(3/4\eta_1+3/2\zeta_\perp)\text{(iii)}\notag \\
  & +(9/4\eta_1 -4\eta_2 +3/2\zeta_\perp+3\zeta_\parallel )\text{(iv)} +(-\eta_2 )\text{(v)} +(-\eta_4)\text{(vi)} \notag \\
  & +(-\eta_3+\eta_4)\text{(vii)},
\end{align}
with the $\eta'$s being the shear viscosities and the $\zeta'$s the bulk viscosities.

Substituting \eqref{eq:r4ViscoTensMag} into \eqref{onsagercondition} we find the following viscous tensor
\begin{align}\label{eq:ViscTensMag}
\Pi_{\mu\nu} &= -2\eta_0\left(w_{\mu\nu}-\Delta_{\mu\nu}\frac{\theta}{3} \right)-\eta_1\left(\Delta_{\mu\nu}-\frac{3}{2}\Xi_{\mu\nu} \right)\left(\theta-\frac{3}{2}\phi\right)+2\eta_2\left(b_\mu\Xi_{\nu\alpha}b_{\beta}+b_\nu\Xi_{\mu\alpha}b_{\beta}\right)w^{\alpha\beta} \notag \\
    & +\eta_3\left(\Xi_{\mu\alpha}b_{\nu\beta}+\Xi_{\nu\alpha}b_{\mu\beta}\right)w^{\alpha\beta}-2\eta_4\left(b_{\mu\alpha}b_{\nu}b_{\beta}+b_{\nu\alpha}b_{\mu}b_{\beta}\right)w^{\alpha\beta}-\frac{3}{2}\zeta_{\perp}\Xi_{\mu\nu}\phi - 3\zeta_{\parallel}b_{\mu}b_{\nu}\varphi,
\end{align}
where $w_{\mu\nu}=\frac{1}{2}\left(D_{\mu}u_{\nu}+D_{\nu}u_{\mu}\right)$, $D_{\mu}=\Delta_{\mu\alpha}\nabla^{\alpha}$, $\Xi_{\mu\nu}\equiv\Delta_{\mu\nu}-b_\mu b_\nu$ (orthogonal projector), $\theta=\nabla_\mu u^\mu$, $\phi\equiv\Xi_{\mu\nu}w^{\mu\nu}$ and $\varphi\equiv b_{\mu}b_{\nu}w^{\mu\nu}$. Note that the derivative operator $D_\mu$ is given in terms of the covariant derivative, i.e. we are generalizing the viscous tensor to a curved spacetime; this will be essential to extract the Kubo formulas, once they come from gravity fluctuations.

\subsection{Kubo formulas for viscous magnetohydrodynamics}

With the expression for the viscous tensor $\Pi_{\mu\nu}$ \eqref{eq:ViscTensMag} at hand, it is time to derive the Kubo formulas that relate the viscosity coefficients to the retarded Green's functions. In this sense, this subsection is the generalization of what we did in Sec. \ref{Chap2.3.1} by computing the Kubo formulas for an isotropic and homogeneous fluid. We remark that Ref. \cite{Huang:2011dc} also derived the Kubo formulas though using the Zubarev formalism.

Let us resume then the procedure developed in Sec. \ref{Chap2.3.1} to obtain the Kubo formulas for the viscosity: adopting the Minkowski background, we perform small gravity perturbations assuming that they are all homogeneous, which means that we can work only with the spatial indices, i.e.  $g_{ij}= \eta_{ij}+h_{ij}(t)$, with $h_{00}=h_{0i}=0$. Also, we work in the rest frame of the fluid where $u^{\mu}=(1,0,0,0)$\footnote{In other words, we will work in the Landau-Lifshitz frame where $u_\mu\Pi^{\mu\nu}=0$, and all the information about the viscosities are in the components $\lbrace i, j, k, l \rbrace$ of the retarded Green function.}, . The novelty here is the presence of magnetic field, which  is assumed to be constant along the $z-$direction, i.e. $b^{\mu}=(0,0,0,1)$.

Thus, we have the variation for the viscous tensor\footnote{Note that:
\begin{equation}
\delta\Xi_{\mu\nu}=h_{\mu\nu}, \ \ \ \delta\theta= \frac{1}{2}\partial_t h^{\lambda}_{\lambda}, \ \ \ \delta\varphi=\frac{1}{2}\partial_t h_{zz}. \notag
\end{equation}}
\begin{align}
\delta\Pi_{ij}=\delta(i)+\delta(ii)+\delta(iii)+\delta(iv)+\delta(v)+\delta(vi)+\delta(vii),
\end{align}
where
\begin{align}
\delta(i) &=  -\eta_0\left(\partial_t h_{ij}-\frac{1}{3}\delta_{ij}\partial_t h^{k}_{k}\right),
\end{align}
\begin{align}
\delta(ii) &=-\frac{1}{4} \eta_1 \left[ (\delta_{ij}-3b_i b_j)\frac{1}{2}\partial_t h^{k}_{k} -3(\delta_{ij}-3b_i b_j)\frac{1}{2}\partial_t h_{zz}\right],
\end{align}
\begin{align}
\delta(iii) &=  \eta_2  \left[ b_i b^k\partial_t h_{jk}+b_j b^k\partial_t h_{ik} - 2b_i b_j b^k b^l\partial_t h_{kl} \right],
\end{align}
\begin{align}
\delta(iv) &=\eta_3 \left( \delta_{ik}+b_i b_k \right)\epsilon_{jlz}\partial_t h^{kl},
\end{align}
\begin{align}
\delta(v) &= -2\eta_4 \left(\epsilon_{ikz}b_j b_k+\epsilon_{jlz}b_i b_k  \right)\partial_t h^{kl},
\end{align}
\begin{align}
\delta(vi) &=-\frac{3}{4}\zeta_\perp\left(\delta_{ij}-b_ib_j \right) \left(\partial_t h^{k}_{k}+\partial_t h_{zz}\right),
\end{align}
\begin{align}
\delta(vii) &=  -\frac{3}{2}\zeta_\parallel b_ib_j\partial_t h_{zz}.
\end{align}

The next step is to write the variations above in Fourier space ($h_{ij}\sim e^{-i\omega t}$), which gives us the following expressions
\begin{align}
\delta(i) &= \frac{i\omega}{2}h_{kl}(\omega)\left[ \eta_0 \left( \delta^{k}_{i}\delta^{l}_{j} +\delta^{l}_{i}\delta^{k}_{j}-\frac{2}{3}\delta_{ij}\delta^{kl}\right) \right]
\end{align}
\begin{align}
\delta(ii) &= \frac{i\omega}{2}h_{kl}(\omega)\frac{1}{4} \eta_1 \left[ (\delta_{ij}-3b_i b_j)\delta^{kl} -3(\delta_{ij}-3b_i b_j)\delta^{k}_{z}\delta^{l}_{z}\right],
\end{align}
\begin{align}
\delta(iii) &= -\frac{i\omega}{2}h_{kl}(\omega)\left[ \frac{1}{2} \eta_2 \left( b_{i}b^{k}\delta^{l}_{j}+b_{i}b^{l}\delta^{k}_{j}+b_{j}b^{k}\delta^{l}_{i}+b_{j}b^{l}\delta^{k}_{i}-4b_{i}b_{j}b^{k}b^l \right)\right],
\end{align}
\begin{align}
\delta(iv) &= -\frac{i\omega}{2}h_{kl}(\omega)\left[ 2\eta_3 \epsilon_{j \ z}^{ \ l} \left( \delta_{ik}+b_i b_k \right)\right],
\end{align}
\begin{align}
\delta(iv) &= \frac{i\omega}{2}h_{kl}(\omega)\left[ 4\eta_4  \left(\epsilon_{i \ z}^{ \ k}b_i b_k +\epsilon_{j \ z}^{ \ l}b_i b_k \right)\right],
\end{align}
\begin{align}
\delta(v) &= \frac{i\omega}{2}h_{kl}(\omega)\left[ \frac{3}{2}\zeta_\perp\left(\delta_{ij}\delta^{kl}+\delta_{ij}\delta_{z}^{k}\delta_{z}^{l}-b_ib_j\delta^{kl}-b_ib_j\delta_{z}^{k}\delta_{z}^{l} \right)  \right],
\end{align}
\begin{align}
\delta(vii) &=\frac{i\omega}{2}h_{kl}(\omega)\left[ 3\zeta_\parallel b_ib_j\delta_{z}^{k}\delta_{z}^{l}  \right].
\end{align}

Collecting all the variations above in Fourier space, we write
\begin{align}
\delta\Pi_{ij}(\omega)&=\frac{i\omega}{2}h_{kl}(\omega)\left[ \eta_0\left(\delta^{k}_{i}\delta^{l}_{j}+\delta^{l}_{i}\delta^{k}_{j}-\frac{2}{3}\delta_{ij}\delta^{kl}\right) +\frac{1}{4} \eta_1 \left[ (\delta_{ij}-3b_i b_j)\delta^{kl} -3(\delta_{ij}-3b_i b_j)\delta^{k}_{z}\delta^{l}_{z}\right]  \right. \notag \\
 &\left.  -\frac{1}{2} \eta_2 \left( b_{i}b^{k}\delta^{l}_{j}+b_{i}b^{l}\delta^{k}_{j}+b_{j}b^{k}\delta^{l}_{i}+b_{j}b^{l}\delta^{k}_{i}-4b_{i}b_{j}b^{k}b^l \right) -2\eta_3 \epsilon_{j \ z}^{ \ l} \left( \delta_{ik}+b_i b_k \right)   \right.\notag \\
   &\left. + 4\eta_4  \left(\epsilon_{i \ z}^{ \ k}b_i b_k +\epsilon_{j \ z}^{ \ l}b_i b_k \right) +\frac{3}{2}\zeta_\perp\left(\delta_{ij}\delta^{kl}+\delta_{ij}\delta_{z}^{k}\delta_{z}^{l}-b_ib_j\delta^{kl}-b_ib_j\delta_{z}^{k}\delta_{z}^{l} \right)  +3\zeta_\parallel b_ib_j\delta_{z}^{k}\delta_{z}^{l}  \right],
\end{align}
which allows us to express the retarded Green's function as a function of the viscosities,
\begin{align}\label{eq:greenvisco}
 - \lim_{\omega\rightarrow 0}\frac{1}{\omega}\text{Im}\,G^{R, \, kl}_{ij}(\omega) = &   \eta_0\left(\delta^{k}_{i}\delta^{l}_{j}+\delta^{l}_{i}\delta^{k}_{j}-\frac{2}{3}\delta_{ij}\delta^{kl}\right) +\frac{1}{4} \eta_1 \left[ (\delta_{ij}-3b_i b_j)\delta^{kl} -3(\delta_{ij}-3b_i b_j)\delta^{k}_{z}\delta^{l}_{z}\right]   \notag \\
 & -\frac{1}{2} \eta_2 \left( b_{i}b^{k}\delta^{l}_{j}+b_{i}b^{l}\delta^{k}_{j}+b_{j}b^{k}\delta^{l}_{i}+b_{j}b^{l}\delta^{k}_{i}-4b_{i}b_{j}b^{k}b^l \right) -2\eta_3 \epsilon_{j \ z}^{ \ l} \left( \delta_{ik}+b_i b_k \right)   \notag \\
   & + 4\eta_4  \left(\epsilon_{i \ z}^{ \ k}b_i b_k +\epsilon_{j \ z}^{ \ l}b_i b_k \right) +\frac{3}{2}\zeta_\perp\left(\delta_{ij}\delta^{kl}+\delta_{ij}\delta_{z}^{k}\delta_{z}^{l}-b_ib_j\delta^{kl}-b_ib_j\delta_{z}^{k}\delta_{z}^{l} \right) \notag \\
   &  +3\zeta_\parallel b_ib_j\delta_{z}^{k}\delta_{z}^{l}.
\end{align}

The final stage is to isolate the viscosities and obtain their associated Kubo formulas. For such a task, we only need to select specific components of $G^{R}_{ij,kl}$. For instance, if we take $i=k=x$ and $j=l=y$ in \eqref{eq:greenvisco}, we have
\begin{equation}
\eta_0 = - \lim_{\omega\rightarrow 0}\frac{1}{\omega}G^{R}_{T_{xy}T_{xy}}(\omega),
\end{equation}
and so forth.

In the end, we have the following Kubo formulas:
\begin{align}
\eta_0 &= - \lim_{\omega \to 0}\frac{1}{\omega}\text{Im}\, G^{R}_{T_{xy}T_{xy}}(\omega)\label{eq:eta0} , \\
\eta_1 &= -\frac{4}{3}\eta_0+ 2\lim_{\omega \to 0}\frac{1}{\omega}\text{Im}\, G^{R}_{P_{\parallel}P_{\perp}}(\omega), \label{eq:eta1} \\
\eta_2 &= -\eta_0 - \lim_{\omega \to 0}\frac{1}{\omega}\text{Im}\, G^{R}_{T_{xz}T_{xz}}(\omega), \\
\eta_3 &= - \lim_{\omega\rightarrow 0}\frac{1}{\omega}G^{R}_{P_{\perp}T_{12}}(\omega), \label{eq:eta3} \\
4 \eta_4 &= - \lim_{\omega \to 0}\frac{1}{\omega}\text{Im}\, G^{R}_{T_{xz}T_{yz}}(\omega), \label{eq:eta4} \\
\zeta_\perp &= -\frac{2}{3}\lim_{\omega \to 0} \frac{1}{\omega}\left[ \text{Im}G^{R}_{P_\perp P_\perp}(\omega) + \text{Im}G^{R}_{P_\parallel, P_\perp}(\omega) \right], \label{eq:kubo1} \\
\zeta_\parallel &= -\frac{4}{3}\lim_{\omega \to 0}\frac{1}{\omega}\left[ \text{Im}G^{R}_{P_\perp  P_\parallel}(\omega) + \text{Im}G^{R}_{P_\parallel, P_\parallel}(\omega) \right],\label{eq:kubo2}
\end{align}
where
\begin{equation}
P_{\perp} \equiv\frac{1}{2}T^{a}_{\ a}=\frac{1}{2}(T^{x}_{\ x}+T^{y}_{\ y}), \ \ P_{\parallel} \equiv \frac{1}{2} T^{z}_{\ z}.
\end{equation}

At first sight, the Kubo formulas obtained here seem different from the ones obtained in Ref. \cite{Huang:2011dc}. The reason is that the formulas written in \cite{Huang:2011dc} are in a fully covariant way. However, if we use the following identity
\begin{equation}\label{eq:KuboTrick}
\left\langle \left[\int d^3x T^{00}, A \right]\right\rangle = \langle\left[H, A \right]\rangle = i\left\langle\frac{\partial A}{\partial t}\right\rangle = 0,
\end{equation}
where $A$ is a generic operator and $H$ is the Hamiltonian, we get rid of the term $\hat{\epsilon} \sim T^{00}$ - recall that the mean values $\langle \cdots \rangle$ for the Kubo formulas are aways related with the equilibrium state\footnote{Apparently, however, we found an overall factor disagreement in $\eta_3$ and $\eta_4$, though this will not influence the results since these coefficients vanish trivially for the magnetic brane.}. Furthermore, when we recover isotropy, i.e. $B=0$, the formulas of both bulk viscosities, $\zeta_{\perp}$ and $\zeta_{\parallel}$, return to the well-known isotropic formula. Moreover, due to the structure of the Kubo formulas for the bulk viscosity, we have the relation
\begin{equation}\label{eq:zetas-relation}
\zeta =\frac{2}{3}\zeta_\parallel +\frac{1}{3}\zeta_\perp,
\end{equation}
where $\zeta$ is the isotropic bulk viscosity obtained by the well-known Kubo formula \eqref{eq:Bulk-Kubo-iso}.

Following the usual convention, we define\footnote{Note that this notation is different from the one considered in \cite{Jain:2014vka, Jain:2015txa}.}
\begin{equation}
\eta_{\perp}\equiv \eta_0, \ \ \eta_{\parallel}\equiv \eta_0+\eta_2.
\end{equation}

Another common way to write the formulas for the shear viscosities is
\begin{equation}
\eta_{ijkl} = - \lim_{\omega \to 0} \frac{1}{\omega} \text{Im} \ G_{T_{ij}T_{kl}}^{R} (\omega, \vec{k} = 0) \ \ \text{with} \ i, j,k, l = x, y, z.
\end{equation}
For example, in the above notation the isotropic shear viscosity $\eta_0$ is
\begin{equation}
\eta_0=\eta_{xyxy}=\eta_{\perp}.
\end{equation}

We finish this subsection emphasizing that the Kubo formulas for $\eta_1$, $\eta_3$ and $\eta_4$ vanishes trivially in the context of the magnetic brane. For example, the Kubo formula for $\eta_3$ \eqref{eq:eta3} depends on the operators $P_\perp$ and $T^{xy}$; however, the dual bulk fields of these operators, $h_{xx}$ and $h_{xy}$ respectively, are decoupled on the fluctuated on-shell action, which causes the vanishing of the two-point function.

\subsection{From the kinetic theory}
\label{Chap4.3.1}

Let us mention briefly how we include the effects of an external magnetic field on the framework of the kinetic theory. In this case, the expression for the relativistic transport equation \eqref{eq:Rel-Boltz} is generalized to \cite{Cerci}
\begin{equation}
p^\mu \partial_\mu f + \frac{\partial (fzeF^{\mu\nu} u_\nu)}{\partial p^\mu} =C[f],
\end{equation}
where $ze$ is the charge of the species of the plasma (e.g: quarks), and $F^{\mu\nu}$ is the usual electromagnetic tensor.

Ref. \cite{Tuchin:2011jw} calculated the anisotropic shear viscosities \eqref{eq:eta0}-\eqref{eq:eta4} within the relativistic kinetic theory framework. The final result for the anisotropic viscosities scales with respect to the magnetic field as follows
\begin{equation}
\eta_{1, \, 2} \sim \frac{1}{(zeB)^2}, \ \ \  \eta_{3, \ 4}\sim \frac{1}{zeB}.
\end{equation}

The same result holds for the non-relativistic case done in $§59$ of Ref. \cite{LandauKine}. Furthermore, the author from \cite{Tuchin:2011jw} argues that this could enhance the anisotropic flow $v_2$, though the weakly interacting kinetic model seems somewhat unrealistic for the QGP near the crossover region where $T\sim 150-250 \  \text{MeV}$.

\chapter{The anisotropic shear viscosity from the magnetic brane}
\label{Chap5.0}

This chapter initiates some novel applications of the gauge/gravity duality (see Chapter \ref{Chap 3.0}) for the calculation of observables in strongly coupled non-Abelian plasmas, such as the QGP described in Chapter \ref{Chap1.0}. More specifically, this chapter is concerned with the anisotropic shear viscosity that comes up when we introduce strong magnetic fields - see Chapter \ref{Chap 2} for a detailed discussion about the isotropic viscosity. 

\hspace{1cm}

The strong magnetic fields that are generated in heavy ion collisions (cf. Chap. \ref{Chap4.0}) break the spatial $SO(3)$ rotational symmetry to a $SO(2)$ invariance about the magnetic field axis and this type of magnetic field-induced anisotropic relativistic hydrodynamics has more transport coefficients than the more symmetric case in order to distinguish the dynamics along the magnetic field direction from that on the plane orthogonal to the field. In fact, this means that the number of independent transport coefficients in the shear viscosity tensor $\eta_{ijkl}$ increases from 1 (in the isotropic case) to 5 in the presence of the magnetic field while there are 2 bulk viscosity coefficients \cite{Huang:2011dc,LandauKine,LandauEla,Tuchin:2011jw} as mentioned in the previous chapter. Therefore, one needs to know how this ``Zeeman-like'' splitting of the different viscosity coefficients depends on the external magnetic field to correctly assess the phenomenological consequences of strong fields on the hydrodynamic response of the QGP formed in heavy ion collisions.

Since one no longer has $SO(3)$ invariance, one may expect that some of the different shear viscosities could violate the universal result $\eta/s = 1/(4\pi)$ valid for isotropic Einstein geometries \cite{Buchel:2003tz,Kovtun:2004de}, which would then constitute an example of the violation of the viscosity bound that is of direct relevance to heavy ion collisions. As mentioned in Sec. \ref{Chap3.4}, previous examples involving the violation of the viscosity bound include: anisotropic deformations of $\mathcal{N} = 4$ Super-Yang-Mills (SYM) theory due to a $z$-dependent axion profile \cite{Mateos:2011ix}  computed in \cite{Rebhan:2011vd} where $\eta_{\parallel}/s< 1/(4\pi)$ along the direction of anisotropy; anisotropic holographic superfluids with bulk $SU(2)$ non-Abelian fields which present universality deviation for $\eta_{\parallel}/s$ \cite{Natsuume:2010ky,Erdmenger:2010xm,Erdmenger:2012zu}; and a dilaton-driven anisotropic calculation recently shown in \cite{Jain:2014vka}. We remark, however, that the first examples of viscosity bound violation were found in ($SO(3)$ invariant) theories with higher order derivatives in the gravity dual \cite{Kats:2007mq,Brigante:2007nu,Brigante:2008gz,Buchel:2008vz}.

In this chapter we evaluate two components of the shear viscosity tensor, namely $\eta_{\perp} \equiv \eta_{xyxy}$ and $\eta_{\parallel} \equiv \eta_{xzxz}=\eta_{yzyz}$, in a strongly coupled non-Abelian plasma in the presence of an external magnetic field using the gauge/gravity duality. These calculations are done using the membrane paradigm \cite{Iqbal:2008by,Thorne:1986iy}. The holographic model we consider is simple Einstein gravity (with negative cosmological constant) coupled with a (prescribed) Maxwell field, which correspond to strongly coupled $\mathcal{N} = 4$ SYM subjected to an external constant and homogenous magnetic field \cite{DK1,DK2,DK3}, discussed at lengthy in Sec. \ref{sec:magbranes}. We examine the role played by the anisotropy introduced by the external field searching for a violation of the viscosity bound in $\eta_{\parallel}/s$. A study of the behavior of $\eta_{\parallel}/s$ is also of phenomenological interest for the modeling of the strongly coupled QGP under strong magnetic fields.

This chapter is organized as follows. In Section \ref{sec:shear}, after a preliminary discussion about the computation of $\eta/s$ from the membrane paradigm in isotropic theories, we show that metric fluctuations in this background parallel and transverse to the external magnetic field result in scalar field fluctuations with two different couplings. This result can then be used in the context of the membrane paradigm to evaluate the shear viscosity coefficients $\eta_\perp$ and $\eta_\parallel$. We finish the chapter in Section \ref{sec:conc} with a discussion of our results.

\section{Anisotropic shear viscosity due to an external magnetic field}
\label{sec:shear}

\subsection{The membrande paradigm and the isotropic shear viscosity}

In this subsection we shall complete the discussion initiated in Chapter \ref{Chap 3.0} where we calculated the shear viscosity from various methods. Here we shall discuss how to obtain the shear viscosity using the so-called membrane paradigm; the results presented for the isotropic shear viscosity serve as guidance for the anisotropic calculation. 

Let us start, once more, with linear response theory. The viscosity tensor for an anisotropic theory is given by the Kubo formula
\begin{equation}
\label{eq:sheartensor}
\eta_{ijkl} = - \lim_{\omega \to 0} \frac{1}{\omega} \text{Im} \ G_{T_{ij}T_{kl}}^{R} (\omega, \vec{k} = 0) \ \ \text{with} \ i, j,k, l = x, y, z
\end{equation}
where $G_{T_{ij}T_{kl}}^{R}(\omega, \vec{k})$ is the Fourier space retarded Green's function given by
\begin{equation}
G_{ij,kl}^{R}(\omega, \vec{k}) = -i \int d^4x \, e^{-i k \cdot x} \theta(t)\left\langle \left[ \hat{T}_{ij}(x), \hat{T}_{kl}(0) \right] \right\rangle,
\end{equation}
while $\hat{T}_{ij}$ is the stress energy tensor operator in the quantum field theory.

For an isotropic theory of hydrodynamics in the absence of other conserved currents, there are only two transport coefficients associated with energy and momentum at the level of relativistic Navier-Stokes theory, namely the isotropic shear viscosity $\eta$ and the bulk viscosity $\zeta$. The computation of $\eta$ in strongly coupled gauge theories using the gauge/gravity duality, in the case of isotropic gauge theories with two derivative gravitational duals, gives a universal value \cite{Policastro:2001yc,Kovtun:2004de} reviewed already in Sec. \ref{Chap3.4}
\begin{equation}
\label{eq:etasiso}
\frac{\eta}{s} = \frac{1}{4\pi}.
\end{equation}

A convenient method that can be used to derive this result is the membrane paradigm \cite{Iqbal:2008by}. In this framework, if we want to compute the transport coefficient $\chi$ of a scalar operator $\hat{O}$ given by the Kubo formula
\begin{equation}
\chi = - \lim_{\omega \to 0} \frac{1}{\omega} \text{Im} \ G^{R} (\omega, \vec{k} = 0),
\end{equation}
where $G^R$ is the retarted correlator associated with the scalar operator $\hat{O}$
\begin{equation}
G^{R}(\omega, \vec{k}) = -i \int d^4x \, e^{-i k \cdot x} \theta(t)\langle \left[ \hat{O} (x), \hat{O} (0) \right] \rangle,
\end{equation}
one needs to look for fluctuations $\phi$ of the associated bulk field in dual gravity theory, in accordance with the gauge/gravity dictionary \cite{Gubser:1998bc,Son:2002sd}. In the case that the action for the fluctuations is given by a massless scalar field with an $r$ dependent coupling $\mathcal{Z}(r)$,
\begin{equation}
S_{fluc} = - \int d^5x \, \sqrt{-g} \frac{1}{2 \mathcal{Z}(r)} (\partial \phi)^2 ,
\end{equation}
the transport coefficient $\chi$ is given by the corresponding transport coefficient $\chi_{mb}$ of the stretched membrane of the black brane horizon \cite{Iqbal:2008by}
\begin{equation}
\label{eq:chi}
\chi = \chi_{mb} = \frac{1}{\mathcal{Z}(r_h)}.
\end{equation}
In the case of the isotropic shear viscosity $\eta$, we must consider the fluctuations $h_{xy}$ of the metric component $g_{xy}$ since the energy-momentum tensor operator in the gauge theory $\hat{T}_{\mu \nu}$ is dual to the bulk metric $g_{\mu \nu}$ of the gravity dual. Given that in isotropic backgrounds the mixed fluctuation $h_{x}^y$ can be described as the fluctuation of a massless scalar field with $\mathcal{Z}(r) = 16 \pi G_5$ \cite{Kovtun:2004de}, then $\eta = 1/(16 \pi G_5)$. The universal result in \eqref{eq:etasiso} follows from identifying the entropy density with the area of the horizon via the Bekenstein formula \cite{Bek,Hawk}.

\subsection{Metric fluctuations and anisotropic shear viscosity}
\label{MetricFluAniso}

Let us now consider metric fluctuations around the background \eqref{eq:background}, which is a solution of the Einstein-Maxwell system \eqref{eq:action}. In a fluid with axial symmetry about an axis due to an external magnetic field there are, in principle, 7 independent transport coefficients in the full viscosity tensor $\eta_{ijkl}$ defined in \eqref{eq:sheartensor}, five of which are shear viscosities and the other two bulk viscosities \cite{LandauKine,Huang:2011dc} - the complete discussion regarding the structure of the viscosity tensor was done in Sec. \ref{Chap4.3}. However, as also argued in Sec. \ref{Chap4.3}, out of the five shear viscosities, three of them are identically zero for the class of anisotropic diagonal backgrounds given by Eq. \eqref{eq:background}, which reduces the total number of independent components of the shear tensor from 7 to 4 (anisotropic superfluids have 5 transport coefficients \cite{LandauEla,Erdmenger:2010xm}). Therefore, we end up with the following shear components of $\eta_{ijkl}$,
\begin{equation}
 \eta_{xyxy} = \eta_{\perp}, \ \ \ \text{and} \ \ \ \eta_{yzyz}= \eta_{xzxz} =\eta_{\parallel}\,.
\end{equation}

The magnetic field breaks the $SO(3)$ rotational invariance of background to only a $SO(2)$ rotation invariance about the $z$ axis. Thus, as expected, it is possible to show that linearized $\phi(t,r) = h_x^y(t,r)$ fluctuations obey
\begin{equation}
\label{eq:gravityflucfin}
\delta S = -\frac{1}{32 \pi G_5} \int d^5x \sqrt{-g}\, (\partial \phi)^2,
\end{equation}
which means that the shear viscosity $\eta_{xyxy} \equiv \eta_{\perp}$ is still given by \eqref{eq:etasiso} and this shear coefficient saturates the viscosity bound.

However, $h_{zx}$ (or, equivalently, $h_{zy}$) fluctuations are not protected by the remaining rotation invariance of the background. In fact, in the context of the membrane paradigm, we must first show that the fluctuation $h_{zx}(t,r)$ obeys the equation of a massless scalar field in order to apply \eqref{eq:chi}. However, the coupling in the action may differ from \eqref{eq:gravityflucfin} and, thus, $\eta_{\parallel} \neq \eta_{\perp}$.

Consider then a fluctuation of the form $g_{zx} \to g_{zx} + h_{zx}$ \footnote{One can show that homogeneous fluctuations of the $\mathrm{U(1)}$ bulk field $A_{\mu}$ decouple from the corresponding fluctuations $h_{xy}$ and $h_{zx}$.}. In order to have a scalar-like action with just the kinetic term (and possibly an $r$ dependent coupling), we choose the mode $\psi(t,r) \equiv h_{y}^{z}(t,r) $, rather than $h_z^y $ for example. Inserting this fluctuation into the action and keeping only quadratic terms one can show that
\begin{align}
\delta S = & \frac{1}{16 \pi G_5}  \int \sqrt{-g} \left\{ \psi^2 \left[ \frac{\square p}{f} -\frac{p}{f^2}\square f -\frac{3}{2f^2}\partial_\mu f\partial^\mu p +\frac{3p}{2f^3}(\partial f)^2  \right]  \right. + \nonumber \\
+ & \left[ \frac{2p}{f}\psi\square\psi - \frac{3p}{2f^2}\partial_\mu f\partial^\mu\psi^2+\frac{2}{f}\partial_\mu p\partial^\mu\psi^2 \right] + \nonumber \\
+ & \left[-\frac{3p}{2f}\frac{(\partial_t\psi)^2}{U} + \frac{3p}{2f}U(\partial_r\psi)^2 = \frac{3p}{2f}\partial_\mu\psi\partial^\mu\psi \right] + \\
- & \left.\left[\left(R+\frac{12}{L^2} -F^2 \right)\frac{p}{2f}\psi^2 +\frac{p}{f}F^2\psi^2 \right] \right\}, \nonumber
\end{align}
where the d'Alembertian is 
\begin{equation}
\Box = -\frac{1}{U}\partial_t^2 + U\partial_r^2+\left(U'+\frac{Uf'}{f}+\frac{Up'}{2p}  \right)\partial_r\,.
\end{equation}
Now, using that the trace of the Einstein's equations gives $R + 20/L^2 = F^2/3$ and, integrating by parts the $\psi \Box \psi$ term, we obtain
\begin{align}
\label{eq:maxfluc}
\delta S &  = \frac{1}{16 \pi G_5}  \int d^5x \, \sqrt{-g} \left[- \frac{p}{2f}\partial_\mu\psi\partial^\mu\psi  - \frac{p}{2f^2}\partial_\mu f\partial^\mu\psi^2+\frac{1}{f}\partial_\mu p\partial^\mu\psi^2 + \right. \nonumber \\ & \left. + \psi^2 \left( \frac{\square p}{f} -\frac{p}{f^2}\square f -\frac{3}{2f^2}\partial_\mu f\partial^\mu p +\frac{3p}{2f^3}(\partial f)^2 \right) + \left(\frac{4p}{fL^2}\psi^2+\frac{F^2}{3}\frac{p}{f}\psi^2  \right) -\frac{p}{f}F^2\psi^2  \right].
\end{align}
We now use the unperturbed Einstein's equations. One needs the $zz$ equation
\begin{equation}
\label{eq:einsteinzz}
\frac{4p}{fL^2} = \frac{\square p}{2f} - \frac{(\partial p)^2}{2pf} - \frac{F^2}{3}\frac{p}{f}
\end{equation}
and also the $yy$ equation,
\begin{equation}
\label{eq:einsteinyy}
-\frac{1}{2}\square p +\frac{(\partial p)^2}{2p} = -\frac{4p}{L^2}-\frac{F^2}{3}p.
\end{equation}
Using the $zz$ \eqref{eq:einsteinzz} equation in \eqref{eq:maxfluc} and integrating by parts once again, noting that
\begin{equation}
\frac{1}{f}\partial_\mu p \partial^\mu\psi^2 = \nabla_\mu\left( \frac{\partial^\mu p}{f}\psi^2 \right)+\frac{1}{f^2}\partial_\mu f\partial^\mu p \psi^2 -\psi^2\frac{\square p}{f} \quad \quad \mathrm{and}
\end{equation}
\begin{equation}
-\frac{p}{2f^2}\partial_\mu f\partial^\mu\psi^2 = - \nabla_\mu\left(\psi^2\frac{p}{2f}\partial^\mu f \right) +\frac{\psi^2}{2f^2}\partial_\mu p\partial^\mu f-\frac{p}{f^3}\psi^2(\partial f)^2 +\frac{p}{2f^2}\psi^2 \square f,
\end{equation}
we arrive at
\begin{align}
\label{eq:maxfluc2}
\delta S = & \frac{1}{16\pi G_5}  \int d^5x \, \sqrt{-g}\left[- \frac{p}{2f}\partial_\mu\psi\partial^\mu\psi +\frac{p}{f}\psi^2 \right. \nonumber + \\ & + \left. \frac{p}{f} \psi^2 \left( \frac{1}{2}\frac{\square p}{p} -\frac{1}{2f}\square f  +\frac{1}{2f^2}(\partial f)^2 - \frac{(\partial p)^2}{2p^2} \right) -\frac{p}{f}F^2\psi^2 \right]\,.
\end{align}
Finally, from \eqref{eq:einsteinzz} and \eqref{eq:einsteinyy}
\begin{equation}
\frac{1}{2}\frac{\square p}{p} -\frac{1}{2f}\square f  +\frac{1}{2f^2}(\partial f)^2 - \frac{(\partial p)^2}{2p^2} = F^2,
\end{equation}
one can show that the action for the fluctuations \eqref{eq:maxfluc2} becomes
\begin{equation}
\label{eq:maxfluc3}
\delta S =  - \frac{1}{16\pi G_5}  \int d^5x \, \sqrt{-g}\left(\frac{p(r)}{2f(r)}\partial_\mu\psi\partial^\mu\psi \right).
\end{equation}
Therefore, we have a massless scalar field with an $r$ dependent coupling $\mathcal{Z}(r) = 16\pi G_5f(r)/p(r)$. These functions were found in the previous section to determine the thermodynamic properties of this system and, thus, in the next section we shall evaluate $\eta_\parallel$. 

\subsection{Viscosity bound violation due to an external magnetic field}

\begin{figure}
\centering
  \includegraphics[width=.65\linewidth]{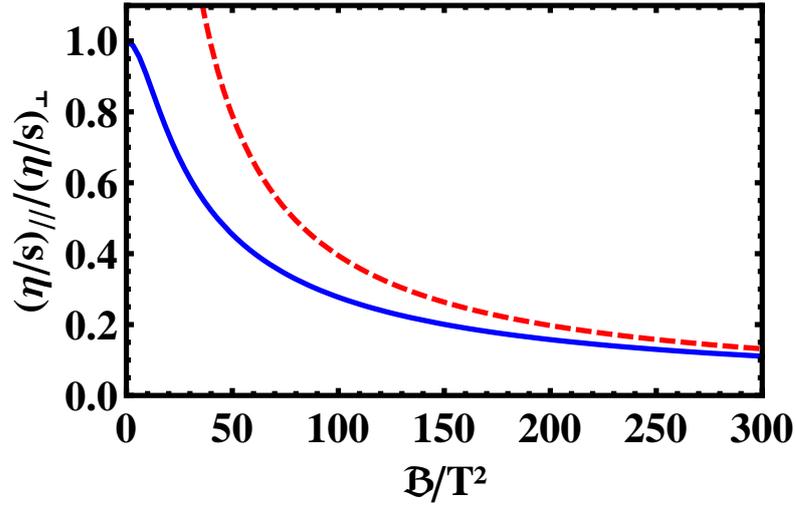}
  \caption{ The ratio of shear viscosities $(\eta/s)_{\parallel}/(\eta/s)_{\perp}$ as a function of $\mathcal{B}/T^2$. The solid blue line is the numerical result from $(\eta/s)_{\parallel}/(\eta/s)_{\perp} = w/v$; the dashed red line is the asymptotic result valid only when $\mathcal{B} \gg T^2$. \eqref{eq:high}}
  \label{fig:etaaniso}      
\end{figure}

From the result of the previous section, it follows that we can also apply the membrane paradigm to \eqref{eq:maxfluc3} to evaluate $\eta_{\parallel}$, using \eqref{eq:chi}. We then have
\begin{equation}\label{eq:EtaParDK}
\frac{\eta_{\parallel}}{s} = \frac{1}{4\pi} \frac{p(r_h)}{f(r_h)}.
\end{equation}
In terms of the numerical, rescaled geometry described in \eqref{eq:resc.metric}, we then obtain
\begin{equation}
\frac{\eta_{\parallel}}{s} = \frac{1}{4\pi} \frac{v(b)}{w(b)}.
\end{equation}
Thus, the ratio $(\eta/s)_{\parallel}/(\eta/s)_{\perp}$ is given by $v/w$. Using this result, we can then evaluate the degree of anisotropy of the shear viscosities as a function of $\mathcal{B}/T^2$; we show the results in Fig.\ \ref{fig:etaaniso}. One can see that for $\mathcal{B}/T^2 \ll 1$, $\eta_{\parallel} \to \eta_{\perp}$, reflecting the fact that at high temperatures we recover the isotropic strongly coupled SYM plasma limit. The asymptotic behavior in the opposite limit, $\mathcal{B}/T^2 \gg 1$, can be understood by looking at the BTZ metric \eqref{eq:BTZ}, which is the relevant geometry in this case. Evaluating $\eta_{\parallel}$ in this limit, one obtains the asymptotic behavior
\begin{equation}
\label{eq:high}
\frac{\eta_{\parallel}}{s} \sim \pi \frac{T^2}{\mathcal{B}}, \ \ \ (\mathcal{B} \gg T^2),
\end{equation}
which is also shown in Fig.\ \ref{fig:etaaniso}. We should note that in this model, $\eta_{\parallel}/s < 1/(4\pi)$ whenever $\mathcal{B}>0$. This gives another example in which the viscosity bound in a gravity dual is violated due to anisotropy. The formula above indicates that $\eta_{\parallel}/s$ can become much smaller than $1/(4\pi)$ for sufficiently strong fields. However, it is conceivable that in this limit other constraints must be imposed to obtain a well defined theory. In fact, it was found in \cite{Brigante:2007nu,Brigante:2008gz} that causality in the gauge theory constituted an important constraint that was used to set a lower value for $\eta/s$ in that particular case involving higher order derivatives in the gravity dual. This matter deserves further study and we hope to address this question in the future. For now, we remark that Ref. \cite{Jain:2014vka} did not detect instabilities for a dilaton driven anisotropy, which has qualitative similarities with the magnetic system treated here.

\section{Conclusions of the chapter}
\label{sec:conc}

Motivated by the recent studies involving the effects of electromagnetic fields on the strongly coupled plasma formed in heavy ion collisions, in this chapter we used the holographic correspondence to compute two anisotropic shear viscosity coefficients of a strongly coupled $\mathcal{N}=4$ SYM plasma in the presence of a magnetic field. As expected, the shear viscosity that describes the dynamics in the plane transverse to the magnetic field, $\eta_\perp$ is not affected by the field and, thus, it still saturates the viscosity bound, i.e., $\eta_\perp/s=1/(4\pi)$. On the other hand, the shear viscosity coefficient along the axis parallel to the external field, $\eta_\parallel$, violates the bound when $\mathcal{B}>0$. In fact, we find $\eta_\parallel/s < 1/(4\pi)$. These results are qualitatively similar to those found in \cite{Rebhan:2011vd} for the case of an anisotropic plasma created by a spatial dependent axion profile \cite{Mateos:2011ix}. Indeed, after the publication of this work, a novel paper appeared \cite{Jain:2015txa} summarizing this ``universal'' behavior of anisotropic branes.   However, the source of anisotropy in our case (the magnetic field) is arguably more directly connected to heavy ion phenomenology than the one used in \cite{Rebhan:2011vd,Jain:2014vka}. 

Plasmas in the presence of magnetic fields usually experience instabilities and it would be interesting to investigate whether there are instabilities induced by strong magnetic fields in the strongly coupled plasma studied in this paper. In fact, one could compute the spectral functions and the quasi-normal modes associated with $\eta_\parallel$ and check if there is any sudden change in their behavior at strong fields. Also, instabilities in homogeneous magnetic media can sometimes be resolved by the formation of magnetic domains and, thus, it would be interesting to investigate whether this is the case for the theory considered in this paper.

Our results for the magnetic field dependence of $\eta_\parallel/s$ show that this ratio only deviates significantly from $1/(4\pi)$ when $\mathcal{B}/ T^2 \gg 1$. Taking the typical temperature at the early stages of heavy ion collisions to be $T \sim 2 m_\pi$, we see that $4\pi\eta_\parallel/s \sim 0.9$ when $\mathcal{B} \sim 40 m_\pi^2$. This value of magnetic field may be too large for heavy ion phenomenology and, thus, our results suggest that anisotropic shear viscosity effects in strongly coupled plasmas are minimal and the isotropic approximation is justified. Alternatively, one could also study the effects of strong magnetic fields on the weak coupling calculations of \cite{Arnold:2000dr,Arnold:2003zc} following the general procedure to compute transport coefficients of relativistic hydrodynamics from the Boltzmann equation proposed in \cite{Denicol:2011fa}.

\chapter{The anisotropic bulk viscosity from magnetic branes}
\label{Chap6.0}

This chapter is the natural sequence of the last one. The main goal here is to calculate the two bulk viscosities, $\zeta_\perp$ \eqref{eq:kubo1} and $\zeta_{\parallel}$ \eqref{eq:kubo2}, that arise when in the presence of an external magnetic field, as explained in Chapter \ref{Chap4.0}, in the context of the magnetic brane solution.

At first one may think that the magnetic brane solution is conformal, with the vanishing of the stress-energy tensor $T^{\mu}_{\mu}=0$, so the both bulk viscosities $\zeta_\parallel$ \eqref{eq:kubo1} and $\zeta_{\perp}$ \eqref{eq:kubo2} would vanish too (see the discussion in Appendix \ref{appB}). However, as shown in Ref. \cite{DK-applications4}, the trace of the stress-energy tensor does not vanish in the presence of an external magnetic field. On contrary, a trace anomaly proportional to $\mathcal{B}^2$ in the rest frame is found.

After we realized that the magnetic brane solution induces a trace anomaly, we computed the bulk viscosities $\zeta_\parallel$ and $\zeta_{\perp}$, whose Kubo formulas are \eqref{eq:kubo1} and \eqref{eq:kubo2}, respectively. It turns out that we found that both are identically zero, which is an odd result, since non-conformal theories usually possess bulk viscosities. Furthermore, as we emphasized in Sec. \ref{Chap1.2.1}, bulk viscosity plays an important role in the dynamics of the QGP, and an understanding of this transport coefficient is certainly important. The anisotropic bulk viscosity, in the context of the dense matter inside neutrons stars was calculated in Ref. \cite{Huang:2009ue}. In Ref. \cite{Agasian:2013wta} we have the calculation of the (isotropic) bulk viscosity of the QGP in a magnetic field. For a calculation of the dependence of the (isotropic) bulk viscosity with respect to the magnetic field in the context of the HRG model, see Ref. \cite{Kadam1}. For holographic calculations of the isotropic bulk viscosity, without a magnetic field, see \cite{gubser1, Benincasa:2005iv,Mas:2007ng, Buchel:2007mf, Buchel:2008uu,Eling:2011ms,Hoyos:2013cba};.

The plan for the rest of this chapter is as follows. Section \ref{Chap6.1} is dedicated to show that the trace of the stress-energy does not vanish for the magnetic brane solution. In Sec. \ref{Chap6.2} we define precisely the dual bulk fields of the relevant operators from the Kubo formulas \eqref{eq:kubo1}-\eqref{eq:kubo2}, necessary to proceed  with a generalization of the recipe developed in Sec. \ref{Chap3.1.2} which gives the two-point functions. Section \ref{Chap6.3} is dedicated to explain how we numerically solved the coupled set of differential equations for the metric fluctuations. We present our conclusions about this study in Sec. \ref{Chap6.4}.

\section{The expectation value of the stress-energy tensor}
\label{Chap6.1}

The bulk viscosity is related with the non-vanishing part of the stress-energy tensor. Consequently, for a conformal theory ($T^{\mu}_{\ \mu}=0$) such as thermal $\mathcal{N}=4$ SYM, the bulk viscosity is identically zero. We shall now compute explicitly how the magnetic field induces an external scale that breaks conformality, albeit in a rather different way than introducing an intrinsic energy scale (e.g. $\Lambda_{QCD}$) for the field theory. The details can be found in \cite{DK-applications4} - and also in \cite{DK-applications5}, which was the first paper to calculate $\langle T_{ij} \rangle$ for the magnetic brane background.

The standard procedure to extract holographically the expectation value of the stress-energy tensor of the field theory requires the near boundary behavior of the metric and the formula for the stress-energy tensor is given by \cite{ren1, ren2, ren3}
\begin{equation}\label{eq:stress-energy-tensor}
\langle T_{ij} \rangle = \frac{1}{4\pi G_5}\left[ g^{(4)}_{ij}-g^{(0)}_{ij}\text{tr}g^{(4)} -(\log(\mu)+\mathcal{C})h^{(4)}_{ij}\right],
\end{equation}
with the metric above being expressed in Fefferman-Graham coordinates\footnote{We use the Fefferman-Graham coordinates \emph{only} in this section of the work. Thus, one should not be confused with the numerical coordinates employed in the rest of the dissertation.}, defined as
\begin{equation}
ds^2_{FG}=\frac{1}{u^2}\left[ du^2+g_{ij}(u,x)dx^idx^j \right]=\frac{1}{u^2}\left[ du^2+g_{tt}(u)dt^2+g_{xx}(u)(dx^2+dy^2)+g_{zz}(u)dz^2\right].
\end{equation}
In any asymptotic AdS$_5$ space one has the following asymptotic behavior of the metric
\begin{equation}
g_{ij}(u\rightarrow 0) = g_{ij}^{(0)}+g_{ij}^{(2)}u^2+g_{ij}^{(4)}u^4+h_{ij}^{(4)}u^4\log u^2+ \mathcal{O}(u^5).
\end{equation}
Note that the conformal boundary is  now located at $u\rightarrow 0$. In \eqref{eq:stress-energy-tensor}, there is also an energy scale $\mu$ and a scheme-dependent renormalization constant $\mathcal{C}$. However, since the trace of $h_{ij}^{(4)}$ vanishes as we will show below, we do not bother to set precise values to $\mu$ and $\mathcal{C}$.\footnote{For instance, in \cite{DK-applications4, DK-applications5}, they choose $\mathcal{C}=-1/4$ whilst the energy scale $\mu$ is proportional to $\sqrt{\mathcal{B}}$ at low temperatures. }

Working out the equations of motion obtained from \eqref{eq:eom1} in the near boundary region, one finds
\begin{align}\label{eq:FG-near-bdry}
g_{tt} &=-1+\left(2g^{(4)}_{xx}+g^{(4)}_{zz}-B^2/6\right)u^4-\frac{B^2}{2}u^4\log u +\mathcal{O}(u^5),\notag \\
g_{xx} &=1+g^{(4)}_{xx}u^4-\frac{B^2}{2}u^4\log u +\mathcal{O}(u^5),\notag \\
g_{tt} &=1+g^{(4)}_{zz}u^4+\frac{B^2}{2}u^4\log u +\mathcal{O}(u^5),
\end{align}
where $g^{(4)}_{xx}$ and $g^{(4)}_{zz}$ are the two free parameters.

Substituting \eqref{eq:FG-near-bdry} into \eqref{eq:stress-energy-tensor}, and taking the trace of the former, we obtain the trace of the stress-energy tensor of the field theory
\begin{equation}\label{eq:Trace-mag-bran}
\langle T^{\mu}_{\ \mu} \rangle=-\frac{\mathcal{B}^2}{24\pi G_{5}},
\end{equation}
where we have used the definition of the physical magnetic field, i.e. $\mathcal{B}=\sqrt{3}B$. The result shows that the trace anomaly is a function of the magnetic fiel, and, since the absence of the bulk viscosity relies on the vanishing of $\langle T^{\mu}_{\ \mu} \rangle$, we expect a non-zero bulk viscosity for the $\mathcal{N}=4$ SYM plasma in presence of a magnetic field.

\section{Dual operators, metric fluctuations and the Green's function}
\label{Chap6.2}

We know, from the holographic calculation of the bulk viscosity done in Sec. \ref{Chap3.5}, that we need to perform small fluctuations of the diagonal part of the metric field in order to extract the holographic bulk viscosity. However, there are some subtleties that arise in the anisotropic case; we have to state precisely what is the field/operator map in this anisotropic case. Thus, we begin by rewriting the Kubo formulas for the anisotropic bulk viscosities,
\begin{equation}\label{eq:kubo-bulk-perp}
\zeta_\perp = -\frac{2}{3}\lim_{\omega \to 0} \frac{1}{\omega}\left[ \text{Im}G^{R}_{P_\perp P_\perp}(\omega) + \text{Im}G^{R}_{P_\parallel, P_\perp}(\omega) \right],
\end{equation}
\begin{equation}\label{eq:kubo-bulk-para}
\zeta_\parallel = -\frac{4}{3}\lim_{\omega \to 0}\frac{1}{\omega}\left[ \text{Im}G^{R}_{P_\perp  P_\parallel}(\omega) + \text{Im}G^{R}_{P_\parallel, P_\parallel}(\omega) \right],
\end{equation}
where
\begin{equation}\label{eq:operators}
P_{\perp} \equiv\frac{1}{2}T^{a}_{\ a}=\frac{1}{2}(T^{x}_{\ x}+T^{y}_{\ y}), \ \ P_{\parallel} \equiv \frac{1}{2} T^{z}_{\ z}.
\end{equation}
and
\begin{equation}\label{eq:Green-def}
G^{R}_{AB}(\omega, \vec{k}) = -i \int d^4x \, e^{-i q \cdot x} \theta(t) \left\langle \left[ \hat{A}(x), \hat{B}(0) \right] \right\rangle,
\end{equation}
as usual.

The question that appears involves identification of the dual bulk fields associated with the operators $P_\perp$ and $P_\parallel$, defined in Eq. \eqref{eq:operators}. Recall that the interaction between gravity and matter (in a linearized level) has the form
\begin{equation}
S_{int} = \frac{1}{2}\int_x \delta g_{\mu\nu}T^{\mu\nu} \supset \int_x \left[ \delta g_{xx}\underbrace{\frac{1}{2}( T^{xx}+T^{yy})}_{=P_\perp}+\delta g_{zz}\underbrace{\frac{1}{2}T^{zz}}_{=P_\parallel} \right],
\end{equation}
where we assumed a SO(2) symmetry to set $\delta g_{yy}=\delta g_{zz}$. Therefore, we see that the dual bulk fields of the operators $P_{\perp}$ and $P_{\parallel}$ are $\delta g_{xx} (=\delta g_{yy})$ and $\delta g_{zz}$, respectively\footnote{The subtlety here is that this is valid as long as one works in the so-called radial gauge $\delta g_{r\mu}=0$, which is the one used in this work. If we choose another gauge, as in \cite{gubser1}, then we ought show the equivalence between this new gauge and the radial gauge by analyzing the behavior of the fluctuations near the boundary to conclude that the gauge choice did not affect their values there.}.

Knowing the dual bulk fields for the retarded Green's functions, such as the ones in Eqs. \eqref{eq:kubo-bulk-perp} and \eqref{eq:kubo-bulk-para}, the holographic dictionary tells us to perform small fluctuations of the metric and the gauge field around the background,
\begin{align}
g'_{\mu\nu} &= g_{\mu\nu} + \delta g_{\mu\nu}, \notag \\
a'_{\mu} &= a_{\mu} + \delta a_\mu, 
\end{align}
where $g_{\mu\nu}$ and $a_{\mu}$ are the background fields in the magnetic brane \eqref{eq:background}. Moreover, we assume that all the fluctuations have an harmonic profile $X(r,t,\vec{x})=\tilde{X}(r)e^{-i\omega t+i\vec{k}\cdot x}$ for $X \in \lbrace\delta g, \delta a\rbrace$. Because the Green's functions require only $\vec{k}\rightarrow\vec{0}$, we can now set $\vec{k}=0$ and recover the SO(2) symmetry on the plane $(x,y)$.

For the specific case of the bulk viscosity, we keep track only of the diagonal fluctuations of the metric, which are the dual bulk field of the operators \eqref{eq:operators} as discussed above. Due to symmetry SO(2) one can show, by looking at the linearized field equations for instance, that the diagonal perturbations decouple from the fluctuations of the U(1) field. This fact is obvious since the fluctuations $\delta g_{xx}+\delta g_{yy}$ are scalars under the SO(2) group, whereas $\delta a_{i}$ are vectors under the SO(2) group. Hence, the only non-vanishing fluctuations here are
\begin{equation}\label{eq:diag-flu}
\delta g_{\mu\nu}=\text{diag}\left\lbrace -U H_{tt},\, \frac{H_{rr}}{U},\, \frac{e^{2V}}{v} H_{xx},\,  \frac{e^{2V}}{v}H_{xx},\,\frac{e^{2W}}{w} H_{zz} \right\rbrace,
\end{equation}
where $H_{tt}\equiv \delta g^t_t$, $H_{rr}\equiv \delta g^r_r$, $H_{xx}\equiv \delta g^x_x$ and $H_{zz}\equiv \delta g^z_z$. Due to SO(2) symmetry, we have already set $H_{xx}=H_{yy}$. Also, we take the radial gauge $H_{r\mu}=0$, which means that $H_{rr}=0$ from now on.

From the Kubo formulas given by \eqref{eq:kubo-bulk-perp} and \eqref{eq:kubo-bulk-para}, we expect somehow a mixing between the fluctuations  $H_{xx}$ and $H_{zz}$ in their equations of motion, and indeed, this is exactly what happens as is shown below. With the mixing of the operators in the bulk, we have to employ the generalization of the holographic recipe to calculate the retarded Green's function \cite{Son:2002sd, Herzogcorre}, which is worked in full detail in \cite{Kaminski}\footnote{Enlightening discussions about the mixing issue can also be found in \cite{Kim:2014bza}.}. As we proceed with the calculations, we shall review the main ingredients of this generalization.

In this scenario, the retarded Green's function will have the following form (schematically)
\begin{equation}
\mathbb{G}^{R}(\omega) \equiv \left(
\begin{array}{ccc}
 G^{R}_{T^{t}_{t}T^{t}_{t}}(\omega) & G^{R}_{T^{t}_{t}P_{\perp}}(\omega)  & G^{R}_{T^{t}_{t}P_{\parallel}}(\omega) \\
G^{R}_{P_{\perp}T^{t}_{t}}(\omega)  &  G^{R}_{P_{\perp}P_{\perp}}(\omega) &  G^{R}_{P_{\perp}P_{\parallel}}(\omega)  \\
 G^{R}_{P_{\parallel}T^{t}_{t}}(\omega)& G^{R}_{P_{\parallel}P_{\perp}}(\omega) & G^{R}_{P_{\parallel}P_{\parallel}}(\omega) \\
\end{array}
\right).
\end{equation}
Then, by looking at the Kubo formulas for the bulk viscosities given by \eqref{eq:kubo-bulk-perp} and \eqref{eq:kubo-bulk-para}, we see that the relevant entries of $\mathbb{G}^{R}(\omega)$ to calculate $\zeta_{\perp}$ are $\left[\mathbb{G}^{R}(\omega)\right]_{xx}=G^{R}_{P_{\perp}P_{\perp}}(\omega)$ and $\left[\mathbb{G}^{R}(\omega)\right]_{zx}=G^{R}_{P_{\parallel}P_{\perp}}(\omega)$. On the other hand, to calculate $\zeta_{\parallel}$, we need the entries $\left[\mathbb{G}^{R}(\omega)\right]_{zz}=G^{R}_{P_{\parallel}P_{\parallel}}(\omega)$ and $\left[\mathbb{G}^{R}(\omega)\right]_{xz}=G^{R}_{P_{\perp}P_{\parallel}}(\omega)$. After this brief digression about the retarded Green's function within the mixing operator framework, we move on to the calculation of the quadratic fluctuated action, which is obtained from plugging \eqref{eq:diag-flu} into \eqref{eq:action}.

For the fluctuated on-shell action, we adopt a notation similar to the one used in \cite{gubser1} to calculate the fluctuated gravity action. But, unlike what occurs in \cite{gubser1}, the fluctuations of the metric are coupled and so are the retarded Green's functions.

Using the the numerical coordinates of the magnetic brane \eqref{eq:resc.metric}\footnote{As discussed before, we do not use the tildes here.}, the fluctuated action acquires the form
\begin{equation}
S = \frac{1}{16\pi G_5}\int_{\mathcal{M}_5}d^5x \mathcal{L}
\end{equation}
\begin{equation}
\mathcal{L}=\hat{\mathcal{L}}+\partial_t\hat{\mathcal{L}}^t +\partial_r\hat{\mathcal{L}}^r,
\end{equation}
where $\hat{\mathcal{L}}$ is the ``improved'' Lagrangian, whose structure is
\begin{equation}
\hat{\mathcal{L}} = \frac{1}{2}\partial_t\vec{H}^T \mathfrak{M}^{tt}\partial_t\vec{H}+\frac{1}{2}\partial_r\vec{H}^T \mathfrak{M}^{rr}\partial_r\vec{H}+\frac{1}{2}\vec{H}^T \mathfrak{M}\vec{H}+\partial_r\vec{H}^T \mathfrak{M}^{r}\vec{H},
\end{equation}
with
\begin{align}\label{eq:bulk-fluc-perp}
\vec{H}&=\left(\begin{array}{c}
H_{tt} \\
H_{xx} \\  
H_{zz}
\end{array} \right), \ \ \
 \mathfrak{M}^{tt} =-\frac{e^{2V+W}}{Uv\sqrt{w}}\left(
 \begin{array}{ccc}
  0 & 0 & 0 \\
  0 & 1 & 1 \\
  0 & 1 & 0 \\
 \end{array}
 \right), \ \ \
 \mathfrak{M}^{rr}= \frac{Ue^{2V+W}}{v\sqrt{w}}\left(
  \begin{array}{ccc}
   0 & 1 & 1/2 \\
   1 & 1 & 1 \\
   1/2 & 1 & 0 \\
  \end{array}
  \right), \notag \\
\mathfrak{M} &= \frac{e^{2V+W}}{12v\sqrt{w}}\left(
\begin{array}{ccc}
 -2 b^2 e^{-4 V}+\Sigma  & 28b^2 e^{-4 V}-2\Sigma  & 2b^2 e^{-4 V}-\Sigma  \\
 28 b^2 e^{-4 V}-2\Sigma  & -48b^2 e^{-4 V} & 28 b^2 e^{-4 V}-2\Sigma  \\
 2 b^2 e^{-4 V}-\Sigma  & 28 b^2 e^{-4 V}-2\Sigma  &  -2b^2 e^{-4 V}+\Sigma \\
\end{array}
\right), \notag \\
\mathfrak{M}^{r} &=\frac{e^{2V+W}}{4v\sqrt{w}}\left(
\begin{array}{ccc}
 U' & -2 U' & -U' \\
 -4 U V' & 0 & -4 U V' \\
 -2 U W' & -4 U W' & 2 U W' \\
\end{array}
\right),
\end{align}
and
\begin{align}
\Sigma &\equiv -3 \left(U' \left(2 V'+W'\right)+2 U V' \left(V'+2 W'\right)-20\right).
\end{align}

The boundary terms, $\partial_t\hat{\mathcal{L}}^t$ and $\partial_r\hat{\mathcal{L}}^r$, will not play any role in the calculation of the bulk viscosity; indeed, they are closely related to the Gibbons-Hawking-York action \cite{ghy1, ghy2} and the counter-term action \cite{ren1, ren2, ren3, ren4, ren5}. Therefore we bypass the calculation of boundary terms because any imaginary part of a retarded Green's function is free from the divergences that we encounter on the on-shell action. For this reason we did not make any attempt to simplify \eqref{eq:bulk-fluc-perp} by introducing boundary terms.

The next step is to complexify the field $\vec{H}(t,r)$, i.e. we promote $\vec{H}^{T}$ to $\vec{H}^{\dagger}$. Now we have a real-valued Lagrangian density, denoted by $\hat{\mathcal{L}}_C$, which is given by a set of complex-valued fields contained in  $\vec{H}$. Noticing that $\partial_t\vec{H}=-i\omega\vec{H}$, we write $\hat{\mathcal{L}}_C$ as
\begin{align}
2\hat{\mathcal{L}}_C = \partial_r\vec{H}^{\dagger}\mathfrak{M}^{rr}\partial_r\vec{H}+\partial_r\vec{H}^{\dagger}\mathfrak{M}^{r}\vec{H}+\vec{H}^{\dagger}\mathfrak{M}^{r \dagger}\partial_r\vec{H}+\vec{H}^{\dagger}\mathfrak{K}\vec{H},
\end{align}
where $\mathfrak{K}=\omega^2\mathfrak{M}^{tt}+\mathfrak{M}$.

We now substitute this new improved Lagrangian into the fluctuated action and integrate by parts. The result is 
\begin{align}
S = \frac{V_3}{32\pi G_5}\int_{-\infty}^{\infty} dt & \left[ \vec{H}^{\dagger}\left(\mathfrak{M}^{rr}\partial_r + \mathfrak{M}^{r}\right) \vec{H} \bigg\vert_{r=r_h}^{r=\infty}  + \right. \nonumber \\ & \left.  + \int dr \vec{H}^{\dagger}\left(-\partial_r\left[\left(\mathfrak{M}^{rr}\partial_r +\mathfrak{M}^{r}\right)\vec{H}\right]+\mathfrak{M}^{r\dagger}\partial_r\vec{H}+\mathfrak{K}\vec{H} \right) \right],
\end{align}
where $V_3=\int dxdydz$ is the 3 volume. Notice that the factor multiplying $\vec{H}^{\dagger}$ in the integrand of $r$ is the equation of motion of $\vec{H}$, which is obtained by varying the action with respect to $\vec{H}^{\dagger}$, so this factor vanishes when we take the on-shell action.

Since we want to work with the fluctuations in momentum space, we Fourier transform them,
\begin{equation}\label{eq:fourier-trans-H}
\vec{H}(t,r)=\int_{-\infty}^{\infty}\frac{d\omega}{2\pi}e^{-i\omega t}\mathfrak{D}(r,\omega)\vec{h}_{\omega}(r).
\end{equation}
The matrix $\mathfrak{D}(r,\omega)$ is introduced in such a way that all the fluctuations $\vec{h}_\omega (r)$ go to constants at the boundary, which will then be regarded as the sources of the operators in the dual field theory. To obtain the explicit form of $\mathfrak{D}(r,\omega)$ we recall that near the boundary the components of $\vec{H}$ goes like, $\vec{H}_{i}(r\rightarrow\infty) \sim r^{-\Delta_{-}^{i}}A_{i}+r^{-\Delta_{+}^{i}}B_{i}$, where $i\in \lbrace t, x, z \rbrace$, and $\Delta_{-}^{i}$ is the smallest exponent. Therefore, $\mathfrak{D} = \text{diag}\left(r^{-\Delta_{-}^{t}},r^{-\Delta_{-}^{x}},r^{-\Delta_{-}^{z}} \right)$.

Inserting \eqref{eq:fourier-trans-H} into the on-shell action, we end up with (after discarding the horizon contribution as required by the holographic prescription)
\begin{equation}\label{eq:bulk1-onshell}
S_{on-shell}=\frac{V_3}{32\pi G_{5}}\int_{-\infty}^{\infty}\frac{d\omega}{2\pi}\,\vec{h}_{-\omega}^T(r)\left(\bar{\mathfrak{M}}^{rr}\partial_r+\bar{\mathfrak{M}}^{r}\right)\vec{h}_{\omega}(r) \bigg\vert_{r\rightarrow\infty},
\end{equation}
where
\begin{align}
\bar{\mathfrak{M}}^{rr} &\equiv \mathfrak{D}^{\dagger}\mathfrak{M}^{rr} \mathfrak{D}, \notag \\
\bar{\mathfrak{M}}^{r} &\equiv \mathfrak{D}^{\dagger}\mathfrak{M}^{r} \mathfrak{D}+\mathfrak{D}^{\dagger}\mathfrak{M}^{rr} \partial_r\mathfrak{D}.
\end{align}

The general solution for the fluctuations $\vec{h}$ can be written as
\begin{equation}\label{eq:gen-solu-1}
\vec{h}_\omega(r) = \mathcal{H}(r,\omega)\vec{J},
\end{equation}
where $\mathcal{H}$ is the matrix formed by three linearly independent (LI) solutions of the fluctuations with the Dirichlet boundary condition,
\begin{equation}\label{eq:Dirichlet}
\lim_{r\rightarrow\infty}\mathcal{H} = \mathsf{1}_{3\times 3},
\end{equation}
which allows us to consider $\vec{J}$ as the value of the fields at the boundary. We can write $\mathcal{H}$ and $\vec{J}$ as
\begin{equation}
\mathcal{H} = \left(\vec{h}_1 \ \vec{h}_2 \ \vec{h}_3 \right), \ \ \ \vec{J} = \left(H_{tt}^{far}, H_{xx}^{far}, H_{zz}^{far}\right)^T,
\end{equation}
with $\vec{h}_1, \vec{h}_2, \vec{h}_3$ being three linearly independent solutions of the system \eqref{eq:httDK} - \eqref{eq:hrtDK} for the fluctuations contained in $\vec{h}_\omega$.

In practice, what we do to achieve the Dirichlet condition \eqref{eq:Dirichlet}, is to write the general solution as
\begin{equation}\label{eq:relation-S-H}
\mathcal{H}(r,\omega)=\mathcal{S}(r,\omega)\mathcal{S}^{-1}(\infty ,\omega),
\end{equation} 
where $S(r,\omega)$ is some generic solution generated by giving initial conditions near the horizon. Notice, from the relation above, that whenever $\text{det}\,\mathcal{S}=0$ the matrix $\mathcal{H}$ is ill-defined and, as detailed in \cite{Kaminski}, this happens when we encounter a quasinormal mode. Another reason to have $\text{det}\,\mathcal{S}=0$, is that we are not really with a complete linear independent (LI) basis of solutions - we shall find this last situation since the incoming wave solution is not enough to construct the complete basis of solutions. We discuss how to circumvent this problem in the next section using the prescription given in \cite{Kim:2014bza} where a similar situation was encountered.

Returning to the fluctuated action and plugging \eqref{eq:gen-solu-1} in \eqref{eq:bulk1-onshell} the on-shell action will be reduced to
\begin{equation}
S_{on-shell}=\frac{V_3}{2}\int_{-\infty}^{\infty}\frac{d\omega}{2\pi}\,\vec{J}^T_{-\omega}\mathcal{F}(\omega,r)\vec{J}_\omega\, \bigg\vert_{r\rightarrow\infty}.
\end{equation}
where we have defined the matrix flux in the same spirit of \cite{Kaminski},
\begin{equation}
\mathcal{F} = \frac{1}{16\pi G_{5}}\left( \mathcal{H}^\dagger\bar{\mathfrak{M}}^{rr}\partial_r\mathcal{H}+\mathcal{H}^{\dagger}\bar{\mathfrak{M}}^{r}\mathcal{H}\right),
\end{equation}
with
\begin{equation}\label{eq:flux-eq}
\partial_r \left(\mathcal{F}-\mathcal{F}^{\dagger}\right) = 0.
\end{equation}
To demonstrate the above property, we have to use the equations of motion for the fluctuations and its hermitian conjugated version, replacing $\vec{H}$ by $\mathcal{H}$,
\begin{equation}\label{eq:eom-fluc}
-\partial_r\left[\left(\mathfrak{M}^{rr}\partial_r +\mathfrak{M}^{r}\right)\mathcal{H}\right]+\mathfrak{M}^{r\dagger}\partial_r\mathcal{H}+\mathfrak{K}\mathcal{H}=0, 
\end{equation}
\begin{equation}\label{eq:eom-fluc-hc}
-\partial_r\left[\partial_r\mathcal{H}^{\dagger}\mathfrak{M}^{rr}+\mathcal{H}^{\dagger}\mathfrak{M}^{r\,\dagger} \right]+\partial_r\mathcal{H}^{\dagger}\mathfrak{M}^{r}+\mathcal{H}^{\dagger}\mathfrak{K}=0.
\end{equation}
Then, by performing the subtraction $\mathcal{H}^{\dagger}$\eqref{eq:eom-fluc}-\eqref{eq:eom-fluc-hc}$\mathcal{H}$, one arrives at the desired result \eqref{eq:flux-eq}. Thus, even in this more general case the imaginary part of the retarded Green's function is closely connected with conserved fluxes from graviton scatterings.

Finally, the expressions above allow us to extract the following Green's function
\begin{equation}\label{eq:green-real}
\mathbb{G}^{R}(\omega) =\lim_{r\rightarrow\infty} \mathcal{F}(r,\omega)=  \frac{1}{16\pi G_{5}}\lim_{r\rightarrow\infty}\left( \bar{\mathfrak{M}}^{rr}\partial_r\mathcal{H}+\bar{\mathfrak{M}}^{r}\right),
\end{equation}
along with the two bulk viscosities given by the formulas \eqref{eq:kubo-bulk-perp} and \eqref{eq:kubo-bulk-para}
\begin{align}
\zeta_{\perp} &=-\frac{2}{3}\lim_{\omega\rightarrow 0}\left[\text{Im}\left(\mathbb{G}^{R}(\omega)\right)_{xx} + \text{Im}\left(\mathbb{G}^{R}(\omega)\right)_{zx}  \right],  \label{eq:GreenBulkPerp} \\  
\zeta_{\parallel} &=-\frac{4}{3}\lim_{\omega\rightarrow 0}\left[\text{Im}\left(\mathbb{G}^{R}(\omega)\right)_{zz} + \text{Im}\left(\mathbb{G}^{R}(\omega)\right)_{xz}  \right]. \label{eq:GreenBulkPara}
\end{align}

Note that the Green's function defined in Eq. \eqref{eq:green-real} is divergent. Nonetheless, this is not a problem since we want its imaginary part, obtained from the conserved flux \eqref{eq:flux-eq}, which is divergence free. Moreover, we have a symmetry between the non-diagonal part of the Green's function, i.e. $\left(\mathbb{G}^{R}(\omega)\right)_{zx}=\left(\mathbb{G}^{R}(\omega)\right)_{xz}$.

Now that all the cards are on the table, it is just a matter of solving numerically the equations of motion for the fluctuations to get the matrix $S(r,\omega)$ and, consequently, the matrix $\mathcal{H}(r,\omega)$ \eqref{eq:relation-S-H} so that we can calculate both bulk viscosities. We do this analysis in the next section.

\section{Towards the numerical solution for $\zeta_\perp$ and $\zeta_\parallel$}
\label{Chap6.3}

The calculation of the retarded Green's function \eqref{eq:green-real} that gives the two bulk viscosities $\zeta_\perp$ and $\zeta_\parallel$ relies on solving the linearized equations of motion of gravity under the fluctuation \eqref{eq:diag-flu}, whose influence in \eqref{eq:green-real} is given by the matrix $\mathcal{H}$. Although we just need the near boundary coefficients of $\mathcal{H}$, these coefficients depend of the whole bulk geometry, such as the near horizon behavior of the fluctuations. For this reason, an analytical solution of $\lim_{r\rightarrow\infty}\mathcal{H}$ can be achieved for some simple cases only (using the matching procedure for instance); hence, as detailed below, we proceed to numerically obtain the retarded Green's function.

The linearized components $(tt)$, $(xx)$, and $(zz)$ of the Einstein-Maxwell field equations \eqref{eq:eom1} with respect to the diagonal fluctuation \eqref{eq:diag-flu}, in the numerical coordinates \eqref{eq:resc.metric} and in momentum space \eqref{eq:fourier-trans-H}, are respectively given by (as usual, the prime denotes $\partial_r$)\footnote{The matrix $\mathfrak{D}(r,\omega)$ in \eqref{eq:fourier-trans-H} is the identity matrix since the fluctuation $\vec{h}$ already goes to a constant vector at the boundary.}
\begin{align}\label{eq:httDK}
 \frac{h_{xx}  \left(8 b^2 U e^{-4 V}+6 \omega
   ^2\right)}{3 U^2}  +\frac{\omega ^2 h_{zz} }{U^2}+\frac{U'
   h_{xx} '}{U}+\frac{U' h_{zz} '}{2 U} 
    +h_{\text{tt}}'\left(\frac{3 U'}{2 U}+2 V'+W'\right)+h_{\text{tt}}''= 0, 
\end{align}
\begin{align}\label{eq:hxxDK}
\frac{h_{xx}  \left(3 \omega
   ^2-16b^2 U e^{-4 V}\right)}{3 U^2} & +V'h_{\text{tt}}'+V'h_{zz}' 
   +h_{xx} ' \left(\frac{U'}{U}+4 V'+W'\right)+h_{xx} ''= 0,  
\end{align}
\begin{align}\label{eq:hzzDK}
\frac{8b^2 e^{-4 V} h_{xx}
   }{3 U}+\frac{\omega ^2 h_{zz}
   }{U^2}+W' h_{\text{tt}}'+2 W'h_{xx}' 
   +h_{zz} '\left(\frac{U'}{U}+2 \left(V'+W'\right)\right)+h_{zz} ''=0,
\end{align}
where $H_{ii}(r,t)=h_{ii}(r)e^{-i\omega t}$. The equations above form a set of three linear independent equations, which is required to solve for the fluctuations $h_{tt}$, $h_{xx}$, and $h_{zz}$ as functions of the radial coordinate. Besides these equations, we also have two constraint equations (CE); the first CE is just the component $(rt)$ of \eqref{eq:eom1},
\begin{align}\label{eq:hrtDK}
h_{xx}  \left(\frac{U'}{U}-2V'\right)+h_{zz}  \left(\frac{U'}{2 U}-W'\right)-2 h_{xx} '-h_{zz}'=0.
\end{align}
The second CE is obtained by combining the $(rr)$ component with the sum \eqref{eq:httDK}+\eqref{eq:hxxDK}+\eqref{eq:hzzDK}. The result is
\begin{align}\label{eq:system-constraint}
\frac{4 h_{xx}  \left(\omega
   ^2-b^2 U e^{-4 V}\right)}{U^2} & +2 h_{\text{tt}}' \left(2
   V'+W'\right)+\frac{2\omega ^2 h_{zz} }{U^2} \notag \\
   &+ h_{xx} ' \left(\frac{2 U'}{U}+4
   \left(V'+W'\right)\right)+h_{zz} ' \left(\frac{U'}{U}+4 V'\right)=0.
\end{align}
In summary, the task now is to solve the system of three coupled differential equations \eqref{eq:httDK}-\eqref{eq:hzzDK}, along with the CE \eqref{eq:hrtDK} and \eqref{eq:system-constraint}. Furthermore, we have 2N boundary conditions to determine, where N is the number of independent fluctuations; thus, we have to fix 6 boundary conditions in our case. 

For the sake of clarity, we shall first analyse the solution of the system \eqref{eq:httDK}-\eqref{eq:hzzDK} in the absence of the magnetic field, $b=0$. In this situation, isotropy is restored ($h_{xx}=h_{zz}$, $W=V$) and the background metric is the usual AdS$_5$-Schwarzschild\footnote{In this case, the background metric is given by: $U(r)=r^2(1-r_h^4/r^4)$ and $e^{2V(r)}=r^2$, with $r_h$ being the radius of the black hole.}. The solution for the fluctuations in this scenario is simple, even analytical, with the expressions for the fluctuations being
\begin{align}\label{eq:sol-flut-iso}
h_{tt}(r;b=0)&= \frac{C_1}{2\sqrt{U}r^3}\left[2(r^4+r_h^4)+r^2\omega^2 \right] +C_2\notag \\
h_{xx}(r;b=0) &= C_1\frac{\sqrt{U}}{r}
\end{align}
where $C_1$ and $C_2$ are two arbitrary constants. However, it happens that the solution \eqref{eq:sol-flut-iso} is a pure gauge solution, i.e. it can be obtained via an infinitesimal diffeomorphism transformation\footnote{In the context of general relativity, given a smooth manifold M, a diffeomorphism transformation is regarded to be a isomorphic map $\phi:M\rightarrow M$ such that the pullback metric $(\phi^\star g)_{\mu\nu}$, and the pullback energy tensor $(\phi^{\star}T)_{\mu\nu}$, satisfy the Einstein's equations if $g_{\mu\nu}$ and $T_{\mu\nu}$ satisfy them.}
\begin{align}
x^\mu &\rightarrow x^\mu+\xi^\mu, \notag \\
g_{\mu\nu} &\rightarrow \pounds_\xi g_{\mu\nu}=g_{\mu\nu}-\nabla_\mu\xi_\nu-\nabla_\nu\xi_\mu,
\end{align}
where $\pounds_\xi$ represents the Lie derivative along $\xi^{\mu}$, and
\begin{equation}
\xi_{\mu}(t,r)= -e^{-i \omega t}\left(-i\omega C_1 \frac{\sqrt{U}}{r}+C_2 ,\frac{C_1}{\sqrt{U}} , 0, 0, 0 \right).
\end{equation}
In this work, we consider that all the diffeomorphism changes will affect only the fluctuated part of the metric $\delta g_{\mu\nu}$. For instance, we can write $\xi_\mu$ as
\begin{equation}
\xi_\mu = \xi^{(0)}_{\mu} + \lambda\xi_{\mu}^{(1)}+\mathcal{O}(\lambda^2),
\end{equation}
where $\lambda$ is the order of the fluctuation. Thus, we consider only $\xi_{\mu}^{(1)}$ different from zero.

It is then obvious that we can perform a gauge choice and eliminate the solution \eqref{eq:sol-flut-iso}. This makes sense because for $b=0$ we have a conformal theory and the bulk viscosity vanishes. Also, the solution \eqref{eq:sol-flut-iso} is the same pure gauge solution obtained in \cite{Policastro1} (with $\vec{q}=\vec{0}$) when studying the so-called sound channel.

Now, let us give a step further and include the magnetic field. In this case, we shall also encounter two LI pure gauge solutions, along with one incoming wave solution near the horizon. The pure gauge solutions should not be discarded otherwise the matrix $\mathcal{S}$ would not be invertible. One of the reasons for the inclusion of pure gauge solutions, as detailed also in \cite{Kim:2014bza}, is because the Einstein's equations have \emph{constraint equations} (CE) (two in our case \eqref{eq:hrtDK}-\eqref{eq:system-constraint}, and one in \cite{Kim:2014bza}) which tie the initial conditions of each fluctuation.\footnote{One way to circumvent these issues is to work with variables invariant under diffeomorphism.}

\begin{figure}
\centering
  \includegraphics[width=.66\linewidth]{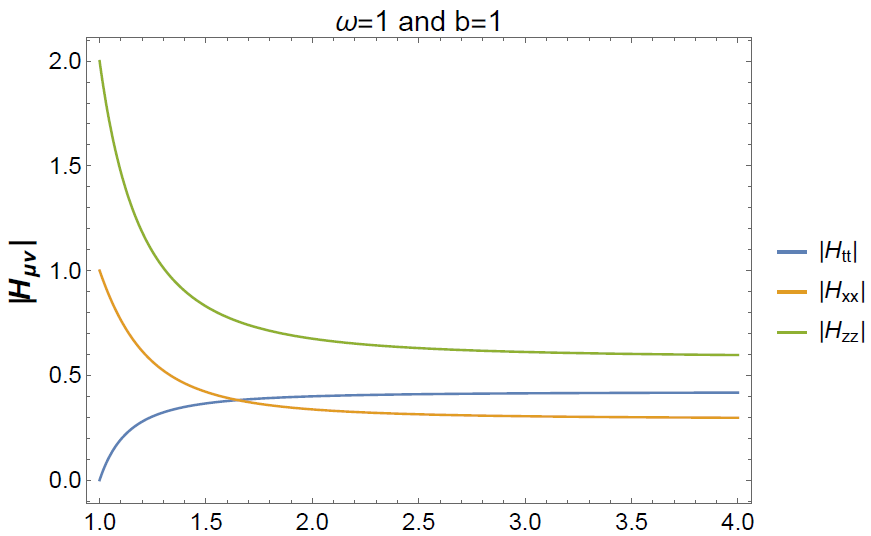}
  \caption{An example for the incoming wave solution of the diagonal fluctuations of the metric.}
  \label{fig:SolH}      
\end{figure}

To exemplify the above statement, let us first analyze the incoming wave solution of the system \eqref{eq:httDK}-\eqref{eq:hzzDK}. Since this system displays of a regular singular point at the horizon, the natural Ansatz for the fluctuations is $(r-1)^{\alpha}F(r)$, where $F(r)$ is a regular function near the horizon ($r_h=1$ in the numerical coordinates) and $\alpha$ is the characteristic exponent. Hence, for $r\rightarrow 1$, we write
\begin{align}\label{eq:inc-wave-sol}
h_{tt}^{inc}(r) &= (r-1)^{\pm\frac{i\omega}{4\pi T}}\left( h_{tt}^{(0)}+h_{tt}^{(1)}(r-1)+\cdots \right), \notag \\
h_{xx}^{inc}(r) &= (r-1)^{\pm\frac{i\omega}{4\pi T}}\left( h_{xx}^{(0)}+h_{xx}^{(1)}(r-1)+\cdots \right), \notag \\
h_{zz}^{inc}(r) &= (r-1)^{\pm\frac{i\omega}{4\pi T}}\left( h_{zz}^{(0)}+h_{zz}^{(1)}(r-1)+\cdots \right).
\end{align}
As we want the incoming wave solution, we select the minus sign in the exponents; this fixes half of the boundary conditions. The coefficients $h_{tt}^{(1)}$, $h_{xx}^{(1)}$, $h_{zz}^{(1)}$, and all the higher coefficients are functions of the parameters $h_{tt}^{(0)}$, $h_{xx}^{(0)}$, $h_{zz}^{(0)}$, and $b$. However, these later parameters are not independent and, in fact, they obey the relations
\begin{align}\label{eq:inc-wave-constraints}
h_{tt}^{(0)}=0, \ \ \ \text{and} \ \ \ 2h_{xx}^{(0)}+h_{zz}^{(0)}=0,
\end{align}
so we have a free parameter, $h_{xx}^{(0)}$ for instance, which is another boundary condition. The other two boundary conditions will come from the pure gauge solutions. So, if we take some value for $h_{xx}^{(0)}$, we determine the next coefficients in terms of $h_{xx}^{(0)}$ and $b$ and use them to seed the \verb+NDSolve+ of the \verb+Mathematica+. However, this procedure will not ensure that $h_{xx}=h_{zz}$ for $b=0$. We circumvent this problem using $h_{xx}^{(0)}\rightarrow h_{zz}^{(0)}+ h_{xx}^{(0)}f(b)$, for some smooth function $f(b)$ such that $f(b=0)=0.$ In Fig. \ref{fig:SolH} we give an example of the incoming wave solution for a generic value of $\omega$ and $b$.

The constraints  \eqref{eq:inc-wave-constraints} shall not allow us to write the general solution only in terms of the incoming wave solution \eqref{eq:inc-wave-sol}. To see this, we write the general solution (near the horizon) as 
\begin{equation}
\mathcal{S} = \left(\vec{h}_1 \ \vec{h}_2 \ \vec{h}_3 \right) = \left(\begin{array}{ccc}
h_{tt}^{I \,(0)} & h_{xx}^{I \,(0)} & h_{zz}^{I \,(0)} \\ 
h_{tt}^{II \,(0)} & h_{xx}^{II \,(0)} & h_{zz}^{II \,(0)} \\ 
h_{tt}^{III \,(0)} & h_{xx}^{III \,(0)} & h_{zz}^{III \,(0)}
\end{array} \right),
\end{equation}
where $I$, $II$ and $III$ denotes three solutions with distinct $h_{xx}^{(0)}$. Clearly the determinant of the above matrix is zero if we use the constraints \eqref{eq:inc-wave-constraints} since the lines are proportional to each other. Thus, as done in \cite{Kim:2014bza}, we resort to the pure gauge solutions to complete the solution basis.

The choice of the radial gauge, $H_{\mu r}=0$ does not fix the gauge completely and we still have a residual gauge freedom. Demanding the condition $H_{\mu r}=0$, we can only perform a diffeomorphism transformation if
\begin{equation}\label{eq:Killing eq}
\nabla_r\xi_\mu + \nabla_\mu\xi_r=0.
\end{equation}
Therefore, the pure gauge solutions must come from the Killing equation \eqref{eq:Killing eq} in order to satisfy the initial radial gauge choice.\footnote{We could also derive these pure gauge solutions in a similar way to the incoming wave solution \eqref{eq:inc-wave-sol}. Actually, this was done for the case where $b=0$.}  The one form $\xi_\mu$ which  satisfies \eqref{eq:Killing eq} is given by
\begin{equation}
\xi_\mu = e^{-i\omega t}\left( i\omega U C_1 \left(\int \frac{1}{U(r)^{3/2}} \,
   dr\right)+C_2,\frac{C_1}{\sqrt{U}},0,0,0 \right),
\end{equation}
which produces the following pure gauge solutions
\begin{align}\label{eq:pure-gauge-sol}
h_{tt}(r)&= -\frac{2 C_1  \omega ^2 U \left(\int \frac{1}{U^{3/2}} \, dr\right)-C_1 
   \sqrt{U} U'}{2 U}+C_2, \notag \\
h_{xx}(r) &= C_1 \sqrt{U} V', \notag \\
h_{zz}(r) &= C_1 \sqrt{U} W',
\end{align}
where $C_1$ and $C_2$ are the two last boundary conditions. Also, notice that  \eqref{eq:pure-gauge-sol} reduces to \eqref{eq:sol-flut-iso} for $b=0$.

We end up with three free parameters: $h_{xx}^{(0)}$, $C_1$, and $C_2$. Then, to generate the three LI solutions for each value of $b$, we take:
\begin{align}
& h_{xx}^{(0)}=1, C_1=0, C_2=0 \Rightarrow \vec{S}_1 = \left(\begin{array}{c}
h_{tt}^{inc}(r) \\ 
h_{xx}^{inc}(r) \\ 
h_{zz}^{inc}(r)
\end{array} \right),\label{eq:b1}  \\
& h_{xx}^{(0)}=0, C_1=1, C_2=0 \Rightarrow \vec{S}_2 = \left(\begin{array}{c}
-\frac{2  \omega ^2 U \left(\int \frac{1}{U^{3/2}} \, dr\right)- 
   \sqrt{U} U'}{2 U} \\ 
\sqrt{U} V' \\ 
\sqrt{U} W'
\end{array} \right), \label{eq:b2} \\
& h_{xx}^{(0)}=0, C_1=0, C_2=1 \Rightarrow \vec{S}_3 = \left(\begin{array}{c}
1 \\ 
0 \\ 
0
\end{array} \right).\label{eq:b3}
\end{align}

We are close now to obtain the Green's functions \eqref{eq:GreenBulkPerp} and \eqref{eq:GreenBulkPara}. With the above set of LI solutions, we construct the general solution $\mathcal{S}=(\vec{S}_1 \ \vec{S}_2 \ \vec{S}_3)$ using Eqs. \eqref{eq:b1}, \eqref{eq:b3}, and \eqref{eq:b3}. Thus, we are able to calculate the matrix $\mathcal{H}$ via the relation \eqref{eq:relation-S-H}, the missing ingredient to obtain $\mathbb{G}^R$ in \eqref{eq:green-real}.

However, the numerical results led to
\begin{equation}
\text{Im}\left(\mathbb{G}^{R}(\omega)\right)_{xx} = \text{Im}\left(\mathbb{G}^{R}(\omega)\right)_{zz} = - \text{Im}\left(\mathbb{G}^{R}(\omega)\right)_{xz}.
\end{equation}
The obvious consequence is the vanishing of both bulk viscosities, $\zeta_\perp$ and $\zeta_\parallel$. This result leads us to conjecture that the trace anomaly induced by the magnetic field does not give rise to a bulk viscosity. A basic numerical check, in order to detect some inconsistency, was to confirm the Ward identity, i.e. $k^{\mu} G_{\mu\nu\alpha\beta}^R(\omega)=\omega G_{t\nu\alpha\beta}^R(\omega)=0\Rightarrow G_{t\nu\alpha\beta}^R(\omega)=0$, which is consistent with our results.

\section{Conclusions of the chapter}
\label{Chap6.4}

The presence of a constant magnetic field in a strongly coupled non-Abelian plasma breaks the original SO(3) rotation symmetry down to SO(2). The consequence, as discussed in detail in Chapter \ref{Chap4.0} is the proliferation of the viscosity coefficients; in particular we have two bulk viscosities, $\zeta_\perp$ and $\zeta_\parallel$, which are related to the trace of the stress-energy tensor. 

Using the magnetic brane set-up, which is the dual theory of magnetic $\mathcal{N}=4$ SYM theory, we performed a calculation of the two anisotropic bulk viscosities. We found that both viscosities vanish, even though there is a trace anomaly induced by the magnetic field. Another approach that we could have used is the calculation of the quasinormal modes, in which one defines (on the AdS side) gauge invariant fields and calculate their eigenvalues; by comparing these eigenvalues with the modes of the hydrodynamic theory, we can extract the bulk viscosity \cite{Kovtun:2005ev}. 

We remark once again, though, that the tracelessness of the stress-energy tensor does not guarantee existence of a bulk viscosity. Indeed, the Bag Model has a non-vanishing trace for the stress-energy tensor, i.e. $T^{\text{Bag} \ \mu}_{ \ \ \mu}=\varepsilon-3P=4B_{\text{Bag}}$  where $B_{\text{Bag}}$ is the bag energy, and the bulk viscosity for this theory does vanish. It is also important to note that the trace \eqref{eq:Trace-mag-bran} is the same for a vacuum geometry - again, this is also true for the Bag Model; consequently, this trace is insensitive about temperature effects, and it is the temperature that gives sense to many-body physics phenomena and their respective properties, such as transport coefficients. 

A similar result is found in the context of kinetic theory in Abelian plasmas. Although four of the five shear viscosities acquire a magnetic field dependent profile, the two bulk viscosities are zero if they vanish in the limit $B=0$, which is the case of ideal gases treated in Sec. \ref{Chap 2.2}. For the complete discussion see $§58$ and $§59$ of Ref. \cite{LandauKine}.

\chapter{The anisotropic heavy quark potential in strongly coupled $\mathcal{N}=4$ SYM in  a magnetic field}
\label{Chap7.0}

The holographic correspondence \cite{Maldacena:1997re,Gubser:1998bc,Witten:1998qj} is a powerful nonperturbative tool that has been widely used to investigate the properties of strongly coupled non-Abelian gauge theories with a large number of colors. In fact, its relevance to the physics of the strongly-coupled quark gluon plasma formed in relativistic heavy ion collisions  became evident after the discovery \cite{Kovtun:2004de} that strongly coupled (spatially isotropic) plasmas that can be described by holographic methods behave as nearly perfect fluids where the shear viscosity to entropy density ratio, $\eta/s$, is close to the estimates obtained within relativistic hydrodynamic modeling of heavy ion collisions. Other applications of the correspondence to the physics of the Quark-Gluon Plasma (QGP) have been reviewed in \cite{solana}. 

Given the recent interest regarding the effects of strong electromagnetic fields in the physics of strong interactions, as discussed in Chapter \ref{Chap4.0}, it is natural to investigate whether holography can also be as insightful in this case. For instance, it has been shown in Chapter \ref{Chap5.0} that in a presence of a magnetic field, $\mathcal{B}$, the shear viscosity tensor of strongly coupled $\mathcal{N}=4$ SYM theory becomes anisotropic and the shear viscosity coefficient in the direction of the magnetic field violates the $\eta/s = 1/(4\pi)$ result \cite{Kovtun:2004de}.

Motivated by the recent lattice work on the effects of strong external (Abelian) magnetic fields on the QCD heavy quark potential at zero temperature done in \cite{Bonati:2014ksa}, in this chapter we study the effect of a constant magnetic field on the heavy quark potential in strongly coupled $\mathcal{N}=4$ SYM theory both at zero and nonzero temperature $T$. The magnetic field distinguishes the different orientations of the $Q\bar{Q}$ pair axis with respect to direction of the magnetic field (defined here to be $z$ axis) and, thus, there is now a perpendicular potential, $V_{Q\bar{Q}}^{\perp}$, for which the pair's axis is on the transverse plane $xy$ and also a parallel potential, $V_{Q\bar{Q}}^{\parallel}$, for which the $Q\bar{Q}$ axis coincides with that of the magnetic field. Clearly, other orientations are possible but here we shall focus only on these two cases.

These heavy quark potentials (both at zero and nonzero temperature) in the gauge theory are defined in this work via their corresponding identification involving the appropriate Wilson loops
\begin{equation}
\lim_{\mathcal{T}\to\infty}\langle W(\mathcal{C}_{\parallel})\rangle  \sim e^{ i V_{Q\bar{Q}}^{\parallel} \mathcal{T}   }\, \qquad \lim_{\mathcal{T}\to\infty}\langle W(\mathcal{C}_{\perp})\rangle  \sim e^{ i V_{Q\bar{Q}}^{\perp} \mathcal{T}   }\,,
\end{equation} 
where $\mathcal{C}_{\parallel}$ is a rectangular time-like contour of spatial length $L_\parallel$ in the $z$ direction and extended over $\mathcal{T}$ in the time direction while $\mathcal{C}_{\perp}$ is the corresponding contour of spatial length $L_\perp$ in the $x$ direction\footnote{Due to the matter content of $\mathcal{N} = 4$ SYM theory, the Wilson loop also contains the coupling to the six $SU(N)$ adjoint scalars $X^I$. In this work we shall neglect the dynamics of the scalars and the holographic calculation of the Wilson loop is defined in 5 dimensions.}. We shall follow D'Hoker and Kraus' construction of the holographic dual of $\mathcal{N}=4$ SYM theory in the presence of a magnetic field \cite{DK1,DK2,DK3} and perform the calculations of the loops defined above in the background given by the asymptotic $AdS_5$ holographic Einstein-Maxwell model to be reviewed below. 

This chapter is organized as follows. In the next section we review the necessary details about the holographic dual of $\mathcal{N}=4$ SYM theory in the presence of a magnetic field at zero temperature and perform the calculation of the parallel and perpendicular potentials and forces in this case. The effects of the breaking of $SO(3)$ spatial invariance induced by the magnetic field on the heavy quark potential and the the interquark force at nonzero temperature are studied in Sec.\ \ref{wsec3.1}. Our conclusions of the chapter are presented in Sec.\ \ref{wconclusion} and other minor details of the calculations can be found in the Appendices \ref{AppD} and \ref{AppE}.

\section{The holographic setup at zero temperature}
\label{wsec2}

In this section we review the properties of the asymptotic $AdS_5$ background corresponding to the holographic dual of strongly coupled $\mathcal{N}=4$ SYM theory in a magnetic field worked out by D'Hoker and Kraus in \cite{DK1,DK2,DK3}. We shall focus here on the $T=0$ properties of the model.

The holographic model involves the Einstein-Maxwell action defined and explained in Sec. \ref{sec:magbranes}. However, we have treated only the background metric with temperature, so we still need to construct the metric field for the vanishing temperature case. For such task, we begin by writing the metric Ansatz on the light-cone gauge\footnote{See Appendix \ref{AppD}.} \cite{DK2}
\begin{align}
ds^2=\frac{dr^2}{P^2(r)}+2P(r)dudv+e^{2W(r)}(dx^2+dy^2),\,\,\,F=Bdx\wedge dy,
\label{w2.2}
\end{align}
where the boundary of the asymptotically $AdS_5$ space is located at $r\rightarrow\infty$. A simple gauge choice for the Maxwell field giving the electromagnetic field strength tensor specified above is $A=Bx\,dy$. Maxwell's equations, $\nabla_\mu F^{\mu\nu}=0$, are then automatically satisfied.

The set of linearly independent components of Einstein's equations is given by the $rr$-, $uv$- and $xx$-components of \eqref{eq:eom1}, respectively
\begin{align}
W''+\frac{P''}{2P}+W'\,^2+\frac{P'\,^2}{4P^2}+\frac{P'W'}{P}-\frac{1}{6P^2}\left(12+2B^2e^{-4W}\right) &=0,\label{w2.8}\\
\frac{P''}{2P}+\frac{P'\,^2}{2P^2}+\frac{P'W'}{P}-\frac{1}{3P^2}\left(12+2B^2e^{-4W}\right) &=0,\label{w2.9}\\
W''+2W'\,^2+\frac{2P'W'}{P}-\frac{1}{3P^2}\left(12-4B^2e^{-4W}\right) &=0,\label{w2.10}
\end{align}
where the prime denotes the derivative with respect to the radial direction, $r$.

It is useful to recast the above equations of movement in the following manner
\begin{align}
P''+2P'W'+4P(W''+W'\,^2) &=0,\label{w2.12}\\
\frac{3P^2W''}{2}+2P^2W'\,^2-\frac{P'\,^2}{4}+PP'W'+B^2e^{-4W} &=0,\label{w2.13}\\
(P^2e^{2W})'' &=24e^{2W}\,.\label{w2.14}
\end{align}
We shall use the coupled ODE's (\ref{w2.12}) and (\ref{w2.13}) to obtain the numerical solutions for $W(r)$ and $P(r)$. For this sake, we also need to specify the initial conditions to start the numerical integration of these ODE's. We are going to work with infrared boundary conditions which we shall specify in a moment. First, notice we can formally solve (\ref{w2.14}) for $P^2$ as follows
\begin{align}
P^2(r)=24e^{-2W(r)}\int_0^r d\xi\int_0^\xi d\lambda e^{2W(\lambda)},
\label{w2.15}
\end{align}
where we fixed the integration constants by imposing that in the infrared $P^2(0)=(P^2)'(0)=0$ \cite{DK2}. Besides (\ref{w2.15}), another equation that will be useful in the determination of the parameters of the infrared expansions we shall take below for $W(r)$ and $P(r)$ is given by the combination\footnote{We note that $P(r)$ enters in this equation only through $P^2$ and $(P^2)'=2PP'$, which can be immediately read off from (\ref{w2.15}).} $2[$(\ref{w2.12})$+$(\ref{w2.13})$]$
\begin{align}
3\left[(P^2)'W'+P^2(W''+2W'\,^2)\right]-12+B^2e^{-4W}=0\,.
\label{w2.16}
\end{align}

\begin{figure}[tbp]
\begin{tabular}{cc}
\includegraphics[width=0.45\textwidth]{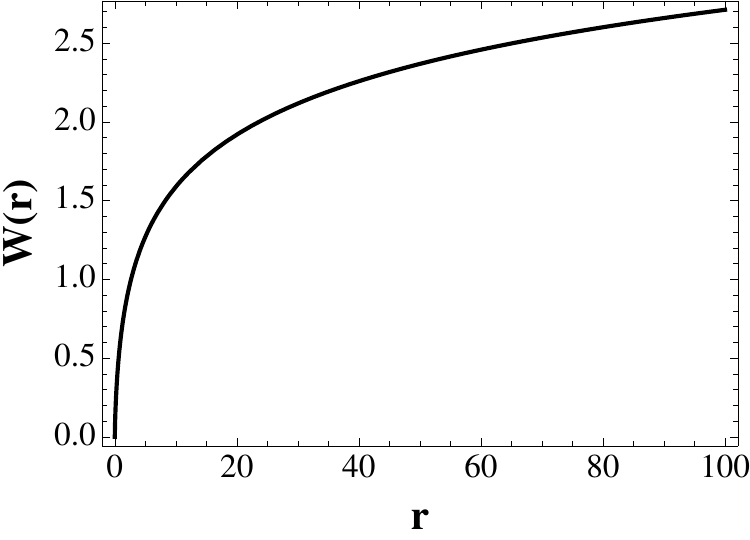} & %
\includegraphics[width=0.45\textwidth]{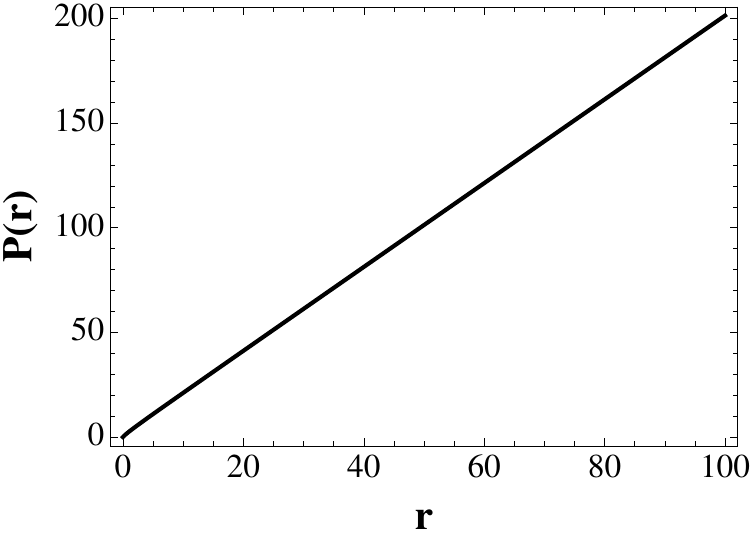} \\
&
\end{tabular}%
\caption{Numerical solution for the functions $W(r)$ and $P(r)$ that appear in the background metric at zero temperature (\ref{w2.2}), which interpolates between $AdS_3\times\mathbb{R}^2$ in the infrared (small $r$) and $AdS_5$ in the ultraviolet (large $r$).}
\label{wfig1}
\end{figure}

Let us now work out the infrared expansions for $W(r)$ and $P(r)$. Following \cite{DK2}, we are interested in numerical solutions of the dynamical ODE's (\ref{w2.12}) and (\ref{w2.13}) that interpolate between $AdS_3\times\mathbb{R}^2$ for small $r$ in the infrared and $AdS_5$ for large $r$ in the ultraviolet. As discussed in \cite{DK1}, this corresponds to a renormalization group flow between a CFT in $(1+1)$-dimensions in the infrared and a CFT in $(3+1)$-dimensions in the ultraviolet, which is the expected behavior of SYM theory in the presence of a constant magnetic field \cite{DK1}. Then, for small $r$ we can take the following infrared expansions
\begin{align}
W(r) &= r^a+\omega r^{2a}+\mathcal{O}(r^{3a}),\label{w2.17}\\
P^2(r) &\approx 12r^2\left[1-2r^a+(2-2\omega)r^{2a}\right]\left[1+\frac{4r^a}{2+3a+a^2} +\frac{2(1+\omega)r^{2a}}{1+3a+2a^2}\right],\label{w2.18}
\end{align}
where (\ref{w2.18}) was obtained by substituting (\ref{w2.17}) into (\ref{w2.15}). Now we substitute (\ref{w2.17}) and (\ref{w2.18}) into (\ref{w2.16}) and set to zero the coefficients of each power of $r$ in the resulting expression, obtaining
\begin{align}
\mathcal{O}(r^0)&:\,\,\,B=2\sqrt{3},\label{w2.19}\\
\mathcal{O}(r^a)&:\,\,\,9a^2+9a-B^2=0\Rightarrow a=a_+\approx 0.758,\label{w2.20}\\
\mathcal{O}(r^{2a})&:\,\,\,\omega\approx -0.634,\label{w2.21}
\end{align}
where we have chosen the positive root in (\ref{w2.20}) in order to obtain a finite $W(0)$ and used (\ref{w2.19}) and (\ref{w2.20}) to obtain (\ref{w2.21}). Substituting (\ref{w2.20}) and (\ref{w2.21}) into (\ref{w2.17}) and (\ref{w2.18}), we determine the first terms in the infrared expansions for $W(r)$, $W'(r)$, $P(r)$ and $P'(r)$, which are enough to initialize the numerical integration of the coupled ODE's (\ref{w2.12}) and (\ref{w2.13}). We start the integration in the deep infrared at some small $r=r_{\textrm{min}}$ and integrate up to some large $r=r_{\textrm{max}}$ near the boundary. The numerical results for the metric functions $W(r)$ and $P(r)$ appearing in (\ref{2.2}) are shown in Fig.\ \ref{wfig1} (these results match those in \cite{DK2}).

The ultraviolet asymptotics for this numerical solution is given by: $\left(e^{2W(r_\textrm{max})},P(r_\textrm{max})\right)$ $\approx\left(1.12365,1.00002\right)\times 2r_\textrm{max}$. Therefore, in order to have an asymptotically $AdS_5$ space at the ultraviolet cutoff, $r=r_\textrm{max}$, we rescale $\left(e^{2W(r)},P(r)\right)\mapsto \left(e^{2\bar{W}(r)},\bar{P}(r)\right)$, where $e^{2\bar{W}(r)}=e^{2W(r)}/1.12365$ and $\bar{P}(r)=P(r)/1.00002$. With this metric rescaling, the physical constant magnetic field in the gauge theory reads\footnote{This rescaling changes the $x$- and $y$-coordinates in (\ref{w2.2}) as follows: $\left(x,y\right)\mapsto\left(x,y\right)/\sqrt{1.12365}$. Furthermore, as discussed after eq. (\ref{eq:Non-Phys-Mag}), the extra factor of $\sqrt{3}$ relates the bulk magnetic field and the magnetic field in the gauge theory.}: $\mathcal{B}=\sqrt{3}B/1.12365\approx 5.34$.

These results were originally obtained in Ref.\ \cite{DK2}. In the following we use them to evaluate the parallel and perpendicular heavy quark potential at zero temperature in the presence of a constant magnetic field.

\section{Holographic Wilson loop $\parallel \mathcal{B}$ at $T=0$}
\label{wsec2.1}

Now we determine the parallel heavy quark potential from the VEV of a rectangular Wilson loop defined by a contour $\mathcal{C}_\parallel$ with its spatial length along the magnetic field direction. We follow the holographic prescription proposed in \cite{maldacena,rey,sonne} (see also \cite{yaffe,jorge} and references therein for more recent discussions) to evaluate the rectangular loops in SYM in the strong t'Hooft coupling limit, $\lambda \gg 1$, with a large number of colors, $N \to \infty$, in terms of a classical Nambu-Goto action in the background discussed in the previous section. 

For this sake, it is better to recast the rescaled version of the metric (\ref{w2.2}) as follows\footnote{See Appendix \ref{AppD}.}
\begin{align}
ds^2=\frac{dr^2}{\bar{P}^2(r)}+\bar{P}(r)(-dt^2+dz^2)+e^{2\bar{W}(r)}(dx^2+dy^2),
\label{w2.22}
\end{align}
where $\bar{P}(r)$ and $\bar{W}(r)$ are the rescaled numerical functions discussed in the previous section. For the sake of notation simplicity, since in the remaining of this section we are going to use only these rescaled functions, we shall omit  from now on the bars in their notation.

The rectangular Wilson loop at the boundary of the asymptotically $AdS_5$ space (\ref{w2.22}) parallel to the magnetic field  is extended along the time direction by $\mathcal{T}$ and has spatial length $L^{\parallel}$, which denotes the heavy quark-antiquark spatial separation in the direction of the magnetic field (we take $\mathcal{T}\gg L^{\parallel}$). We choose to place the probe quark $Q$ at $-\hat{z}L^{\parallel}/2$ and the $\bar{Q}$-probe charge at $+\hat{z}L^{\parallel}/2$. Attached to each of the probe charges in the pair there is a string that sags in the interior of the bulk of the space (\ref{w2.22}). As usual \cite{maldacena,rey,sonne}, in the limit $\mathcal{T}\to \infty$ we consider a classical U-shaped configuration that extremizes the Nambu-Goto action and has a minimum at some value $r_0$ of the radial coordinate in the interior of the bulk. 

The parametric equation of the 2-dimensional string worldsheet swept out in the 5-dimensional bulk is formally given by
\begin{align}
X^\mu &: \textrm{Internal Space} \rightarrow \textrm{Target Space (Bulk)}\nonumber\\
      & \,\,\,\,\,\,\,\,\,\,\,\,\,\,\,\,\,\,\,\,\,\,\,\,\,\,\,\,\, (\tau,\sigma) \mapsto X^\mu(\tau,\sigma) = x^\mu,
\label{w2.23}
\end{align}
and, in static gauge $\tau\rightarrow t,\,\,\, \sigma\rightarrow z$, the target space coordinates over the string worldsheet become
\begin{align}
X^r(t,z)=r,\,\,\,X^t=t,\,\,\,X^x=0,\,\,\,X^y=0,\,\,\,X^z=z,
\label{w2.25}
\end{align}
where $X^r(t,z)=r$ is a constraint equation. For loops where $\mathcal{T} \gg L^{\parallel}$, the static string configuration is invariant under translations in time and one can write $X^r(t,z)=X^r(z)=r$. For the sake of notation simplicity, we take a slight abuse of language and write simply $r=r(z)$ for this constraint equation. Therefore, the static gauge condition can be summarized as follows
\begin{align}
(\tau,\sigma)\rightarrow (t,z)\Rightarrow X^\mu(t,z)=(r(z),t,0,0,z).
\label{w2.26}
\end{align}

The pullback or the induced metric over the string worldsheet in the numerical background (\ref{w2.22}) is defined by
\begin{align}
\gamma_{ab}=g_{\mu\nu}\partial_a X^\mu\partial_b X^\nu,\,\,\,a,b\in\{\tau,\sigma\},
\label{2.27}
\end{align}
with components 
\begin{align}
\gamma_{tz}&=\gamma_{zt}=0,\label{w2.28}\\
\gamma_{tt}&=-P(r(z)),\label{w2.29}\\
\gamma_{zz}&=\frac{\dot{r}^2(z)}{P^2(r(z))}+P(r(z)),\label{w2.30}
\end{align}
where the dot denotes the derivative with respect to $z$. The square root of minus the determinant of the induced metric reads
\begin{align}
\sqrt{-\gamma}=\sqrt{\frac{\dot{r}^2(z)}{P(r(z))}+P^2(r(z))},
\label{w2.31}
\end{align}
and, therefore, the Nambu-Goto action for this $Q\bar{Q}$-configuration is
\begin{align}
S_{\textrm{NG}}=\frac{1}{2\pi\alpha'}\int d^2\sigma \sqrt{-\gamma}= \frac{\mathcal{T}}{2\pi \alpha'} \int_{-L^{\parallel}/2}^{L^{\parallel}/2} dz \sqrt{\frac{\dot{r}^2(z)}{P(r(z))}+P^2(r(z))}\,,
\label{w2.32}
\end{align}
where $\alpha'=\ell_s^2$ and $\ell_s$ is the string length.

Since the integrand in (\ref{w2.32}), $L_{\textrm{NG}}$, does not depend explicitly on $z$, $H_{\textrm{NG}}$ defined below is a \emph{constant of motion} in the z direction
\begin{align}
H_{\textrm{NG}}\equiv \frac{\partial L_{\textrm{NG}}}{\partial\dot{r}}\dot{r}-L_{\textrm{NG}}=\frac{-P^2(r(z))}{\sqrt{\frac{\dot{r}^2(z)}{P(r(z))}+P^2(r(z))}}=C\,.
\label{w2.33}
\end{align}
We may determine $C$ by evaluating (\ref{w2.33}) at the minimum of $r(z)$ where the U-shaped string configuration has a minimum in the interior of the bulk, $r(z=0)=r_0$, where $\dot{r}(0)=0$ and find
\begin{align}
C = \frac{-P^2(r_0)}{\sqrt{P^2(r_0)}}\,.
\label{w2.34}
\end{align}
Substituting (\ref{w2.34}) into the square of (\ref{w2.33}) and solving for $\dot{r}(z)$, one obtains 
\begin{align}
\dot{r}(z)=\frac{dr(z)}{dz}=\sqrt{P^3(r(z))\left[\frac{P^2(r(z))}{P^2(r_0)}-1\right]},
\label{w2.35}
\end{align}
which implies that
\begin{align}
L^\parallel(r_0)= 2\int_{r_0}^{\infty} \frac{dr}{\sqrt{P^3(r)\left[\frac{P^2(r)}{P^2(r_0)}-1\right]}},
\label{w2.36}
\end{align}
where we used that for the U-shaped string configuration described before, $r(\pm L^{\parallel}/2)\rightarrow\infty$, since the probe charges are localized at the boundary of the space (\ref{w2.22}), and we also took into account the fact that the U-shaped contour of integration in the $rz$-plane is symmetric with respect to the $r$-axis, with $r(z=0)=r_0$.

The bare parallel heavy quark potential for this static $Q\bar{Q}$-configuration reads
\begin{align}
V_{Q\bar{Q},\textrm{bare}}^{\parallel}(r_0)=\frac{S_{\textrm{NG}}}{\mathcal{T}}\biggr|_{\textrm{on-shell}} &=\frac{1}{2\pi\alpha'}\int_{-L^{\parallel}/2}^{L^{\parallel}/2} dz \sqrt{\frac{P^4(r(z))}{P^2(r(0))}}\nonumber\\
&=\frac{1}{\pi \alpha'}\int_{r_0}^\infty dr\sqrt{\frac{P(r)}{P^2(r)-P^2(r_0)}},
\label{w2.37}
\end{align}
where we used (\ref{w2.35}) to evaluate the on-shell Nambu-Goto action (\ref{w2.32}). Now we need to regularize (\ref{w2.37}) by subtracting the divergent self-energies of the infinitely heavy probe charges $Q$ and $\bar{Q}$. These contributions correspond to strings stretching from each probe charge at the boundary to the deep interior of the bulk and, in practice, one identifies the ultraviolet divergences to be subtracted by looking at the dominant contribution in the integrand of (\ref{w2.37}) in the limit $r\rightarrow\infty$
\begin{align}
\sqrt{\frac{P(r)}{P^2(r)-P^2(r_0)}}\stackrel{r\rightarrow\infty}{\longrightarrow} \frac{1}{\sqrt{P(r)}}\biggr|_{r\rightarrow\infty}\sim\frac{1}{\sqrt{2r}}\,.
\label{w2.38}
\end{align}
Therefore, the sum of the self-energies of the probe charges is given by
\begin{align}
2\times V_0=2\times\frac{1}{2\pi \alpha'}\int_0^\infty\frac{dr}{\sqrt{2r}},
\label{w2.39}
\end{align}
and the renormalized parallel heavy quark potential is
\begin{align}
V_{Q\bar{Q}}^{\parallel}(r_0)=V_{Q\bar{Q},\textrm{bare}}^{\parallel}(r_0)-2V_0= \frac{1}{\pi \alpha'}\left[ \int_{r_0}^\infty dr\left(\sqrt{\frac{P(r)}{P^2(r)-P^2(r_0)}}-\frac{1}{\sqrt{2r}}\right) - \int_0^{r_0}\frac{dr}{\sqrt{2r}}\right]\,.
\label{w2.40}
\end{align}

In order to obtain the curve $V_{Q\bar{Q}}^{\parallel}(L^{\parallel})$, one may construct a table with pairs of points $(L^{\parallel}(r_0),V_{Q\bar{Q}}^{\parallel}(r_0))$ by taking different values of the parameter $r_0$ in Eqs.\ (\ref{w2.36}) and (\ref{w2.40}), and then numerically interpolate between these points. Before doing this, let us first obtain the corresponding expressions for the perpendicular potential $V_{Q\bar{Q}}^\perp(L^\perp)$. After that, we will make a comparison between the heavy quark potentials and forces obtained in the presence of the magnetic field and the standard isotropic SYM results discussed in \cite{maldacena}.

\section{Holographic Wilson loop $\perp \mathcal{B}$ at $T=0$}
\label{wsec2.2}

Now we consider a rectangular Wilson loop with spatial length $L^\perp$ located in the plane perpendicular to the magnetic field direction at the boundary of the space (\ref{w2.22}). We place the $Q$-probe charge at $-\hat{x}L^{\perp}/2$ and the $\bar{Q}$-probe charge at $+\hat{x}L^{\perp}/2$. For this $Q\bar{Q}$-configuration, it is convenient to define the following static gauge
\begin{align}
(\tau,\sigma)\rightarrow (t,x)\Rightarrow X^\mu(t,x)=(r(x),t,x,0,0)\,.
\label{w2.41}
\end{align}

Following the same general steps discussed in detail in the previous section, one obtains
\begin{align}
L^{\perp}(r_0)&= 2\int_{r_0}^{\infty} \frac{dr}{\sqrt{P^2(r)e^{2W(r)}\left[\frac{P(r)e^{2W(r)}}{P(r_0)e^{2W(r_0)}}-1\right]}},\label{w2.42}\\
V_{Q\bar{Q}}^{\perp}(r_0)&= \frac{1}{\pi \alpha'}\left[ \int_{r_0}^\infty dr\left(\sqrt{\frac{e^{2W(r)}}{P(r)e^{2W(r)}-P(r_0)e^{2W(r_0)}}}-\frac{1}{\sqrt{2r}}\right) - \int_0^{r_0}\frac{dr}{\sqrt{2r}}\right].\label{w2.43}
\end{align}
We note that both the (renormalized) parallel and perpendicular potentials are regularized by the same subtraction term, $2\times V_0$, in Eq.\ (\ref{w2.39}).

\begin{figure}[tbp]
\begin{center}
\begin{tabular}{cc}
\includegraphics[width=0.45\textwidth]{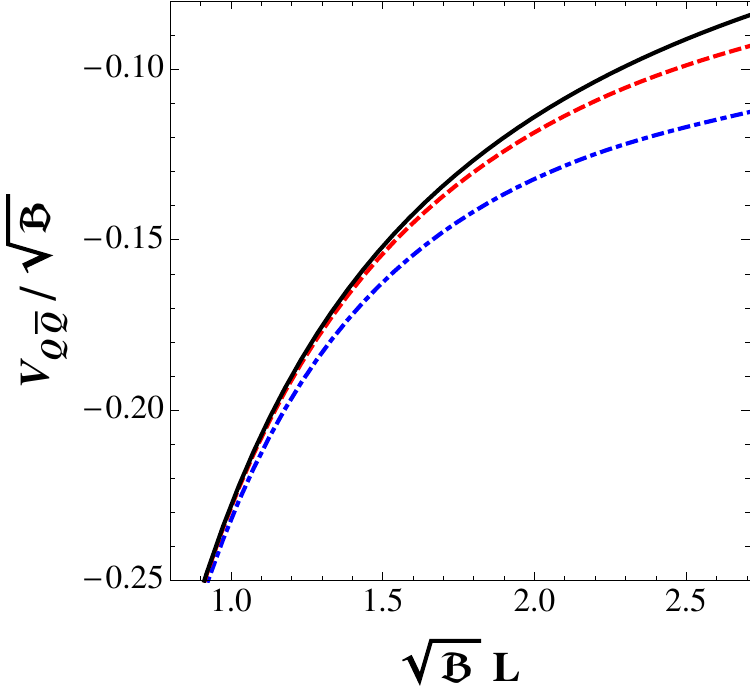} & %
\includegraphics[width=0.45\textwidth]{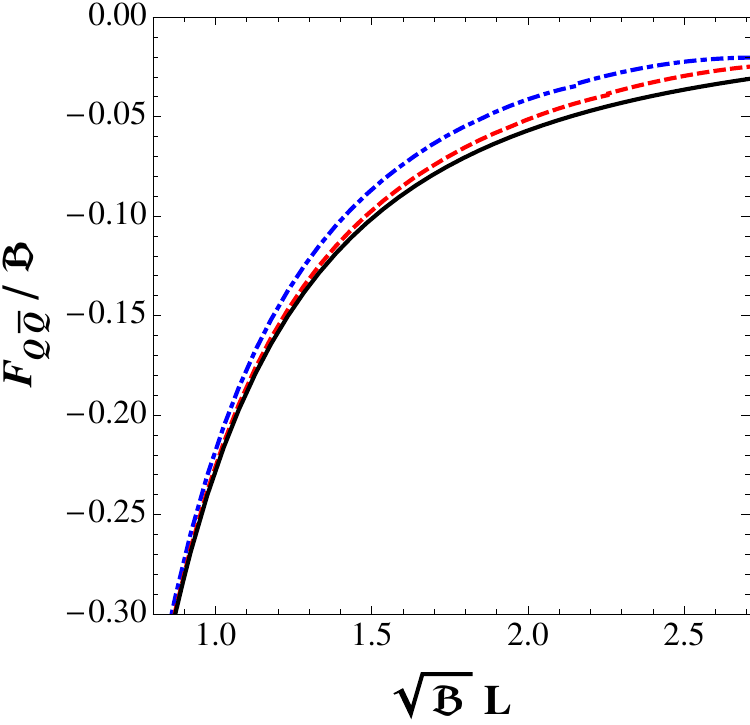} \\
&
\end{tabular}%
\caption{ Anisotropy induced by a magnetic field $\mathcal{B}$ in $\mathcal{N}=4$ SYM at $T=0$ (in this plot $\alpha'=1$) in the heavy quark potential (left panel) and the corresponding force (right panel). The solid black lines denote the isotropic result $\sim -0.228/L$ \cite{maldacena}, the dashed red lines correspond to the perpendicular potential $V_{Q\bar{Q}}^\perp$ and force $F_{Q\bar{Q}}^\perp=-dV_{Q\bar{Q}}^\perp/dL$, and the dotted-dashed blue lines correspond to the parallel potential $V_{Q\bar{Q}}^\parallel$ and force $F_{Q\bar{Q}}^\parallel=-dV_{Q\bar{Q}}^\parallel/dL$.}
\label{wfig2}
\end{center}
\end{figure}

In practice, for the numerical integrations to be performed in Eqs.\ (\ref{w2.36}), (\ref{w2.40}), (\ref{w2.42}), and (\ref{w2.43}), the boundary at $r\rightarrow\infty$ is numerically described by $r_{\textrm{max}}$, in accordance with the numerical solution obtained for the metric (\ref{w2.22}). Our plots for the parallel and perpendicular potentials at $T=0$ are shown on the left panel of Fig.\ \ref{wfig2}. One can see that for the $T=0$ anisotropic holographic setup considered in this section the magnitudes of both the parallel and perpendicular potentials at nonzero $\mathcal{B}$ are enhanced with respect to the $\mathcal{B}=0$ isotropic case (given by $\sim -0.228/L$ \cite{maldacena}), though the parallel potential is more affected by the magnetic field. Also, for very short distances $\sqrt{\mathcal{B}}\, L \ll 1$, both potentials converge to the isotropic potential \cite{maldacena} since the effects from the magnetic field become negligible in this limit. On the right panel of Fig.\ \ref{wfig2} we show the forces associated with these potentials. One can see that the magnetic field generally decreases the magnitude of the attractive force between the quarks in comparison to the isotropic scenario and that the force experienced by the quarks becomes the weakest when the pair axis is parallel to the direction of the magnetic field. 

Moreover, in the absence of any other scale in the theory besides $\mathcal{B}$ (and the interquark distance $L$), the actual value of $\mathcal{B}$ is immaterial. This situation changes when one switches on the temperature and, in this case, there is a new dimensionless scale given by the ratio $\mathcal{B}/T^2$. In fact, we shall see in the next section that in this case one is able to tune the anisotropy in the heavy quark potential by varying the value of the magnetic field.  

\section{Anisotropic heavy quark potential for $T\neq 0$}
\label{wsec3.1}

The holographic calculation of the $T\neq 0$ Wilson loops used in the definition of the parallel and perpendicular potentials follows the same procedure done before in the case where $T=0$. The boundary conditions for each string configuration are the same as before and the overall shape of the string in the bulk is the U-shaped profile \cite{sonne}. The only difference is that when $T\neq 0$ the background metric to be used is the numerically found anisotropic black brane in Eq. \eqref{eq:background} according to the discussion above. Therefore, it is easy to show that 
the interquark separation and (renormalized) heavy quark potential for the parallel case are 
\begin{align}
L^{\parallel}(r_0)&= 2\int_{r_0}^{\infty} \frac{dr}{\sqrt{U(r)e^{2W(r)}\left[\frac{U(r)e^{2W(r)}}{U(r_0)e^{2W(r_0)}}-1\right]}},\label{w3.20}\\
V_{Q\bar{Q}}^{\parallel}(r_0)&= \frac{1}{\pi\alpha'}\left[ \int_{r_0}^\infty dr\left(\sqrt{\frac{U(r)e^{2W(r)}}{U(r)e^{2W(r)}-U(r_0)e^{2W(r_0)}}}-1\right) - \int_0^{r_0}dr\right]\,
\label{w3.21}
\end{align}
while for the perpendicular setup one finds
\begin{align}
L^{\perp}(r_0)&= 2\int_{r_0}^{\infty} \frac{dr}{\sqrt{U(r)e^{2V(r)}\left[\frac{U(r)e^{2V(r)}}{U(r_0)e^{2V(r_0)}}-1\right]}},\label{w3.22}\\
V_{Q\bar{Q}}^{\perp}(r_0)&= \frac{1}{\pi\alpha'}\left[ \int_{r_0}^\infty dr\left(\sqrt{\frac{U(r)e^{2V(r)}}{U(r)e^{2W(r)}-U(r_0)e^{2V(r_0)}}}-1\right) - \int_0^{r_0}dr\right]\,,
\label{w3.23}
\end{align}
where $r_0$ is the point in the bulk where the U-shaped configuration has its minimum. Note that we used the same (temperature independent) subtraction scheme employed at $T=0$ to define the renormalized potentials at finite temperature. These potentials are proportional to the (regularized) area of the Nambu-Goto worldsheet and they are interpreted in the strongly coupled gauge theory as the difference in the total free energy of the system due to the addition of the heavy $Q\bar{Q}$-pair \cite{Noronha:2010hb}.  While one can may argue that one should remove an ``entropy-like" contribution from this free energy difference \cite{mocsy,Noronha:2009ia}, in this work we shall not perform such a subtraction and, for simplicity, we define this free energy difference (which equals the regularized Nambu-Goto action) in each case to be the corresponding heavy quark potential at finite temperature. 

As done before, in the numerical integrations to be performed in Eqs.\ (\ref{w3.20}), (\ref{w3.21}), (\ref{w3.22}), and (\ref{w3.23}), the boundary at $r\rightarrow\infty$ is numerically described by $r_{\textrm{max}}$. At finite temperature, there is a maximum value of $LT$ above which there are other string configurations that may contribute to the evaluation of the Wilson loops at finite temperature \cite{yaffe} besides the semi-classical U-shaped string configuration. This implies that one cannot compute the potentials with the setup described here when $LT$ is large. In fact, one can show that the inclusion of the magnetic field makes this problem worse, as it is shown in Fig.\ \ref{wfig4} below. In this plot we show $LT$ as a function of  the appropriate rescaled horizon $y_H$ (see Appendix \ref{AppE} for the definition of this variable) for the isotropic case (solid black line) and for the parallel (dotted-dashed blue line) and perpendicular (dashed red curve) cases computed using $\mathcal{B}/T^2=50$ (left panel) and $\mathcal{B}/T^2=1000$ (right panel). When $y_h \to 0$ the curves follow the isotropic SYM case while one can see that the maximum of $LT$ is considerably decreased if the magnetic field is sufficiently intense and this effect is stronger for the perpendicular configuration. This implies that the region of applicability of the U-shaped string worldsheet decreases with the applied magnetic field and, thus, other string configurations must be taken into account when computing the string generating functional for sufficiently large $LT$ \cite{yaffe}. This problem was investigated in the case of an isotropic $\mathcal{N}=4$ SYM plasma in \cite{Grigoryan:2011cn} but the extension of these calculations to the anisotropic scenario studied here will be left as a subject of a future study. Nevertheless, for the values of $LT$ in which the U-shaped configuration is dominant our results for the potential are trustworthy and we shall discuss them below.

Also, the fact that the maximum of $LT$ decreases with the applied magnetic field implies that the imaginary part of the potential, computed for instance within the worldsheet fluctuation formalism \cite{Noronha:2009da,Finazzo:2013rqy}, may be enhanced by the magnetic field and this would affect the thermal width of heavy quarkonia in a strongly coupled plasma.

\begin{figure}[tbp]
\begin{center}
\begin{tabular}{cc}
\includegraphics[width=0.45\textwidth]{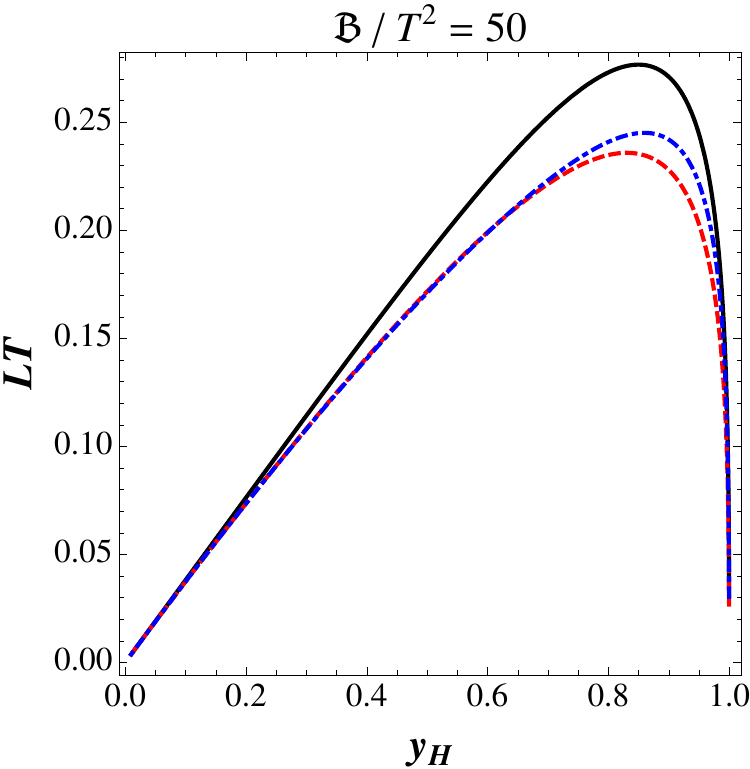} & %
\includegraphics[width=0.45\textwidth]{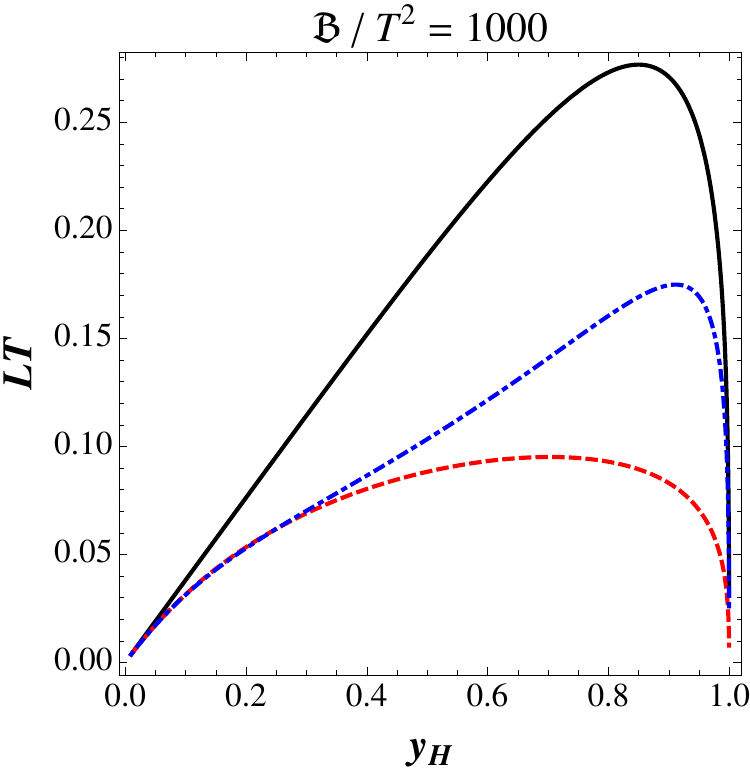} \\
&
\end{tabular}
\caption{ Interquark separation $LT$ versus the rescaled horizon $y_H$ (see Appendix \ref{AppE}). In the left panel $\mathcal{B}/T^2=50$ while for the right panel $\mathcal{B}/T^2=1000$. For both panels the solid black line corresponds to the isotropic SYM case while the dashed red line (dotted-dashed blue line) corresponds to the case of anisotropic SYM with $Q\bar{Q}$ axis perpendicular (parallel) to the magnetic field axis.}
\label{wfig4}
\end{center}
\end{figure}

\begin{figure}[tbp]
\begin{center}
\begin{tabular}{cc}
\includegraphics[width=0.45\textwidth]{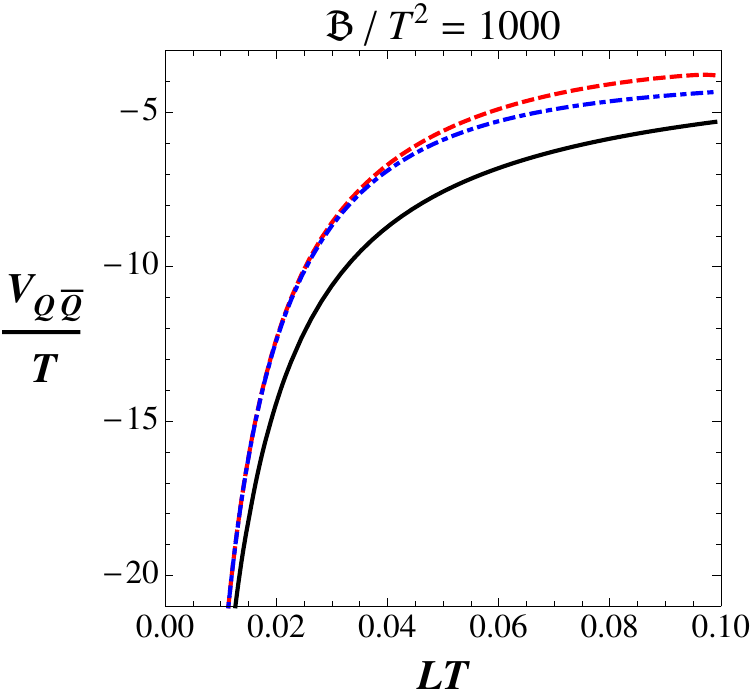} & %
\includegraphics[width=0.45\textwidth]{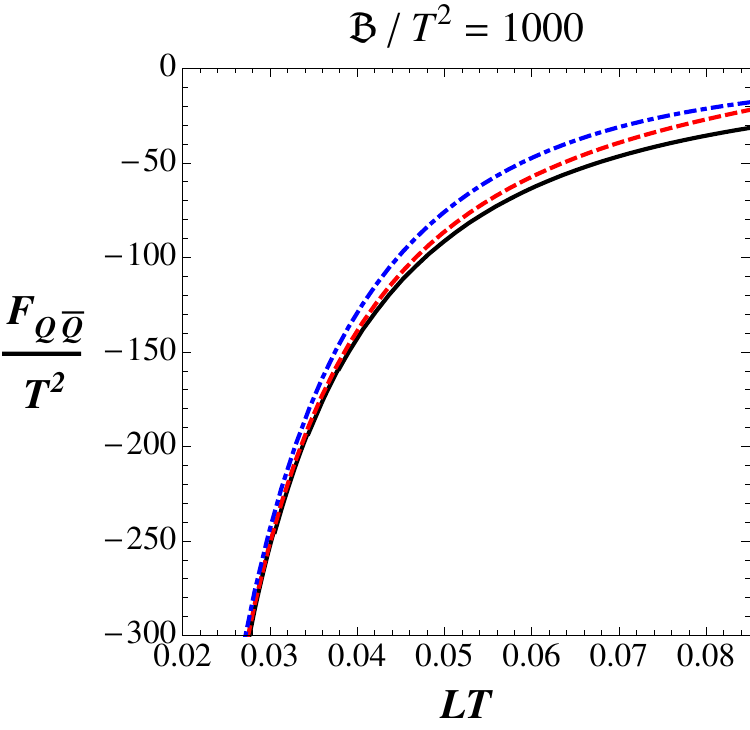} \\
&
\end{tabular}%
\caption{ Anisotropy induced by a strong magnetic field $\mathcal{B}/T^2=1000$ in the heavy quark potential (left panel) and the corresponding force (right panel) experience by a $Q\bar{Q}$ pair in a strongly-coupled $\mathcal{N}=4$ SYM plasma. The solid black lines correspond to the isotropic result $V_{Q\bar{Q}}^{\mathcal{B}=0}$ and isotropic force $F_{Q\bar{Q}}^{\mathcal{B}=0}=-dV_{Q\bar{Q}}^{\mathcal{B}=0}/dL$, the dashed red lines correspond to the perpendicular potential $V_{Q\bar{Q}}^\perp$ and perpendicular force $F_{Q\bar{Q}}^\perp=-dV_{Q\bar{Q}}^\perp/dL$, and the dotted-dashed curves correspond to the parallel potential $V_{Q\bar{Q}}^\parallel$ and force $F_{Q\bar{Q}}^\parallel=-dV_{Q\bar{Q}}^\parallel/dL$. In this plot $\alpha'=1$. }
\label{wfig5}
\end{center}
\end{figure}

The combined effects from nonzero temperature and magnetic field on the heavy quark potential (left panel) and the corresponding force between the quarks (right panel) can be seen in Fig.\  \ref{wfig5}. We found that the anisotropy in the heavy quark potential (and the force) induced by the magnetic field only becomes relevant for very large values of the field. In fact, in Fig.\ \ref{wfig5} we have set $\mathcal{B}/T^2=1000$ to better illustrate the effects. The solid black lines correspond to the isotropic result for the potential $V_{Q\bar{Q}}^{\mathcal{B}=0}$ and its respective force, the dashed red lines correspond to the perpendicular potential $V_{Q\bar{Q}}^\perp$ and perpendicular force, and the dotted-dashed curves correspond to the parallel potential $V_{Q\bar{Q}}^\parallel$ and parallel force (in this plot $\alpha'=1$). By comparing Fig.\ \ref{wfig5} and Fig.\ \ref{wfig2} one can see that, roughly, the overall effect of the temperature is to shift the parallel and perpendicular potentials upwards with respect to the isotropic result. However, the pattern found at $T=0$ regarding the corresponding forces between the quarks is maintained, i.e., the force experienced by the quarks is the weakest when the pair axis is aligned with the magnetic field. Therefore, at least in the case of strongly coupled $\mathcal{N}=4$ SYM, we find that the inclusion of a magnetic field generally weakens the attraction between heavy quarks in the plasma.

\section{Conclusions of the chapter}
\label{wconclusion}

In this chapter we have studied how the inclusion of a constant magnetic field $\mathcal{B}$ affects the interaction between heavy $Q\bar{Q}$ pairs in strongly coupled $\mathcal{N}=4$ SYM theory both at zero and finite temperature by computing rectangular Wilson loops using the holographic correspondence. The magnetic field makes the heavy quark potential and the corresponding force anisotropic and we found that the attraction between the heavy quarks weakens in the presence of the magnetic field (both at $T=0$ and $T\neq 0$). One may see this as indication that in a strongly coupled plasma deconfinement is facilitated by the inclusion of a magnetic field. Although, in practice, in the model considered here this effect only becomes relevant when $\mathcal{B}/T^2$ is extremely large \cite{DK-applications1,DK-applications2}. 

We note that Ref.\ \cite{Chernicoff:2012bu} studied the anisotropy in the heavy quark potential induced by a nontrivial axion field in the bulk \cite{Mateos:2011ix} and found a reduction in the binding energy of the $Q\bar{Q}$ pair. This result is consistent with ours even though the source of anisotropy used in \cite{Chernicoff:2012bu} is different than the one used here (the constant magnetic field). This agreement between different anisotropic holographic models has also been found to hold in the case of transport coefficients since the shear viscosity coefficient along the direction of anisotropy computed in the axion-induced model \cite{Rebhan:2011vd} and in Ref.\ \cite{DK-applications2} display the same qualitative behavior.

One may think that results in this paper give support to the idea that in a strongly coupled plasma deconfinement is facilitated by the inclusion of a magnetic field. However, such a conclusion may only be properly drawn in the case where the underlying gauge theory is not conformal at $T=0$ and $B=0$. In fact, the lattice results of Ref.\ \cite{Bonati:2014ksa} show that in QCD in a magnetic field at $T=0$ the absolute value of the Coulomb coupling in the direction of the magnetic field is enhanced with respect to its vacuum value while this coupling is suppressed in the case perpendicular to the magnetic field. On the other hand, the string tension perpendicular to the field is enhanced with respect to its vacuum value while the string tension parallel to the field is suppressed. This illustrates how complicated the effects of a magnetic field-induced anisotropy can be in a gauge theory with a mass gap.

It would be interesting to study modifications of the current setup and consider systems that are not conformal at $T=0$. For instance, consider a confining theory at $T=0$ with confinement scale $\Lambda$. In this case, there is already a relevant dimensionless ratio $\mathcal{B}/\Lambda^2$ and, for instance, one can study how the mass gap of the theory is affected by the presence of the magnetic field and also how the area law of the rectangular Wilson loop becomes anisotropic and can be used to define a string tension for the heavy quark potential that depends on the angle between the $Q\bar{Q}$ pair and the magnetic field direction. 

Such a model could be easily constructed following the bottom up studies in \cite{ihqcd-1,ihqcd-2,ihqcd-veneziano,GN1,Noronha:2009ud} this time involving a dynamical metric, a scalar field, and a vector field in the bulk. The parallel and perpendicular potentials computed in this non-conformal model could be more easily compared to the lattice QCD study of Ref.\ \cite{Bonati:2014ksa}.Indeed, the next chapter is devoted to build this holographic model that can be fruitful for the phenomenology of the QGP formed in heavy ion collisions.

\chapter{Magnetic non-conformal strongly coupled plasma}
\label{Chap8.0}

A top-down holographic dual for $\mathcal{N}=4$ super Yang-Mills theory (SYM) in the presence of an external constant magnetic field was studied  in Sec. \ref{sec:magbranes} \cite{DK1,DK2,DK3} and calculations for different physical observables in this scenario were carried out, such as the anisotropic viscosity and the anisotropic heavy quark potential - see also \cite{DK-applications1,DK-applications4} for some other applications. However, the QGP formed in heavy ion collisions \cite{reviewQGP1,reviewQGP2} probes the temperature region within which the QCD plasma is highly nonconformal \cite{latticedata1} (when $T \sim 150-300$ MeV). Therefore, in order to make contact with realistic heavy ion collision applications, one needs to develop holographic models that are able to capture some of the relevant aspects of the physics of the strongly coupled QGP near the QCD crossover \cite{Aoki:2006we}. One possible way to accomplish this within holography is to deform the boundary quantum field theory by turning on a dynamical scalar field in the bulk whose boundary value sources a relevant operator in the gauge theory. Near the boundary the scalar field approaches zero and conformal invariance is recovered in the ultraviolet. In the infrared, however, the holographic dual gauge theory generated by such deformation behaves very differently than a conformal plasma and may be tuned to display some of the properties of QCD in the strong coupling regime.

In this chapter we construct a nonconformal anisotropic bottom-up holographic model that is suited for the study of a QCD-like plasma at nonzero magnetic field and vanishing chemical potential(s). Our model is built up on classical nonconformal anisotropic black brane solutions to the Einstein-Maxwell-Dilaton (EMD) model defined with a negative cosmological constant and in the presence of an external constant magnetic field. This constitutes a sequel to the studies of strongly coupled nonconformal plasmas via black brane solutions initiated by \cite{GN1,GN2} in the case of finite temperature, zero magnetic field, and vanishing chemical potential\footnote{These nonconformal solutions can also be adapted to study the vacuum properties of the gauge theory, as recently discussed in \cite{stefanovacuum}. This was studied in detail earlier in \cite{ihqcd-1,ihqcd-2,hot-ihqcd} in the case of similar bottom-up models at zero and finite temperature concerning pure glue Yang-Mills theory.}, which was later extended in \cite{gubser1,gubser2} and also \cite{finitemu} to take into account the presence of a nonzero baryon chemical potential at zero magnetic field\footnote{See also \cite{ihqcd-veneziano} for a bottom-up holographic model at finite temperature, nonzero chemical potential, and zero magnetic field in the Veneziano limit \cite{Veneziano}.}. This type of nonconformal model has been used in the last years to investigate how different observables of phenomenological relevance to the QGP and the physics of heavy ion collisions vary near the QCD crossover transition. In fact, after the original calculations in \cite{GN1,GN2}, which included the evaluation of the bulk viscosity at zero baryon chemical potential and zero magnetic field \cite{GN2}, a series of other quantities were computed within this type of holographic model such as the heavy quark free energy \cite{Noronha:2009ud,Noronha:2010hb}, the energy loss of highly energetic probes \cite{Ficnar:2010rn,Ficnar:2011yj,Ficnar:2012yu}, the Debye screening mass \cite{stefanovacuum}, the electric conductivity \cite{conductivity}, a large set of first and second order viscous hydrodynamic transport coefficients \cite{Finazzo:2014cna}, the spectrum of quasinormal modes \cite{Janik:2015waa} and the thermal photon production rate \cite{yang-muller}. In the context of the holographic models developed in \cite{gubser1,gubser2} and \cite{finitemu} as extensions of the original models \cite{GN1,GN2}, taking into account the presence of a nonvanishing baryon chemical potential, we mention the calculation of the holographic critical point in the $(T,\mu_B)$-plane and the associated critical exponents \cite{gubser1}, the evaluation of the holographic equation of state, the heavy quark drag force, the Langevin diffusion coefficients, the jet quenching parameter, the energy loss of light quarks and an estimate of the equilibration time in the baryon-rich strongly coupled QGP \cite{finitemu}, the evaluation of the bulk viscosity \cite{gubser2}, as well as the baryon susceptibility, baryon conductivity, thermal conductivity, baryon diffusion \cite{baryondiff}, and the thermal photon and dilepton production rates \cite{Finazzo:2015xwa} at finite baryon chemical potential and zero magnetic field. Here we add one more entry to this family of nonconformal black hole solutions by taking into account, for the first time, the presence of a magnetic field in the nonconformal, QCD-like gauge theory.

Our model is a bottom-up holographic setup in which the dilaton potential and the Maxwell-Dilaton gauge coupling are dynamically fixed in order to describe lattice data at zero chemical potential(s) and vanishing magnetic field, which should be contrasted with top-down models coming from compactifications of known string theory solutions. Although in bottom-up models the holographic dual is not precisely known, the fact that these models may be constructed using some phenomenological input from QCD makes it possible that at least part of the physics of the boundary gauge field theory resembles, even at the quantitative level, QCD in the strong coupling limit. Thus, one may regard such constructions as holographic effective theories that are engineered to model some specific aspects of QCD phenomenology, e.g. the correct thermodynamics around the crossover. Once the model parameters are fixed, these theories can be used to make predictions about observables that are currently beyond the scope of lattice calculations, such as most of the second order hydrodynamic coefficients \cite{Finazzo:2014cna}.

This chapter is organized as follows. In Section \ref{sec2} we describe in detail the construction of our holographic model and how the dilaton potential and the Maxwell-Dilaton gauge coupling can be determined by lattice data for the $(2+1)$-flavor lattice QCD equation of state and magnetic susceptibility at zero magnetic field, respectively. With the holographic model parameters fully specified, we proceed in Section \ref{sec3} to obtain the holographic equation of state at nonzero magnetic field and present results for the temperature and magnetic field dependence of the entropy density and the pressure. We find that the deconfinement temperature in our holographic model decreases with an increasing magnetic field, as recently observed on the lattice. Moreover, our model results for the pressure and the crossover temperature are in quantitative agreement with current lattice data up to $eB \lesssim 0.3$ GeV$^2$, which is the relevant range of magnetic fields for heavy ion collisions. In Sec. \ref{SecAnisoShear} we present the calculation of the anisotropic shear viscosity for this EMD model and we compare the result with the one obtained from the magnetic brane in Chapter \ref{Chap5.0}. We present our conclusions in Section \ref{conclusion} where we also point out other applications to be pursued in the near future using the anisotropic nonconformal holographic model developed here.

\section{The holographic model}
\label{sec2}

Assuming as usual that charm quarks are not relevant in the crossover transition, in QCD there are three different chemical potentials associated with three independent globally conserved charges. These different chemical potentials are the three lighter quark chemical potentials $\mu_u$, $\mu_d$, $\mu_s$ or, equivalently, the baryon chemical potential $\mu_B$, the electric charge chemical potential $\mu_Q$, and the strangeness chemical potential $\mu_S$. For each nonzero chemical potential in the gauge theory there must be a nonzero temporal component of the associated gauge field in the bulk. It is also clear that an Abelian magnetic field $B$ in the gauge theory should come from a nonzero spatial component of the gauge potential in the electric charge sector.\footnote{Recall our discussion of the three Cartan subgroups when we defined the magnetic brane in Sec. \ref{sec:magbranes}.}

In the present work we solely focus on the electric charge sector at $B \neq 0$ with $\mu_Q=\mu_B=\mu_S=0$, which may be described by the following EMD action
\begin{align}
S&=\frac{1}{16\pi G_5}\int_{\mathcal{M}_5}d^5x\sqrt{-g}\left[R-\frac{1}{2}(\partial_\mu\phi)^2-V(\phi) -\frac{f(\phi)}{4}F_{\mu\nu}^2\right] +S_{\textrm{GHY}}+S_{\textrm{CT}},
\label{2.1}
\end{align}
where $S_{\textrm{GHY}}$ is the Gibbons-Hawking-York action \cite{ghy1,ghy2} needed to establish a well-posed variational problem with Dirichlet boundary condition for the metric, and $S_{\textrm{CT}}$ is the counterterm action that can be constructed using the holographic renormalization procedure \cite{ren1,ren2,ren3,ren4,ren5}. These two boundary terms contribute to the total on-shell action but not to the equations of motion and, since we shall not need to compute the total on-shell action in the present work, we do not need to worry about their explicit form here. Also, as we are going to discuss in detail in Section \ref{sec2.4}, we shall dynamically fix the gravitational constant $G_5$, the dilaton potential $V(\phi)$, and the Maxwell-Dilaton gauge coupling $f(\phi)$, by solving the equations of motion for the EMD fields with the requirement that the holographic equation of state and magnetic susceptibility at zero magnetic field match the corresponding lattice QCD results.

In \eqref{2.1}, the metric field in the bulk is dual to the stress-energy tensor of the boundary field theory while the dilaton field is introduced in order to dynamically break the conformal symmetry of the gauge theory in the infrared. The Abelian gauge field in the bulk is employed here to introduce an external magnetic field at the boundary, which we take to be constant and uniform in the $\hat{z}$-direction and, as stated before, in the present work we set all the chemical potentials to zero. The constant and uniform magnetic field breaks the $SO(3)$ rotational invariance of the gauge theory down to $SO(2)$ rotations around the $\hat{z}$-axis implying that the Ansatz for the bulk metric must be anisotropic and translationally invariant. Also, at zero temperature this Ansatz must be invariant under boosts in the $(t,z)$-plane though this symmetry is not present at nonzero temperature. Based on these symmetry properties, which are phenomenologically dictated by the corresponding symmetry content present in current lattice QCD calculations defined on the $(T,B)$-plane, we take the following black brane Ansatz for the bulk fields in\footnote{As we shall discuss soon, $\mathcal{B}$ is one of the two initial conditions controlling the temperature and the external magnetic field at the boundary quantum field theory. The other initial condition corresponds to the value of the dilaton field evaluated at the black brane horizon, $\phi_0$. The set of initial conditions $(\phi_0,\mathcal{B})$ is nontrivially related to the thermodynamical pair $(T,B)$ in the gauge theory. In Sections \ref{sec2.3} and \ref{sec2.4} we discuss how one can relate $\mathcal{B}$ to the external magnetic field at the boundary gauge theory, $B$.} \eqref{2.1}:
\begin{align}
ds^2&=e^{2a(r)}\left[-h(r)dt^2+dz^2\right]+e^{2c(r)}(dx^2+dy^2)+\frac{e^{2b(r)}dr^2}{h(r)},\nonumber\\
\phi&=\phi(r),\,\,\,A=A_\mu dx^\mu=\mathcal{B}xdy\Rightarrow F=dA=\mathcal{B}dx\wedge dy,
\label{2.2}
\end{align}
where the radial location of the black brane horizon, $r_H$, is given by the largest root of the equation $h(r_H)=0$ and in our coordinates the boundary of the asymptotically AdS$_5$ spacetime is located at $r\rightarrow\infty$. In \eqref{2.2} we have already fixed a convenient gauge for the Maxwell field, which in the present case is a prescribed non-dynamical field. Also, for simplicity, we shall adopt units where the asymptotic AdS$_5$ radius is equal to one.

Using \eqref{2.2}, the equations of motion obtained from \eqref{2.1} may be expressed as follows
\begin{align}
\phi''+\left(2a'+2c'-b'+\frac{h'}{h}\right)\phi'-\frac{e^{2b}}{h} \left(\frac{\partial V(\phi)}{\partial\phi}+\frac{\mathcal{B}^2e^{-4c}}{2}\frac{\partial f(\phi)}{\partial\phi}\right)&=0,\label{2.3}\\
a''+\left(\frac{14}{3}c'-b'+\frac{4}{3}\frac{h'}{h}\right)a' +\frac{8}{3}a'^2+\frac{2}{3}c'^2+\frac{2}{3}\frac{h'}{h}c'
+\frac{2}{3}\frac{e^{2b}}{h} V(\phi)-\frac{1}{6}\phi'^2&=0,\label{2.4}\\
c''-\left(\frac{10}{3}a'+b'+\frac{1}{3}\frac{h'}{h}\right)c' +\frac{2}{3}c'^2-\frac{4}{3}a'^2-\frac{2}{3}\frac{h'}{h}a'
-\frac{1}{3}\frac{e^{2b}}{h} V(\phi)+\frac{1}{3}\phi'^2&=0,\label{2.5}\\
h''+\left(2a'+2c'-b'\right)h'&=0,\label{2.6}
\end{align}
where the prime denotes a derivative with respect to the radial direction. Using these equations of motions one can also derive a useful constraint
\begin{align}
a'^2+c'^2-\frac{1}{4}\phi'^2+\left(\frac{a'}{2}+c'\right)\frac{h'}{h}+4a'c'
+\frac{e^{2b}}{2h}\left(V(\phi)+\frac{\mathcal{B}^2e^{-4c}}{2}f(\phi)\right)=0.
\label{2.7}
\end{align}
The equation of motion for the Maxwell field is automatically satisfied by the Ansatz \eqref{2.2}. Moreover, $b(r)$ has no equation of motion and, thus, it can be freely chosen to take any value due to reparametrization invariance. In the next Section we specify a subsidiary condition for $b(r)$ that defines a convenient gauge for the metric that will be used in the numerical calculations carried out in the present work.

\subsection{Ultraviolet expansions}
\label{sec2.1}

For the calculation of physical observables in the gauge theory one needs to obtain the near-boundary, far from the horizon expansions for the bulk fields $a(r)$, $c(r)$, $h(r)$, and $\phi(r)$. In the present work, we use the domain-wall gauge defined by the subsidiary condition $b(r)=0$. At the boundary the dilaton field goes to zero in such a way that $V(\phi(r\rightarrow\infty)\rightarrow 0)=-12$ (cf. Eq. \eqref{eq:AsymPotGub}) and $f(0)$ is a finite positive constant\footnote{Note that in \eqref{2.1} the Maxwell-Dilaton gauge coupling $f(\phi)$ plays the role of an inverse effective gauge coupling squared and, therefore, it must correspond to a positive-definite function.}. Also, the metric blackening factor, $h(r)$, must go to a constant at the boundary, which we denote by\footnote{This constant is equal to one in the so-called ``standard coordinates'' of the domain-wall gauge, which we shall discuss soon. Here we are considering general coordinates where this constant may be different than one. We shall also see later how to relate these two sets of coordinates.} $h(r\rightarrow\infty)=h_0^{\textrm{far}}$.

Moreover, since we are interested in asymptotically AdS$_5$ solutions to the equations of motion \eqref{2.3}, \eqref{2.4}, \eqref{2.5}, and \eqref{2.6}, at the boundary one finds $a(r\rightarrow\infty)=c(r\rightarrow\infty)$. In the domain-wall gauge $b(r)=0$, the leading order near-boundary expression for $a(r)$ (and also $c(r)$) is linear in $r$ \cite{gubser1,gubser2} such that at lowest order in $\phi(r\rightarrow\infty)\rightarrow 0$ we may consider the following leading order far from the horizon ultraviolet asymptotics
\begin{align}
V(\phi)\approx -12,\,\,\, f(\phi)\approx f(0),\,\,\,h(r)\approx h_0^{\textrm{far}},\,\,\,
a(r)\approx a_0^{\textrm{far}}+a_{-1}^{\textrm{far}}r,\,\,\, c(r)\approx c_0^{\textrm{far}}+c_{-1}^{\textrm{far}}r,
\label{2.8}
\end{align}
where $a_{-1}^{\textrm{far}}=c_{-1}^{\textrm{far}}$, as discussed above. Indeed, by substituting \eqref{2.8} into the equations of motion and taking the asymptotic limit of large $r$ (where the ultraviolet expansions hold), one concludes that
\begin{align}
a_{-1}^{\textrm{far}}=c_{-1}^{\textrm{far}}=\frac{1}{\sqrt{h_0^{\textrm{far}}}}.
\label{2.9}
\end{align}
In order to obtain the next to leading order term for $h(r)$ and also the first terms for $\phi(r)$ in the ultraviolet expansions for the bulk fields, we consider the first backreaction of the near-boundary fields expressed in \eqref{2.8} and \eqref{2.9} on the equations of motion\footnote{This procedure may be repeated to obtain all the other subleading terms in the ultraviolet expansions. However, we only need the first few terms in these expansions to compute the thermodynamical observables.}. In fact, we first consider the next to leading order near-boundary expansion for the dilaton potential
\begin{align}
V(\phi)\approx -12+\frac{m^2}{2}\phi^2,\,\,\, m^2=-\nu\Delta,
\label{2.10}
\end{align}
where $\Delta$ is the ultraviolet scaling dimension of the gauge invariant operator dual to the bulk dilaton field and we defined $\nu=d-\Delta$, where $d=4$ is the dimension of the boundary. We shall see in Section \ref{sec2.4} that a good description of lattice data can be achieved  by taking $\Delta\approx 3$ ($\nu\approx 1$). One can now show that the far from horizon ultraviolet asymptotics for the bulk fields may be written as
\begin{align}
a(r)&\approx\alpha(r)+\cdots,\nonumber\\
c(r)&\approx\alpha(r)+(c_0^{\textrm{far}}-a_0^{\textrm{far}})+\cdots,\nonumber\\
h(r)&\approx h_0^{\textrm{far}}+h_4^{\textrm{far}}e^{-4\alpha(r)}+\cdots,\nonumber\\
\phi(r)&\approx \phi_Ae^{-\nu\alpha(r)}+\phi_Be^{-\Delta\alpha(r)}+\cdots,
\label{2.11}
\end{align}
where we defined $\alpha(r)=a_0^{\textrm{far}}+r/\sqrt{h_0^{\textrm{far}}}$ while $\cdots$ denotes subleading terms. We note that the ultraviolet asymptotics \eqref{2.11} are in agreement with our numerical solutions. By comparing these numerical solutions to \eqref{2.11} one can determine the ultraviolet coefficients $a_0^{\textrm{far}}$, $c_0^{\textrm{far}}$, $h_0^{\textrm{far}}$ and $\phi_A$, which are needed to compute the thermodynamical observables in Sections \ref{sec2.3} and \ref{sec2.4}.

\subsection{Infrared expansions}
\label{sec2.2}

Now we consider the infrared, near-horizon expansions for the bulk fields $a(r)$, $c(r)$, $h(r)$, and $\phi(r)$. Near the horizon all the bulk fields in \eqref{2.2} are assumed to be smooth and we may consider the Taylor expansions
\begin{align}
X(r)=\sum_{n=0}^\infty X_n(r-r_H)^n,
\label{2.12}
\end{align}
where $X=\left\{a,c,h,\phi\right\}$.

In order to numerically solve the equations of motion \eqref{2.3}, \eqref{2.4}, \eqref{2.5}, and \eqref{2.6} we need to specify the boundary conditions $X(r_{\textrm{start}})$ and $X'(r_{\textrm{start}})$, where $r_{\textrm{start}}$ is a value of the radial coordinate that is slightly above the horizon\footnote{The horizon is a singular point of the equations of motion and, thus, we need to initialize the numerical integrations slightly above it.}. In this work we work with Taylor expansions up to second order, which are sufficient to perform the numerical integrations if $r_{\textrm{start}}$ is close enough to $r_H$. Therefore, we must determine 12 Taylor coefficients in order to specify $X(r_{\textrm{start}})$ and $X'(r_{\textrm{start}})$ at second order. One of these 12 coefficients, namely, $\phi_0$, is one of the two initial conditions of the problem\footnote{As discussed before, the other initial condition is $\mathcal{B}$.}. Four of these 12 coefficients, namely, $a_0$, $c_0$, $h_0$, and $h_1$ and also the radial location of the black hole horizon, $r_H$, may be fixed by rescaling the bulk coordinates while taking into account also the fact that $h(r)$ vanishes at the horizon. For definiteness, we adopt here numerical coordinates fixed in such a way that
\begin{align}
r_H=0;\,\,\,a_0=c_0=h_0=0,\,\,\,h_1=1.
\label{2.13}
\end{align}
Note that $r_H=0$ may be obtained by rescaling the radial coordinate while $h_0=0$ comes from the fact that $h(r)$ has a simples zero at the horizon. Also, $h_1=1$ may be obtained by rescaling $t$ while $a_0=0$ may be arranged by rescaling $(t,z)$ by a common factor. Similarly, $c_0=0$ may be arranged by rescaling $(x,y)$ by a common factor. After this, the remaining 7 coefficients in the near-horizon Taylor expansions for the bulk fields can be fixed on-shell as functions of the initial conditions $(\phi_0,\mathcal{B})$ by substituting the second order Taylor expansions into the equations of motion and setting to zero each power of $r_{\textrm{start}}$ in the resulting algebraic equations\footnote{In practice, we set to zero the following 7 terms: $\mathcal{O}(r_{\textrm{start}}^{0})$, $\mathcal{O}(r_{\textrm{start}}^{1})$, and $\mathcal{O}(r_{\textrm{start}}^{2})$ in \eqref{2.6}, $\mathcal{O}(r_{\textrm{start}}^{-1})$ in \eqref{2.7}, $\mathcal{O}(r_{\textrm{start}}^{-1})$ and $\mathcal{O}(r_{\textrm{start}}^{0})$ in \eqref{2.3}, and $\mathcal{O}(r_{\textrm{start}}^{0})$ in \eqref{2.4}.}.

With $X(r_{\textrm{start}})$ and $X'(r_{\textrm{start}})$ determined as discussed above, the equations of motion are numerically integrated from $r_{\textrm{start}}$ near the horizon up to some numerical ultraviolet cutoff $r_{\textrm{max}}$ near the boundary. We used $r_{\textrm{start}}=10^{-8}$ and $r_{\textrm{max}}=10$ to numerically solve the equations of motion. It is important to remark, however, that even before reaching $r_{\textrm{conformal}}=2$ the numerical backgrounds we considered in the present work have already reached the ultraviolet fixed point corresponding to the AdS$_5$ geometry. This fact is used in Section \ref{sec2.3} to reliably obtain the ultraviolet coefficients in \eqref{2.11} and it will be also employed in Section \ref{sec2.4} to properly compute the holographic magnetic susceptibility numerically.

\subsection{Coordinate transformations and thermodynamical observables}
\label{sec2.3}

Let us now introduce the so-called ``standard coordinates'' of the domain-wall metric gauge, $\tilde{b}(\tilde{r})=0$, where variables with $\sim$ refer to quantities evaluated in these standard coordinates where the background reads
\begin{align}
d\tilde{s}^2&=e^{2\tilde{a}(\tilde{r})}\left[-\tilde{h}(\tilde{r})d\tilde{t}^2+d\tilde{z}^2\right]+ e^{2\tilde{c}(\tilde{r})}(d\tilde{x}^2+d\tilde{y}^2)+\frac{d\tilde{r}^2}{\tilde{h}(\tilde{r})},\nonumber\\
\tilde{\phi}&=\tilde{\phi}(\tilde{r}),\,\,\, \tilde{A}=\tilde{A}_\mu d\tilde{x}^\mu=\hat{B}\tilde{x}d\tilde{y}\Rightarrow \tilde{F}=d\tilde{A}=\hat{B}d\tilde{x}\wedge d\tilde{y},
\label{2.14}
\end{align}
and the boundary is at $\tilde{r}\rightarrow\infty$ while the horizon is at $\tilde{r}=\tilde{r}_H$. The ``hat'' in $\hat{B}$ accounts for the fact that this is the magnetic field measured in units of the inverse of the AdS radius squared, while $B$ shall be used to denote the boundary magnetic field measured in physical units, as we shall discuss in Section \ref{sec2.4}. In the standard coordinates, the ultraviolet asymptotics for the bulk fields are given by \cite{gubser1,gubser2} (see also \cite{finitemu})
\begin{align}
\tilde{a}(\tilde{r})&\approx\tilde{r}+\cdots,\nonumber\\
\tilde{c}(\tilde{r})&\approx\tilde{r}+\cdots,\nonumber\\
\tilde{h}(\tilde{r})&\approx 1+\cdots,\nonumber\\
\tilde{\phi}(\tilde{r})&\approx e^{-\nu\tilde{r}}+\cdots.
\label{2.15}
\end{align}

The standard coordinates (in which $h(r)$ goes to one at the boundary) are the coordinates where we obtain standard holographic formulas for the gauge theory's physical observables such as the temperature and the entropy density. However, in order to obtain numerical solutions for the bulk fields one needs to give numerical values for all the infrared near-horizon Taylor expansion coefficients, which in turn requires rescaling these standard coordinates, as discussed in the previous Section. The numerical solutions are obtained in the numerical coordinates described by the Ansatz \eqref{2.2} with the ultraviolet asymptotics \eqref{2.11}, while standard holographic formulas for physical observables are obtained in the standard coordinates described by the background \eqref{2.14} with the ultraviolet asymptotics \eqref{2.15}. One may relate these two sets of coordinates by equating $\tilde{\phi}(\tilde{r})=\phi(r)$, $d\tilde{s}^2=ds^2$ and $\hat{B}d\tilde{x}\wedge d\tilde{y}=\mathcal{B}dx\wedge dy$ and this leads to the following relations\footnote{As mentioned in \cite{gubser1}, if $\phi_A<0$ one must replace $\phi_A\mapsto|\phi_A|$ in these relations.} (by comparing the near-boundary asymptotics \eqref{2.11} and \eqref{2.15} for $r\rightarrow\infty$)
\begin{align}
\tilde{r}&=\frac{r}{\sqrt{h_0^{\textrm{far}}}}+a_0^{\textrm{far}}-\ln\left(\phi_A^{1/\nu}\right),\nonumber\\
\tilde{t}&=\phi_A^{1/\nu}\sqrt{h_0^{\textrm{far}}}t,\nonumber\\
\tilde{x}&=\phi_A^{1/\nu}e^{c_0^{\textrm{far}}-a_0^{\textrm{far}}}x,\nonumber\\
\tilde{y}&=\phi_A^{1/\nu}e^{c_0^{\textrm{far}}-a_0^{\textrm{far}}}y,\nonumber\\
\tilde{z}&=\phi_A^{1/\nu}z;\nonumber\\
\tilde{a}(\tilde{r})&=a(r)-\ln\left(\phi_A^{1/\nu}\right),\nonumber\\
\tilde{c}(\tilde{r})&=c(r)-(c_0^{\textrm{far}}-a_0^{\textrm{far}})-\ln\left(\phi_A^{1/\nu}\right),\nonumber\\
\tilde{h}(\tilde{r})&=\frac{h(r)}{h_0^{\textrm{far}}},\nonumber\\
\tilde{\phi}(\tilde{r})&=\phi(r);\nonumber\\
\hat{B}&=\frac{e^{2(a_0^{\textrm{far}}-c_0^{\textrm{far}})}}{\phi_A^{2/\nu}}\mathcal{B}.
\label{2.16}
\end{align}

The temperature of the plasma is given by the black brane horizon's Hawking temperature
\begin{align}
\hat{T}=\frac{\sqrt{-\tilde{g}'_{\tilde{t}\tilde{t}} \tilde{g}^{\tilde{r}\tilde{r}}\,'}}{4\pi}\biggr|_{\tilde{r}=\tilde{r}_H}= \frac{e^{\tilde{a}(\tilde{r}_H)}}{4\pi}|\tilde{h}'(\tilde{r}_H)|=\frac{1}{4\pi\phi_A^{1/\nu}\sqrt{h_0^{\textrm{far}}}},
\label{2.17}
\end{align}
while the entropy density is obtained via the Bekenstein-Hawking's relation \cite{Bek,Hawk}
\begin{align}
\hat{s}=\frac{S}{V}=\frac{A_H/4G_5}{V}=\frac{\int_{\textrm{horizon}}d^3\tilde{x}\sqrt{\tilde{g}(\tilde{r}=\tilde{r}_H, \,\tilde{t}\,\textrm{fixed})}}{4G_5V} = \frac{2\pi}{\kappa^2}e^{\tilde{a}(\tilde{r}_H)+2\tilde{c}(\tilde{r}_H)} = \frac{2\pi e^{2(a_0^{\textrm{far}}-c_0^{\textrm{far}})}}{\kappa^2\phi_A^{3/\nu}},
\label{2.18}
\end{align}
where we defined $\kappa^2=8\pi G_5$ and used \eqref{2.12}, \eqref{2.13}, and \eqref{2.16}.

One can see from \eqref{2.16}, \eqref{2.17}, and \eqref{2.18} that the only ultraviolet coefficients in the numerical coordinates which we need to fix by fitting the numerical solutions with \eqref{2.11} are $a_0^{\textrm{far}}$, $c_0^{\textrm{far}}$, $h_0^{\textrm{far}}$, and $\phi_A$. The numerical solutions for $h(r)$ converge quickly to their asymptotic values at large $r$ and we may reliably set $h_0^{\textrm{far}}=h(r_{\textrm{conformal}})$. With $h_0^{\textrm{far}}$ fixed in this way, we may fix $a_0^{\textrm{far}}$, $c_0^{\textrm{far}}$, and $\phi_A$, respectively, by employing the fitting functions $a(r)=a_0^{\textrm{far}}+r/\sqrt{h_0^{\textrm{far}}}$, $c(r)=c_0^{\textrm{far}}+r/\sqrt{h_0^{\textrm{far}}}$, and $\phi(r)=\phi_Ae^{-\nu a(r)}$ in the interval $r\in[r_{\textrm{conformal}}-1,r_{\textrm{conformal}}]$. We were able to obtain good fits for the near-boundary behavior of the numerical solutions using this fitting scheme.

\begin{figure}[h]
\begin{centering}
\includegraphics[scale=0.55]{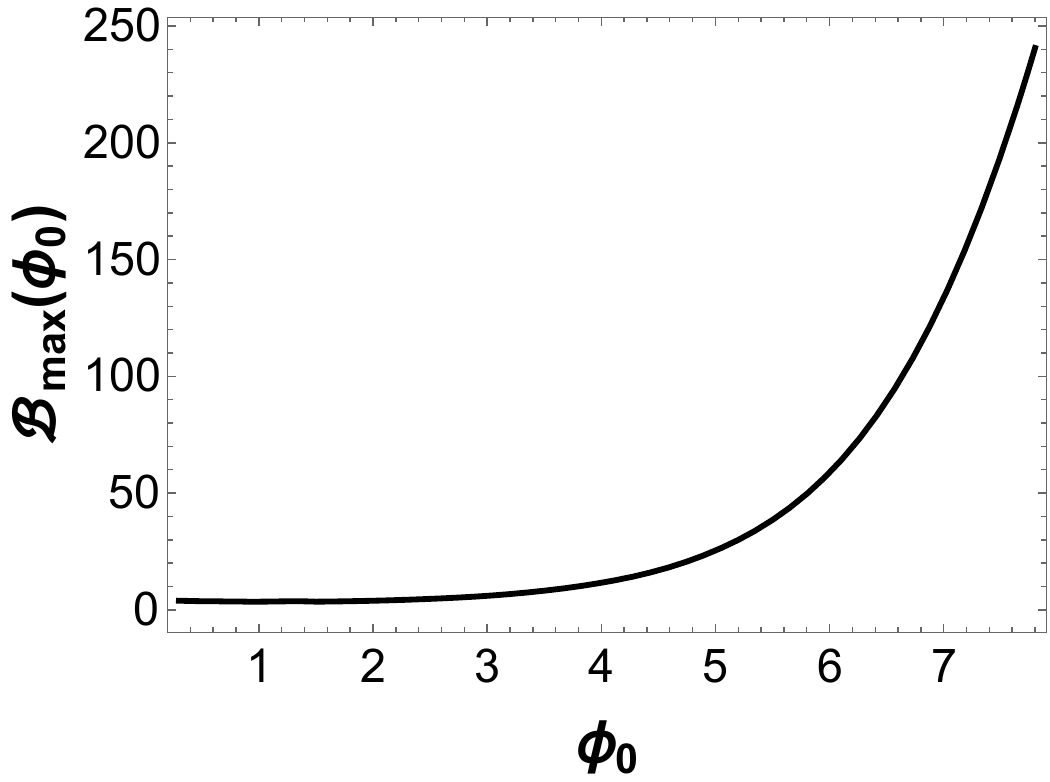}
\par\end{centering}
\caption{The curve corresponds to the upper bound for the initial condition $\mathcal{B}$ as a function of the initial condition for the dilaton $\phi_0$, below which the solutions of the equations of motion are asymptotically AdS$_5$. This curve depends on the chosen profiles for the dilaton potential $V(\phi)$ and gauge coupling function $f(\phi)$ to be discussed in the next Section. \label{fig1}}
\end{figure}

Also, it is important to remark that there is an upper bound on the initial condition $\mathcal{B}$ for a given value of the initial condition for the dilaton $\phi_0$. In fact, for values of $\mathcal{B}$ above this bound, all the numerical backgrounds we generated failed to be asymptotically AdS$_5$. Such a bound, which we denote by $\mathcal{B}\le\mathcal{B}_{\textrm{max}}(\phi_0)$, may be numerically constructed by interpolating a list with pairs of points $\left\{\left(\phi_0^i,\mathcal{B}_{\textrm{max}}^i\right),\,i=1,2,3,\cdots\right\}$ and the corresponding result is presented in Fig.\ \ref{fig1}. 

In the next Section we explain how one can express the thermodynamical quantities $\hat{B}$, $\hat{T}$, and $\hat{s}$ in physical units\footnote{Note from \eqref{2.16}, \eqref{2.17}, and \eqref{2.18} that $\hat{B}$, $\hat{T}$, and $\hat{s}$ are proportional to $\phi_A^{-2/\nu}$, $\phi_A^{-1/\nu}$ and $\phi_A^{-3/\nu}$, respectively. Correspondingly, their counterparts in physical units (without the ``hat'') are given in MeV$^2$, MeV, and MeV$^3$, respectively. This is related to the fact that the leading mode for the dilaton field, $\phi_A$, corresponds to the insertion of a relevant deformation in the quantum field theory, which is responsible for generating an infrared scale that breaks the conformal invariance of the theory at low energies \cite{gubser2}.} using the lattice data for the equation of state and the magnetic susceptibility at zero magnetic field.

\section{Specifying the dilaton potential $V(\phi)$ and the Maxwell-Dilaton gauge coupling $f(\phi)$}

Given the general EMD action \eqref{2.10}, one still needs to specify the dilaton potential $V(\phi)$ and the Maxwell-dilaton gauge coupling $f(\phi)$ in order to solve the Einstein's equations \eqref{2.3}-\eqref{2.6}. The idea, as already sketched in section \ref{Chap3.5}, is to use data from lattice QCD to fix $V(\phi)$ and $f(\phi)$. Let us first discuss how to fix $V(\phi)$.

\subsection{Fixing the dilaton potential using lattice data for the QCD EoS with (2+1) flavors}

Here we follow Refs.\ \cite{Finazzo:2014cna,finitemu} which discuss in detail how to dynamically fix the dilaton potential, $V(\phi)$, and the gravitational constant, $\kappa^2$, using the recent lattice data \cite{latticedata1} for the QCD equation of state with $(2+1)$-flavors.

The first step towards the specification of the dilaton potential is to make a functional Ansatz such that $V(\phi=0)=-12$, i.e. one still has an asymptotically AdS$_5$ space. In particular, we take
\begin{equation}\label{eq:GenPot}
V(\phi) = -12(1+a^4)^{1/4}\cosh\gamma\phi+b_2\phi^2+b_4\phi^4+b_6\phi^6, 
\end{equation}
where $a$, $\gamma$, $b_2$, $b_4$ and $b_6$ are the fit parameters. This functional form is, so far, the most useful one to reproduce the equation of state for the QCD with few parameters.

However, the parameters of the potential \eqref{eq:GenPot} are not entirely free, they must satisfy some constraints  which are shown below:
\begin{itemize}
\item Positivity of $c_s^2$: In the adiabatic approximation (cf. sec. 2 of Ref. \cite{GN1}), one may approximate the speed of sound as follows
\begin{equation}
c_s^2\approx \frac{1}{3} -\frac{V'(\phi)}{V(\phi)},
\end{equation}
which leads us to $c_s^2\approx\frac{1}{3}-\frac{\gamma^2}{2}$ near the boundary\footnote{Recall that $\phi\rightarrow\infty$ near the boundary.}. Therefore, in order to have a positive definite speed of sound, one must have $\gamma\leq \sqrt{2/3}$.

\item Breitenlohner-Freedman (BF) bound: As we remarked in Sec. \ref{Chap3.2.2}, the mass of the scalar field may be a little negative as long as it satisfies Eq. \eqref{eq:BFbound5}. The dilaton's mass may be found expanding the potential near the boundary, where $\phi\rightarrow0$, i.e.
\begin{equation}
V(\phi\rightarrow 0)=-12+\frac{1}{2}m^2\phi^2+\mathcal{O}(\phi^4).
\end{equation}
Thus, with the specific functional form \eqref{eq:GenPot}, the mass of the dilaton field becomes
\begin{equation}
m^2=-6 a+2b_2-3 \gamma ^2\geq - 4,
 \end{equation} 
with the inequality representing the BF bound \eqref{eq:BFbound5}. Furthermore, the dilaton deformation must be a relevant one, i.e. it deforms the IR of our theory and, from Sec. \ref{Chap3.2}, this is achieved whenever the dilaton scaling dimension $\Delta$ has values in the interval
\begin{equation}
 2\leq\Delta<4, \text{where} \ \ \Delta(\Delta-4) =-m^2.
 \end{equation} 
 
 \item Singularity criteria: Large curvatures in the IR are allowed if and only if \cite{Gubser:2000nd}
 \begin{equation}
 V(\phi)\leq -12.
 \end{equation}
\end{itemize}

Bearing in mind all the remarks done above, one can start to generate different geometries for different parametrizations. Since we want a holographic model that mimics the QCD equation of state for (2+1) flavors, we compute the speed of sound,
\begin{equation}
c_s^2 =\frac{d \log T }{d\log s},
\end{equation}
and then we compare the holographic result with the lattice data given in Ref. \cite{latticedata1}. The best parametrization was found to be
\begin{align}
V(\phi)=-12\cosh(0.606\,\phi)+0.703\,\phi^2-0.1\,\phi^4+0.0034\,\phi^6.
\label{2.19}
\end{align}
From the dilaton potential specified above one obtains the dilaton mass $m^2\approx-3$, as anticipated in Section \ref{sec2.1}.

\begin{figure}[h]
\begin{tabular}{cc}
\includegraphics[width=0.48\textwidth]{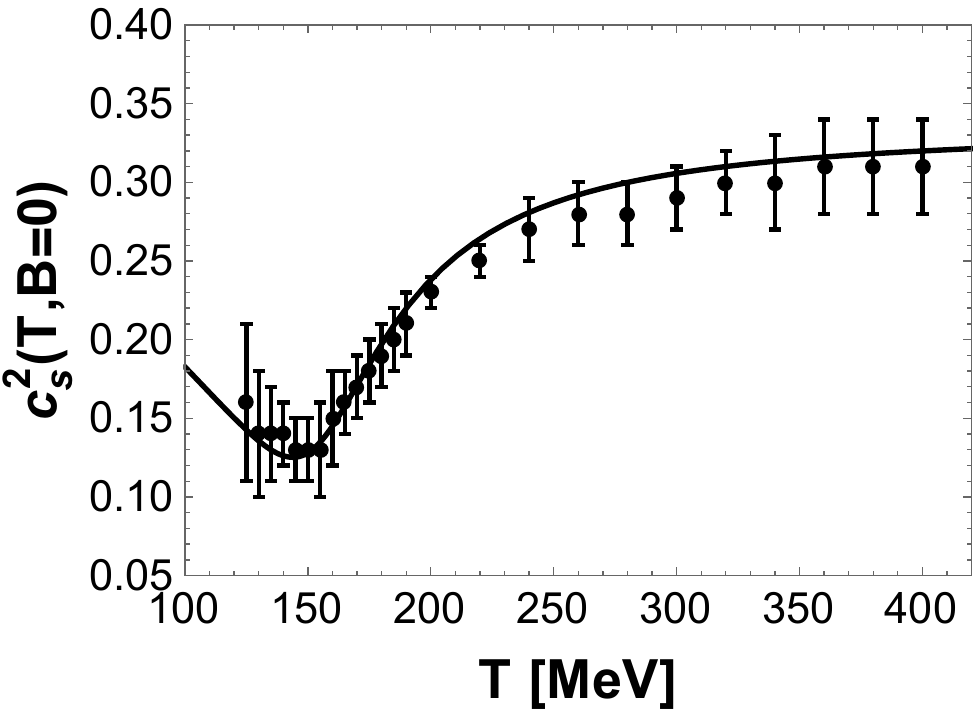} & \includegraphics[width=0.46\textwidth]{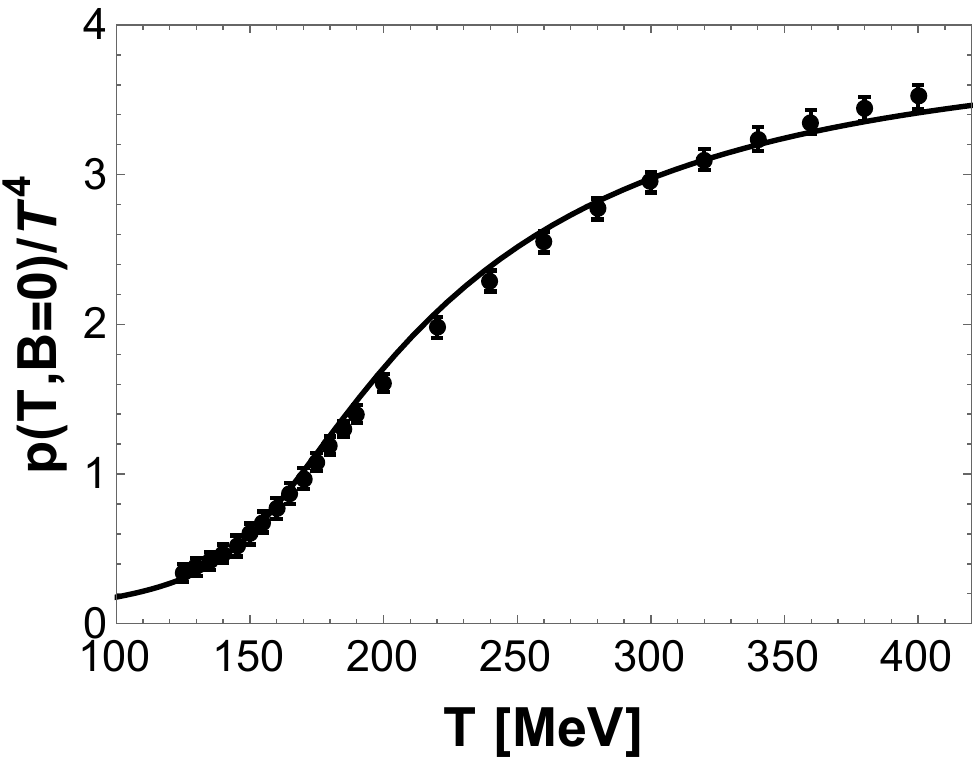} 
\end{tabular}
\caption{Holographic calculation of the speed of sound squared $c_s^2$ and the (normalized) pressure $p/T^4$. The data points correspond to lattice QCD results from \cite{latticedata1} computed at zero magnetic field. \label{fig2}}
\end{figure}

However, we still need to fix the gravitational constant $\kappa^2=8\pi G_5$ once the speed of sound is insensitive to this parameter. To fix $\kappa^2$, we calculate the pressure 
\begin{equation}
p(T)=\int_{T_{ref}}^{T}s(T')dT', \ \ \text{where} \ \ T_{ref}\approx 20 \ \text{MeV},
\end{equation}
which is a sensitive quantity with respect $\kappa^2$, and then we compare with the lattice data \cite{latticedata1}. The result is
\begin{equation}
\kappa^2 = 8\pi G_5 = 12.5.
\end{equation}
The results for the holographic equation of state are given in Fig. \ref{fig2}. 

We remark that, although the present EMD construction does not explicit introduce fundamental flavors at the dual boundary quantum field theory, the dilaton potential in Eq. \eqref{2.19} was adjusted in order to quantitatively \emph{mimic} the $(2+1)$-flavor lattice QCD equation of state and its crossover. This \emph{mimicking} procedure was originally introduced in \cite{GN1} (see also \cite{Yaresko:2015ysa} for more recent discussions), where it was also discussed how different choices for the dilaton potential may \emph{emulate} not only the QCD crossover, as done in the present work, but also first and second order phase transitions, which may be useful for a large variety of different physical systems.

In the present work, we employ the same procedure used in \cite{Finazzo:2014cna,finitemu} to express the holographically determined thermodynamical observables in physical units, i.e., we find the temperature at which our speed of sound squared, $c_s^2$, displays a minimum (at zero magnetic field) and match it to the corresponding lattice QCD result \cite{latticedata1}
\begin{align}
\lambda=\frac{T_{\textrm{min.}\,c_s^2}^{\textrm{lattice}}}{T_{\textrm{min.}\,c_s^2}^{\textrm{BH}}}\approx \frac{143.8\,\textrm{MeV}}{0.173} \approx 831\,\textrm{MeV}.
\label{2.20}
\end{align}
In what follows, we relate any black hole thermodynamical observable, $\hat{X}$, with its counterpart in physical units, $X$, with mass dimension [MeV$^p$], by taking $X = \lambda^p \hat{X}$ [MeV$^p$]. This prescription respects the fact that dimensionless ratios, such as $s/T^3$, must give the same result regardless of the units. A comparison between our holographic results for the speed of sound squared, $c_s^2(T,B=0)$, and the (normalized) pressure, $p(T,B=0)/T^4$ (at zero magnetic field) and the corresponding lattice QCD results from \cite{latticedata1} is shown in Fig.\ \ref{fig2}. One can see that the holographic model provides a good description of the lattice data in the absence of an external magnetic field.  

Although our potential \eqref{2.19} mimics the thermodynamics of the QCD and, consequently, has a crossover as phase transition, one could use another parametrization to obtain a different phase transition. Table \ref{TabV} summarizes this idea showing a couple of parametrizations that gives us different types of phase transition. We denote by $V_{2nd}$ the holographic model that displays a second order phase transition, whilst the holographic model with $V_{1st}$ displays a first order phase transition. The dilaton potential $V_{IHQC}(\phi)$ is the one used in the improved holographic QCD \cite{ihqcd-1,ihqcd-2}, which describes the pure glue sector of the QCD near the crossover temperature.

\begin{table}[h]
\begin{center}
\caption{Different parametrizations of $V(\phi)$ which give different types of phase transitions.}
\begin{tabular}{c|c|c|c|c|c|c}\label{TabV}
Potential & $a$ & $\gamma$ & $b_2$ &  $b_4$ & $b_6$ & $\Delta$ \\ 
\hline
$V_{(2+1)N_f}$ \eqref{2.19} & 0 & 0.606 & 0.703 & -0.1 & 0.0034 & 3.00 \\ 
$V_{2nd}$ \cite{GN1} & 0 & $1/\sqrt{2}$ & 1.942 & 0 & 0 & 3.37 \\ 
$V_{1st}$ \cite{GN1} & 0 & $\sqrt{7/12}$ & 2.0 & 0 & 0 & 3.00 \\ 
$V_{IHQC}$ \cite{ihqcd-1,ihqcd-2} & 1 & $\sqrt{2/3}$ & 6.25 & 0 & 0 & 3.58 \\ 
\hline 
\end{tabular} 
\end{center}
\end{table}

\textit{Note added:} After the finish of this dissertation we generated another parametrization for $V(\phi)$ that greatly improves the agreement with the lattice data. This new parametrization may be found in Ref. \cite{Finazzo:2016mhm}.

\subsection{Fixing the Maxwell-Dilaton gauge coupling using lattice data for the magnetic susceptibility at zero magnetic field}
\label{sec2.4}

In order to fully determine our holographic model and include the effects from a magnetic field we also need to dynamically fix the Maxwell-Dilaton gauge coupling $f(\phi)$. This can be done using the recent lattice data \cite{latticedata2} for the magnetic susceptibility of QCD with $(2+1)$-flavors evaluated at zero magnetic field. In order to compute the magnetic susceptibility in our holographic model we follow the same general steps discussed in \cite{donos}: we substitute the Ansatz \eqref{2.2} into the action \eqref{2.1} and calculate the second derivative of the on-shell action with respect to the magnetic field, dividing the result by the entire spacetime volume of the boundary. In order to obtain the bare magnetic susceptibility we plug the on-shell numerical solutions into the expression obtained in the previous step\footnote{As mentioned in footnote 7 of \cite{donos}, the Euclidean action has the opposite sign of the Lorentzian action.},
\begin{align}
\chi_{\textrm{bare}}(T,B)=-\frac{\partial^2 f_{\textrm{bare}}}{\partial B^2}=-\frac{1}{V_{\textrm{bdy}}}\frac{\partial^2 S_{E,\,\textrm{bare}}^{\textrm{on-shell}}[B]}{\partial B^2} &=
\frac{1}{V_{\textrm{bdy}}}\frac{\partial^2 S_{\textrm{bare}}^{\textrm{on-shell}}[B]}{\partial B^2}\nonumber\\
& = -\frac{1}{2\kappa^2}\int_{\tilde{r}_H}^{\tilde{r}^{\textrm{fixed}}_{\textrm{max}}}d\tilde{r} f(\tilde{\phi}(\tilde{r})) e^{2(\tilde{a}(\tilde{r})-\tilde{c}(\tilde{r}))}\biggr|^{\textrm{on-shell}},
\label{2.21}
\end{align}
where $f_{\textrm{bare}}$ is the bare free energy density and, formally, one should take the limit $\tilde{r}^{\textrm{fixed}}_{\textrm{max}}\rightarrow\infty$. However, in numerical calculations, $\tilde{r}^{\textrm{fixed}}_{\textrm{max}}$ must be a fixed ultraviolet cutoff for all the geometries in order to ensure that the ultraviolet divergence in \eqref{2.21} is independent of the temperature. Since we are interested here in calculating the magnetic susceptibility at zero magnetic field where $a(r)=c(r)$, one obtains from \eqref{2.21}
\begin{align}
\chi_{\textrm{bare}}(T,B=0)= -\frac{1}{2\kappa^2}\int_{\tilde{r}_H}^{\tilde{r}^{\textrm{fixed}}_{\textrm{max}}}d\tilde{r} f(\tilde{\phi}(\tilde{r}))\biggr|^{\textrm{on-shell}}.
\label{2.22}
\end{align}
In order to regularize \eqref{2.22} we follow the same procedure adopted on the lattice \cite{latticedata2} and subtract from \eqref{2.22} the vacuum contribution at zero temperature. Clearly, this removes the ultraviolet divergences since those are temperature independent. More precisely, we subtract the geometry corresponding to $(T_{\textrm{small}},B)\approx(0.005\,\textrm{MeV},0)$, which is generated by the initial conditions $(\phi_0,\mathcal{B})=(7.8,0)$; this is the asymptotically AdS$_5$ geometry with the lowest temperature and zero magnetic field which we could reach in our numerical computations\footnote{Note that $\phi_0=7.8$ corresponds to the local minimum of our dilaton potential \eqref{2.19}. For $\phi_0>7.8$, our dilaton potential becomes non-monotonic and, in practice, we took $\phi_0=7.8$ as the upper bound for the initial condition $\phi_0$ in our numerical calculations to avoid complications with extra singular points in the equations of motion.}. Therefore, we obtain the following holographic formula for the magnetic susceptibility at zero magnetic field (which is valid for any EMD model of the kind considered here)
\begin{align}
\chi(T,B=0)&=\chi_{\textrm{bare}}(T,B=0)-\chi_{\textrm{bare}}(T_{\textrm{small}},B=0)\nonumber\\
&=-\frac{1}{2\kappa^2}\left[\left(\int_{\tilde{r}_H}^{\tilde{r}^{\textrm{fixed}}_{\textrm{max}}}d\tilde{r} f(\tilde{\phi}(\tilde{r}))\right)\biggr|_{T,B=0}-\left(\textrm{same}\right)\biggr|_{T_{\textrm{small}},B=0} \right]^{\textrm{on-shell}}
\nonumber\\
&=-\frac{1}{2\kappa^2}\left[\left(\frac{1}{\sqrt{h_0^{\textrm{far}}}}\int_{r_{\textrm{start}}}^{r^{\textrm{var}}_{\textrm{max}}} dr f(\phi(r))\right)\biggr|_{T,B=0}-\left(\textrm{same}\right)\biggr|_{T_{\textrm{small}},B=0} \right]^{\textrm{on-shell}},
\label{2.23}
\end{align}
where $\tilde{r}^{\textrm{fixed}}_{\textrm{max}}$ must be chosen in such a way that the upper limits of integration in the numerical coordinates satisfy $r_{\textrm{conformal}}\le r^{\textrm{var}}_{\textrm{max}}=\sqrt{h_0^{\textrm{far}}}\left[\tilde{r}^{\textrm{fixed}}_{\textrm{max}}- a_0^{\textrm{far}}+\ln\left(\phi_A^{1/\nu}\right)\right]\le r_{\textrm{max}}$ for all the geometries considered. We found that for $\tilde{r}^{\textrm{fixed}}_{\textrm{max}} \sim 33$ such requirement is met. We also checked that one can vary the value of the ultraviolet cutoff  $\tilde{r}^{\textrm{fixed}}_{\textrm{max}}$ and the results for the holographic magnetic susceptibility do not change, which confirms the stability of our numerical procedure.

\begin{figure}[h]
\begin{centering}
\includegraphics[scale=0.55]{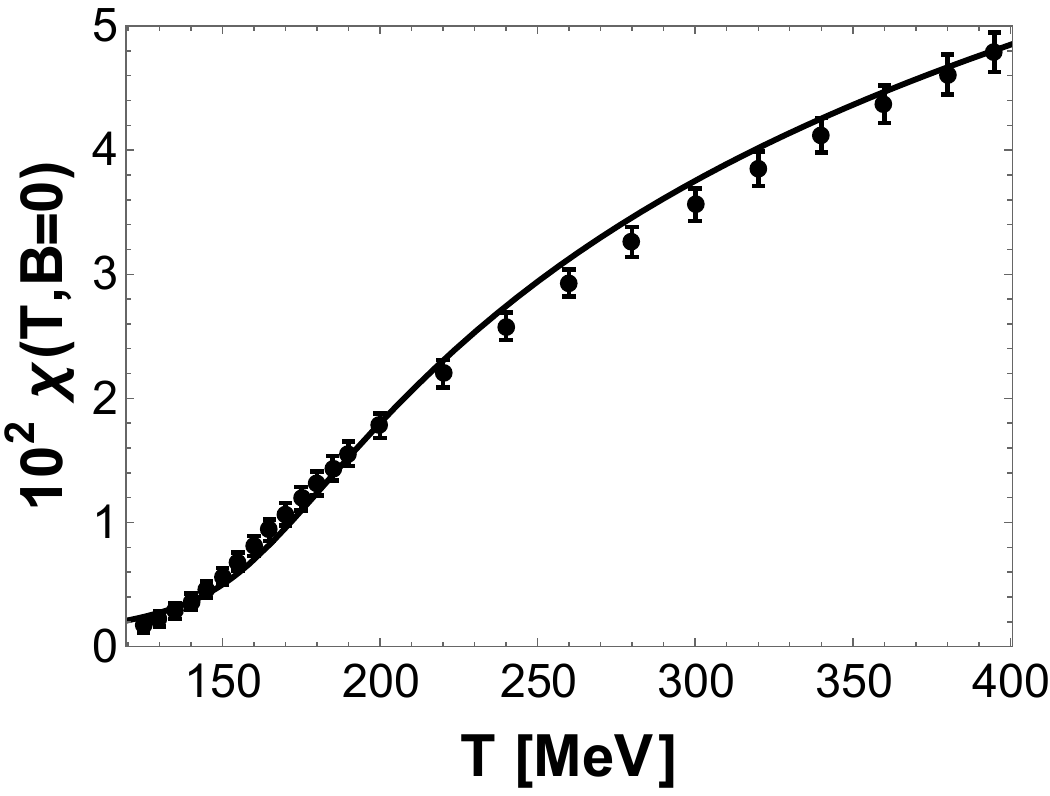}
\par\end{centering}
\caption{Holographic calculation of the magnetic susceptibility at zero magnetic field and comparison with lattice data from \cite{latticedata2} (we consider 10.9 times the data available in table III in \cite{latticedata2}, which corresponds to the magnetic susceptibility in natural units - see footnote 1 in \cite{latticedata2}). \label{fig3}}
\end{figure}

We can now use many different trial profiles for $f(\phi)$ to evaluate \eqref{2.23} over the zero magnetic field background solutions, trying to holographically fit the recent lattice data from \cite{latticedata2} for the magnetic susceptibility of $(2+1)$-flavor QCD at zero magnetic field. We found that a good description of the lattice data can be obtained by fixing
\begin{align}
f(\phi)=1.12\,\textrm{sech}(1.05\,\phi-1.45),
\label{2.24}
\end{align}
with the corresponding results displayed in Fig.\ \ref{fig3}.

With the dilaton potential \eqref{2.19} and the Maxwell-Dilaton gauge coupling \eqref{2.24} dynamically fixed by the description of adequate lattice data at zero magnetic field, our holographic model is now fully determined. This setup may be employed to investigate the physics of the dual quantum field theory at finite temperature and nonzero magnetic field with vanishing chemical potential(s). 

We finish this Section by mentioning some limitations of the holographic model presented here:
\begin{itemize}
\item The model cannot describe phenomena directly related to chiral symmetry and its breaking/restoration (such as $T=0$ magnetic catalysis \cite{Gusynin:1994re,Gusynin:1995nb,Miransky:2002rp}). This could be studied by adding flavor D-branes in the bulk (see, for instance, Ref. \cite{ihqcd-veneziano});\\
\item The model cannot properly describe hadron thermodynamics (which sets in at low temperatures, below $T \sim 150 $ MeV) and the effects of magnetic fields at low temperatures (for a study of the hadron resonance gas in a magnetic field see \cite{Endrodi:2013cs}). Moreover, in this holographic model asymptotic freedom is replaced by conformal invariance at sufficiently high temperatures. Furthermore, for high enough magnetic fields the nonlinear nature of the DBI action for the D-branes should be taken into account \cite{vacilao};\\
\item In Appendix \ref{appF}, we present a brief discussion on the behavior of electric field response functions in the present EMD model, which indicates that this simple model is not versatile enough to simultaneously cover in a quantitative way both the magnetic and electric sectors of the QGP.
\end{itemize}
With these limitations in sight, we expect that the present bottom-up holographic model will be mostly useful to understand the effects of magnetic fields on the QGP within the range $T \sim 150-400$ MeV and $eB\lesssim 1$ GeV$^2$.

\section{Holographic QCD thermodynamics at nonzero magnetic field}
\label{sec3}

\begin{figure}[h]
\begin{tabular}{cc}
\includegraphics[width=0.48\textwidth]{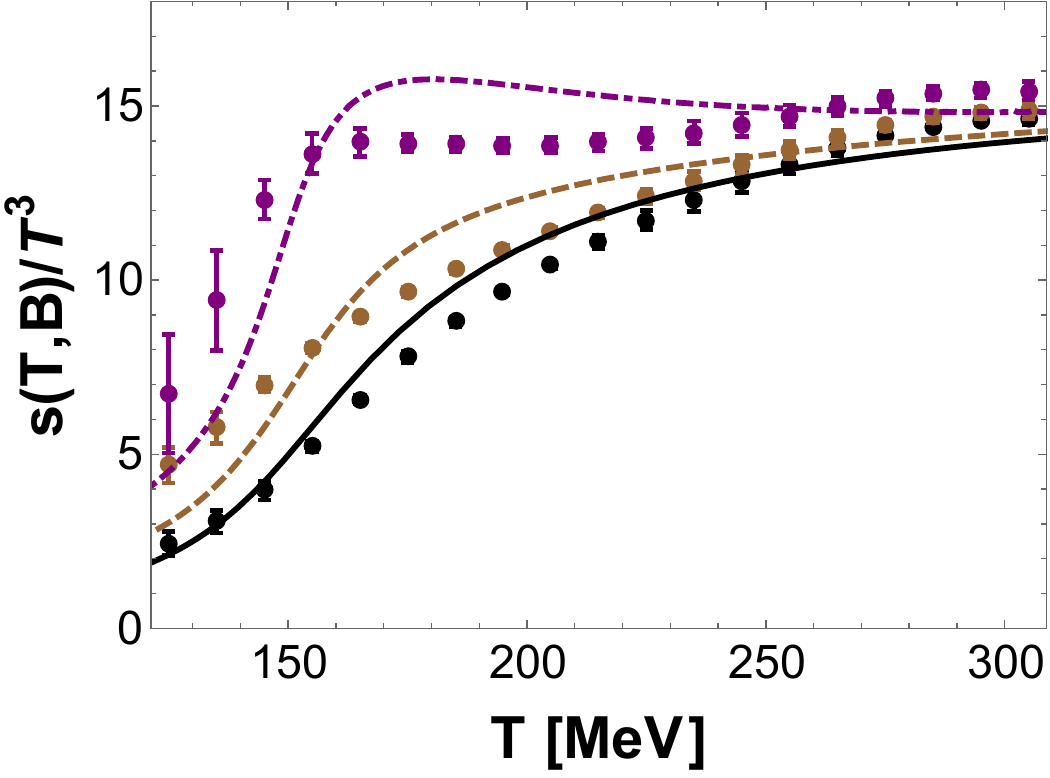} & \includegraphics[width=0.48\textwidth]{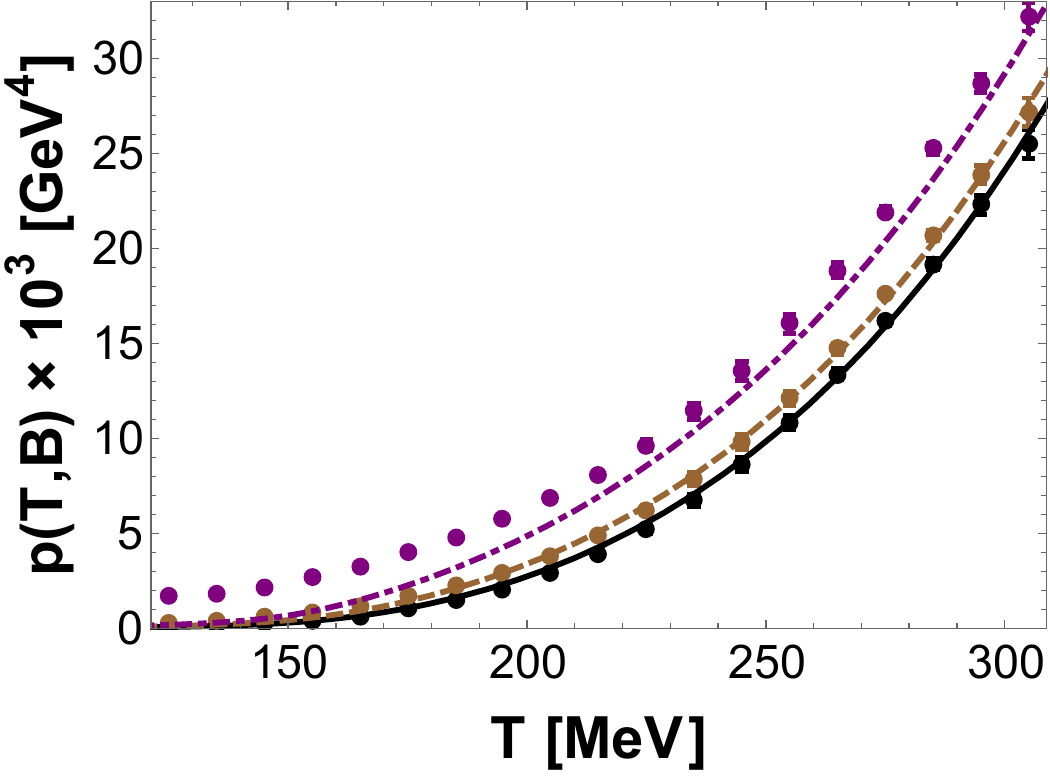} 
\end{tabular}
\caption{ Holographic calculation for the normalized entropy density, $s/T^3$, and pressure, $p$, in the presence of an external magnetic field. The solid, dashed, and dot-dashed curves correspond to magnetic fields $eB=0$, $0.3$, and $0.6$ GeV$^2$, respectively. The data points correspond to the lattice calculations for these quantities performed in \cite{latticedata3}.}
\label{fig4}
\end{figure}

In this Section the results for the holographic equation of state at nonzero magnetic field are presented. The formulas needed to compute the observables shown below were presented in the last Section. Here, we define the pressure as the temperature integral of the entropy density performed while keeping the magnetic field fixed\footnote{As discussed in detail in Section 2 of \cite{latticedata3} this corresponds to the isotropic pressure in the so-called ``$B$-scheme'' where the magnetic field is kept fixed during compression. Also, this corresponds to the anisotropic pressure in the direction of the magnetic field in the so-called ``$\Phi$-scheme'' where the magnetic flux is kept fixed during compression.}
\begin{align}
p(T,B)=\int_{T_{\textrm{ref}}}^T dT' s(T',B),
\label{3.1}
\end{align}
where we took a low reference temperature, $T_{\textrm{ref}}=22$ MeV, in agreement with what was done in \cite{Finazzo:2014cna,finitemu} to obtain the fit for the dilaton potential and the gravitational constant \eqref{2.19}. By doing so, the holographic curves for the pressure in Fig. \ref{fig4} (and also Fig. \ref{wfig2}) actually correspond to differences with respect to reference pressures calculated at $T_{\textrm{ref}}$ for each value of the magnetic field.

In Fig.\ \ref{fig4} we show our holographic results for the normalized entropy density, $s/T^3$, and pressure, $p$, and compare them to recent lattice data \cite{latticedata3} for $eB = 0$, $0.3$, and 0.6 GeV$^2$. It is important to remark, however, that the above convention to calculate the pressure is not exactly the same used in \cite{latticedata3} since in \eqref{3.1} the pressure (difference) vanishes at $T=T_{\textrm{ref}}=22$ MeV while in the calculation carried out in \cite{latticedata3} the pressure goes like $\sim\mathcal{O}\left((eB)^4\right)$ for $T\rightarrow 0$ and, therefore, one should expect that the differences between these two calculations\footnote{Note that in \cite{latticedata3} the pressure was obtained from the renormalized free energy density. Here, we could have done the analogous holographic procedure by calculating the free energy density from the holographically renormalized on-shell action for the EMD model. This is, however, a much more laborious calculation than the one we have carried out here where we first calculated the entropy density using the Bekenstein-Hawking's relation \eqref{2.18} and then we calculated the pressure (difference) using Eq. \eqref{3.1}.} become more pronounced at low temperatures and large magnetic fields, as seen in Fig.\ \ref{fig4}. However, even for $eB=0.6$ GeV$^2$, we do find a reasonable agreement for the pressure at large temperatures ($T > 200$ MeV).

On the other hand, when it comes to the ratio $s/T^3$, the agreement between our holographic results and the lattice is only at the qualitative level. This is in part due to the uncertainties in the holographic description of this observable already at $B=0$: the holographic model parameters were chosen to describe the lattice data for the pressure and the speed of sound squared at $B=0$ and not\footnote{Probably a better agreement with $B\neq 0$ lattice data may be obtained by improving the choice of the model parameters through a global fit to $B=0$ lattice data for the pressure, the entropy density, the speed of sound, and the trace anomaly. See Ref. \cite{Finazzo:2016mhm} for this implementation.} $s/T^3$. In any case, one can see that $s/T^3$ increases with an increasing magnetic field, which is the general behavior observed on the lattice \cite{latticedata3}. Moreover, note that the curve $s/T^3$ becomes steeper near the transition region for increasing values of the magnetic field, which is again in agreement with the general trend observed on the lattice \cite{Endrodi:2015oba}.

\begin{table}[h]
 \begin{center}
  \begin{tabular}{| c || c |}
    \hline
    $eB$ $[$GeV$^2]$ & $T_c(eB)$ $[$MeV$]$ \\
    \hline
    \hline
    0   & 158.2 \\
    \hline
    0.1 & 157.6 \\
    \hline
    0.2 & 154.9 \\
    \hline
    0.3 & 153.2 \\
    \hline
    0.4 & 151.3 \\
    \hline
    0.5 & 149.9 \\
    \hline
  \end{tabular}
 \caption{Deconfinement temperature (defined by the inflection point of $s/T^3$) for different values of the magnetic field in the bottom-up holographic model. \label{tab1}}
 \end{center}
\end{table}

\begin{figure}[h]
\begin{centering}
\includegraphics[scale=0.55]{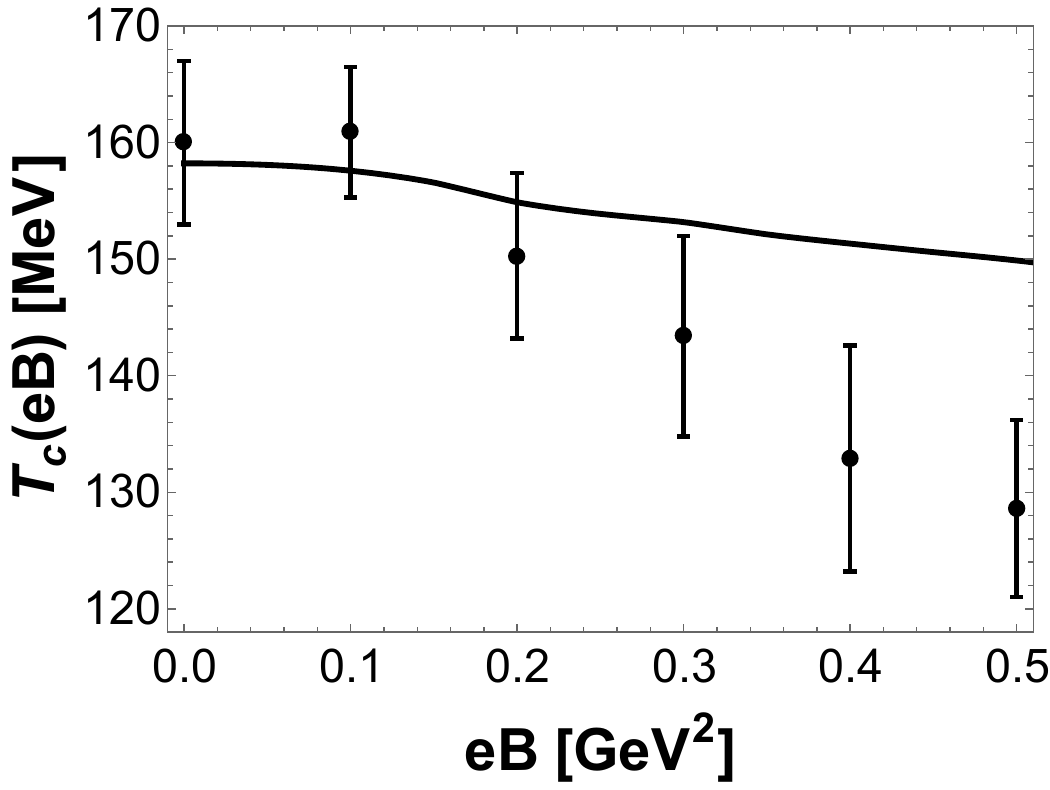}
\par\end{centering}
\caption{Deconfinement temperature (defined by the inflection point of $s/T^3$) for different values of the magnetic field in the bottom-up holographic model. The data points correspond to the lattice calculation performed in \cite{latticedata3}.}
\label{TcB}
\end{figure}

As discussed in \cite{latticedata3}, the inflection point of $s/T^3$ may be used to characterize the crossover temperature as a function of the magnetic field\footnote{Since the crossover is not a genuine phase transition, the free energy is analytic in the region where the degrees of freedom change from a hadron gas to a deconfined plasma. Thus, the definition of the crossover temperature $T_c$ depends on the observable one uses to characterize it. Different observables can give in principle different values for $T_c$ and one may use them to obtain a band defining the crossover region \cite{latticedata0,latticedata3}.}. Correspondingly, the peak in $T\partial_T(s/T^3)$ may be used to estimate the crossover temperature as a function of the magnetic field in our holographic model. We used our results for $s/T^3$ to find how the crossover temperature changes with a magnetic field and the results are displayed in table \ref{tab1} and in Fig.\ \ref{TcB}. One can see in Fig.\ \ref{TcB} that in our model the crossover temperature decreases with an increasing magnetic field, as found on the lattice \cite{latticedata0,latticedata3}, but a quantitative agreement with the data from \cite{latticedata3} occurs only for $eB \lesssim 0.3$ GeV$^2$.

Some general comments regarding the crossover found in our holographic model are in order at this point. Depending on the chosen dilaton potential, the black hole solutions may or may not have a minimum temperature, as detailed discussed, for instance, in Refs. \cite{GN1,stefanovacuum,Yaresko:2015ysa,Charmousis:2010zz}. In the case there is some minimum temperature below which the black hole solutions do not exist, the system generally features a first order Hawking-Page phase transition \cite{Hawking:1982dh} to the thermal gas phase at some critical temperature a little bit higher than the minimum temperature for the existence of the black hole solutions. Also, in this case, the black hole solutions are not unique and there is at least one unstable branch of black hole solutions above this minimum temperature. But for some choices of the dilaton potential the temperature of the black hole solutions may monotonically decrease as a function of the radial position of the horizon until going to zero, in which case the black hole solutions are unique and thermodynamically preferred over the thermal gas solution and the system does not feature any phase transition at nonzero temperature (at least at zero magnetic field and vanishing chemical potentials): this is the case realized in our EMD model. Note also this is indeed the adequate situation to mimic the QCD crossover instead of the pure Yang-Mills first order phase transition. In fact, by analyzing our dilaton potential according to the general criteria discussed in \cite{Charmousis:2010zz}, one notes that in the deep infrared our dilaton potential goes like $V(\phi\to\infty)\sim -e^{0.606\phi}$, in which case at each finite value of temperature (at zero magnetic field and vanishing chemical potentials) there exists a unique black hole solution and this corresponds to the true ground state of the system, having a larger pressure than the thermal gas solution. Moreover, since within the region of the $(T,B)$-phase diagram analyzed in our manuscript the pressure of the plasma increases with $B$ (as also seen on the lattice, see Fig. \ref{fig4}), within this region the black hole solutions are always thermodynamically preferred and do correspond to the true ground state of the system.

As a technical detail, in order to obtain the curves in Fig.\ \ref{fig4} we used a large grid of initial conditions with 720,000 points taking 900 equally spaced points in the $\phi_0$-direction starting from $\phi_0=0.3$ and going up to $\phi_0=7.8$, and 800 equally spaced points in the $\frac{\mathcal{B}}{\mathcal{B}_{\textrm{max}}(\phi_0)}$-direction starting from $\frac{\mathcal{B}}{\mathcal{B}_{\textrm{max}}(\phi_0)}=0$ and going up to $\frac{\mathcal{B}}{\mathcal{B}_{\textrm{max}}(\phi_0)}=0.99$. A large number of points was required to obtain sufficiently smooth curves for $s/T^3$ that allowed for the extraction of the crossover temperature and its dependence on the magnetic field. However, smooth curves for $p$ could be obtained using much smaller (and faster) grids.

\section{The anisotropic shear viscosity}
\label{SecAnisoShear}

We now calculate the shear viscosity coefficients in this novel EMD model that emulates effects of an external magnetic field. Actually, as discussed at length in Chapter \ref{Chap4.0}, we have seven viscosity coefficients, being five shear viscosities and two bulk viscosities. In this section we shall calculate the shear viscosities, in the same spirit of Chapter \ref{Chap5.0}.

For the sake of completeness, we will show how to obtain the same result for the anisotropic shear viscosity \eqref{eq:EtaParDK} from Chapter \ref{Chap5.0} for this QCD-like theory using a matching procedure \cite{GN2}. Although this bottom-up model has a dilaton field, the anisotropic shear mode $h_{xz}$ is still decoupled; the consequence is that its equations of motion remains the same. Assuming the usual harmonic fluctuation $\psi\equiv h^{z}_{x} = h^{z}_{x}(r)e^{-i\omega t} $, we write its equation of motion explicitly,
\begin{equation}\label{eq:psieq}
\psi ''+\psi' \left(4a'-b'+\frac{h'}{h}\right)+\frac{\omega ^2 e^{2b-2 a}}{h^2}\psi=0,
\end{equation}
where primes denote $\partial_r$ derivatives. The conserved flux for the differential equation above is
\begin{equation}
\text{Im}\, \mathcal{F} = he^{4a-b}\,\text{Im}(\psi^* \psi').
\end{equation}

We relate the imaginary part of the retarded Green's function with $\text{Im}\mathcal{F}$ by (cf. discussion in Sec. \ref{Chap3.4})
\begin{equation}\label{eq:kubo-rel-EMD}
\text{Im}\, G^{R}_{T^{xz}T^{xz}}(\omega) = -\frac{\text{Im}\, \mathcal{F}}{16\pi G_5}.
\end{equation}
The task now is to fully determine the flux $\text{Im}\, \mathcal{F}$. However, as we shown below, only the near horizon region will not be enough, because there will be an undetermined constant. We will circumvent this issue by analyzing the solution with $\omega=0$ and comparing it with the near horizon solution (matching).

\begin{figure}
\begin{center}
\begin{tabular}{c}
\includegraphics[width=0.66\textwidth]{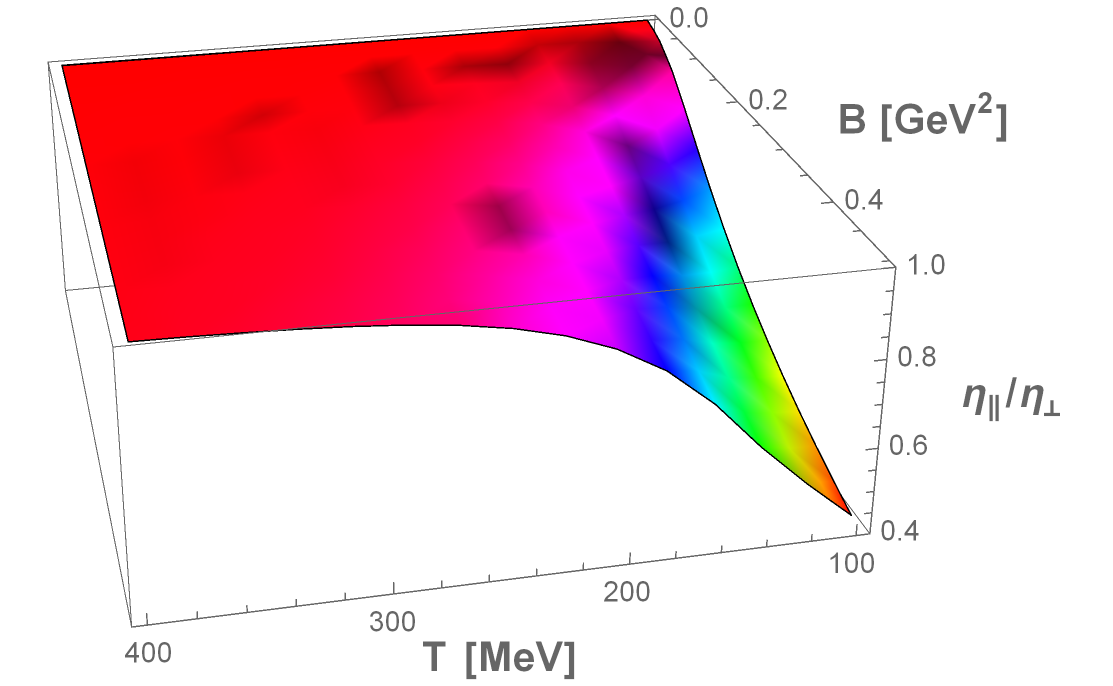} 
\end{tabular}
\begin{tabular}{c}
\includegraphics[width=0.6\textwidth]{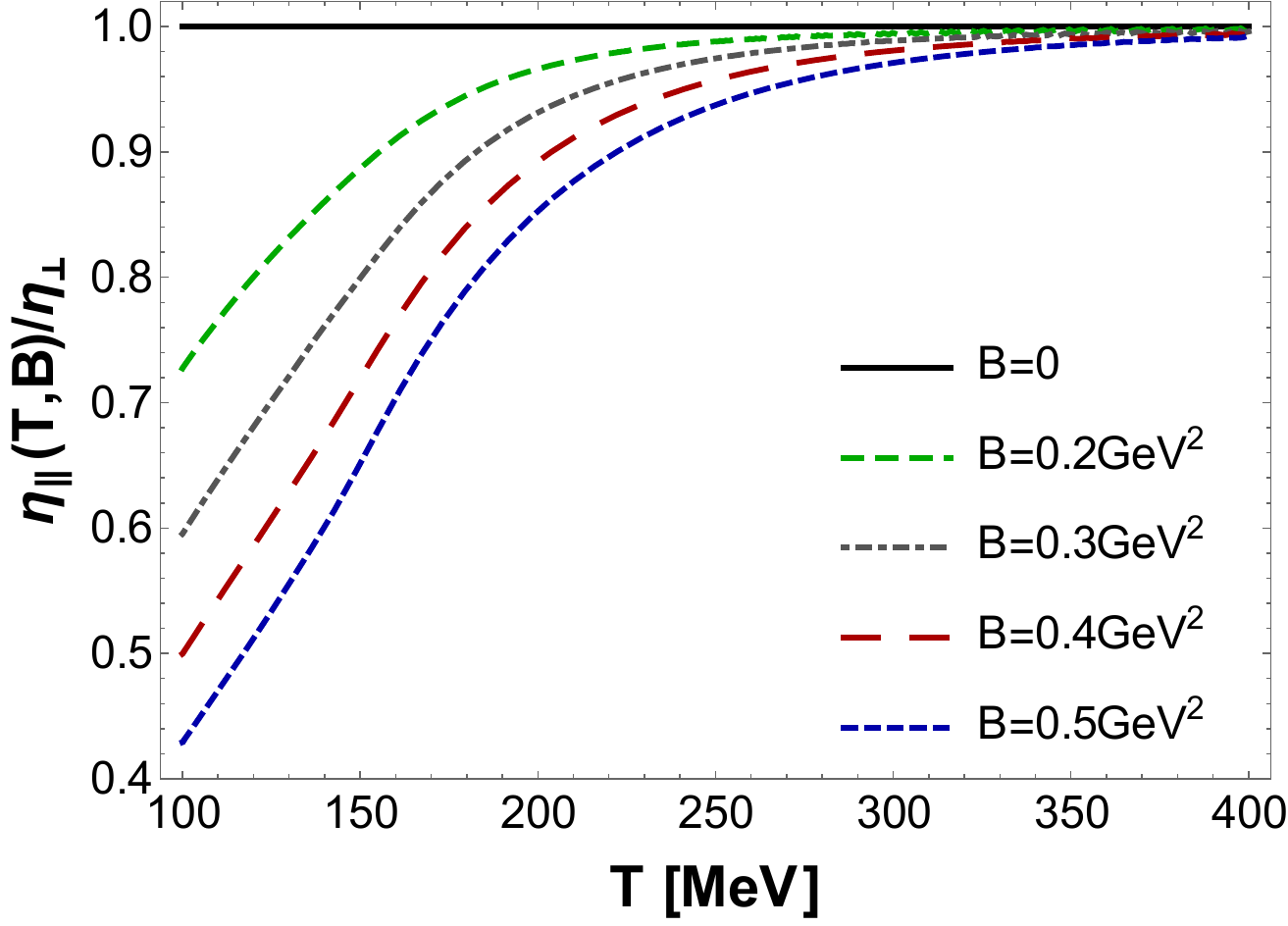} 
\end{tabular}
\begin{tabular}{c}
\includegraphics[width=0.6\textwidth]{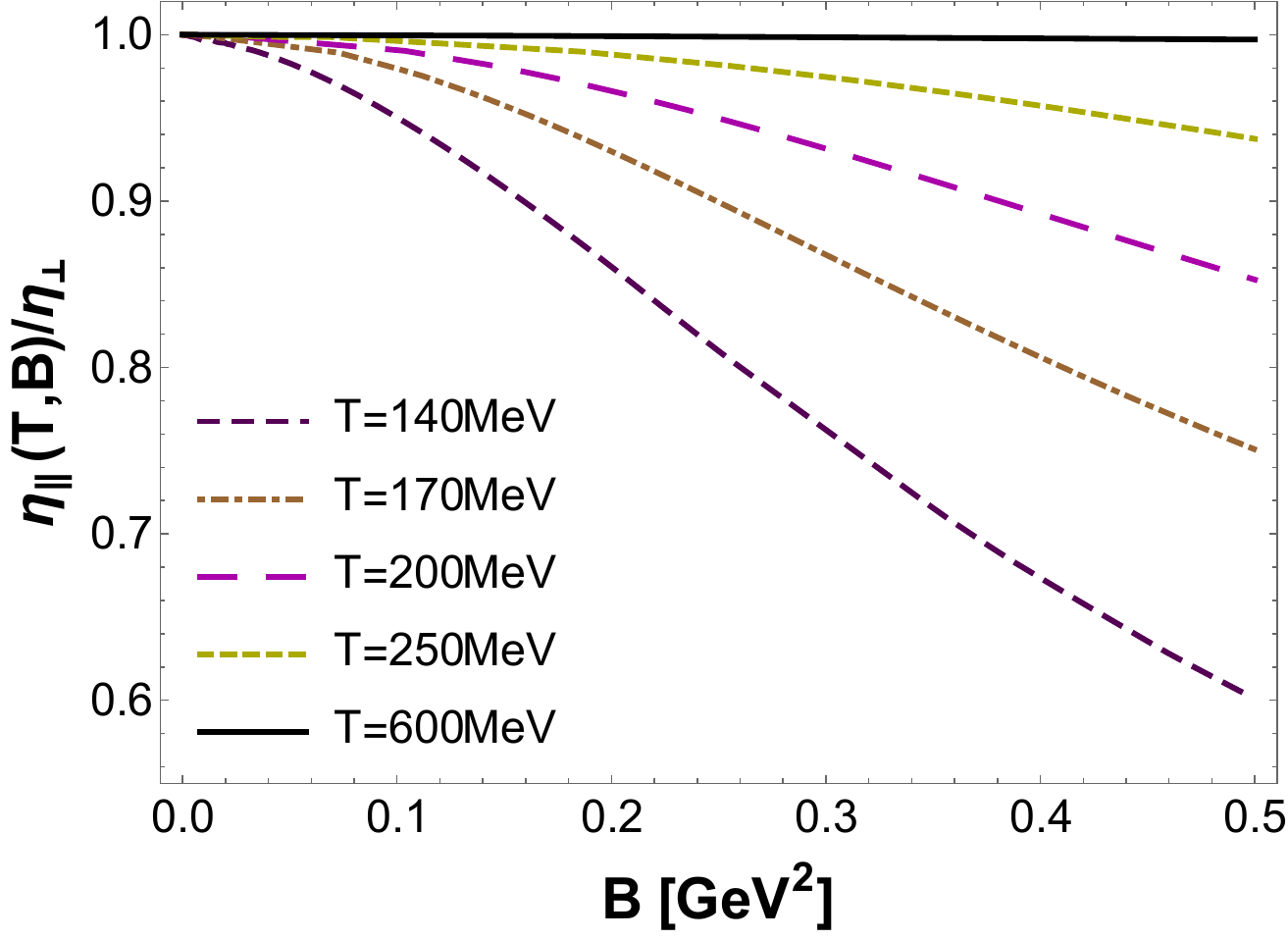} 
\end{tabular}
\end{center}
\caption{Numerical results for the anisotropic shear viscosity \eqref{eq:EtaParEMD} of the EMD model with an external magnetic field.}
\label{fig:shearnonconfor}
\end{figure}

In the near horizon limit, $r\rightarrow r_h$, Eq. \eqref{eq:psieq} is reduced to
\begin{equation}
\psi ''+\psi'\frac{1}{r-r_h}+\frac{\omega ^2 e^{2b-2 a}}{h'(r_h)^2(r-r_h)^2}\psi=0,
\end{equation}
whose solution is given by
\begin{equation}\label{eq:psinearh}
\psi(r\rightarrow r_h) \approx c_+(r-r_h)^{\frac{i\omega}{4\pi T}} + c_-(r-r_h)^{-\frac{i\omega}{4\pi T}},
\end{equation}
where $T=\frac{e^{a(r_h)-b(r_h)}}{4\pi}h'(r_h)$ is the Hawking temperature and $c_{+/-}$ are constants. Since we want the retarded Green's function we discard the outgoing term of the solution above, i.e. $c_+=0$. On the other hand, the solution of \eqref{eq:psieq} around $\omega=0$ is given by
\begin{equation}\label{eq:psinearw}
\psi(r) = a_1 +a_2\int_{r}^{\infty} \frac{e^{-4a(r')+b(r')}}{h(r')}dr',
\end{equation}
where $a_1$ and $a_2$ are another constants. The matching technique to solve \eqref{eq:psieq} comes about when we expand \eqref{eq:psinearh} around $\omega=0$ and \eqref{eq:psinearw} aroung $r_h$, giving us the following relation
\begin{equation}
c_-\left[ 1-\frac{i\omega}{4\pi T}\log\left(r-r_h\right) \right] \approx a_1 +  a_2\frac{e^{-4a(r_h)+b(r_h)}}{h'(r_h)}\log\left(r-r_h\right).
\end{equation}
Thus, we have the result
\begin{equation}
a_1 = c_-, \ \ \ a_2 = -i\omega e^{3a(r_h)}c_-,
\end{equation}
which is valid for small values of $\omega$. The condition at the boundary tells us that $\psi(r\rightarrow \infty)=1$ and, therefore, we have that $c_-=1$. Bearing this in mind, we can now write the flux $\mathcal{F}$ using the solution of the near horizon geometry, which is the more convenient region
\begin{align}
\text{Im}\, \mathcal{F} &= h(r_h)e^{4a(r_h)-b(r_h)}\,\text{Im}(\psi^*(r_h) \psi'(r_h)) \notag \\
            &= \omega e^{3a(r_h)}.
\end{align}
Plugging this result into the Kubo relation \eqref{eq:kubo-rel-EMD}, we have that
\begin{equation}
\eta_{\parallel} = \frac{e^{3a(r_h)}}{16\pi G_5}.
\end{equation}
Dividing the formula above of the parallel shear viscosity by the entropy density, we obtain
\begin{equation}\label{eq:EtaParEMD}
\frac{\eta_{\parallel}}{s} =  \frac{1}{4\pi}e^{2a(r_h)-2c(r_h)},
\end{equation}

We show the numerical result for the anisotropic shear viscosity \eqref{eq:EtaParEMD} in Fig. \ref{fig:shearnonconfor}. Note that the result is in qualitative agreement with the calculation done in the magnetic brane context (compare with Fig. \ref{fig:etaaniso}). Furthermore, the result for the anisotropic viscosity is also in qualitative agreement with the kinetic result \cite{Tuchin:2011jw} (mentioned in Sec. \ref{Chap4.3.1}). However, as one can see in Fig. \ref{fig:shearnonconfor}, it would be hard to detect some effect due to the anisotropic viscosity in a heavy ion collision because at early times, where $B$ is relevant, the QGP temperature is high, which decreases the effect of a magnetic field on the viscosity.

\section{Conclusions of the chapter}
\label{conclusion}

In this chapter we developed, for the first time, a bottom-up holographic model that provides a quantitative description of the crossover behavior observed in the equation of state and in the magnetic susceptibility of a QCD plasma with $(2+1)$-flavors at zero magnetic field. We employed this model to study how an Abelian magnetic field $B$ affects the thermodynamic properties of this strongly coupled plasma (at zero chemical potentials). In the presence of the magnetic field the plasma becomes anisotropic and we used the inflection point of the holographically calculated $s/T^3$ curve to determine how the crossover temperature is affected by the external magnetic field. We found that the crossover temperature decreases with an increasing magnetic field, which agrees with the general behavior recently observed on the lattice. Our model calculations display some level of quantitative agreement with the lattice data for values of the magnetic field up to $eB \lesssim 0.3$ GeV$^2$, which is the expected range achieved in ultrarelativistic heavy ion collisions.

We believe that this agreement with the lattice data can be further improved toward larger values of $eB$ if one tries to carefully match the lattice thermodynamic calculations at $B=0$ by simultaneously taking into account different observables such as the pressure and the speed of sound squared, as we have done in the present approach, with the addition of the entropy density and the trace anomaly in a global fit; in this sense, our choice for the holographic model parameters (fixed at $B=0$) may be systematically improved.

An interesting feature of our holographic model that distinguishes it from other constructions (such as \cite{Ballon-Bayona:2013cta,Mamo:2015dea}) is that the suppression of the crossover temperature with the external magnetic field found here is directly tied to a \emph{quantitative description} of near crossover lattice QCD thermodynamics at $B=0$. It would be desirable to generalize the present holographic model by taking into account the contribution of the chiral condensate. Moreover, motivated by the recent studies in Refs.\ \cite{Endrodi:2015oba,cohen}, one could also investigate if this model indicates the existence of a critical point in the $(T,B)$-plane at higher values of the magnetic field\footnote{Note from Fig. \ref{fig4} that for the values of $B$ considered here we only have a smooth analytical crossover, as also seen on the lattice.}.

The holographic setup constructed here may be employed to obtain estimates for the magnetic field dependence of many other physical observables relevant to the strongly coupled QGP. For instance, in Sec. \ref{SecAnisoShear} we calculated the anisotropic viscosities and how they vary as we increase the value of a magnetic field; we also learned that $\eta_{\parallel}<\eta_{\perp}$ for any nonzero value of magnetic field, which is in qualitative agreement with the calculation done in Chapter \ref{Chap5.0}.

Recently, the effects of an external magnetic field on the equilibration dynamics of strongly coupled plasmas have been studied using holography \cite{DK-applications4,Mamo:2015aia}. In this context, it would be interesting to see how the quasinormal mode spectrum in our nonconformal plasma varies with an external magnetic field. Given that our model can capture the nonconformal behavior of the QGP near the crossover transition, with and without the external magnetic field, a detailed study of the quasinormal modes in this model may shed some light on the thermalization process that takes place in an anisotropic nonconformal strongly magnetized QGP. We hope to report results in this direction in the near future.

\chapter{Conclusions and outlook}
\label{Chap9}

Throughout the first part of this dissertation, which comprehends Chapters \ref{Chap1.0}, \ref{Chap 2}, and \ref{Chap 3.0}, we have reviewed how we can heat up hadronic matter (``melt'' the hadrons) and create the so-called quark-gluon plasma. After we summarize some important properties of this new state of matter, emphasizing how small its viscous effects (in the sense that $\eta/s$ is small) are, we began, in Chapter \ref{Chap 2}, a diligent study of the shear viscosity and the bulk viscosity; we also had a glimpse on why a relativistic viscous theory of hydrodynamics is not easy to construct. We have ended the second chapter by introducing the linear response formalism, which is the standard way to compute the transport coefficients in a strongly coupled plasma (and on the weakly coupled systems as well). After that, we studied the basics of the gauge/gravity duality \cite{Maldacena:1997re,Gubser:1998bc,Witten:1998qj}, and why it is so ``simple'' to calculate transport coefficients within this framework - such as the viscosity calculation in Sections \ref{Chap3.4} and \ref{Chap3.5}. The studies in these chapters served as a preparation for the rest of the dissertation.

Motivated by the recent indications of the presence of very strong magnetic fields in heavy ion collisions \cite{noncentralB1,noncentralB2,noncentralB3,noncentralB4,noncentralB5,noncentralB6}, we began, in Chapter \ref{Chap4.0}, to explore effects of strong magnetic fields on the QGP - the main focus of this work. We also discussed the holographic top-down construction of the magnetic brane \cite{DK1,DK2,DK3}, dual to the magnetic strongly coupled $\mathcal{N}=4$ SYM, which is the background used in the subsequent chapters. Moreover, in Sec. \ref{Chap4.3} we discussed in detail the emergence of anisotropic viscosities due to the anisotropy induced by an external magnetic field (we now have five shear viscosities and two bulk viscosities).

In Chapter \ref{Chap5.0} we gathered the knowledge developed in previous chapters and obtained the anisotropic shear viscosities of the magnetic brane setup, i.e. $\eta_{\perp}$ and $\eta_{\parallel}$ \footnote{We recall that the other three shear viscosities trivially vanish  for the magnetic brane case.}. We observed another violation of the result $\eta/s=1/4\pi$ for anisotropic theories \cite{Erdmenger:2012zu,Rebhan:2011vd,Jain:2015txa}, with $\eta_{\parallel}$ decreasing with increasing magnetic field. Besides calculating only the transport coefficients, it was necessary to tackle the dynamical problem of relativistic magnetohydrodynamics of strongly coupled theories, i.e. its equations of motion. For a such task, one could try to extend the fluid/gravity paradigm \cite{Hubeny:2011hd} in presence of magnetic fields in (2+1) dimensions \cite{Blake:2015hxa}, where the magnetic field is a (pseudo) scalar and the theory is isotropic, to the (3+1)-dimensional case, which would certainly shed some light on this matter. 

Naturally, the shear viscosity is one among many other transport coefficients. Thus, given the suitability of the gauge/gravity duality to calculate real-time phenomena, we can exploit other directions as well, such as the effects of magnetic fields on heavy quark diffusion \cite{Fukushima:2015wck,CasalderreySolana:2006rq}.

We also presented a preliminary discussion in Chapter \ref{Chap6.0} about the anisotropic bulk viscosities for the magnetic $\mathcal{N}=4$ SYM, $\zeta_{\perp}$ and $\zeta_{\parallel}$. These coefficients were shown to vanish in this theory, even though there is a trace anomaly induced by the magnetic field.

In Chapter \ref{Chap7.0} we studied the anisotropic heavy $Q\bar{Q}$ potential using the magnetic brane solution.  We have found that at zero and finite temperature, the inclusion of the magnetic field decreases the attractive force between heavy quarks with respect to its $\mathcal{B}=0$ value and the force associated with the parallel potential is the least attractive force. Qualitatively, the same result is found in Ref. \cite{Chernicoff:2012bu} in the context of an anisotropy induced by the axion \cite{Mateos:2011ix}.

When we developed the EMD model to mimic the strongly coupled QGP embedded in a magnetic field in Chapter \ref{Chap8.0}, following the previous holographic constructions \cite{GN1,gubser1,gubser2}, we demonstrated that is completely feasible to construct realistic holographic models that include a magnetic field. The toll we paid, though, was to relinquish some firm connections with the string theory. Furthermore, the lack of breaking/restoration of the chiral symmetry on the EMD model must be fixed. A promising way to remedy the chiral issue, aside the D-brane embedding \cite{vacilao}, is outlined in Refs. \cite{Chelabi:2015cwn,Fang:2015ytf}.

We end this dissertation commenting the lack of studies about dense stars (neutron stars) via gauge/gravity duality. Inside these compact stars, it is likely that quarks and gluons are deconfined in a strongly interacting regime. As an example, we conjecture that an extension of the EMD model developed in Chapter \ref{Chap8.0} that accommodates the chemical potential sector\footnote{Neutron stars have large values of baryonic chemical potential.} $\mu_B$ \cite{gubser1,finitemu}  suffices for a first approach. This is certainly a new branch of study that deserves further investigations.

\renewcommand{\chaptermark}[1]{\markboth{\MakeUppercase{\appendixname\ \thechapter}} {\MakeUppercase{#1}} }
\fancyhead[RE,LO]{}

\appendix

\chapter{Maxwell's result for the shear viscosity of a diluted gas}
\label{appA}

The goal of this appendix is to show, in a rather simple but effective way (see \cite{LeStat}), how one calculates the shear viscosity of a diluted gas. The results of these calculations had profound consequences and it is evidently within the subject of this dissertation.

To calculate $\eta$, we again resort to the simple laminar flow (for a gas, obviously) illustrated in Fig. \ref{fig:lami-flow}. In this flow, we have the relation
\begin{equation}\label{eq:FAppendix}
\frac{F}{A}=\eta \frac{du_x}{dy}\cong \eta\frac{u}{d},
\end{equation}
with the boundary conditions: $u_x(0)=0$, and $u_{x}(d)=u$. The relation above was obtained in Sec. \ref{Chap 2.1} from macroscopic arguments, i.e. the Navier-Stokes equation. We have to arrive at a similar expression using the microscopic nature of the gas, and, by comparing with Eq. \eqref{eq:FAppendix}, extract the shear viscosity.

\begin{figure}[h]
\centering
\includegraphics[width=12cm,height=4.5cm]{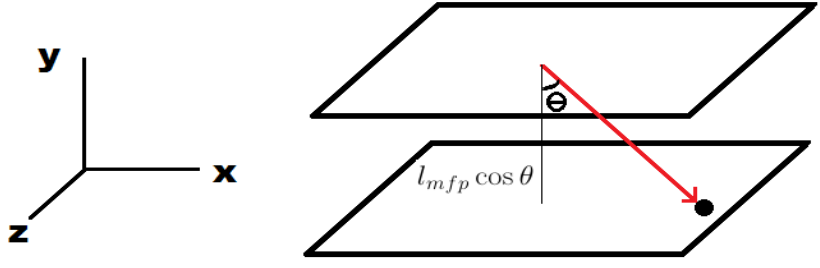}
\caption{The path of a molecule's gas between two consecutive collisions. The distance traveled by this molecule - the red arrow, is given by the mean free path $l_{mfp}$.}
\label{fig:moleculepath}
\end{figure}

The strategy is to relate the momentum diffusion with the consecutive collisions of the gas molecules. More specifically, we track what happens with \emph{one} molecule in a limited region, and afterwards we use some distribution function to take the average. To better visualize this, we depicted this situation in Fig. \ref{fig:moleculepath}. 

We begin by looking at the variation of the momentum $\Delta p$ (not the pressure!) of the molecule drawn in Fig. \ref{fig:moleculepath},
\begin{equation}
\Delta p = m\left[ u_x(y+\Delta y)-u_x(y) \right] \cong m\frac{du_x}{dy}\Delta y.
\end{equation}
Notice also that what happens in the $x-$axis and $z-$axis is not important since the momentum diffusion occurs only on the $y-$axis.

The next step is take the average of $\Delta p$. This can be done once we know the momentum flux density, which is the number of particles per unit of time and area crossing some section in the $xz-$plane, such as the planes drawn in Fig.  \ref{fig:moleculepath}. This number is given by $N v_y f(\vec{v})d^3x$\footnote{We can directly derive this number from the expression $\Pi_{ij}=m\int d^3v v_i v_j f(\vec{v})$, which is the momentum flux. In this case, we are interested on the $\Pi_{xy}$ component.}. Therefore, the force per unit of area is
\begin{align}\label{eq:FAgas}
\frac{F}{A}&=-\frac{1}{A}\frac{\Delta p}{\Delta t} = -N\int d^3v \Delta p\, v_y f(\vec{v}) \notag \\
        &=-mN\frac{du_x}{dy}\int d^3v  v_z f(\vec{v}).
\end{align}

We cannot solve directly the integral above because $f(\vec{v})$ is not the usual Maxwell distribution function. Instead, it is some anisotropic version of it, where $\langle v_{y}\rangle=\langle v_{z}\rangle=0$, and $\langle v_{x}\rangle\neq0$. The clever way to deal with this integral is to assume $u \ll \langle v\rangle$ (slightly anisotropic distribution function), so that the following approximation
\begin{equation}\label{eq:ApproxIntegralDistrib}
\int d^3v\, v f(\vec{v})Q(cos\theta) \approx \frac{1}{2}\langle v\rangle\int_{-1}^{1} d(\cos\theta)\cos\theta Q(cos\theta),
\end{equation}
is legitimate for some generic function $Q(cos\theta)$. 

Hence, by using the approximation \eqref{eq:ApproxIntegralDistrib} in Eq. \eqref{eq:FAgas}, and knowing that $v_y=-v\cos\theta$, we obtain
\begin{equation}
\frac{F}{A}=\frac{1}{3}mN\langle v\rangle l_{mfp}\frac{du_x}{dy}.
\end{equation}

Now, we just compare the above result with expression \eqref{eq:FAppendix}. This gives us the following expression for the shear viscosity of a gas
\begin{equation}
\eta = \frac{1}{3}mN\langle v\rangle l_{mfp}.
\end{equation}

Note that this expression derived for the shear viscosity of a gas does not depend of its density $N$ (cf. Eq. \eqref{eq:visc-gas-dens}). This is a remarkable feature and helped to consolidate the kinetic theory since this result was a successful theoretical prediction derived by Maxwell in 1860 \cite{Brush}.

\chapter{The conformal symmetry}
\label{appB}

Throughout this dissertation we mentioned conformal field theories ($\mathcal{N}=4$ SYM) and how they were important in establishing  the foundations of the gauge/gravity duality (cf. Chap. \ref{Chap 3.0}). In addition, we emphasized the effect of conformal invariance on the bulk viscosity, i.e. $\zeta=0$ for conformal theories. Therefore, this appendix is dedicated to show more explicitly these results. The canonical reference for conformal field theory is Ref. \cite{FranCFT}

We start with the \emph{conformal transformation}. In a a conformal transformation the coordinates change as $x\rightarrow \tilde{x}(x)$, whose effect on the metric is
\begin{equation}
g_{\mu\nu}\rightarrow e^{2\Omega(x)}g_{\mu\nu},
\end{equation}
where $e^{2\Omega(x)}$ is the conformal factor. These conformal transformations are part of the so-called \emph{conformal group}.

The conformal group contains translations, rotations, and boosts (for Lorentzian manifolds). Besides them, we have two more transformations which characterizes the conformal group, the \emph{scale transformation}
\begin{equation}
x\rightarrow \lambda x,
\end{equation}
and the \emph{special conformal transformation} (SCT),
\begin{equation}
x\rightarrow \frac{x^\mu+x^2b^\mu}{1+2x\cdot b+x^2b^2},
\end{equation}
which can be seen as an inversion-translation-inversion transformation.

For a $D$-dimensional space, the conformal group has $1/2(D+2)(D+1)$ generators, which is the same number of the generators of the rotation group in $D+2$ dimensions. Therefore, the conformal group is related to the group $SO(D+2)=SO(1+p,1+q)$, for $p+q=D$.

The generators of the conformal group are
\begin{align}
&\text{Translations:} \ \ P_\mu \equiv -i\partial_\mu \\
&\text{Rotations:} \ \ M_{\mu\nu}\equiv i(x_\mu\partial_\nu - x_\nu \partial_\mu) \\
&\text{Dilatations:} \ \ D\equiv -ix_\mu\partial^\mu \\
&\text{SCT:} \ \ \mathcal{K}\mu \equiv i(x^2\partial_\mu-2x_\mu x_\nu\partial^\nu)  .
\end{align}

The algebra of the conformal group is given by (the non-vanishing commutators)
\begin{align}
[D,P_\mu]& =iP_\mu , \\
[D,\mathcal{K}_\mu] &=-i\mathcal{K}_\mu , \\
[\mathcal{K}_\mu,P_\nu] &= 2i\eta_{\mu\nu}D-2iM_{\mu\nu},  \\
[P_\sigma,M_{\mu\nu}] &= i(\eta_{\sigma\mu}P_\nu-\eta_{\sigma\nu}P_\mu), \\
[M_{\mu\nu},M_{\mu\nu}] &= i(\eta_{\nu\rho}M_{\mu\sigma}-\eta_{\mu\sigma}M_{\nu\rho}- (\mu \leftrightarrow\nu)).
\end{align}

After this short prelude, let us check that $T^{\mu}_{\mu}=0$ in conformal theories and what consequences this fact brings. The simplest way at arrive in this result is to consider the following (constant) scale transformation
\begin{equation}
\delta g_{\mu\nu}=\epsilon g_{\mu\nu}.
\end{equation}
Now, we compute the effect of this transformation on the action,
\begin{equation}
\delta S = \int d^D x \frac{\delta S}{\delta g^{\mu\nu}}\delta g^{\mu\nu} = -\frac{1}{2}\int d^D x \,\epsilon\, T^{\mu}_{\mu},
\end{equation}
where we have used the usual formula for the energy-stress tensor in the last equality,
\begin{equation}
T_{\mu\nu}=\frac{-2}{\sqrt{-g}} \frac{\delta S}{\delta g^{\mu\nu}}.
\end{equation}

Since the scale transformation is a symmetry of the system, we have
\begin{equation}
\delta S=0 \Rightarrow T^{\mu}_{\mu}=0,
\end{equation}
which is an essential feature of any conformal field theory. The converse is not true though, i.e. $T^\mu_\mu=0$ does not imply that the theory is conformal.

In order to investigate the consequences of $T^{\mu}_{\mu}=0$, we just rewrite the Kubo formula for the bulk viscosity of a D-dimensional theory derived in Sec. \ref{Chap2.3.1}. The formula is
\begin{equation}
\zeta = \frac{1}{(D-1)^2}\lim_{\omega\rightarrow 0}\frac{1}{\omega}\int d^D x e^{ik\cdot x} \theta(t)\langle [T^{\mu}_{\mu}(x),T^{\mu}_{\mu}(0)]\rangle,
\end{equation}
where we have used the property of Eq. \eqref{eq:KuboTrick} to have this explicit covariant expression for the bulk viscosity. Hence, it is obvious from the above Kubo relation that the bulk viscosity $\zeta$ \emph{must} vanish in conformal theories.

Regarding the trace of the stress-energy tensor of the strong interactions, it is well known that, at the classical level, the non-Abelian Yang Mills theory is a conformal theory in four dimensions, i.e. $T^{YM \, \mu}_{\mu}=0$. However, quantum effects prevent this theory from being conformal. Indeed, the trace of the stress-energy tensor of QCD is
\begin{equation}\label{eq:TraAnQCD}
T^{QCD \, \mu}_{\mu}=\sum_{q}^{N_f} m_q \bar{q}q +\frac{\beta(g)}{2g^3}G^{a\mu\nu}G_{\mu\nu}^{a},
\end{equation}
where $\beta(g)$ is the beta function defined in Eq. \eqref{eq:betaQCD}. Obviously, the masses of the quarks induces an energy scale, though very small. The important contribution comes from the Yang-Mills anomaly. \\

A final important warning is that, in general, the trace of the stress-energy tensor of some conformal field theory may vanish, which is an \emph{apparent} puzzle (cf. the discussion in 4.3.3 of \cite{FranCFT}). For example, let us take simplest case of the free massless scalar field in $D-$dimensions. In this case its stress-energy tensor is given by
\begin{equation}
T^{\mu\nu} = \partial_\mu\phi\partial_\nu\phi-\frac{1}{2}g^{\mu\nu}(\partial\phi)^2,
\end{equation}
so its trace is
\begin{equation}
T^{\mu}_{\mu} = (1-D/2)(\partial\phi)^2,
\end{equation}
which is zero only for $D=2$. However, we can easily fix it by introducing the improved stress-energy tensor,
\begin{equation}
T^{\mu\nu}_{I}=T^{\mu\nu}+\partial_\alpha J^{\alpha \mu\nu},
\end{equation}
where $J^{\alpha \mu\nu}=-J^{\alpha \nu\mu}$ in order to respect the Noether charges; in other words, we add total derivatives on the action to get $T^{\mu}_{I\ \mu}=0$. For the scalar field, we add $\partial_\alpha J^{\alpha \mu\nu}=-1/6(\partial_\mu\partial_\nu-g_\mu\nu)\phi^2$. 

The above procedure ~can be similarly used to obtain the symmetric stress-energy tensor, i.e. the Belinfante tensor.

\chapter{Universality of the low energy limit absorption of a scalar field by a spherical symmetric black hole}
\label{appC}

In this appendix we show in detail the proof regarding the low energy limit absorption of a scalar field by a spherical symmetric black hole \cite{DasAbs}, i.e. $\lim_{\omega\rightarrow 0}\sigma (\omega)=A_h$, where $\sigma(\omega)$ is the absorption cross section of a (massless) scalar field wave scattered by a spherical black hole, $\omega$ is the frequency (energy) of the wave, and $A_h$ is the black hole's horizon. As mentioned in Section \ref{Chap3.4}, this theorem was of crucial importance to arrive at the ration $\eta/s=1/4\pi$\footnote{See also \cite{EmparanAbs} for the absorption cross section of extended branes.}. Naturally, the method for computing two-point functions from absorption cross sections is outdated once we acquired powerful tools to compute them (e.g. the membrane paradigm).

First, one takes a generic metric with isotropic coordinates\footnote{For instance, in isotropic coordinates, the metric of the Schwarzschild black hole is given by $ds^2=-\left(\frac{1-M/2r}{1+M/2r} \right)dt^2+(1+M/2r)^4\left[dr^2 + r^2 d\Omega_{2}^{2}  \right]$.} in $p+2$ dimensions,
\begin{equation}
ds^2 = - f(r) dt^2 + g(r) \left[dr^2 + r^2 d\Omega_{p}^{2}  \right]
\end{equation}
with $\lim_{r \to \infty} f(r) = 1$ and  $\lim_{r \to \infty} g(r) = 1$ (asymptotically flatness), and $d\Omega_{p}^{2}= d\phi_{1}^2 + \sin^2\phi_{1} d\phi_{2}^2+ \dots + \prod_{k=1}^{d-1} \sin^2\phi_{k} d\phi_{k}^2$ being the metric of a p-sphere with unity radius.

Using that $\sqrt{-g} = \sqrt{r^{2p}fg^{p+1}} = r^p f^{1/2}g^{\frac{p+1}{2}}$, the equation of motion for the massless scalar field becomes
\begin{align}
\frac{1}{\sqrt{-g}} \partial_{\mu} \left( \sqrt{-g}g^{\mu\nu}\partial_{\nu} \Phi \right) &= -\frac{\partial^2_{t}\Phi}{f} + \frac{1}{r^p f^{1/2}g^{\frac{p+1}{2}}} \partial_{r} \left[ \frac{r^p f^{1/2}g^{\frac{p+1}{2}}}{g} \partial_r \Phi \right] + \Delta_{\Omega_{p}^{2}} \Phi   =0
\end{align}
where $\Delta_{\Omega_{p}^{2}}$ is the Laplacian of the p-sphere, whose eigenfunctions are the spherical harmonics in $p$ dimensions. More precisely, $\Delta_{\Omega_{p}^{2}} Y_{lm...} = -l(l+p-3) Y_{lm...}$. Since we are interested in the low energy limit, we discard the excitations of the scalar field with $l\geq 1$. 

Adopting the plane-wave Ansatz, $\Phi(t,r,\theta) = \phi_{w}(r)e^{-iwt}$, we have
\begin{align}
 \frac{w^2 \phi}{f} +  \frac{1}{r^p f^{1/2}g^{\frac{p+1}{2}}} \partial_{r} \left[ r^p f^{1/2}g^{\frac{p-1}{2}} \partial_r \phi \right] = 0
\end{align}
or
\begin{equation} \label{eq:eq. mov. camp. esc. absorcao}
\left[( r^p f^{1/2}g^{\frac{p-1}{2}} \partial_r )^2  + w^2 r^{2p}g^p \right] \phi_{w}(r) = 0
\end{equation}

The strategy to solve \eqref{eq:eq. mov. camp. esc. absorcao} to use the matching procedure (see Sec. \ref{SecAnisoShear} for another calculation using this method). In this procedure, we will slice the space-time in three regions, always focusing the low energy limit $w \rightarrow 0$. The regions to be considered are:
\begin{description}
\item[I] Near the black hole's horizon, i.e., $r \rightarrow r_H$;

\item[II] An intermediate region, i.e., $r>>M$ (the mass of the black hole) and $rw<<1$.

\item[III] A region at far spatial infinity, far from the horizon, i.e., $rw>>1$.
\end{description}

We will consider the solutions in the regions \textbf{I} and \textbf{III}, and then we will extrapolate both solutions to the intermediate region \textbf{II} so that we determine all the unknown constants.

Since we are handling a scattering problem, it is natural to think that, at spatial infinity, we have a mixture between the incoming wave and the scattered (reflected) wave. On the other hand, in the near horizon region, we should have only the transmitted (absorbed) component. Thus,
\begin{equation} \label{eq: cond. cont}
\phi_w(r) \sim  e^{-iwr} + R(w) e^{iwr}  \ \ r \rightarrow \infty,
\end{equation}
where $R(w)$ is the reflection amplitude. Naturally, the absorption probability associated with the scalar field is $\Gamma = 1- |R(w)|^2$. 

Now we define $\xi $ such that
\begin{equation}
d\xi= \frac{dr}{r^p f^{1/2}g^{\frac{p-1}{2}}}
\end{equation}
and we simplify \eqref{eq:eq. mov. camp. esc. absorcao} to
\begin{equation}\label{eq: eq. dif. radial simplif campo absorcao}
\left[\partial_\xi^2  + w^2 r^{2p}g^p \right] \phi_{w}(r) = 0.
\end{equation}

The area of the black hole is given by
\begin{align}
A_H &=\int\limits_{r=r_H} \sqrt{\prod_{i=1}^{p}g_{\theta_i \theta_i}} d\theta_i 
        =r_{H}^{p}[g(r_H)]^{\frac{p}{2}} \int d\Omega_p 
       = r_{H}^{p}[g(r_H)]^{\frac{p}{2}} \Omega_p \equiv R_{H}^{p}\Omega_p,
\end{align}
where $\Omega_p = \frac{2\pi^{(p+1)/2}}{\Gamma \left( \frac{p+1}{2} \right)}$ is the area of a p-sphere with radius equals unity.

In the near horizon limit, we can consider $r^2g(r) \sim  R_{H}^{2} $ as being a constant. Then, the boundary condition given in \eqref{eq: cond. cont} becomes
\begin{equation}\label{sol. perto hor. absor}
\phi_w(r) = T(w) e^{-iwR_{H}^{p}\xi},  \ \ \ r \rightarrow r_H,
\end{equation}
where $T(w)$ is the constant to be related with the transmission amplitude. Consequently, for distances $r>>M$ (region \textbf{II}), i.e., $g(r)\sim 1$ and $f(r) \sim 1$, but $rw<<1$, the solution  \eqref{sol. perto hor. absor} is approximately given by
\begin{equation} \label{aprox. reg. 1 em 2}
\phi_w (r) \sim T(1 - iwR_{H}^{p}\xi), \   \  \   \xi \sim -\frac{T}{p-1}r^{-p+1}.
\end{equation}

For the case where $rw>>1$ (region \textbf{III}), Eq. \eqref{eq: eq. dif. radial simplif campo absorcao}  for the scattered wave is given by
\begin{equation}
\lbrace r^p \partial_r (r^p \partial_r) +w^2r^{2p}  \rbrace \phi_w(r) = 0.
\end{equation}
Making a change of variables, $\rho=rw$, one obtains
\begin{equation}\label{eq. dif. espalha. reg. 3}
\lbrace \partial_{\rho}^{2} +p/\rho \partial_\rho +1 \rbrace \phi_w(\rho) = 0,
\end{equation}
whose solutions are given in terms of Bessel functions. From now on, let us assume that $p$ is an even number (the procedure for $p$ odd is the same); thus, the solution for \eqref{eq. dif. espalha. reg. 3} has the following form
\begin{equation} \label{sol. onda reg. 3}
\phi_w(\rho) = \rho^{\frac{1-p}{2}} \left[ AJ_{\nu} (\rho) + BJ_{-\nu} (\rho)  \right], \ \ \ \nu= (p-1)/2
\end{equation}

To extrapolate the solution of region \textbf{III} to the intermediate region \textbf{II}, we use the approximation of Bessel's functions for small arguments, i.e., $J_\nu \approx \frac{1}{\Gamma(\nu+1)} \left( \frac{x}{2} \right)^\nu$ if $|x|\ll 1$. Then, for $rw\ll 1$, 
\begin{equation}\label{aprox. regi. 3 em 2}
\phi_w(r) \approx \frac{2^{\frac{-p+1}{2}} A}{\Gamma\left(\frac{p+1}{2} \right)} + \frac{2^{\frac{p-1}{2}} w^{1-p}}{\Gamma\left(\frac{3-p}{2} \right)} \frac{B}{r^{p-1}}
\end{equation}

Comparing, in the sense of the power series in $r$, the coefficients of \eqref{aprox. regi. 3 em 2} with the coefficients of \eqref{aprox. reg. 1 em 2},we have
\begin{equation}
T =\frac{2^{\frac{-p+1}{2}} A}{\Gamma\left(\frac{p+1}{2} \right)} ; \ \ \ iwR_{H}^{p}\frac{1}{p-1}T =  \frac{2^{\frac{p-1}{2}} w^{1-p}}{\Gamma\left(\frac{3-p}{2} \right)} B,
\end{equation}
\begin{equation}
\therefore \frac{B}{A} = i \frac{2^{-p+1} (wR_H)^p}{p-1} \frac{\Gamma\left( \frac{3-p}{2} \right)}{\Gamma \left( \frac{p+1}{2} \right)}.
\end{equation}

For the calculation of the absorption cross section of the scalar field by the black hole, we have to find the absorption probability $\Gamma$ of a spherical wave with $l=0$, defined by $\Gamma = |T(w)|^2 = 1 - |R(w)|^2 $. In this problem, we will use $T(w)$. However, one cannot extract such information from the solution \eqref{sol. onda reg. 3} and, because of this, we have to use its limit when $r \rightarrow \infty$. Knowing that $J_\nu(x) \approx \sqrt{\frac{2}{\pi x}}\cos\left(x - \nu\pi/2 +\pi/4  \right)$ for $|x|\gg 1$, one obtains
\begin{equation}
\phi_w(r) \approx \sqrt{\frac{2}{\pi (rw)^p}} \left[ A \cos\left(rw - (p-1)\pi/4 +\pi/4  \right) + B\cos\left(rw + (p-1)\pi/4 +\pi/4  \right) \right].
\end{equation}
Now, we must write the formula above in terms of complex numbers so we can extract the transmission coefficients of the waves going towards the black hole:
\begin{align}
\cos\left(rw - (p-1)\pi/4 +\pi/4  \right) &= e^{- i(p-1)\pi/4+i\pi/4} \left( \frac{e^{irw} + e^{-irw+ i(p-1)\pi/2 -i\pi/2 }}{2} \right) \notag \\
                                                    &= e^{i\beta} \left( \frac{e^{irw} - ie^{-irw+ i\alpha }}{2} \right),
\end{align}
with $\alpha\equiv (p-1)\pi/2 $, e $\beta \equiv -(p-1)\pi/4+\pi/4$. Similarly,
\begin{align}
\cos\left(rw + (p-1)\pi/4 +\pi/4  \right) &= e^{- i(p-1)\pi/4+i\pi/4} \left( \frac{e^{irw+i(p-1)\pi/2} + e^{-irw-i\pi/2 }}{2} \right) \notag \\
                                                    &= e^{i\beta} \left( \frac{e^{irw+i\alpha} - ie^{-irw }}{2} \right),
\end{align}
which results in
\begin{align}\label{eq: prob. absor.}
\phi_w(r) &\approx C(r) \left[ (e^{irw} - ie^{-irw+ i\alpha }) + B(e^{irw+i\alpha} - ie^{-irw } ) \right] \notag \\
             &\approx C'(r) \left[ e^{-iwr} + i\left( \frac{1+\frac{B}{A}e^{i\alpha}}{1+ \frac{B}{A}e^{-i\alpha}} \right) e^{-i\alpha} e^{iwr}  \right],
\end{align} 
where $C(r)$ and $C'(r)$ are two undetermined functions of $r$. Notice that, from the expression \eqref{eq: prob. absor.}, we can extract the transmission coefficient $T(w)$ sine it is accompanied by the exponential term $e^{iwr}$. Therefore, the absorption probability is
\begin{align}
\Gamma = 1 - \left\vert  \frac{1+\frac{B}{A}e^{i\alpha}}{1+ \frac{B}{A}e^{-i\alpha}}   \right\vert ^2.
\end{align}
Also, if  $B/A= i\gamma$, with $\gamma = \frac{2^{-p+1} (wR_H)^p}{p-1} \frac{\Gamma\left( \frac{3-p}{2} \right)}{\Gamma \left( \frac{p+1}{2} \right)} $,  then,
\begin{align}
\Gamma &= 1 - \left\vert  \frac{1+i\gamma\cos2\alpha - \gamma\sin2\alpha}{1+i\gamma\cos2\alpha +\gamma\sin2\alpha}  \right\vert ^2 \notag \\
              &= \frac{4\gamma\sin2\alpha}{1+\gamma^2 +2\gamma\sin 2\alpha}.
\end{align} 

In the low limit frequency $\omega\rightarrow 0$ (low energies), we can approximate the denominator of the above expression for $\Gamma$ to the unity; ergo, one obtains
\begin{equation}
\Gamma = 4 \frac{2^{-p+1}}{p-1} (wR_H)^p \sin\left[\pi(p-1)/2 \right]\frac{\Gamma\left( \frac{3-p}{2} \right)}{\Gamma \left( \frac{p+1}{2} \right)}.
\end{equation}

For the calculation of the absorption cross section, we still have to extract the plane wave from the spherical wave that will be scattered by the black hole\footnote{This is entirely analogous to standard textbook computation of scatterings in quantum mechanics (in four dimensions) where one decomposes the radial wave in terms of the Bessel's functions. However, in higher dimensions, one has to use a generalization of the Bessel's functions, i.e. the Gegenbauer polynomials \cite{BesselF}.}. As we are considering only terms with $l = 0$ for the plane wave, it is useful to write
\begin{equation}
e^{-i\omega z}= K \frac{e^{-iwr}}{r^{p/2}} Y_{00...} + \text{partial waves with higher} \ l + \text{reflected waves},
\end{equation}
where $Y_{00...}$ is the analogue of the spherical harmonic for higher dimensions. Given the standard normalization $\int |Y|^2 d\Omega_p=1$, we conclude that $Y_{00...}=\Omega_{p}^{-1/2}$.

One way to determine the constant $K$ is to integrate both sides over the solid angle of the p-sphere, 
\begin{equation} \label{eq: extr. onda plana}
 \int e^{-iwz} d\Omega_p = K \frac{e^{-iwr}}{r^{p/2}} Y_{00...} \int d\Omega_p =  K \frac{e^{-iwr}}{r^{p/2}} \Omega_p^{1/2}.
 \end{equation} 
To integrate the left-hand side of the above expression, we recall that
 \begin{equation}
 \Omega_p =  \int_{\phi_{p}=0}^{2\pi}\int_{\phi_{p-1}=0}^{\pi} . . . \int_{\phi_1 =0}^{\pi} \sin^{p-1}\phi_1 \sin^{p-2}\phi_2 ...\sin\phi_{p-1} d\phi_1...d\phi_{p-1}d\phi_{p},
 \end{equation}
and we choose $z$ such that $z= r\cos\phi_1$. Thus,
 \begin{equation}
  \int e^{-iwz} d\Omega_p = \Omega_{p-1} \int_{0}^{\pi} \sin^{p-1}\phi \; e^{-iwrcos\phi} d\phi,
 \end{equation}
using the identity
 \begin{equation}
 \int \sin^{k}\phi \; e^{-iwrcos\phi} d\phi = \sqrt{\pi} \left( \frac{2}{wr} \right)^{k/2} \Gamma\left( \frac{k+1}{2} \right) J_{\frac{k}{2}}(wr),
 \end{equation}
we are lead to
 \begin{equation}\label{eq: integral p-esfera}
 \int e^{-iwz} d\Omega_p = \Omega_{p-1} \sqrt{\pi} \left( \frac{2}{wr} \right)^{(p-1)/2} \Gamma\left( \frac{p}{2} \right) J_{\frac{p-1}{2}}(wr).
 \end{equation}
 
However, the Bessel function does not give us information about the coefficient of the spherical wave going towards the black hole; we circumvent this by taking the asymptotic limit of the Bessel function, which has the appropriate functional form of an ingoing wave. With this trick, the integral \eqref{eq: integral p-esfera} becomes
 \begin{equation}
  \int e^{-iwz} d\Omega_p \cong \frac{2^{\frac{p}{2}}}{2w^{\frac{p}{2}}} \Omega_{p-1} \Gamma \left( \frac{p}{2} \right) \frac{e^{-iwr +i\theta}}{r^{\frac{p}{2}}},
 \end{equation}
 where $\theta$ is just a phase. Comparing this result with \eqref{eq: extr. onda plana}, we conclude that
 \begin{equation}
 |K|^2 = \frac{2^{p}}{4w^p}\Omega_{p}^{-1}\Omega_{p-1}^{2} \left[ \Gamma (p/2) \right]^2.
 \end{equation}
We can use that $\Omega_{p-1}^2 = \frac{4\pi^p}{\left[ \Gamma (p/2) \right]^2}$, to obtain
 \begin{equation}
|K|^2 = \frac{1}{\Omega_{p}} \left( \frac{2\pi}{w} \right)^p.
  \end{equation} 
  
 Finally, the absorption cross section $\sigma_{abs}(\omega)$, in the low energy limit $\omega\rightarrow 0$, is given by
 \begin{equation}
 \sigma_{abs}(0) = |K|^2 \Gamma.
 \end{equation}

Obviously, such result is valid only for $l=0$. For sake of completeness, we just quote here the general result for $\sigma_{abs}(\omega)$, valid for any value of $l$, given in \cite{gubserAbs}
\begin{equation}
\sigma^{l}_{abs}(w) = \frac{2^{p-1}\pi^{(p-1)/2}}{w^{p}} \Gamma((p-1)/2)  (l+(p-1)/2) \binom {l+p-2} {l} \Gamma_{l}(w).
\end{equation}

Returning to our case, we have that
 \begin{align}
 \sigma_{abs}(0) = \frac{R_{H}^{p}}{\Omega_p} \frac{8\pi^p}{p-1} \sin\left[\pi(p-1)/2  \right] \frac{\Gamma\left( \frac{3-p}{2} \right)}{\Gamma \left( \frac{p+1}{2} \right)},
 \end{align}
and applying Euler's reflection formula,
 $$\Gamma(1-z)\Gamma(z) = \frac{\pi}{\sin\left(\pi z  \right)} \Rightarrow \sin\left[\pi(p-1)/2  \right]\Gamma\left( \frac{3-p}{2} \right) = \frac{\pi}{\Gamma \left( \frac{p-1}{2} \right)},$$
we simplify it to
 \begin{align}
  \sigma_{abs}(0) &= \frac{R_{H}^{p}}{\Omega_p} \frac{4\pi^{p+1}}{\frac{(p-1)}{2}\Gamma \left( \frac{p-1}{2} \right)\Gamma \left( \frac{p+1}{2} \right)} \notag \\
                     &= \frac{1}{\Omega_p} \frac{4\pi^{p+1}}{\left[ \Gamma \left( \frac{p+1}{2} \right) \right]^2} R_{H}^{p}.
 \end{align}
Noting the term $\Omega_p^2$ we finally arrive at
 \begin{equation}
 \sigma_{abs}(0) = A_H.
 \end{equation}
Which demonstrates the initial statement.

\chapter{Coordinate transformations}
\label{AppD}

In this Appendix we list the different coordinate systems used on Chapter \ref{Chap7.0} and how one may write the metric of $AdS_5$ spacetime in each one of them. A common way of expressing the $AdS_5$ metric in the context of the holographic correspondence is through the explicitly conformal coordinate system below
\begin{align}
ds^2=\frac{L^2}{U^2}(dU^2-dt^2+dx^2+dy^2+dz^2),
\label{a1}
\end{align}
where the boundary of the $AdS_5$ space is at $U=0$. Defining the coordinate transformation
\begin{align}
\bar{r}:=\frac{L^2}{U},
\label{a2}
\end{align}
one may rewrite the $AdS_5$ metric as follows
\begin{align}
ds^2=\frac{L^2}{\bar{r}^2}d\bar{r}^2+\frac{\bar{r}^2}{L^2}(-dt^2+dx^2+dy^2+dz^2),
\label{a3}
\end{align}
where the boundary of the $AdS_5$ space is now at $\bar{r}\rightarrow\infty$. This coordinate system is the one used in \cite{DK1} and in Sec.\ \ref{sec:magbranes} to obtain the finite temperature solutions.

Also, through the coordinate transformation
\begin{align}
r:=\frac{\bar{r}^2}{2L}=\frac{L^3}{2U^2},
\label{a4}
\end{align}
one can write the $AdS_5$ metric as
\begin{align}
ds^2=\frac{L^2}{4r^2}dr^2+\frac{2r}{L}(-dt^2+dx^2+dy^2+dz^2),
\label{a5}
\end{align}
where the boundary is at $r\rightarrow\infty$. We can further define the light-cone coordinates
\begin{align}
u:=\frac{z+t}{\sqrt{2}},\,\,\,v:=\frac{z-t}{\sqrt{2}},
\label{a6}
\end{align}
in terms of which (\ref{a5}) is rewritten as follows
\begin{align}
ds^2=\frac{L^2}{4r^2}dr^2+\frac{4r}{L}dudv+\frac{2r}{L^2}(dx^2+dy^2)\,.
\label{a7}
\end{align}
This coordinate system is the one used in \cite{DK2} and in Sec.\ \ref{wsec2} to study the zero temperature solution of the magnetic brane.

\chapter{Wilson loops in strongly coupled $\mathcal{N}=4$ SYM}
\label{AppE}

For the sake of completeness, in this Appendix we give a brief review of the holographic computation of rectangular Wilson loops in SYM at finite temperature \cite{sonne,rey} without magnetic fields. We shall closely follow the discussions in Section 5.1 of Ref.\ \cite{jorge}. At finite $T$ and $B=0$, the background giving an holographic description of thermal SYM is the $AdS_5$-Schwarzschild metric
\begin{align}
ds^2=\frac{L^2}{\bar{r}^2f(\bar{r})}d\bar{r}^2-\frac{\bar{r}^2f(\bar{r})}{L^2}d\bar{t}^2+ \frac{\bar{r}^2}{L^2}(d\bar{x}^2+d\bar{y}^2+d\bar{z}^2), \,\,\,f(\bar{r})=1-\frac{\bar{r}_H^4}{\bar{r}^4},
\label{b1}
\end{align}
where the boundary is at $\bar{r}\rightarrow\infty$ and the horizon is at $\bar{r}=\bar{r}_H$. Rescaling $\bar{r}=:4\bar{r}_H(r-3/4)$, $(\bar{t},\bar{x},\bar{y},\bar{z})=:(t,x,y,z)/4\bar{r}_H$ and adopting units where $L=1$, one rewrites (\ref{b1}) as follows
\begin{align}
ds^2=\frac{dr^2}{\left(r-\frac{3}{4}\right)^2f(r)}-\left(r-\frac{3}{4}\right)^2f(r)dt^2+\left(r-\frac{3}{4}\right)^2 d\vec{x}^{\,2}, \,\,\,f(r)=1-\frac{1}{\left[4\left(r-\frac{3}{4}\right)\right]^4},
\label{b2}
\end{align}
where the boundary is at $r\rightarrow\infty$ and the horizon is now at $r=1$. From (\ref{b2}), the Hawking temperature reads
\begin{align}
T=\frac{\sqrt{-g'_{tt}\,g^{rr}\,'}}{4\pi}\biggr|_{r=1}=\frac{1}{4\pi},
\label{b3}
\end{align}
which is the same constant temperature obtained before for the magnetic backgrounds. Indeed, one can check numerically that the magnetic backgrounds at finite temperature derived in Sec.\ \ref{sec:magbranes} converge for the metric (\ref{b2}) in the limit of zero magnetic field, as it should be.

The formal expressions for the interquark distance and the heavy quark potential (we set $\alpha'=1$ below) as functions of the parameter $r_0$ are given by\footnote{For instance, one may obtain these expressions by replacing $U(r)\rightarrow (r-3/4)^2f(r)$ and $e^{2V(r)}\rightarrow (r-3/4)^2$ in Eqs.\ (\ref{w3.22}) and (\ref{w3.23}).}
\begin{align}
L_{Q\bar{Q}}^{(T\neq 0)}(r_0)&= 32\sqrt{(4r_0-3)^4-1}\int_{r_0}^{\infty} \frac{dr}{\sqrt{[(4r-3)^4-1][(4r-3)^4-(4r_0-3)^4]}},\label{b4}\\
V_{Q\bar{Q}}^{(T\neq 0)}(r_0)&= \frac{1}{\pi}\left[ \int_{r_0}^\infty dr\left(\sqrt{\frac{(4r-3)^4-1}{(4r-3)^4-(4r_0-3)^4}}-1\right) - \int_0^{r_0}dr\right].\label{b5}
\end{align}
Defining the new integration variable $R:=4r-3$ and also the constant $R_0:=4r_0-3$, we rewrite (\ref{b4}) and (\ref{b5}) as follows
\begin{align}
L_{Q\bar{Q}}^{(T\neq 0)}(R_0)&= 8\sqrt{R_0^4-1}\int_{R_0}^{\infty} \frac{dR}{\sqrt{(R^4-1)(R^4-R_0^4)}},\label{b6}\\
V_{Q\bar{Q}}^{(T\neq 0)}(R_0)&= \frac{1}{4\pi}\left[ \int_{R_0}^\infty dR\left(\sqrt{\frac{R^4-1}{R^4-R_0^4}}-1\right) -R_0-3\right]\,.\label{b7}
\end{align}
Defining now $y:=R/R_0$ and also $y_H:=1/R_0$, one rewrites (\ref{b6}) and (\ref{b7}) as follows
\begin{align}
L_{Q\bar{Q}}^{(T\neq 0)}(y_H)&= 8y_H\sqrt{1-y_H^4}\int_{1}^{\infty} \frac{dy}{\sqrt{(y^4-y_H^4)(y^4-1)}},\label{b8}\\
V_{Q\bar{Q}}^{(T\neq 0)}(y_H)&= \frac{1}{4\pi y_H}\left[ \int_{1}^\infty dy\left(\sqrt{\frac{y^4-y_H^4}{y^4-1}}-1\right) - 1 - 3y_H\right]\,.\label{b9}
\end{align}
Let us denote the integrals in (\ref{b8}) and (\ref{b9}) by $\mathcal{I}_1$ and $\mathcal{I}_2$, respectively. In what follows, we are going to express these integrals in terms of Gaussian hypergeometric functions, by using the following integral representation
\begin{align}
_2F_1(a,b;c;z)=\frac{\Gamma(c)}{\Gamma(b)\Gamma(c-b)}\int_0^1 dx\, x^{b-1}\, (1-x)^{c-b-1}\, (1-zx)^{-a},
\label{b10}
\end{align}
which is valid for $\textrm{Re}[c]>\textrm{Re}[b]>0$ and $|z|<1$.

Defining the new integration variable $x:=y^{-4}$, one obtains for the integral in (\ref{b8})
\begin{align}
\mathcal{I}_1=\int_{1}^{\infty} \frac{dy}{\sqrt{(y^4-y_H^4)(y^4-1)}}&= \frac{1}{4}\int_0^1 dx\, x^{-1/4}\, (1-x)^{-1/2}\, (1-xy_H^4)^{-1/2}\nonumber\\
&=\frac{1}{4}\frac{\Gamma(3/4)\Gamma(1/2)}{\Gamma(5/4)}\,_2F_1\left(\frac{1}{2},\frac{3}{4};\frac{5}{4};y_H^4\right)\nonumber\\
&\approx 0.599\,_2F_1\left(\frac{1}{2},\frac{3}{4};\frac{5}{4};y_H^4\right).
\label{b11}
\end{align}
For the integral in (\ref{b9}),
\begin{align}
\mathcal{I}_2=\int_{1}^\infty dy\left(\sqrt{\frac{y^4-y_H^4}{y^4-1}}-1\right) = \frac{1}{4}\int_0^1 dx\, x^{-5/4} \left[ (1-x)^{-1/2}\, (1-xy_H^4)^{1/2}-1\right],
\label{b12}
\end{align}
we employ the following regularization scheme\footnote{Note this regularization procedure involves commuting the limit $\lambda\rightarrow 0$ with the integral.} in order to allow the use of the integral representation (\ref{b10})
\begin{align}
\mathcal{I}_2^{\textrm{reg}}&=\frac{1}{4}\lim_{\lambda\rightarrow 0}\int_0^1 dx\, x^{-5/4+\lambda} \left[ (1-x)^{-1/2}\, (1-xy_H^4)^{1/2}-1\right]\nonumber\\
&=\frac{1}{4}\lim_{\lambda\rightarrow 0}\left[ \frac{\Gamma(-1/4+\lambda)\Gamma(1/2)}{\Gamma(1/4+\lambda)}\, _2F_1(-\frac{1}{2},-\frac{1}{4}+\lambda;\frac{1}{4}+\lambda;y_H^4) + \frac{4}{1-4\lambda} \right]\nonumber\\
&\approx -0.599\,_2F_1\left(-\frac{1}{2},-\frac{1}{4};\frac{1}{4};y_H^4\right)+1\,.
\label{b13}
\end{align}

Substituting (\ref{b11}) into (\ref{b8}) and (\ref{b13}) into (\ref{b9}), one obtains, respectively
\begin{align}
L_{Q\bar{Q}}^{(T\neq 0)}(y_H)&\approx 4.792\, y_H\sqrt{1-y_H^4}\,_2F_1\left(\frac{1}{2},\frac{3}{4};\frac{5}{4};y_H^4\right),\label{b14}\\
V_{Q\bar{Q}}^{(T\neq 0)}(y_H)&\approx -\frac{0.048}{y_H}\,_2F_1\left(-\frac{1}{2},-\frac{1}{4};\frac{1}{4};y_H^4\right)- \frac{3}{4\pi},\label{b15}
\end{align}
with $|y_H^4|<1$. Eqs.\ (\ref{b14}) and (\ref{b15}) were employed to obtain numerically the parametric SYM curve in Fig.\ \ref{wfig5}.

As a final remark, we mention that the values of $y_H$ considered in the parametric plots shown in Fig.\ \ref{wfig5} were restricted to values below $y_H^{\textrm{max}}$, which is the value of $y_H$ where $LT$ reaches its maximum value.

In the zero temperature case discussed in \cite{maldacena} the potential can be obtained analytically as we now briefly review. The formal expressions for the interquark separation and the interquark potential as functions of the parameter $r_0$ are given by\footnote{For instance, one may obtain these expressions by replacing $U(r)\rightarrow r^2$ and $e^{2W(r)}\rightarrow r^2$ in eqs. (\ref{w3.22}) and (\ref{w3.23}).}:
\begin{align}
L_{Q\bar{Q}}^{(T=0)}(r_0)&= 2r_0^2\int_{r_0}^{\infty} \frac{dr}{r^2\sqrt{r^4-r_0^4}},\label{b16}\\
V_{Q\bar{Q}}^{(T=0)}(r_0)&= \frac{1}{\pi}\left[ \int_{r_0}^\infty dr\left(\frac{r^2}{\sqrt{r^4-r_0^4}}-1\right) - \int_0^{r_0}dr\right].\label{b17}
\end{align}
Defining the new integration variable $y:=r/r_0$, we rewrite (\ref{b16}) and (\ref{b17}) as follows:
\begin{align}
L_{Q\bar{Q}}^{(T=0)}(r_0)&= \frac{2}{r_0}\int_1^{\infty} \frac{dy}{y^2\sqrt{y^4-1}}=\frac{2\sqrt{\pi}\,\Gamma(3/4)}{r_0\,\Gamma(1/4)},\label{b18}\\
V_{Q\bar{Q}}^{(T=0)}(r_0)&= \frac{r_0}{\pi}\left[ \int_1^\infty dr\left(\frac{y^2}{\sqrt{y^4-1}}-1\right) - 1\right]=\frac{r_0}{\pi}\left[-\textrm{E}(-1)+(2-i)\textrm{K}(-1)-\textrm{K}(2)\right],\label{b19}
\end{align}
where $\textrm{K}(x)$ and $\textrm{E}(x)$ are the complete elliptic integrals of the first and second kinds, respectively. From (\ref{b18}) one obtains $r_0$ as an explicitly function of the interquark separation
\begin{align}
r_0=\frac{2\sqrt{\pi}\,\Gamma(3/4)}{\Gamma(1/4)}\frac{1}{L_{Q\bar{Q}}^{(T=0)}}\approx\frac{1.198}{L_{Q\bar{Q}}^{(T=0)}}.
\label{b20}
\end{align}
Taking into account that $\left[-\textrm{E}(-1)+(2-i)\textrm{K}(-1)-\textrm{K}(2)\right]\approx -0.599$ and substituting (\ref{b20}) into (\ref{b19}), one gets the following analytical result \cite{maldacena}
\begin{align}
V_{Q\bar{Q}}^{(T=0)}(L_{Q\bar{Q}}^{(T=0)})\approx -\frac{0.228}{L_{Q\bar{Q}}^{(T=0)}}.
\label{b21}
\end{align}
This result was used to plot the SYM curve in Fig. \ref{wfig2}.

\chapter{Electric susceptibility and conductivity for different coupling functions}
\label{appF}

In order to check further limitations of the present EMD model (some of which have been discussed at the end of Section \ref{sec2.4}), we compare in this Appendix the results for the magnetic susceptibility, and also the electric susceptibility and DC electric conductivity for two different profiles of the Maxwell-Dilaton electric coupling function $f(\phi)$. The first profile is given in Eq. \eqref{2.24}, which was fixed by fitting lattice data \cite{latticedata2} for the magnetic susceptibility at $B=0$, as discussed before. The second profile was fixed in Ref. \cite{Finazzo:2015xwa} by fitting lattice data \cite{Borsanyi:2011sw} for the electric susceptibility also at $B=0$,
\begin{align}
f(\phi)=0.0193\,\textrm{sech}(-100\,\phi)+0.0722\,\textrm{sech}(10^{-7}\,\phi).
\label{A1}
\end{align}

\begin{figure}
\begin{center}
\begin{tabular}{c}
\includegraphics[width=0.45\textwidth]{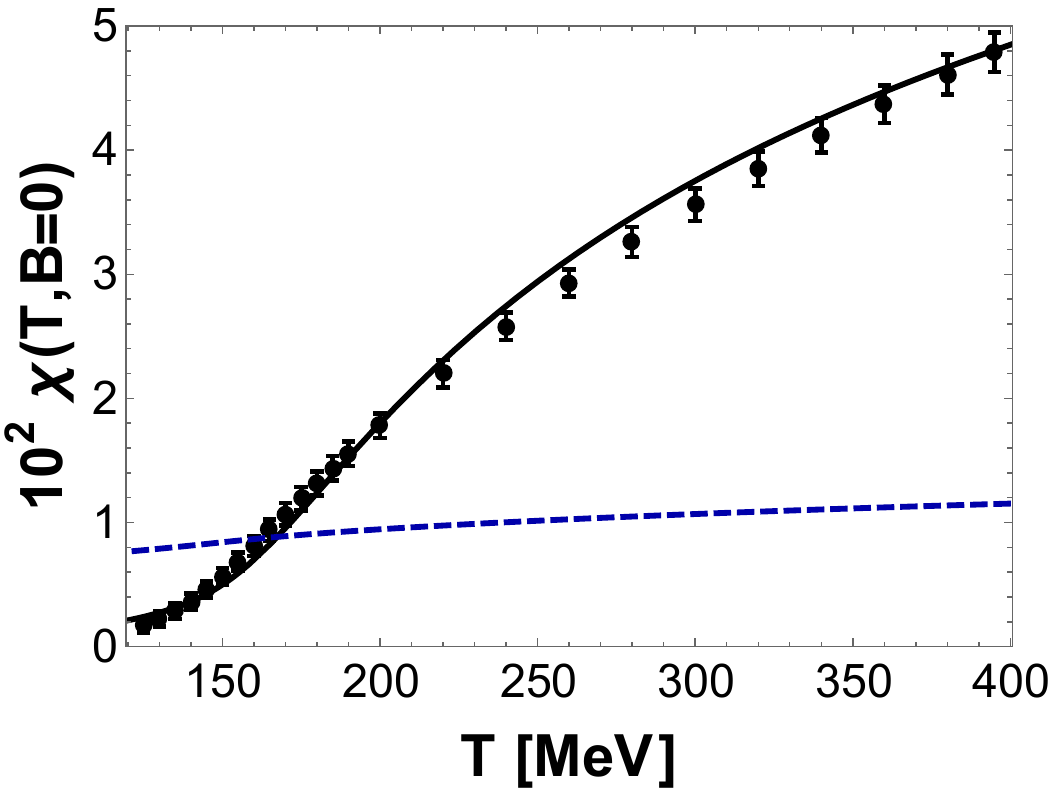} 
\end{tabular}
\begin{tabular}{c}
\includegraphics[width=0.45\textwidth]{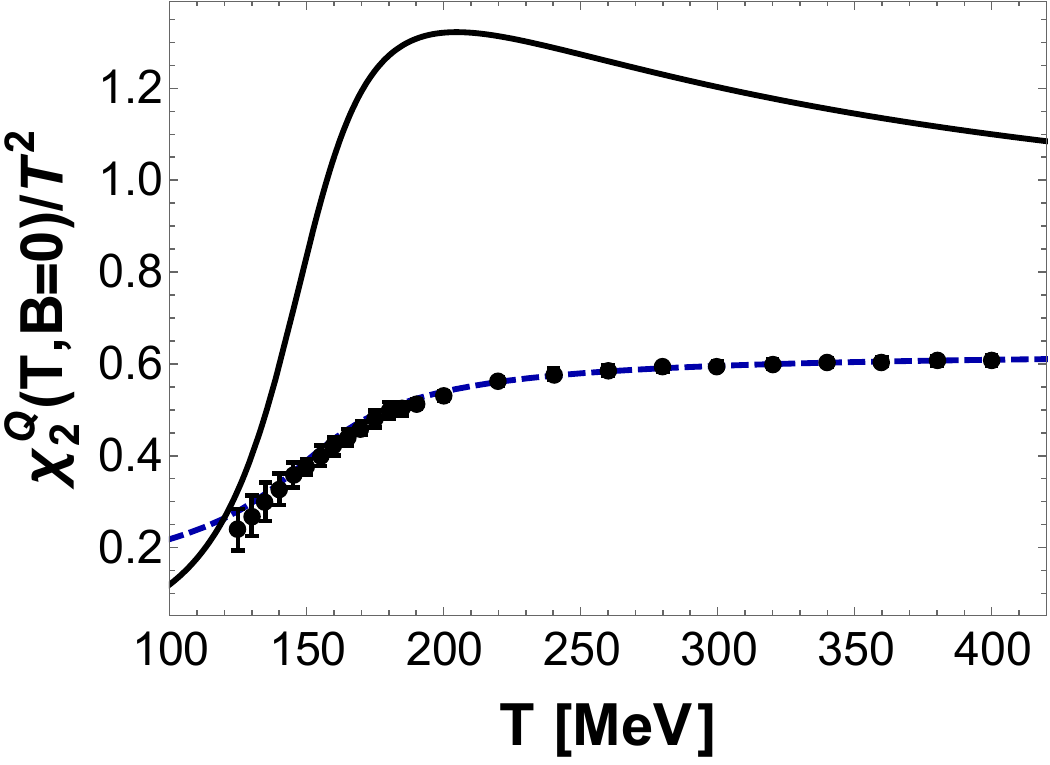} 
\end{tabular}
\begin{tabular}{c}
\includegraphics[width=0.45\textwidth]{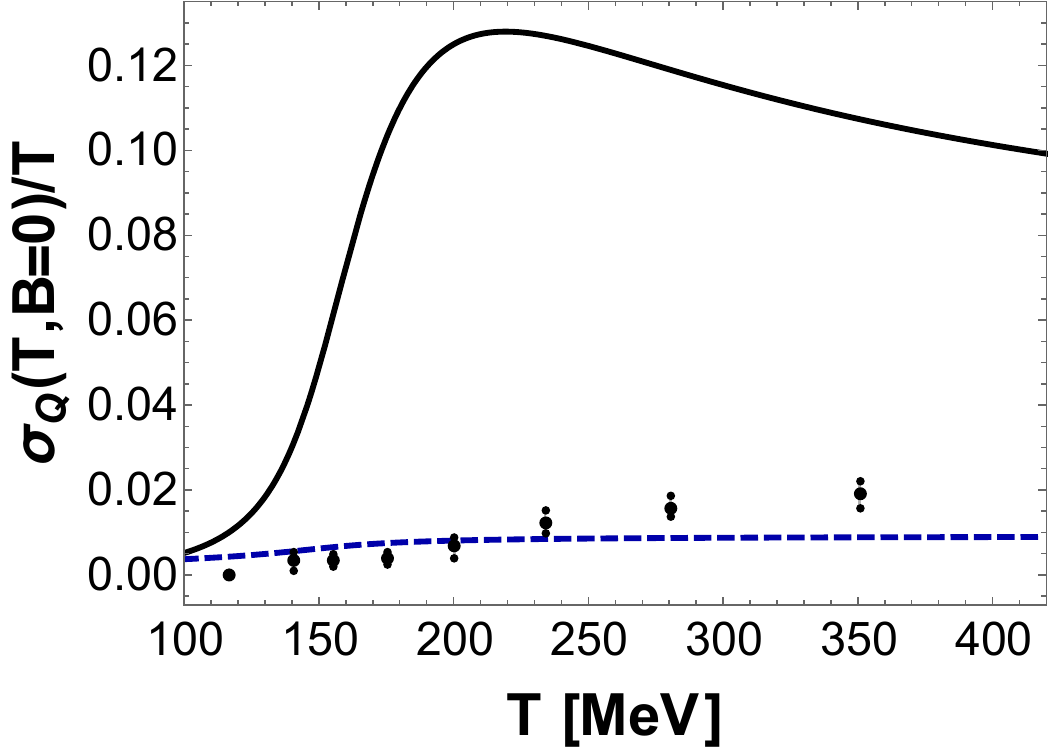} 
\end{tabular}
\end{center}
\caption{{\small  EMD magnetic susceptibility (top left), electric susceptibility (top right) and DC electric conductivity (bottom) for two different choices of the Maxwell-Dilaton electric coupling function $f(\phi)$: the full curves were obtained by using $f(\phi)$ given in Eq. \eqref{2.24}, while the dashed curves were obtained by employing $f(\phi)$ given in Eq. \eqref{A1}. All the lattice data displayed in these plots refer to $(2+1)$-flavor QCD (lattice data for the electric conductivity are taken from \cite{Aarts:2014nba}).}
\label{figlimitations}}
\end{figure}

At $B=0$, the holographic formulas for the electric susceptibility and the DC electric conductivity are given respectively by,
\begin{align}
\frac{\chi_2^Q}{T^2}&=\frac{1}{16\pi^2} \frac{s}{T^3} \frac{1}{f(0)\int_{r_H}^\infty dr\, e^{-2a(r)}f^{-1}(\phi(r))},\label{A2}\\
\frac{\sigma_Q}{T}&=\frac{2\pi\sqrt{h_0^{\textrm{far}}}f(\phi_0)}{\kappa^2},\label{A3}
\end{align}
and we refer the reader to consult Ref. \cite{Finazzo:2015xwa} for a discussion on the derivation of these formulas\footnote{We use that $a(r_H)=a_0=0$ and $\phi(r_H)=\phi_0$.}.

One can see from the results shown in Fig \ref{figlimitations} that a simple EMD holographic model cannot give simultaneously good quantitative descriptions of electric and magnetic field response functions: by adjusting the electric coupling $f(\phi)$ in order to fit the magnetic susceptibility at $B=0$, one is able to attain a good description of the QCD thermodynamics at finite $B$, as shown in Section \ref{sec3}, but response functions to an applied electric field are not well described in a quantitative way within such prescription. On the other hand, if one adjusts the electric coupling $f(\phi)$ in order to match the electric susceptibility, one is not able to obtain a good quantitative agreement with lattice data for the magnetic susceptibility. It would be certainly interesting to think about the construction of some holographic model versatile enough to quantitatively cover the entire electric-magnetic sector of the QGP, which is something that our simple EMD model is not able to do. We must remark, however, that up to now, our EMD model is the only holographic approach available in the literature which is able to match in a quantitative way the behavior of many magnetic field related observables calculated on the lattice.

\backmatter \singlespacing   

\bibliographystyle{unstr}



\end{document}